\documentclass[11pt,letterpaper]{article}

\usepackage[section]{placeins}
\usepackage[letterpaper,margin=1in]{geometry}
\usepackage[T1]{fontenc}
\usepackage{lmodern}
\usepackage{amsmath,amssymb,amsthm,mathtools}
\usepackage{aliascnt}
\usepackage{booktabs,array,longtable}
\usepackage{microtype}
\usepackage{graphicx}
\usepackage{flafter}
\usepackage{enumitem}
\usepackage[dvipsnames]{xcolor}
\usepackage{tikz}
\usetikzlibrary{arrows.meta,positioning}
\usepackage{pgfplots}
\pgfplotsset{compat=1.17}
\usepgfplotslibrary{groupplots}
\definecolor{OursBlue}{HTML}{2A78D6}
\definecolor{MRSOrange}{HTML}{EB6834}
\definecolor{AVAqua}{HTML}{1BAF7A}
\usepackage[colorlinks=true,linkcolor=MidnightBlue,citecolor=OliveGreen,urlcolor=MidnightBlue]{hyperref}
\usepackage[capitalise,nameinlink]{cleveref}
\hypersetup{
  pdftitle={The Exact Approximation Ratio of k-means++},
  pdfauthor={Aravind Reddy},
  pdfsubject={Exact worst-case expected approximation ratio of k-means++ seeding for every k},
  pdfkeywords={k-means++, D2 sampling, seeding, clustering, approximation ratio, harmonic numbers},
}

\allowdisplaybreaks
\setlist{itemsep=2pt,topsep=4pt}

\newtheorem{theorem}{Theorem}[section]
\newcommand{\sharedthm}[2]{\newaliascnt{#1}{theorem}\newtheorem{#1}[#1]{#2}\aliascntresetthe{#1}}
\sharedthm{lemma}{Lemma}
\sharedthm{proposition}{Proposition}
\sharedthm{corollary}{Corollary}
\theoremstyle{definition}
\sharedthm{definition}{Definition}
\sharedthm{remark}{Remark}
\crefname{lemma}{Lemma}{Lemmas}
\crefname{proposition}{Proposition}{Propositions}
\crefname{corollary}{Corollary}{Corollaries}
\crefname{definition}{Definition}{Definitions}
\crefname{remark}{Remark}{Remarks}
\newcounter{algo}
\crefname{algo}{Algorithm}{Algorithms}
\crefname{figure}{Figure}{Figures}
\Crefname{figure}{Figure}{Figures}
\crefname{appendix}{Appendix}{Appendices}

\newcommand{\R}{\mathbb{R}}
\newcommand{\E}{\mathbb{E}}
\newcommand{\cost}{\mathrm{cost}}
\newcommand{\OPT}{\mathrm{OPT}}
\newcommand{\Occ}{\mathcal{O}}
\newcommand{\Del}{\mathcal{D}}
\newcommand{\Pre}{\mathcal{P}}
\newcommand{\Cl}{\operatorname{cl}}
\newcommand{\Dl}{\operatorname{dl}}
\newcommand{\Var}{\operatorname{Var}}

\title{The Exact Approximation Ratio of $k$-means++}

\author{
Aravind Reddy\\
\small Centre for Responsible AI, Wadhwani School of Data Science \& AI\\
\small Indian Institute of Technology Madras\\
\small \href{mailto:aravind@cerai.in}{\texttt{aravind@cerai.in}}
}

\date{29 November 2026}

\begin{document}

\maketitle

\begin{abstract}
The $k$-means++ algorithm is the standard initialization for $k$-means clustering, one of the most widely used methods in machine learning and data analysis, adopted both for its empirical performance and its worst-case approximation guarantee. Given a finite Euclidean point set, $k$-means seeks $k$ centers minimizing the sum of squared distances from each point to its nearest center. $k$-means++ picks the first center uniformly at random from the points and each subsequent center with probability proportional to its squared distance to the nearest chosen center. Let $R_k$ be the smallest factor such that, on every input, the expected cost of these $k$ centers is at most $R_k$ times the optimal cost. In their seminal paper introducing $k$-means++, Arthur and Vassilvitskii (SODA 2007) showed that $2\ln k \le R_k \le 8(\ln k + 2)$, and Makarychev, Reddy and Shan (NeurIPS 2020) lowered the upper bound to $5(\ln k + 2)$. We determine $R_k$ exactly for all $k$ and discover a striking phase transition. For $k \ge 19$,
\[
R_k = 2H_{k-1} + 2,
\]
where $H_j = 1 + \frac12 + \frac13 + \dots + \frac1j$ is the $j$th harmonic number. For $2 \le k \le 18$, $R_k$ is the maximum of an explicit rational function of one variable and strictly exceeds $2H_{k-1} + 2$, with $R_2 = 5$ and $R_3 = 185/32$. In particular, $2\ln k + 2 \le R_k \le 2\ln k + 4$ for every $k \ge 1$. When a cluster is first sampled, previous analyses bound its expected cost by a constant times its optimal cost and charge that cost for the rest of the run. We instead track what happens to a cluster after its first hit. At each later draw, the probability of sampling that cluster equals its share of the total cost. A poor first choice can therefore make a second hit more likely, and that second hit can bring the cluster's cost back down. The full result is formally verified in Lean 4, and the formalization is included as ancillary files.
\end{abstract}

\vspace{0.5em}

\section{Introduction}\label{sec:intro}

Given $n$ points in $\R^d$ and a number $k$, the $k$-means problem asks for $k$ centers that minimize the sum, over all points, of the squared distance from the point to its nearest center. Each optimal center sits at the mean of the points it serves, and the points served by one center form a \emph{cluster}. The problem is the most widely used formulation of clustering in machine learning and data analysis, and it is computationally hard: finding an optimal solution, or even approximating it within some constant factor greater than $1$, is NP-hard~\cite{ACKS15}. In practice, it is attacked with Lloyd's algorithm~\cite{Llo82}, which alternately assigns each point to its nearest center and moves each center to the mean of its points. Lloyd's algorithm never increases the cost and has polynomial smoothed running time (expected polynomial time under small relative Gaussian perturbations)~\cite{AMR11}, but it can stop far from the optimum, and where it stops depends on where it starts.

The $k$-means++ seeding procedure of Arthur and Vassilvitskii~\cite{AV07} chooses the starting centers at random, favoring points that are far from the centers already chosen. The first center is a data point chosen uniformly at random. Each further center is a data point chosen with probability proportional to its squared distance to the nearest center already chosen; this rule is called \emph{$D^2$ sampling}. The procedure is simple and fast, and it has become the standard initialization for Lloyd's algorithm (serving as the default in widely used software such as MATLAB~\cite{MATLABkmeans} and Julia's Clustering.jl~\cite{ClusteringJL}), adopted both for its empirical performance and because it comes with a worst-case approximation guarantee that holds for every input.

\paragraph{The guarantee.} Because the centers are random, so is their cost. Arthur and Vassilvitskii proved that its expected value is at most $8(\ln k+2)$ times the optimal $k$-means cost, for every input, and that no bound below $2\ln k$ is possible~\cite[Theorem~4.1]{AV07}. Makarychev, Reddy and Shan~\cite{MRS20} lowered the upper bound to $5(\ln k+2)$. The constant $5$ has a precise origin: when $D^2$ sampling first selects a point of a cluster, the cluster's expected cost immediately afterward is at most five times its optimal cost, and this factor is sharp. Let $R_k$ denote the best possible guarantee of this kind: the smallest number such that, on every input, the expected cost of the $k$ seeds is at most $R_k$ times the optimal cost. Equivalently, $R_k$ is the supremum over all inputs of the ratio of expected seeding cost to optimal cost (\cref{sec:main}); the expectation is over the random choices of the algorithm, and the input is arbitrary. These results place $R_k$ between $2\ln k$ and $5(\ln k+2)$.

\paragraph{This paper.} We determine $R_k$ exactly for every $k$. For $k\ge19$,
\[
R_k=2H_{k-1}+2=2\Bigl(1+\frac12+\frac13+\dots+\frac1{k-1}\Bigr)+2\approx2\ln k+3.15,
\]
and for $2\le k\le18$, $R_k$ is the maximum of an explicit rational function of one variable and strictly exceeds $2H_{k-1}+2$ (\cref{thm:exact-rk}, \cref{tab:values}, \cref{fig:ratio}). For instance, $R_2=5$, $R_3=\frac{185}{32}$, and $R_{10}\approx7.897$: on every data set, in every dimension, $k$-means++ with $k=10$ centers has expected cost at most $7.897$ times the optimal cost, and there are inputs on which the ratio comes arbitrarily close to $R_{10}$. In particular, $2\ln k+2\le R_k\le 2\ln k+4$ for every $k\ge1$ (\cref{cor:sharp-additive}). The answer exhibits a phase transition between $k=18$ and $k=19$: two different families of inputs approach the worst case, and which of them dominates switches at $k=19$ (\cref{sec:surprise}).

\subsection{Model and main result}\label{sec:main}

Throughout, $X\subset\R^d$ is a finite set of $|X|\ge k$ distinct, equally weighted points in any dimension $d\ge1$, with $k\ge1$. For a finite nonempty $C\subset\R^d$ let $\cost(Y,C)=\sum_{y\in Y}\min_{c\in C}\|y-c\|^2$, and let $\OPT_k(X)$ be the minimum of $\cost(X,C)$ over all $C\subset\R^d$ with $|C|=k$; optimal centers need not be data points. The algorithm (\cref{alg:kmpp}) keeps every center it chooses and outputs $k$ centers $C_k$, of cost $\Phi_k=\cost(X,C_k)$ before any Lloyd iteration, with expectation $\E\,\Phi_k$ over the algorithm's random choices. Write $H_j=\sum_{i=1}^j1/i$ ($H_0=0$) and
\begin{equation}\label{eq:Rk}
R_k=\sup\Bigl\{\frac{\E\,\Phi_k}{\OPT_k(X)}\;:\;d\ge1,\ X\subset\R^d\text{ finite, distinct},\ |X|\ge k,\ \OPT_k(X)>0\Bigr\}.
\end{equation}
Thus $\E\,\Phi_k\le R_k\,\OPT_k(X)$ for every input, and no smaller constant has this property. For $k\ge2$, our two extremal families approach their limiting ratios as $|X|$ grows (with fixed dimension $k-1$ for the hub inputs of \cref{sec:hub}, and growing dimension for the reservoir inputs of \cref{sec:reservoir}). If $\OPT_k(X)=0$, distinctness forces $|X|=k$ and $\Phi_k=0$, so the bound holds trivially. For $k=1$ the single center is uniform, and the centroid identity gives $\E\,\Phi_1=2\,\OPT_1(X)$~\cite[Lemma~3.1]{AV07}; hence $R_1=2$, attained whenever $\OPT_1>0$.

The answer for $2\le k\le18$ is expressed through one explicit rational function. For $m\ge0$ and a \emph{price} $p>0$, let $\pi_m(p)=\prod_{i=1}^m\frac{ip}{1+ip}$ ($\pi_0=1$), the probability that $m$ independent rate-$p$ exponential clocks all ring before one rate-$1$ clock. Put $G_0(p)=5$ and, for $m\ge1$,
\begin{equation}\label{eq:G}
G_m(p)=\frac{1+H_m+mp\,G_{m-1}(p)+4p+4(1-p)\,\pi_m(p)}{1+mp}.
\end{equation}
\Cref{sec:hub} shows that $G_m(p)$ is the limiting ratio of an explicit input family with $k=m+2$ in which $p$ tunes the distance of the competing clusters, with~\eqref{eq:G} arising as its first-step recurrence. With $c=1/p$ and $\pi_m=\prod_{i=1}^m\frac i{i+c}$, $G_m(1/c)=1+H_m+3(1-\pi_m)/c+4\pi_m-4(c-1)\,d\pi_m/dc$, and partial fractions give
\begin{equation}\label{eq:G-binomial}
G_m(p)=5+\sum_{i=1}^m(-1)^{i-1}\binom mi\frac{5p-4+(1/i)}{(1+ip)^2}.
\end{equation}
For $k\ge2$ let $J_k=\sup_{p>0}G_{k-2}(p)$. Thus $J_2=5$; for $k\ge3$ the supremum is attained at a unique price $p^*_{k-2}$ (\cref{thm:laplace}), e.g.\ $p^*_1=11/5$ and $J_3=185/32$.

\begin{theorem}[Exact ratio]\label{thm:exact-rk}
The exact approximation ratio of $k$-means++ is
\[
R_k=\begin{cases}2,&k=1,\\ J_k,&2\le k\le18,\\ 2H_{k-1}+2,&k\ge19.\end{cases}
\]
In particular, $R_k=\max\{J_k,\,2H_{k-1}+2\}$ for every integer $k\ge2$.
\end{theorem}

\begin{corollary}[Sharp logarithmic bounds]\label{cor:sharp-additive}
For every integer $k\ge1$,
\[
2\ln k+2\;\le\;R_k\;\le\;2\ln k+5-2\ln2.
\]
Both additive constants are best possible for the worst-case ratio $R_k$: equality holds in the lower bound at $k=1$ ($R_1=2$) and in the upper bound at $k=2$ ($R_2=5$). In particular, $R_k\le 2\ln k+4$ for every $k\ge1$.
\end{corollary}

\begin{proof}
\emph{Lower bound.} At $k=1$, \cref{thm:exact-rk} gives $R_1=2=2\ln1+2$ (equality). For $k\ge2$, \cref{thm:exact-rk} gives $R_k\ge2H_{k-1}+2$, and integral comparison for the decreasing function $x\mapsto1/x$ yields $H_{k-1}=\sum_{j=1}^{k-1}\frac1j\ge\int_1^k\frac{dx}{x}=\ln k$, hence $R_k\ge2\ln k+2$.

\emph{Upper bound.} At $k=1$, $\ln(1+x)\le x$ gives $\ln2<1<\frac32$, so $R_1=2<5-2\ln2$. At $k=2$, \cref{thm:exact-rk} gives $R_2=5=2\ln2+5-2\ln2$ (equality). For $3\le k\le18$, since $\frac{d}{dx}\bigl(\ln(1+x)-\frac{2x}{2+x}\bigr)=\frac{x^2}{(1+x)(2+x)^2}>0$ on $x>0$, we have $\ln\frac{j+1}{j}>\frac{2}{2j+1}$ for every $j\ge1$; summing over $j=2,\dots,k-1$ gives
\[
2\ln k+5-2\ln2=5+2\ln\frac k2>5+\sum_{j=2}^{k-1}\frac{4}{2j+1}.
\]
At $k=3$, $5+\frac45=\frac{29}5>\frac{185}{32}=R_3$ (margin $\frac{3}{160}$). For $4\le k\le18$, direct exact rational comparison against the rigorous upper enclosures $J^+_k\ge J_k=R_k$ of \cref{tab:bin-brackets} verifies $5+\sum_{j=2}^{k-1}\frac{4}{2j+1}>J^+_k$ at every $k\in\{4,\dots,18\}$ (minimum margin $>0.065$ at $k=4$, increasing to $>0.483$ at $k=18$). For $k\ge19$, \cref{thm:exact-rk} gives $R_k=2H_{k-1}+2$, and integral comparison yields
\[
H_{k-1}=1+\frac12+\sum_{j=3}^{k-1}\frac1j\le\frac32+\int_2^{k-1}\frac{dx}{x}=\frac32+\ln\frac{k-1}{2}<\frac32+\ln\frac k2,
\]
so $R_k\le5+2\ln\frac{k-1}{2}<2\ln k+5-2\ln2$.

\emph{Sharpness and rounded bound.} Because equality holds at $k=1$ and $k=2$, no uniform lower constant can exceed $2$ and no uniform upper constant can be smaller than $5-2\ln2$. Finally, setting $x=1$ in $\ln(1+x)>\frac{2x}{2+x}$ gives $\ln2>\frac23>\frac12$, whence $5-2\ln2<\frac{11}3<4$ and $R_k\le2\ln k+4$ for every $k\ge1$.
\end{proof}

\begin{theorem}[The worst case is not attained]\label{thm:strict}
For every $k\ge2$, every input $X$ with $\OPT_k(X)>0$ satisfies $\E\,\Phi_k<R_k\OPT_k(X)$. For $2\le k\le18$, more precisely,
\begin{equation}\label{eq:theta-refined}
\frac{\E\,\Phi_k}{\OPT_k(X)}\;\le\;(1-\theta)\,R_k+\theta\,(2H_{k-1}+2)\;<\;R_k,
\end{equation}
where $\theta\ge2/|X|$ is the expected share of the optimal cost carried by the cluster of the first center (\cref{eq:theta}).
\end{theorem}

For $k\ge19$ the right side of~\eqref{eq:theta-refined} equals $R_k$, and strictness comes instead from a small gain in the restoration argument (\cref{cor:strict}). The extremal families show that $R_k$ is a supremum.

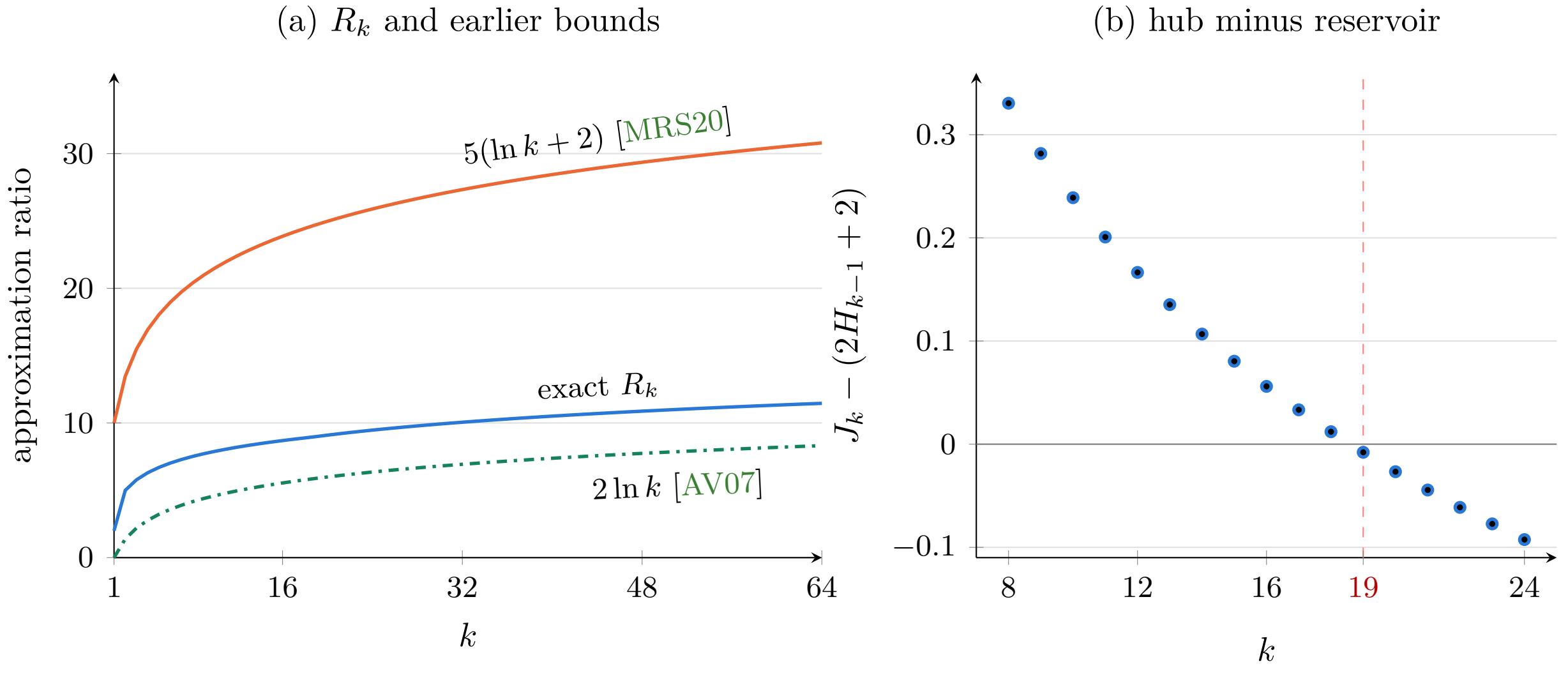
\begin{figure}[!htb]
\centering
\begin{tikzpicture}
\begin{groupplot}[
  group style={group size=2 by 1, horizontal sep=1.6cm},
  height=0.40\linewidth, clip=false,
  tick label style={font=\footnotesize}, label style={font=\small},
  every axis plot/.append style={line width=1pt},
  axis lines=left, ymajorgrids, grid style={black!12},
]
\nextgroupplot[width=0.54\linewidth, xmin=1, xmax=64, ymin=0, ymax=36,
  xlabel={$k$}, ylabel={approximation ratio},
  xtick={1,16,32,48,64}, ytick={0,10,20,30},
  title={\small (a) $R_k$ and earlier bounds}]
\addplot[draw=MRSOrange,mark=none] table[x=k,y=mrs]{figures/ratio.dat}
  node[pos=0.72,above,sloped,font=\footnotesize] {$5(\ln k+2)$ \cite{MRS20}};
\addplot[draw=OursBlue,mark=none] table[x=k,y=R]{figures/ratio.dat}
  node[pos=0.70,above,sloped,font=\footnotesize] {exact $R_k$};
\addplot[draw=AVAqua!75!black,dash dot,mark=none] table[x=k,y=avlow]{figures/ratio.dat}
  node[pos=0.80,below=2pt,sloped,font=\footnotesize] {$2\ln k$ \cite{AV07}};
\nextgroupplot[width=0.46\linewidth, xmin=7, xmax=25, ymin=-0.11, ymax=0.36, clip=true,
  xlabel={$k$}, ylabel={$J_k-(2H_{k-1}+2)$},
  xtick={8,12,16,24}, ytick={-0.1,0,0.1,0.2,0.3},
  extra x ticks={19}, extra x tick style={grid=major, grid style={red!50,dashed}, tick label style={red!70!black}},
  title={\small (b) hub minus reservoir}]
\addplot[draw=OursBlue,only marks,mark=*,mark size=1.4pt,restrict x to domain=8:24] table[x=k,y=gap]{figures/ratio.dat};
\addplot[draw=black!50,thin,mark=none,domain=7:25] {0};
\end{groupplot}
\end{tikzpicture}
\caption{(a) The exact ratio $R_k$ (\cref{thm:exact-rk}) against the upper bound $5(\ln k+2)$ of~\cite{MRS20} and the lower bound $2\ln k$ of~\cite{AV07}. (b) The advantage of the hub input over the reservoir input. It decreases from $1$ at $k=2$ (the plot shows $8\le k\le24$), is positive through $k=18$, and is negative from $k=19$ on; $R_k$ is the larger of the two values. Plotted values are floating-point evaluations; the rigorous comparison is \cref{thm:brackets}.}
\label{fig:ratio}
\end{figure}

\Cref{tab:values} lists the values and \cref{fig:ratio} plots them. Asymptotically, for $k\ge19$ we have $R_k=2\ln k+2+2\gamma_{\mathrm{E}}-1/k+O(k^{-2})$ (\cref{cor:values-asymptotic}), where $\gamma_{\mathrm{E}}\approx0.5772$ is Euler's constant, so the excess $R_k-2\ln k$ converges to the asymptotic limit $2+2\gamma_{\mathrm{E}}\approx3.1544$. This asymptotic limit is distinct from the uniform sharp additive constants of \cref{cor:sharp-additive}: across all $k\ge1$, $R_k-2\ln k$ attains its minimum $2$ at $k=1$ and its maximum $5-2\ln2\approx3.6137$ at $k=2$, exceeding $2+2\gamma_{\mathrm{E}}$ at small $k$ before approaching $2+2\gamma_{\mathrm{E}}$ from below for $k\ge19$. The leading constant is therefore $2$, matching the lower bound of~\cite{AV07}, and for $k\ge19$ the worst case pays only the factor $2$ of a uniformly placed center together with the harmonic term; the local factor $5$ of~\cite{MRS20} is sharp for a single hit but is not paid in the worst case.

\begin{table}[!ht]
\centering
\small
\begin{tabular}{@{}rcccc@{\qquad}rcccc@{}}
\toprule
$k$ & $J_k$ & $p^*_{k-2}$ & $2H_{k-1}+2$ & $R_k$ & $k$ & $J_k$ & $p^*_{k-2}$ & $2H_{k-1}+2$ & $R_k$\\
\midrule
1 & -- & -- & 2 & 2 & 11 & 8.0587 & 2.574 & 7.8579 & 8.0587\\
2 & 5 & -- & 4 & 5 & 12 & 8.2062 & 2.601 & 8.0398 & 8.2062\\
3 & 5.78125 & 2.200 & 5 & 5.78125 & 13 & 8.3417 & 2.626 & 8.2064 & 8.3417\\
4 & 6.3060 & 2.278 & 5.6667 & 6.3060 & 14 & 8.4670 & 2.649 & 8.3603 & 8.4670\\
5 & 6.7020 & 2.340 & 6.1667 & 6.7020 & 15 & 8.5835 & 2.671 & 8.5031 & 8.5835\\
6 & 7.0205 & 2.392 & 6.5667 & 7.0205 & 16 & 8.6925 & 2.692 & 8.6365 & 8.6925\\
7 & 7.2870 & 2.437 & 6.9000 & 7.2870 & 17 & 8.7948 & 2.711 & 8.7615 & 8.7948\\
8 & 7.5162 & 2.477 & 7.1857 & 7.5162 & 18 & 8.8912 & 2.730 & 8.8791 & 8.8912\\
9 & 7.7175 & 2.512 & 7.4357 & 7.7175 & 19 & 8.9823 & 2.747 & 8.9902 & 8.9902\\
10 & 7.8969 & 2.545 & 7.6579 & 7.8969 & 20 & 9.0688 & 2.764 & 9.0955 & 9.0955\\
\bottomrule
\end{tabular}
\caption{The exact ratio $R_k$ for $k\le20$, rounded to four decimals except where exact. The hub value $J_k$ exceeds the reservoir value $2H_{k-1}+2$ exactly for $2\le k\le18$; the margin falls from $1$ at $k=2$ to $0.0121$ at $k=18$, and is $-0.0079$ at $k=19$. The rigorous brackets used in the proof are in \cref{thm:brackets}.}\label{tab:values}
\end{table}

\subsection{Two worst cases, and the change at \texorpdfstring{$k=19$}{k=19}}\label{sec:surprise}

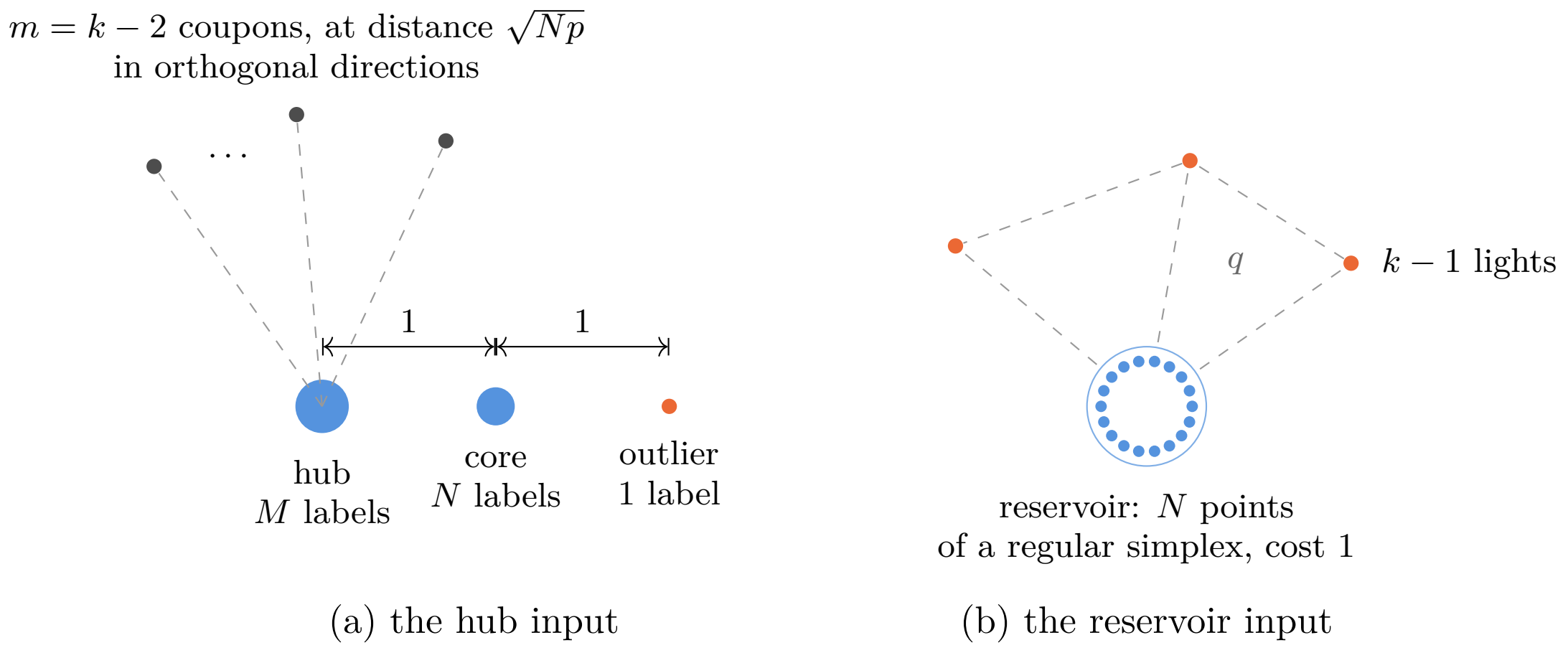
\begin{figure}[!htb]
\centering
\begin{tikzpicture}[scale=1.0,
  lab/.style={font=\footnotesize, align=center},
  heavy/.style={circle, fill=OursBlue!80, inner sep=0pt},
  light/.style={circle, fill=MRSOrange, inner sep=0pt, minimum size=4pt}]
\begin{scope}
\node[heavy, minimum size=14pt] (hub) at (0,0) {};
\node[lab, below=3pt] at (hub.south) {hub\\[-1pt]$M$ labels};
\node[heavy, minimum size=10pt] (core) at (1.6,0) {};
\node[lab, below=3pt] at (core.south) {core\\[-1pt]$N$ labels};
\node[light] (out) at (3.2,0) {};
\node[lab, below=3pt] at (out.south) {outlier\\[-1pt]$1$ label};
\draw[|<->|,thin] (0,0.55) -- node[above,font=\footnotesize]{$1$} (1.6,0.55);
\draw[|<->|,thin] (1.6,0.55) -- node[above,font=\footnotesize]{$1$} (3.2,0.55);
\foreach \a/\n in {65/1,95/2,125/m}{
  \node[light, fill=black!70] (c\n) at (\a:2.7) {};
  \draw[dashed, black!40] (0,0) -- (c\n);
}
\node[lab, above=2pt] at (c2.north) {$m=k-2$ coupons, at distance $\sqrt{Np}$\\[-1pt]in orthogonal directions};
\node[font=\footnotesize] at (110:2.45) {$\cdots$};
\node[font=\small] at (1.4,-2.0) {(a) the hub input};
\end{scope}
\begin{scope}[xshift=7.6cm]
\foreach \i in {1,...,18}{ \node[heavy, minimum size=3pt] at ({0.42*cos(20*\i)},{0.42*sin(20*\i)}) {}; }
\draw[OursBlue!60] (0,0) circle (0.55);
\node[lab, below=2pt] at (0,-0.6) {reservoir: $N$ points\\[-1pt]of a regular simplex, cost $1$};
\foreach \a/\n in {35/1,80/2,140/3}{
  \node[light] (l\n) at (\a:2.3) {};
  \draw[dashed, black!40] (\a:0.6) -- (l\n);
}
\draw[dashed, black!40] (l1) -- (l2) -- (l3);
\node[lab, right=2pt] at (l1.east) {$k-1$ lights};
\node[font=\footnotesize, black!60] at (58:1.55) {$q$};
\node[font=\small] at (0,-2.0) {(b) the reservoir input};
\end{scope}
\end{tikzpicture}
\caption{The two inputs on which $k$-means++ does worst (\cref{sec:lower}); the picture is schematic and not to scale. (a) Many points at the hub make it receive the first center. One cluster consists of a core and a rare outlier at twice the core's distance from the hub; $D^2$ sampling over-samples the outlier, and $m$ distant singletons compete for the remaining draws. (b) A large, nearly uniform cluster and $k-1$ isolated points, all at squared distance $q$ from everything else; every draw that returns to the cluster wastes a center.}
\label{fig:extremal}
\end{figure}

Two different families of inputs approach the worst case, one up to $k=18$ and the other from $k=19$ on, and they exploit different features of $D^2$ sampling (\cref{fig:extremal}). \Cref{sec:lower} constructs both and computes their ratios exactly; here is how they work.

\paragraph{The hub: a biased first hit.} One cluster has a tight core of $N$ points and one rare outlier. For $k=2$, put many points at a \emph{hub}, at distance $1$ from the core and $2$ from the outlier, so that the first center is almost surely at the hub. The cluster made of the core and the outlier then has total rate $N+4$: the core contributes $N$ and the outlier $4$, because its squared distance to the hub is $4$. The one remaining draw lands on the outlier with probability $4/(N+4)\approx4/N$ and leaves the core unserved, at cost about $N$; otherwise it lands in the core and leaves cost about $1$. So the expected cost tends to $1+4=5$ while the optimum tends to $1$. This is the sharp factor $5$ of~\cite{MRS20}, and it is paid once. For larger $k$ the hub adds $m=k-2$ distant singletons, the \emph{coupons}, at squared distance $Np$ from the hub. The outlier now competes with the core and the coupons for the remaining draws. A larger price $p$ makes a missed coupon more expensive but also collects the coupons faster, which shortens the window in which the outlier can strike. The function $G_m(p)$ records this trade-off, and $J_k$ is its optimum.

\paragraph{The reservoir: wasted draws.} A large, nearly uniform cluster of optimal cost $1$ (the vertices of a regular simplex) sits together with $k-1$ isolated points, the \emph{lights}, all at a large squared distance $q$ from everything else. The first center, uniform over the data, lands in the large cluster with probability tending to $1$ and leaves cost about $2$ there, as a uniformly placed center does. Each later draw should find a new light. But with $j$ lights still unfound the cluster is drawn again before all of them are found with probability about $2H_j/q$, and each such return wastes a center and leaves one light, of cost $q$, unserved. So the expected cost tends to $2+2H_{k-1}$. Within the large cluster the first hit is uniform, because all its points are equally far from the lights; the loss comes entirely from wasted draws.

\paragraph{The change at $k=19$.} One might expect the worst case to combine the two effects, paying the factor $5$ on one cluster and the harmonic waste on the others; the classical bounds of~\cite{AV07,MRS20} charge both effects. \Cref{thm:exact-rk} says that the effects do not add: $R_k$ is their maximum. The hub has the larger constant and the reservoir the faster growth. The hub's advantage is $1$ at $k=2$ and shrinks at every step, to $0.0121$ at $k=18$, and it is negative from $k=19$ on (\cref{fig:ratio}(b)). Nothing in either construction singles out $18$ and $19$. The phase transition happens at a moderate $k$ and by margins below one percent, so no asymptotic argument can locate it.

\subsection{Why the examples do not prove the upper bound}\label{sec:must}

Comparing the two constructions locates the change but says nothing about other inputs. The upper bound must hold for every finite input in every dimension, with clusters of arbitrary shape interacting through the sampling probabilities. A proof of the exact value must therefore do four things:
\begin{enumerate}[label=(\roman*),leftmargin=2em]
\item \emph{Rule out mixtures.} No input may combine an outlier-biased first hit with reservoir-type waste at full strength; charging the factor $5$ and the harmonic term separately loses about $3$.
\item \emph{Lose nothing in the reduction to one cluster.} Apart from the first center's cluster, which contributes $\theta(2H_{k-1}+2)$ in~\eqref{eq:theta-refined}, every cluster must be charged no more than the extremal inputs charge it as $\theta\to0$.
\item \emph{Change shape at the right place.} The induction on remaining clusters must follow the hub value through rank $16$ ($k\le18$) and switch at rank $17$ ($k=19$) from a hub-valued hypothesis to the exact reservoir value.
\item \emph{Be sharp near the change.} With relative margins of $0.14\%$ at $k=18$ and $0.09\%$ at $k=19$, the inequalities near the change cannot lose constant factors; a sharpened Euclidean reset (\cref{sec:reset-statement}) and exact finite rational checks supply this precision.
\end{enumerate}

\subsection{The central idea: another hit repairs an expensive first seed}\label{sec:idea}

A bad first hit can correct itself. A biased first hit can leave a cluster $P$ with expected cost close to $5\phi$, because it may land on a rare outlier and leave the bulk of $P$ unserved. At each later draw, $D^2$ sampling hits $P$ again with probability equal to its share of the total cost $T$; whenever this leaves $P$ with a large share of $T$ (larger absolute cost alone need not mean a larger share when outside cost dominates), a repairing return is correspondingly likely. A second hit at any point whose squared distance to the centroid of $P$ is at most the average brings the cost of $P$ back to at most $2\phi$, the expected cost of a uniform center. With \emph{price} $L$ equal to the average current cost of the uncovered clusters, a covered cluster of cost $v$ is charged $vL/T$ at each draw while being hit again with probability $v/T$: for every unit of probability that it is hit again, it pays $L$ (\cref{sec:ov-idea}). The fate of a badly hit cluster is thus a race between repair by a return and collection of all remaining clusters. The hub input tunes this race through $p$; in the reservoir no repair is needed, and the loss is the waste itself.

Two results turn this race into a proof. First, the Euclidean reset (\cref{thm:reset}) sharpens the local factor $5$ to
\[
U(s)=\min\Bigl\{s,\;1+\frac s2,\;\frac{5s}{s+14/5}\Bigr\},
\]
which approaches $5$ only when the pre-hit normalized cost $s$ is large; whether a subsequent return is likely then depends on the cluster's share $v/T$ of the total current cost (not on $s$ alone), which the race comparison tracks directly. Second, an induction on the number of remaining clusters counts actual draws (including wasted ones) and switches invariant at rank $17$. Where the margins near the crossover are too tight for uniform symbolic estimates, explicit analytic reductions convert each claim into finitely many exact rational or polynomial inequalities (\cref{sec:ov-regimes} and Appendices~B--E). \Cref{sec:overview} develops this overview and derives the warm-up bound $R_k\le2H_{k-2}+5$ along the way.

\subsection{What the guarantee means in practice}\label{sec:practice}

\Cref{thm:exact-rk} is a worst-case bound holding for every input, with no dependence on $|X|$, $d$, or clusterability. Because Lloyd iterations never increase cost, the same bound holds after any number of them (though it need not remain tight). By Markov's inequality a single run has cost at most $2R_k\OPT_k$ with probability at least $\frac12$, so the best of $t$ independent runs exceeds $2R_k\OPT_k$ with probability at most $2^{-t}$.

\subsection{Relation to earlier work}\label{sec:related}

\Cref{thm:exact-rk} closes the gap between $2\ln k\le R_k\le5(\ln k+2)$ of~\cite{AV07,MRS20}, determining the leading constant $2$, the asymptotic additive constant $2+2\gamma_{\mathrm{E}}$ (together with the uniform sharp additive constants of \cref{cor:sharp-additive}), and the exact value for every $k$. The proof builds on Dasgupta's potential argument~\cite{Das13,MRS20} in sharpened form (\cref{thm:restoration}) and the tight local factor $5$ of~\cite{MRS20} (\cref{lem:five}), adding post-hit repair tracking via the Euclidean reset $U(s)$ and the actual-draw race comparison.

Ostrovsky, Rabani, Schulman and Swamy~\cite{ORSS12} analyzed a closely related seeding rule. They choose the first center with probability proportional to the sum of its squared distances to all data points, then use the same $D^2$ sampling rule as $k$-means++ for every subsequent center. Under a separation assumption, they prove that variants of Lloyd's method with this initialization quickly find a near-optimal clustering. On the lower-bound side, Brunsch and R\"oglin~\cite{BR13} showed that an $\Omega(\log k)$ ratio occurs with high probability on suitable inputs, and Bhattacharya, Jaiswal and Ailon~\cite{BJA16} gave planar instances.

For fixed $\beta>1$, sampling $\lceil\beta k\rceil$ centers gives a constant-factor bi-criteria guarantee~\cite{ADK09,Wei16}; see also~\cite{MMSW16}. In this bi-criteria setting, Wei~\cite{Wei16} separated the harmonic indices for covered and uncovered clusters. Post-seeding local search gives a constant approximation~\cite{LS19,CGPR20}. Greedy $k$-means++ is analyzed in~\cite{GORT23}, fast approximate seedings in~\cite{BLHK16}, and parallel $k$-means$\|$~\cite{BMVKV12} in~\cite{BLK17,Roz20,MRS20}. For polynomial-time $k$-means approximation~\cite{KMNPSW04,ANSW20}, the factor has improved to $5.912$ in Euclidean space~\cite{CEMN22}, $3+2\sqrt2+\varepsilon$ in general metrics~\cite{CCGGLW25}, $4+\varepsilon$ in Euclidean space~\cite{CCGGLW26}, and $5+\varepsilon$ in general metrics~\cite{BGHLWW26}; approximating Euclidean $k$-means within some constant factor greater than $1$ is NP-hard~\cite{ACKS15}. Recent work also addresses massively parallel~\cite{CKP26}, almost-linear-time~\cite{JJLL26}, and fully dynamic~\cite{BCFJJL26} settings.

\subsection{How to read this paper}\label{sec:reading}

\Cref{sec:intro,sec:overview} assume only basic probability and explain what $k$-means++ guarantees and why. \Cref{sec:lower} constructs both extremal inputs and compares them rigorously. The upper bound follows \cref{fig:roadmap}: the general framework and restoration bridge, already yielding $R_k\le2H_{k-2}+5$ (\cref{sec:framework,sec:restoration}); the race, stopped-price comparison, and Euclidean reset (\cref{sec:race-reset}); the three inductive regimes (\cref{sec:hub-regime,sec:rank17,sec:reservoir-regime}); and the assembly (\cref{sec:assembly}). Where finite exact rational or polynomial computations enter, each section states the exact claim, domain, and analytic reduction; Appendices~B--E present the checking methods, seam coverage, witness parameter tables, and worked representative calculations, while the companion package supplies the complete exact-rational evaluation rules and repeated checks. \Cref{app:notation} (\cref{tab:notation}) summarizes the notation.

\paragraph{Verification and availability.}
An ancillary verification package (referred to throughout as the \emph{companion package} or \emph{companion}) accompanies this paper; it is distributed alongside the submission and documented in its top-level \texttt{README}. It contains both the complete machine-checked formalization of \cref{thm:exact-rk,cor:sharp-additive,thm:strict} mentioned in the abstract and the exact-rational witness data, interval-evaluation rules, and checking routines for every finite inequality used in Sections~3 and~6--9 and Appendices~B--E. All delegated checks (exact rational evaluations, Bernstein-basis polynomial positivity, quadratic discriminants, and finite interval-partition guards) are performed purely in exact rational arithmetic, with no floating-point approximation. Every check can be re-run end to end by following the package \texttt{README}; Appendices~B--E give the analytic reductions, witness parameter tables, and worked representative calculations (including all eight exact reset polynomials in Appendix~B and a full worked main-strip row in Appendix~D.3), while the companion supplies the remaining interval-evaluation rules and exhaustive repeated arithmetic.

\section{Overview and the warm-up argument}\label{sec:overview}

This section explains, without full rigor, why the leading constant is $2$, where the logarithm comes from, and what more the exact bound requires. \Cref{sec:ov-idea} is self-contained and leads to the simpler bound $R_k\le2H_{k-2}+5$; the rest of the section outlines how the exact proof is organized. \Cref{sec:lower} proves the matching lower bounds, and the proof of the upper bound starts in \cref{sec:framework}.

\subsection{The idea in one cluster}\label{sec:ov-idea}

Fix an optimal partition of $X$ into clusters $P_1,\dots,P_k$, with optimal one-center costs $\phi_i=\OPT_1(P_i)$, so that $\sum_i\phi_i=\OPT_k$. At any moment some clusters already contain a center (\emph{covered}) and the others do not (\emph{uncovered}). The final cost is the cost of the covered clusters plus that of the clusters never covered.

\paragraph{Where the logarithm comes from.}
Each draw that lands in a covered cluster is \emph{wasted}: it does not reduce the number of uncovered clusters. Dasgupta's potential argument~\cite{Das13,MRS20}, in the sharpened form of \cref{thm:restoration}, bounds the final cost of the clusters that are still uncovered at the end by the expected sum, over the draws, of
\begin{equation}\label{eq:intro-charge}
\cost(P,C)\cdot\frac{L}{T}
\end{equation}
over covered clusters $P$. Here $C$ is the current set of centers, $T=\cost(X,C)$ is the current total cost, and $L$ is the average current cost of an uncovered cluster, which we call the \emph{price}. We call~\eqref{eq:intro-charge} the \emph{charge} of $P$ at that draw. Since $P$ is drawn next with probability $\cost(P,C)/T$, the charge is the expected price paid if $P$ pays $L$ each time it is drawn again.

The charges of a covered cluster are bounded by a harmonic sum, even though some draws are wasted. Number the draws that follow the first center by $j=0,1,\dots,k-2$. The first center covers one cluster and each later draw covers at most one more, so after $j$ draws at least $k-1-j$ clusters are still uncovered. If $n\ge k-1-j$ clusters are uncovered, their total cost $nL$ is at most $T$, so the charge at that draw is at most $\cost(P,C)/(k-1-j)$. Costs only decrease as centers are added. Summing over $j$, the charges of $P$ total at most its cost when it was first covered times $\frac1{k-1}+\frac1{k-2}+\dots+1=H_{k-1}\approx\ln k$.

\paragraph{Why $5$ appears, and why $2$ should be the truth.}
The \emph{centroid identity} (\cref{sec:prelim}) says that if a center is placed at a \emph{uniformly} random point of $P$, the expected cost of $P$ is exactly $2\phi$. If every cluster had cost $2\phi$ after its first hit, the argument above would give $2\phi\ln k$, that is, leading constant $2$. But the first hit is not uniform. A point $y$ is drawn with probability proportional to its \emph{rate} $q(y)$, the squared distance from $y$ to the current centers, and the conditional expected cost of $P$ after such a hit can be close to $5\phi$; \cref{lem:five} shows that it is at most $5\phi$. Charging that cost for the rest of the run is what gives the constant $5$ of~\cite{MRS20}.

\paragraph{Repeat hits repair the cluster.}
A covered cluster is not frozen. It keeps competing for samples: at a draw where it has cost $v$, it is charged $vL/T$, and it is hit again with probability exactly $v/T$. \emph{So for every unit of probability that the cluster is hit again, it pays exactly the price $L$.} From the current state, the expected total charge through the next hit, or through the last draw if there is no further hit, is at most the current price $L$ (\cref{lem:prefix}). Normalize the cluster to centroid $0$ and mean squared radius $1$, and call its points with squared radius at most $1$ the \emph{inside} points. Serving the whole cluster from a single point $y$ costs exactly $\E_\mu\|Y-y\|^2=1+\|y\|^2$ in these units, by the centroid identity; with the earlier centers kept, the actual cost after a hit at $y$ is at most this. So a hit at an inside point leaves cost at most $2$: inside hits repair the cluster all the way back to the ideal value $2\phi$.

\paragraph{Excess outside, deficit inside.}
Two quantities attached to a point $y$ of the normalized cluster must be kept apart. Its squared radius $\|y\|^2$ describes where $y$ sits in its own cluster. Its rate $q(y)$ describes how far $y$ is from the current centers. The squared radius has mean exactly $1$, so the average excess above $1$ equals the average deficit below $1$:
\[
B:=\E_\mu(\|Y\|^2-1)_+=\E_\mu(1-\|Y\|^2)_+.
\]
Serving the cluster from $y$ alone costs $1+\|y\|^2=2+(\|y\|^2-1)$: an excess of $\|y\|^2-1$ over the ideal $2$ if $y$ is outside the unit ball, and a deficit of $1-\|y\|^2$ if it is inside. If the first hit were uniform, these would cancel exactly. Under $D^2$ sampling the hit has density proportional to $q(y)$ instead, and this weight may favor outside points. The resulting excess must be paid for by repair. The key observation is that the same point $y$ plays two roles: its rate $q(y)$ biases the hit toward it, and its distance from the inside points determines how fast a later inside hit repairs its excess. The triangle inequality ties the two together: the distance $\sqrt{q(y)}$ from $y$ to the centers is at most the distance $\sqrt{q(y')}$ from an inside point $y'$ to the centers plus $\|y-y'\|$. Splitting $\sqrt{q(y)}$ in these two proportions and squaring, one part is paid by the deficit of the inside points and the other by the price of waiting for a repair. That is what keeps the coefficient at $2$.

\paragraph{Paying for the wait.}
Here is how an induction absorbs the price of waiting. Consider a cluster $P$ that is not yet covered, with current cost $v$. Let $n$ be the number of other uncovered clusters and $nL$ their total cost, and $e$ the cost of everything else, so the total cost is $T=v+nL+e$. Let the \emph{clock} $c_{\mathrm{cl}}=\frac1n+\frac1{n-1}+\cdots$ have one term for each draw that will remain after the next one; if the next draw is the last, $c_{\mathrm{cl}}=0$ and nothing remains to be paid. The claim is that the charges of $P$ after its first hit total at most $2\phi\,c_{\mathrm{cl}}$ in expectation. Look at the next draw and weight each outcome by its cost rather than its probability; the weights add up to $T$.
\begin{itemize}[leftmargin=1.5em]
\item If the draw hits $P$ (weight $v$), the repair argument bounds the later charges by $2\phi$ per unit of clock plus a one-time waiting cost: in total at most $2\phi(c_{\mathrm{cl}}\,v+L)$.
\item If it hits one of the $n$ other uncovered clusters (weight $nL$), that cluster becomes covered and, unless the next draw is the last, the clock shrinks by exactly $\frac1n$. By induction this branch costs at most $nL\cdot2\phi(c_{\mathrm{cl}}-\frac1n)=2\phi\,c_{\mathrm{cl}}\,nL-2\phi L$. The saving $2\phi L$ pays back the waiting cost exactly.
\item Any other draw (weight $e$) costs at most $2\phi\,c_{\mathrm{cl}}\,e$.
\end{itemize}
Adding the three branches gives at most $2\phi\,c_{\mathrm{cl}}\,T$, and dividing by $T$ proves the claim (\cref{thm:delayed}). Together with the factor $5$ for the cost right after the first hit, this already gives the warm-up bound $R_k\le2H_{k-2}+5$ (\cref{cor:warmup}), within $3$ of the lower bound $2H_{k-1}+2$.

\subsection{From the whole input to one cluster}\label{sec:ov-reduction}

The exact bound keeps the same bookkeeping but charges each cluster precisely. Every cluster other than the first center's has a \emph{delayed cost}: zero if it is never hit, and otherwise the price at each return after its first hit plus its final cost. All draws are actual $D^2$ draws; exterior draws consume a draw without retiring an uncovered cluster and are handled by uniform thinning (\cref{thm:thinning}), which reclassifies a random live cluster as exterior. Call $\mathsf C_j$ a \emph{stock} for rank $j$ if every cluster $P$ facing $j$ other uncovered clusters with $j+1$ draws left has expected delayed cost at most $\mathsf C_j\phi_P$ from every retained center set. With $\theta=\sum_i\frac{|P_i|}{|X|}\frac{\phi_i}{\OPT_k}\in[\frac2{|X|},1]$, restoration and thinning yield
\[
\frac{\E\,\Phi_k}{\OPT_k}\;\le\;(1-\theta)\,\mathsf C_{k-2}+\theta\,(2H_{k-1}+2),
\]
reducing \cref{thm:exact-rk} to stocks $\mathsf C_j\le J_{j+2}$ for $j\le16$ and $\mathsf C_j=2H_{j+1}+2$ for $j\ge17$.

\subsection{The race and the Euclidean reset}\label{sec:ov-race}

After its first hit, a cluster of normalized cost $\bar u$ faces $j$ uncovered clusters of mean cost $L$: it is drawn again at rate $\bar u$ against a collection rate $iL$ when $i$ clusters remain. Under equal frozen prices and no exterior points, all $j$ clusters would be collected before a return with probability $\pi_j(L/\bar u)$—the exact hub race of~\eqref{eq:G} with $p=L/\bar u$. For the actual process with unequal shrinking prices and exterior draws, a stopped-price comparison (\cref{thm:stopped}) bounds the cost until the first return, and a transfer theorem (\cref{thm:transfer}) pays for the continuation after a return by persistence and the \emph{Euclidean reset} (\cref{sec:reset-statement}): a $D^2$ hit on a cluster of normalized cost $s$ leaves conditional mean cost at most
\[
U(s)=\min\Bigl\{s,\;1+\frac s2,\;\frac{5s}{s+14/5}\Bigr\},
\]
which approaches $5$ only as $s\to\infty$; whether a repairing return is likely after the hit depends on the cluster's resulting share of the total current cost $T$, rather than on $s$ alone.

\subsection{The induction, its regimes, and the role of finite computation}\label{sec:ov-regimes}

\begin{figure}[!htb]
\centering
\begin{tikzpicture}[
  box/.style={draw=black!65, rounded corners=2pt, fill=#1, align=center, font=\footnotesize, inner sep=3.5pt, minimum height=1.05cm, text width=3.05cm},
  box/.default=white,
  fbox/.style={box=#1, densely dashed, line width=0.75pt},
  arr/.style={-{Stealth[length=5pt]}, black!60, line width=0.6pt},
  lbl/.style={font=\scriptsize, fill=white, inner sep=1.5pt}]
\node[box=black!4] (rs) at (0.0,0) {\S\ref{sec:restoration} Restoration;\\ first-seed bridge};
\node[box=black!4] (fw) at (4.6,0) {\S\ref{sec:framework} Charges, clocks;\\ delayed bound};
\node[fbox=black!4] (rr) at (9.2,0) {\S\ref{sec:race-reset} Race, stopped\\ price \& reset};
\node[fbox=OursBlue!10] (hb) at (2.1,-2.4) {\S\ref{sec:hub-regime} Hub regime\\ ranks $0$--$16$};
\node[fbox=OursBlue!10] (r17) at (6.7,-2.4) {\S\ref{sec:rank17} Opening \&\\ rank-$17$ transition};
\node[fbox=OursBlue!10] (rv) at (11.3,-2.4) {\S\ref{sec:reservoir-regime} Reservoir regime\\ ranks $\ge18$};
\node[box=MRSOrange!12] (th) at (6.7,-4.75) {\S\ref{sec:assembly} Proof of\\ \cref{thm:exact-rk,thm:strict}};
\node[fbox=MRSOrange!12] (lb) at (11.3,-4.75) {\S\ref{sec:lower} Extremal inputs\\ \& crossover};
\draw[arr] (fw.south west) -- (hb.north);
\draw[arr] (fw.south east) to[out=-18,in=162] (rv.north west);
\draw[arr] (rr.south west) -- (hb.north east);
\draw[arr] (rr.south) -- (r17.north east);
\draw[arr] (rr.south east) -- (rv.north);
\draw[arr] (hb) -- node[lbl,above=3pt]{$J_{18}<\mathsf P_{16}$} (r17);
\draw[arr] (r17) -- node[lbl,above=3pt]{$\mathsf C_{17}$} (rv);
\draw[arr] (hb.south) -- (th.north west);
\draw[arr] (r17.south) -- (th.north);
\draw[arr] (rv.south) -- (th.north east);
\draw[arr] (rs.south) |- node[lbl,pos=0.72,above=2pt]{bridge} (th.west);
\draw[arr] (lb) -- (th);
\end{tikzpicture}
\caption{How the proof fits together. The top row is shared across regimes; the middle row establishes the stocks rank by rank, handing $J_{18}<\mathsf P_{16}$ into the rank-$17$ transition and $\mathsf C_{17}=2H_{18}+2$ into the reservoir induction; the bridge of \S\ref{sec:restoration} converts stocks into the upper bound, and \S\ref{sec:lower} supplies the matching lower bounds. Solid borders mark purely symbolic arguments; dashed borders mark symbolic reductions combined with exact finite rational or polynomial checks (checking methods, worked examples, and witness tables in Appendices~B--E; independently re-runnable in the ancillary verification package of \cref{sec:reading}).}
\label{fig:roadmap}
\end{figure}
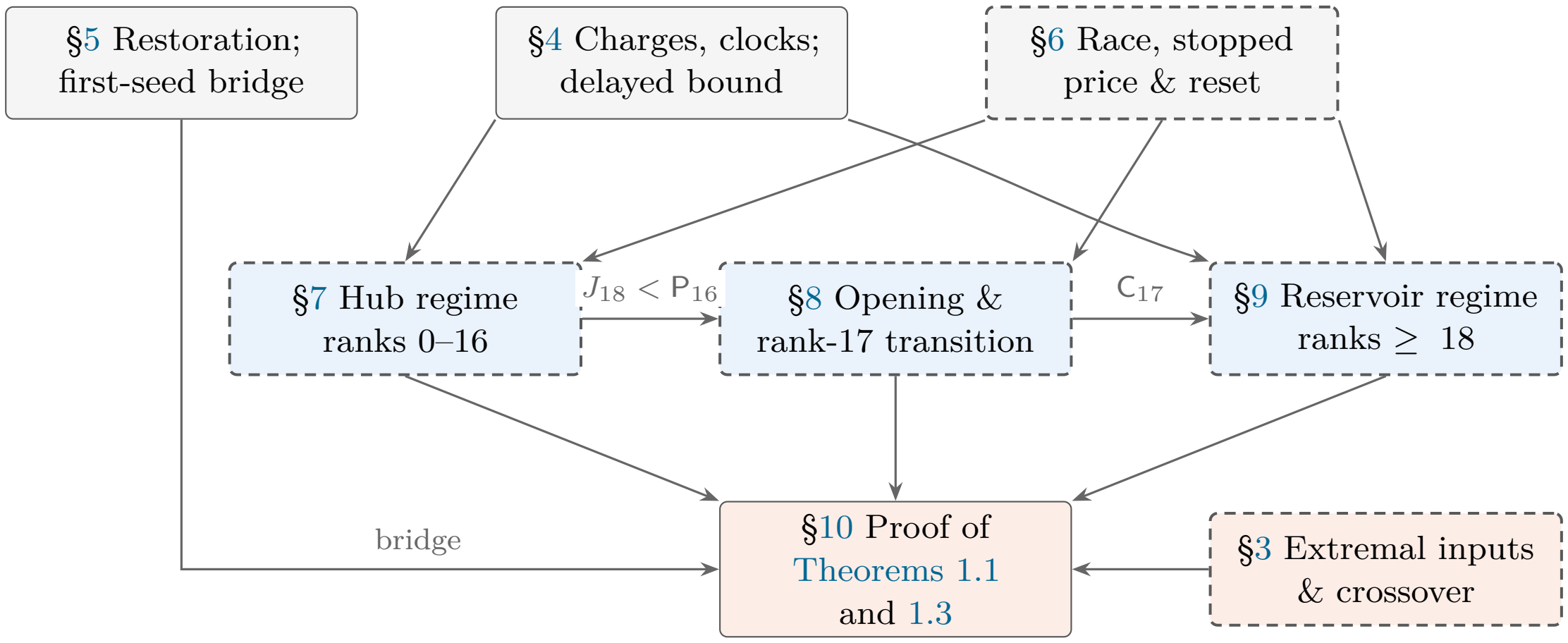

The stocks are proved by induction on the rank $j$, mirroring the crossover at $k=19$ (\cref{fig:roadmap}):
\begin{itemize}[leftmargin=1.5em]
\item \emph{Rank $0$ ($k=2$).} One draw remains, and the reset gives the universal stock $5=J_2$.
\item \emph{Ranks $1$--$16$ ($3\le k\le18$), the hub regime.} Clusters split by their intrinsic worst post-hit mean $M^*(P)\le5$. If $M^*(P)>485883/100000$, a single witness cut decomposes $P$ into a tight bulk and small-mass distant satellites, and a binary comparison bounds it by $J_{j+2}$. Otherwise an opening induction builds explicit stocks $\mathsf C_j<J_{j+2}$.
\item \emph{Rank $17$ ($k=19$), the transition.} From the universal rank-$16$ bound $J_{18}<\mathsf P_{16}=4445581/500000$, an $18$-draw analysis establishes the exact reservoir stock $\mathsf C_{17}=2H_{18}+2$.
\item \emph{Ranks $j\ge18$ ($k\ge20$), the reservoir regime.} One induction upgrades the clock argument of \cref{sec:ov-idea} via an explicit repair kernel and carries $2H_{j+1}+2$ to every larger rank.
\end{itemize}

\paragraph{Finite computation and witness checking.}
Ranks $0$ and $k\ge20$ rely chiefly on symbolic arguments plus a small number of explicit rational inequalities. Ranks $1$--$17$ require larger finite exact checks for a structural reason: at the crossover, the hub and reservoir values differ by $0.14\%$ ($k=18$) and $0.09\%$ ($k=19$), margins too narrow for uniform symbolic estimates. Throughout, we separate \emph{finding witnesses} from \emph{checking them}: thresholds, weights, piecewise envelopes, and interval partitions are rational witnesses whose discovery never enters the proof; correctness relies solely on checking the resulting finite inequalities over their stated domains, including all seams and endpoints. Every analytic reduction from the continuous claim to the finite family is proved in the text. Appendices~B--E state the checking methods in ordinary mathematics—exact rational evaluation, Bernstein-basis positivity, discriminant bounds, and interval partitions—work one representative example per materially different method, and tabulate the witness parameters, while the ancillary verification package (\cref{sec:reading}) supplies the remaining interval-evaluation rules and verifies every delegated check end to end.

\section{The two extremal inputs and the change at \texorpdfstring{$k=19$}{k=19}}\label{sec:lower}

This section proves $R_k\ge\max\{J_k,2H_{k-1}+2\}$ and compares the two lower bounds. \Cref{sec:hub,sec:reservoir} construct the two inputs of \cref{fig:extremal} and compute their ratios from exact finite-$N$ discrete recurrences. \Cref{sec:crossover} compares the two values rigorously and can be skipped on a first reading. The hub input is first described with repeated points (\emph{labels} at a few \emph{sites}); \cref{lem:perturb} separates labels into distinct points while changing the ratio arbitrarily little. All limits below are taken in the stated order.

\begin{lemma}[Separating labels]\label{lem:perturb}
Let $X_0$ be a finite multiset of labels at sites in $\R^d$ ($d\ge1$), with fixed multiplicities, $k\le|X_0|$, and $\OPT_k(X_0)>0$, where cost and sampling treat every label separately. For any family of distinct-point perturbations $X_\varepsilon$ moving every label by at most $\varepsilon$, $\E\,\Phi_k(X_\varepsilon)/\OPT_k(X_\varepsilon)\to\E\,\Phi_k(X_0)/\OPT_k(X_0)$ as $\varepsilon\to0^+$.
\end{lemma}

\begin{proof}
The expectation $\E\,\Phi_k(X_\varepsilon)$ is a finite sum over labelled draw sequences of length $k$. Because $\OPT_k(X_0)>0$, every prefix center set of size at most $k$ has strictly positive cost on $X_0$, so every sampling denominator stays bounded away from $0$ for small $\varepsilon>0$ and each path probability and terminal cost converges to its value on $X_0$. By $1$-Lipschitzness of $D(\cdot,C)$, $\cost(X_\varepsilon,C)\le(1+\varepsilon)\cost(X_0,C)+|X_0|(\varepsilon+\varepsilon^2)$ and symmetrically with $X_0,X_\varepsilon$ swapped, so $\OPT_k(X_\varepsilon)\to\OPT_k(X_0)>0$. Explicit injective perturbations exist in $\R^d$ ($d\ge1$) by shifting label $i$ along coordinate~$1$ by $\frac{i+1}{|X_0|+1}\min(\varepsilon,\Delta_{\min}/3)$, where $\Delta_{\min}>0$ is the minimum distance between distinct sites.
\end{proof}

\subsection{The hub: a rare outlier}\label{sec:hub}

Fix $m=k-2\ge0$, $p>0$, and integers $M,N\ge1$. In $\R^{m+1}$ with orthonormal basis $e_0,\dots,e_m$, put $M$ labels at $0$ (the \emph{hub}), $N$ labels at $e_0$ (the \emph{core}), one label at $2e_0$ (the \emph{outlier}), and one label at $\sqrt{Np}\,e_r$ for each $r=1,\dots,m$ (the \emph{coupons}). The $m+3$ sites have pairwise squared distances at least $\min(1,Np)>0$, so any $k=m+2$ centers merge two distinct sites and $\OPT_k>0$. Placing centers at the hub, the $m$ coupons, and the core--outlier centroid $\frac{N+2}{N+1}e_0$ gives $0<\OPT_k\le N(\frac1{N+1})^2+(\frac N{N+1})^2=\frac N{N+1}<1$. The sites $0,e_0,2e_0$ translate the tight single-hit example of~\cite[Appendix~E, Theorem~E.1]{MRS20} into a full input with competing coupons.

\begin{theorem}[Hub lower bound]\label{thm:hub}
For every $k\ge2$, $R_k\ge J_k$.
\end{theorem}

\begin{proof}
By \cref{lem:perturb} it suffices to bound the ratio of the labelled hub input, which satisfies $\E\,\Phi_k/\OPT_k\ge\E\,\Phi_k$ since $0<\OPT_k\le1$. Dropping all nonnegative outcomes where the uniform first seed is not a hub label (the hub-first event has probability $\frac{M}{M+N+1+m}$) gives
\[
\frac{\E\,\Phi_k}{\OPT_k}\ge\frac{M}{M+N+1+m}\,E_{\mathrm{hub}}(M,N,p,m),
\]
where $E_{\mathrm{hub}}(M,N,p,m)$ is the expected final cost after $m+1$ further $D^2$ draws from $C=\{0\}$.

Whenever $0\in C$ and $C$ is a subset of the $m+3$ sites, exact pairwise distances give: hub cost $0$; each unhit coupon cost $Np$ (its squared distances to $0$, core, outlier, and other coupons are $Np$, $1+Np$, $4+Np$, $2Np$); unhit core cost $N$ (squared distance $1$ to $0$ and to the outlier); and unhit outlier cost $4$ if the core is unhit, or $1$ if the core is hit. Thus from any such $C$ with $j\le m$ unhit coupons and $r$ draws remaining, the expected final cost depends only on $j$, $r$, and whether the core and outlier are hit. Induction on $r$ and $j$ evaluates all four states:
\begin{enumerate}[label=(\roman*),leftmargin=2em]
\item \emph{Both core and outlier hit, $r\le j$ draws left.} Total cost is $jNp$ and every draw hits an unhit coupon, so the expected final cost is $(j-r)Np$.
\item \emph{Core hit, outlier unhit, $j$ draws left.} Total cost is $1+jNp$. Hitting the outlier (rate $1$) enters~(i) with $j-1$ draws and $j$ coupons (cost $Np$); hitting a coupon (rate $jNp$) lowers $j$ by $1$. Hence the expected cost $A_{N,p}(j)$ satisfies $A_{N,p}(0)=1$ and
\[
A_{N,p}(j)=\frac{Np+jNp\,A_{N,p}(j-1)}{1+jNp}\qquad(j\ge1).
\]
Dividing numerator and denominator by $Np$ and inducting on $j$ yields $\lim_{N\to\infty}A_{N,p}(j)=1+H_j$.
\item \emph{Outlier hit, core unhit, $j$ draws left.} Total cost is $N(1+jp)$. Hitting the core (rate $N$) leaves cost $Np$ by~(i); hitting a coupon (rate $jNp$) lowers $j$ by $1$. Hence the expected cost is $N\,K_j(p)$, where $K_0(p)=1$ and $K_j(p)=\frac{p+jp\,K_{j-1}(p)}{1+jp}=p+(1-p)\pi_j(p)$.
\item \emph{Neither hit, $j+1$ draws left.} Total cost is $N+4+jNp$. At $j=0$, one draw yields $C_{N,p}(0)=\frac{5N}{N+4}$; for $j\ge1$, hitting the core (rate $N$), outlier (rate $4$), or a coupon (rate $jNp$) gives
\[
C_{N,p}(j)=\frac{N\,A_{N,p}(j)+4N\,K_j(p)+jNp\,C_{N,p}(j-1)}{N+4+jNp}.
\]
\end{enumerate}
Starting from $C=\{0\}$ gives $E_{\mathrm{hub}}(M,N,p,m)=C_{N,p}(m)$. Dividing numerator and denominator by $N$ and letting $N\to\infty$ inductively yields $C_{N,p}(0)\to5=G_0(p)$ and
\[
\lim_{N\to\infty}C_{N,p}(j)=\frac{1+H_j+4\bigl(p+(1-p)\pi_j(p)\bigr)+jp\lim_{N\to\infty}C_{N,p}(j-1)}{1+jp}=G_j(p)
\]
for every $j\ge1$, matching~\eqref{eq:G}. Sending $M\to\infty$ at fixed $N$, then $N\to\infty$, and taking the supremum over $p>0$ gives $R_k\ge J_k$.
\end{proof}

\paragraph{Exponential-clock intuition.}
As intuition for the closed form of $G_m(p)$ in~\eqref{eq:G-binomial}, in the $N\to\infty$ limit (with time scaled by $N$) the core rings as a rate-$1$ exponential clock $V$, the $m$ coupons as independent rate-$p$ clocks with maximum $t=Y_m/p$, and the outlier at infinitesimal rate $4/N$ before $V$ and $1/N$ after $V$. Whenever the outlier rings before $\max(V,t)$ it displaces the final heavy event (costing $Np$ if $V<t$ and $N$ if $V\ge t$), while otherwise the final cost is $1$; integrating over $V$ and $t$ produces $1+H_m+3(1-\pi_m)/c+4\pi_m-4(c-1)\,d\pi_m/dc$ ($c=1/p$). \Cref{lem:binomial} proves this equivalence directly from~\eqref{eq:G}.

\subsection{The reservoir: a large simplex}\label{sec:reservoir}

\begin{theorem}[Reservoir lower bound]\label{thm:simplex}
For every $k\ge2$ and $N\ge2$ there are inputs with $N+k-1$ distinct points whose ratios tend, as $q\to\infty$, to
\begin{equation}\label{eq:simplex-finite}
2+2H_{k-1}-\frac2{N+k-1}-\frac{2(k-1)}{(N+k-1)N}\Bigl(1-\binom{N+k-2}{k-2}^{-1}\Bigr).
\end{equation}
Hence $R_k\ge2H_{k-1}+2$ for every $k\ge1$.
\end{theorem}

\begin{proof}
Let $m=k-1\ge1$. In $\R^{N+m+1}$, place the $N$ vertices of a centered regular simplex in the first $N$ coordinates with squared norm $1/N$ (vertex $i$ has coordinate $\frac{N-1}{N\sqrt{N-1}}$ at index $i$ and $-\frac1{N\sqrt{N-1}}$ at the other $N-1$ indices), so its one-center cost is $1$ and distinct vertices have squared distance $\frac2{N-1}$. In the remaining $m+1$ orthogonal unit directions $e_0,\dots,e_m$, place the $m$ \emph{lights} $\sqrt{q/2-1/N}\,e_0+\sqrt{q/2}\,e_j$ ($1\le j\le m$) with $q>2$. Every vertex--light and light--light squared distance is $q$. Any cluster containing a light and another point costs at least $q/2>1$, so the unique optimal $k$-partition is the simplex plus $m$ singletons and $\OPT_k=1$.

Send $q\to\infty$ at fixed $N$. Once $a\ge1$ simplex vertices are hit, the $N-a$ remaining vertices have total cost $s_a=\frac{2(N-a)}{N-1}$ ($s_1=2$), while each of the $j$ unhit lights has cost $q$. Let $E_{\mathrm{post}}(a,j,r;q)$ be the expected final cost with $r\le j$ draws left. By induction on $r\ge0$, $E_{\mathrm{post}}(a,j,r;q)-(j-r)q\to s_a(1+H_j-H_{j-r})$: it is exact at $r=0$, and for $r\ge1$,
\begin{multline*}
E_{\mathrm{post}}(a,j,r;q)-(j-r)q
=\frac{s_a}{s_a/q+j}+\frac{s_a}{s_a+jq}\bigl(E_{\mathrm{post}}(a+1,j,r-1;q)-(j-r+1)q\bigr)\\
+\frac{j}{s_a/q+j}\bigl(E_{\mathrm{post}}(a,j-1,r-1;q)-(j-r)q\bigr)
\end{multline*}
tends to $\frac{s_a}{j}+0+s_a(1+H_{j-1}-H_{j-r})=s_a(1+H_j-H_{j-r})$. Setting $a=1$, $r=j$ gives $E_{\mathrm{post}}(1,j,j;q)\to2+2H_j$.

Put $N_{\mathrm{pts}}=N+m$. A uniform first seed hits the simplex with probability $N/N_{\mathrm{pts}}$ (expected cost tending to $2+2H_m$), and otherwise hits a light (probability $m/N_{\mathrm{pts}}$), leaving $h=m-1$ unhit lights and $h+1$ draws. From any all-light state with $h$ unhit lights and $h+1$ draws left, the simplex has rate $Nq$ against $h$ lights at rate $q$, so the expected cost $E_{\mathrm{pre}}(h;q)$ satisfies $E_{\mathrm{pre}}(0;q)=E_{\mathrm{post}}(1,0,0;q)$ and $E_{\mathrm{pre}}(h;q)=\frac N{N+h}E_{\mathrm{post}}(1,h,h;q)+\frac h{N+h}E_{\mathrm{pre}}(h-1;q)$. Induction using $(N+h)\binom{N+h-1}{h-1}=h\binom{N+h}{h}$ gives
\[
\lim_{q\to\infty}E_{\mathrm{pre}}(h;q)=2+2H_h-\frac2N\Bigl(1-\binom{N+h}{h}^{-1}\Bigr).
\]
Averaging both first-seed outcomes with $H_{m-1}=H_m-\frac1m$ yields~\eqref{eq:simplex-finite}. Letting $N\to\infty$ gives $R_k\ge2H_{k-1}+2$ ($R_1=2$ at $k=1$).
\end{proof}

Within each cluster the first hit is uniform, so the loss comes purely from wasted draws. Makarychev, Reddy and Shan~\cite[Appendix~E.2]{MRS20} used a large regular simplex and $k-1$ singletons for a bi-criteria lower bound; equal cross-distances yield the finite-$N$ limit~\eqref{eq:simplex-finite} for $k$ seeds and improve Arthur and Vassilvitskii's $k$-simplex bound $2H_k$~\cite[Theorem~4.1]{AV07} by $2-2/k$ to $2H_{k-1}+2$.

\subsection{Which input wins}\label{sec:crossover}

Write $c=1/p$, $\pi_m=\pi_m(1/c)=\prod_{i=1}^m\frac i{i+c}$, $s_{1,m}(c)=\sum_{i=1}^m\frac1{i+c}$, $s_{2,m}(c)=\sum_{i=1}^m\frac1{(i+c)^2}$, $K_m(1/c)=\pi_m+(1-\pi_m)/c$, and $G'_m(c)=\frac d{dc}G_m(1/c)$. Since $d\pi_m/dc=-\pi_ms_{1,m}$,
\begin{equation}\label{eq:Gprime}
G'_m(c)=3\,\frac{\pi_m-1}{c^2}+\frac{3\pi_ms_{1,m}}c+4(1-c)\,\pi_m\bigl(s_{1,m}^2+s_{2,m}\bigr).
\end{equation}

\begin{lemma}[Two forms of the hub function]\label{lem:binomial}
For $m\ge1$ and $c>0$, the closed form $G_m(1/c)=1+H_m+3(1-\pi_m)/c+4\pi_m-4(c-1)\,d\pi_m/dc$ agrees with the recurrence~\eqref{eq:G} and with~\eqref{eq:G-binomial}, and
\begin{equation}\label{eq:hub-difference}
m\bigl(G_m(1/c)-G_{m-1}(1/c)\bigr)=1+\pi_m\bigl[4(1-c)\,c\,s_{1,m}-1\bigr].
\end{equation}
\end{lemma}

\begin{proof}
Partial fractions of $\pi_m=m!/\prod_{i=1}^m(i+c)$ give $\pi_m=\sum_{i=1}^m(-1)^{i-1}i\binom mi\frac1{i+c}$, $(1-\pi_m)/c=\sum_{i=1}^m(-1)^{i-1}\binom mi\frac1{i+c}$, and $\pi_ms_{1,m}=\sum_{i=1}^m(-1)^{i-1}i\binom mi\frac1{(i+c)^2}$. Substituting these with $1=\sum_{i=1}^m(-1)^{i-1}\binom mi$ and $H_m=\sum_{i=1}^m(-1)^{i-1}\binom mi\frac1i$ into the closed form, $(i+c)^2+i(3-4c)(i+c)-4i^2(1-c)=ic^2\bigl(5p-4+(1/i)\bigr)$ gives~\eqref{eq:G-binomial}. For~\eqref{eq:hub-difference}, use $\pi_m=\frac m{m+c}\pi_{m-1}$, $\pi_{m-1}-\pi_m=\frac c{m+c}\pi_{m-1}$, and $\pi_ms_{1,m}-\pi_{m-1}s_{1,m-1}=\frac{m\pi_{m-1}}{(m+c)^2}-\frac c{m+c}\pi_{m-1}s_{1,m-1}$; rewriting~\eqref{eq:hub-difference} in $p=1/c$ yields~\eqref{eq:G}, matching $G_0=5$ at $m=0$.
\end{proof}

\begin{theorem}[Laplace unimodality]\label{thm:laplace}
Fix $m\ge1$ and $\mathsf v\ge-1$, and put $F(c)=G_m(1/c)+\mathsf v\,K_m(1/c)$.
\begin{enumerate}[label=(\alph*),leftmargin=2em]
\item There is $a\ge(11+2\mathsf v)/8$ such that $c\mapsto e^{ac}F'(c)$ is strictly decreasing on $(0,\infty)$ and eventually negative. If $F'(0^+)>0$ (which holds at $\mathsf v=0$), $F$ has a unique critical point, its strict global maximum.
\item If $G'_m(c_0)>0$, then $J_{m+2}\le G_m(1/c_0)+\frac8{11}G'_m(c_0)$. If moreover $G'_m(c_1)<0$ for some $c_1>c_0$, then the maximizer lies in $(c_0,c_1)$ and $J_{m+2}\le G_m(1/c_0)+(c_1-c_0)G'_m(c_0)$.
\item The maximizers satisfy $\frac{11}5=p^*_1<p^*_2<\dots<p^*_{16}<\frac{273}{100}<\frac{11}4$.
\end{enumerate}
\end{theorem}

\begin{proof}
(a) Let $S(z)=1-(1-e^{-z})^m$ and $f=-S'$. Integration by parts gives $(1-\pi_m)/c=\int_0^\infty e^{-cz}S(z)\,dz$ and $c\,\pi_m'(c)=-\int_0^\infty e^{-cz}(zf(z))'\,dz$, where $(zf)'/f=1-z+(m-1)z/(e^z-1)$. Hence
\[
F(c)=1+H_m+\int_0^\infty e^{-cz}f(z)B(z)\,dz,\qquad B(z)=(3+\mathsf v)\frac{S(z)}{f(z)}+8+\mathsf v-8z+4(m-1)\frac z{e^z-1}.
\]
Since $S/f=\frac1m\sum_{j=0}^{m-1}(1-e^{-z})^{-j}$ decreases weakly to $1$, $3+\mathsf v>0$, and $z/(e^z-1)$ decreases strictly, $B$ decreases strictly to $-\infty$ with a unique zero $a\ge(11+2\mathsf v)/8$ (as $B\ge0$ at $z=(11+2\mathsf v)/8$). Differentiating $e^{ac}F'(c)=-\int_0^\infty ze^{-c(z-a)}f(z)B(z)\,dz$ gives $\int_0^\infty z(z-a)e^{-c(z-a)}f(z)B(z)\,dz<0$, since $(z-a)B(z)\le0$ with equality only at $z=a$. As $c\to\infty$, the $(0,a)$ part dominates and $e^{ac}F'(c)\to-\infty$.

(b) At $\mathsf v=0$, part~(a) gives $G'_m(c)\le e^{-a(c-c_0)}G'_m(c_0)$ for $c\ge c_0$; integrate from $c_0$ to the maximizer using $1/a\le8/11$ or the bracket length $c_1-c_0$.

(c) Differentiating~\eqref{eq:hub-difference} for $c\in(0,1)$ with $u_r=cs_{1,r}$, $w_r=c^2s_{2,r}$ gives $\frac d{dc}[r(G_r-G_{r-1})]=\frac{\pi_ru_r}c[5-8c-4(1-c)(u_r+w_r/u_r)]$. Adding $\Delta=c/(r+1+c)$ to $(u,w)$ increases $u+w/u$ by $\Delta[1+(\Delta-w/u)/(u+\Delta)]>0$ (as $w\le u^2$), so the bracket strictly decreases in $r$. Since the first $m$ derivatives sum to $0$ at $c^*_m$, the $(m+1)$st is negative and $G'_{m+1}(c^*_m)<0$, giving $c^*_{m+1}<c^*_m$ by~(a). Finally $p^*_1=\frac{11}5$ directly, and $G'_{16}(\frac{100}{273})>0$ by \cref{thm:brackets}.
\end{proof}

\begin{theorem}[Brackets at the change]\label{thm:brackets}
Let $\mathsf P_{16}=4445581/500000=8.891162$. Then
\[
2+2H_{17}<G_{16}(\tfrac{11}4)<G_{16}(\tfrac{273}{100})<J_{18}<8.89116164<\mathsf P_{16},\qquad J_{19}<8.988329<2+2H_{18},
\]
and the maximizing price $p^*_{16}$ lies in $(2.7290,2.7300)$.
\end{theorem}

\begin{proof}
For $c\in(0,\frac34)$, $3/c-4+4(1-c)s_{1,m}>0$, so $G_m(1/c)$ decreases in $\pi_m,s_{1,m}$ and $G'_m(c)$ increases in $\pi_m,s_{1,m},s_{2,m}$. \Cref{app:hub-brackets} verifies outward rational brackets for these finite products and sums at $(m,c)=(16,\frac4{11})$, $(17,\frac4{11})$, $(16,\frac{100}{273})$, and $(16,\frac{1000}{2729})$, giving $G_{16}(\frac{11}4)-(2+2H_{17})>0.011973$, $G_{16}(\frac{273}{100})<8.891161578$, $0<G'_{16}(\frac{100}{273})<0.0004485$, and $G'_{16}(\frac{1000}{2729})<-0.0026$. By \cref{thm:laplace}(b), $p^*_{16}\in(2.729,2.73)$ and
\[
J_{18}\le G_{16}(\tfrac{273}{100})+\Bigl(\tfrac{1000}{2729}-\tfrac{100}{273}\Bigr)G'_{16}(\tfrac{100}{273})<8.89116164<\mathsf P_{16},
\]
while $J_{19}\le G_{17}(\frac{11}4)+\frac8{11}G'_{17}(\frac4{11})<8.988329<18358381/2042040=2+2H_{18}$. (Note that $8.891161578$ bounds $G_{16}(\frac{273}{100})$, not $J_{18}$: $G_{16}(\frac{54597}{20000})>8.891161581$. Only $J_{18}<\mathsf P_{16}$ is used later.)
\end{proof}

The following theorem explains why the hub's advantage, once lost, does not return; it is not needed for \cref{thm:exact-rk}, where the sign of $J_k-(2H_{k-1}+2)$ comes from \cref{thm:hub-stocks} for $k\le18$ and from the upper bound for $k\ge19$.

\begin{theorem}[The advantage decreases]\label{thm:gap}
Let $A_m=2+2H_{m+1}$. The optimized hub advantage $J_{m+2}-A_m$ strictly decreases in $m\ge0$; it is positive for $0\le m\le16$ and negative for every $m\ge17$.
\end{theorem}

\begin{proof}
Write $t=1-c$, $u_m=cs_{1,m}$, and $w_m=c^2s_{2,m}$. At $c=c^*_m$, $G'_m(c)=0$ and~\eqref{eq:Gprime} read
\begin{equation}\label{eq:stationary}
\frac1{\pi_m}=1+u_m+\tfrac43t(u_m^2+w_m).
\end{equation}
By~\eqref{eq:hub-difference}, $m(J_{m+2}-J_{m+1})\le m(G_m-G_{m-1})(1/c^*_m)=1+\pi_m(4tu_m-1)$, while $m(A_m-A_{m-1})=1+\frac{m-1}{m+1}$. Thus it suffices that $\pi_m(4tu_m-1)<\frac{m-1}{m+1}$, trivial if $t\le0$; assume $0<t<1$.

\emph{Ranks $1\le m\le4$.} Since $\pi_m^{-1}\ge1+u_m$ and $u_m<cH_m$, it suffices that $(3m+5)u_m-4(m+1)u_m^2/H_m\le2m$, whose maximum is $H_m(3m+5)^2/(16(m+1))$. Exact fractions give $32m(m+1)-H_m(3m+5)^2=0,\frac{21}2,\frac{74}3,\frac{455}{12}$ for $m=1,2,3,4$ (strict at $m=1$ via $u_1<cH_1$).

\emph{Ranks $5\le m\le17$.} Since $H_m\le H_{17}<\frac72$, $t\le1-\frac27u_m$, and by~\eqref{eq:stationary} it suffices that $F(u)=5+(2-12t)u+\frac83tu^2>0$ at $u=u_m$ (giving $\pi_m(4tu_m-1)<\frac23\le\frac{m-1}{m+1}$). For $u\le\frac74$, $21F(u)\ge(105-16u)(1-u)^2+u(16-9u)>0$; for $\frac74\le u\le\frac92$, $t\le\frac12$ gives $F(u)\ge5-4u+\frac43u^2>0$; for $u\ge\frac92$, $F(u)\ge5+2u$.

\emph{Ranks $m\ge17$.} Since $\frac{m-1}{m+1}\ge\frac89$, it suffices that $P_t(u)=17+(8-36t)u+\frac{32}3tu^2>0$. For $t\le\frac29$ all coefficients are nonnegative. For $\frac29\le t\le\frac{19}{20}$ the convex discriminant $(8-36t)^2-\frac{2176}3t$ equals $-\frac{4352}{27}$ and $-\frac{197}{75}$ at the endpoints. For $t>\frac{19}{20}$, $P_t(u)>0$ unless $u_m\in(\frac9{10},\frac{17}{10})$ (containing both roots of $P_1(u)$). There $c<\frac1{20}$, $\ln(1+c/i)\le(1+c)\frac c{i+c}$ gives $\pi_m^{-1}\le e^{(1+c)u_m}<e^{21u_m/20}$, and $e^{9/5}<\frac{61}{10}$ (six Taylor terms plus geometric tail) yields $e^x\le1+x+\frac{55}{54}x^2$ on $[0,\frac95]$. With $u_m>\frac9{10}$ this gives $\pi_m^{-1}<1+u_m+\frac{19}{15}u_m^2$, contradicting~\eqref{eq:stationary} because $\frac43t>\frac{19}{15}$.

\emph{Signs.} \Cref{thm:brackets} gives $J_{18}-A_{16}>0$ and $J_{19}-A_{17}<0$; strict monotonicity fixes every other sign.
\end{proof}

\section{Charges, clocks and the delayed bound}\label{sec:framework}

This section sets up the definitions and elementary bounds shared by all regimes; global notation is summarized in \cref{app:notation} (\cref{tab:notation}). Its main result, \cref{thm:delayed}, bounds a cluster's post-hit repeat charges by twice its optimal cost per unit of harmonic clock, which with \cref{sec:restoration} already yields $R_k\le2H_{k-2}+5$ (\cref{cor:warmup}).

\subsection{Costs, the algorithm and three Euclidean facts}\label{sec:prelim}

For $y\in\R^d$ and a finite nonempty $C\subset\R^d$, write $D(y,C)=\min_{c\in C}\|y-c\|$ and $\cost(S,C)=\sum_{y\in S}D(y,C)^2$ ($S\subseteq X$), and let $\OPT_j(S)=\min_{|C|=j}\cost(S,C)$. For nonempty $S$ with centroid $\bar y_S=\frac1{|S|}\sum_{y\in S}y$, $\OPT_1(S)=\sum_{y\in S}\|y-\bar y_S\|^2$.

\begin{center}
\fbox{\begin{minipage}{0.88\textwidth}
\refstepcounter{algo}\label{alg:kmpp}
\textbf{Algorithm~\thealgo} ($k$-means++ seeding).
\begin{enumerate}[leftmargin=1.5em]
\item Choose $y$ uniformly at random from $X$ and set $C_1=\{y\}$.
\item For $t=1,\dots,k-1$: choose $y\in X$ with probability $D(y,C_t)^2/\cost(X,C_t)$, and set $C_{t+1}=C_t\cup\{y\}$.
\item Output $C_k$; its cost is $\Phi_k=\cost(X,C_k)$.
\end{enumerate}
\end{minipage}}
\end{center}

Since centers are never removed, every cost is nonincreasing in $t$ (\emph{retention}); if $\cost(X,C_t)=0$ then $C_t=X$ and $t=|X|\ge k$, so step~2 is well defined for $t<k$. A \emph{draw} is one execution of step~2, and it \emph{hits} $S$ if the chosen point lies in $S$.

\paragraph{Three Euclidean facts.}
For any finite nonempty $P\subset\R^d$ with centroid $\bar y_P$ and $\phi=\OPT_1(P)$,
\begin{align}
\sum_{y'\in P}\|y'-y\|^2&=\phi+|P|\,\|y-\bar y_P\|^2\qquad(y\in\R^d),\label{eq:centroid}\\
\sum_{y\in P}\sum_{y'\in P}\|y-y'\|^2&=2|P|\,\phi,\label{eq:pairwise}\\
|D(y,C)-D(y',C)|&\le\|y-y'\|\qquad(\text{finite nonempty }C\subset\R^d),\label{eq:lipschitz}
\end{align}
namely the \emph{centroid identity}, its sum over $y\in P$, and $1$-Lipschitzness of distance to a set.

\begin{proposition}[One uniform center and zero optimum]\label{prop:R1}
If $y$ is uniform on $P$, then $\E\,\cost(P,\{y\})=2\phi$. In particular, for every finite nonempty $X\subset\R^d$, $\E\,\Phi_1=2\,\OPT_1(X)$ and $R_1=2$, attained whenever $\OPT_1(X)>0$. For any $1\le k\le|X|$, if $\OPT_k(X)=0$ then $|X|=k$ and $\E\,\Phi_k=0$.
\end{proposition}

\begin{proof}
Averaging~\eqref{eq:centroid} over $y\in P$ gives $2\phi$, yielding $\E\,\Phi_1=2\OPT_1(X)$ and $R_1=2$. If $\OPT_k(X)=0$ with $k\le|X|$, a $k$-point center set covers $X$, forcing $|X|=k$, $C_k=X$, and $\Phi_k=0$.
\end{proof}

\paragraph{Normalized coordinates.}
For a target $P\subseteq X$ with $\phi>0$ and finite nonempty $C\subset\R^d$, apply the affine map $y\mapsto\hat y=\sqrt{|P|/\phi}\,(y-\bar y_P)$ to both $P$ and $C$, writing $\widehat C$ for the image of $C$ and $\mu$ for the uniform measure on the image of $P$ (with independent $Y,Y'\sim\mu$). Then $\E_\mu Y=0$, $\E_\mu\|Y\|^2=1$, and $\E_\mu\|Y-Y'\|^2=2$; every total cost $\Sigma$ of $P$ scales to $\Sigma/\phi$, and the \emph{normalized rate} $q(y)=D(y,\widehat C)^2$ satisfies $\E_\mu q=\cost(P,C)/\phi$.

\begin{lemma}[Local factor five~\cite{MRS20}]\label{lem:five}
Let $P$ have $\phi>0$, let $C\subset\R^d$ be finite nonempty with $v=\cost(P,C)>0$, and let $s=v/\phi$. If $y\in P$ is drawn with probability $D(y,C)^2/v$, then $\E\,\cost(P,C\cup\{y\})\le\min(s,5)\,\phi$. If $\phi=0$, then $|P|=1$ and a hit leaves cost $0$.
\end{lemma}

\begin{proof}
Work in normalized coordinates and set $q(y)=D(y,\widehat C)^2$, $r(y)=\sqrt{q(y)}$, and $w(y)=\E_\mu[\min(q(Y'),\|y-Y'\|^2)]$. Retention gives $w(y)\le s$ and~\eqref{eq:centroid} gives $w(y)\le1+\|y\|^2$, so $Z(y)=(w(y)-1)_+$ obeys $Z\le s$ and $\E_\mu Z\le1$. By~\eqref{eq:lipschitz} and~\eqref{eq:pairwise}, $\Var_\mu(r)=\frac12\E_\mu[(r(Y)-r(Y'))^2]\le1$. Since $(a_1+a_2)^2\le2a_1^2+2a_2^2$ and $(\E_\mu r)^2\le s$,
\[
\E_\mu[qZ]\le2(\E_\mu r)^2\E_\mu Z+2\E_\mu\bigl[(r-\E_\mu r)^2Z\bigr]\le2s+2s\Var_\mu(r)\le4s,
\]
so $\E_\mu[qw]/s\le\E_\mu[q(1+Z)]/s\le5$, and retention gives $\E_\mu[qw]/s\le s$.
\end{proof}

\subsection{Queries, occupations, and clocks}\label{sec:queries}

\begin{definition}[Target, names, state]\label{def:state}
A \emph{target} is a nonempty $P\subseteq X$; a \emph{name family} is a collection $\mathcal Q=\{R_1,\dots,R_j\}$ of pairwise disjoint nonempty subsets of $X\setminus P$. A \emph{state} is a pair $\sigma=(C,\mathcal Q_{\mathrm{live}})$, where $C\subset\R^d$ is a finite nonempty set of retained centers and $\mathcal Q_{\mathrm{live}}\subseteq\mathcal Q$ is the set of \emph{live} names. At $\sigma$, write
\[
v=\cost(P,C),\qquad n=|\mathcal Q_{\mathrm{live}}|,\qquad \ell=\frac1n\sum_{R\in\mathcal Q_{\mathrm{live}}}\cost(R,C)\ \ (\ell=0\text{ if }n=0),\qquad T=\cost(X,C),
\]
and $e=T-v-n\ell\ge0$ for the \emph{exterior cost} (including points of retired names).
\end{definition}

From $T>0$, one $D^2$ \emph{draw} chooses $y\in X$ with probability $D(y,C)^2/T$, adds $y$ to $C$, and removes from $\mathcal Q_{\mathrm{live}}$ any live name containing $y$ (\emph{retiring} it); states with $T=0$ are absorbing. A draw is a \emph{target}, \emph{named}, or \emph{exterior} draw accordingly. Along the resulting Markov chain $(\sigma_t)_{t\ge0}$ from $\sigma_0=\sigma$, write $v_t,\ell_t,n_t,T_t$ for the quantities at $\sigma_t$.

\begin{definition}[Charge and occupations]\label{def:occupation}
The \emph{charge} at $\sigma$ is $\mathrm{chg}(\sigma)=v\ell/T$ ($0$ if $T=0$). Define
\begin{align*}
\Occ_r(\sigma)&=\E\sum_{t=0}^{r-1}\mathrm{chg}(\sigma_t)&&\text{(covered occupation, }0\le r\le n),\\
\Del_h(\sigma)&=\E\sum_{t=0}^{h-1}\mathrm{chg}(\sigma_t)\,\mathbf 1\{\text{one of draws }1,\dots,t\text{ hits }P\}&&\text{(delayed occupation, }0\le h\le n+1),\\
\Pre_r(\sigma)&=\E\sum_{t=0}^{r-1}\mathrm{chg}(\sigma_t)\,\mathbf 1\{\text{none of draws }1,\dots,t\text{ hits }P\}&&\text{(prefix).}
\end{align*}
\end{definition}

\begin{definition}[Covered and delayed queries]\label{def:queries}
At $\sigma$ with $n$ live names, the \emph{covered cost} ($0\le r\le n$) and \emph{delayed cost} ($0\le h\le n+1$) are
\[
Q_r(\sigma)=\Occ_r(\sigma)+\E\,\cost(P,C_r),\quad
Z_h(\sigma)=\Del_h(\sigma)+\E\bigl[\cost(P,C_h)\,\mathbf 1\{\text{one of draws }1,\dots,h\text{ hits }P\}\bigr].
\]
A constant $\mathsf C$ is a \emph{stock for rank $j$} if $Z_{j+1}(\sigma)\le\mathsf C\,\OPT_1(P)$ for every $d\ge1$, finite $X\subset\R^d$, target $P\subseteq X$, name family $\mathcal Q$, and state $\sigma=(C,\mathcal Q_{\mathrm{live}})$ with $|\mathcal Q_{\mathrm{live}}|=j$.
\end{definition}

Thus $\Occ_r=\Pre_r+\Del_r$ and $\Del_0=\Del_1=0$. Because $\mathrm{chg}(\sigma_t)=\ell_t\Pr[\text{draw }t+1\text{ hits }P\mid\sigma_0,\dots,\sigma_t]$ whenever $T_t>0$, the occupations are expectations of pathwise charges $\ell_{t-1}\mathbf 1\{\text{draw }t\text{ hits }P\}$; void draws after $T_t=0$ charge nothing. When $P$ is an optimal cluster and $\mathcal Q$ the remaining uncovered clusters, a covered cluster pays $Q_r$ while an uncovered cluster pays $Z_h$ (and its unhit terminal cost is accounted for by \cref{thm:restoration}). Both sides of a stock bound scale homogeneously, so we normalize by $\phi$ when $\phi>0$ (if $\phi=0$, $Z_h=0$).

\paragraph{Clocks.}
Define the \emph{covered clock} $\Cl(r,n)=H_n-H_{n-r}=\sum_{i=n-r+1}^n\frac1i$ ($0\le r\le n$) and \emph{delayed clock} $\Dl(h,n)=\Cl(h-1,n)$ for $1\le h\le n+1$ ($\Dl(0,n)=0$). For $1\le r\le n$, and for $2\le h\le n+1$ with $m=n-h+2\ge1$,
\begin{align}
\Cl(r-1,n-1)&=\Cl(r,n)-\frac1n,& \Cl(r-1,n)&\le\Cl(r,n),& \Dl(r,n)&\le\Cl(r,n),\label{eq:clock-shift}\\
\Dl(h-1,n-1)&=\Dl(h,n)-\frac1n,& \Dl(h-1,n)&=\Dl(h,n)-\frac1m.\label{eq:delayed-shift}
\end{align}

\subsection{Three elementary bounds}\label{sec:elementary}

\begin{lemma}[Persistence]\label{lem:persistent}
For every state $\sigma$ with target cost $v=\cost(P,C)$, $n$ live names, and $0\le r\le n$, $\Occ_r(\sigma)\le v\,\Cl(r,n)$, and hence $Q_r(\sigma)\le\bigl(1+\Cl(r,n)\bigr)\,v$.
\end{lemma}

\begin{proof}
Each draw retires at most one name, so $n_t\ge n-t$. By retention $v_t\le v$ and $T_t\ge n_t\ell_t$, whence $\mathrm{chg}(\sigma_t)\le v/(n-t)$ (also when $n_t\ell_t=0$). Summing over $0\le t<r$ gives $\Occ_r\le v\,\Cl(r,n)$, and retention gives $\cost(P,C_r)\le v$.
\end{proof}

\begin{lemma}[First-hit prize]\label{lem:prefix}
For every state $\sigma$ with live price $\ell$ and every $r\ge0$, $\Pre_r(\sigma)\le\ell$.
\end{lemma}

\begin{proof}
Let $\tau$ be the first draw hitting $P$ ($\tau=\infty$ if none). Since $\mathrm{chg}(\sigma_t)=\ell_t\Pr[\tau=t+1\mid\sigma_0,\dots,\sigma_t]$ on $\{\tau>t\}$, $\Pre_r(\sigma)=\E[\ell_{\tau-1}\mathbf 1\{\tau\le r\}]$. Set $M_t=\ell_{t\wedge(\tau-1)}$ ($M_0=\ell$). Target draws freeze $M_t$ and exterior draws can only lower live costs by retention. If a draw hits live name $R$ of cost $s_R$ (probability $s_R/T$), the new price is $0$ for $n=1$ and at most $(n\ell-s_R)/(n-1)$ for $n\ge2$, giving conditional drift at most
\[
\sum_{R\in\mathcal Q_{\mathrm{live}}}\frac{s_R}{T}\Bigl(\frac{n\ell-s_R}{n-1}-\ell\Bigr)=\frac{n\ell^2-\sum_Rs_R^2}{T(n-1)}\le0
\]
by Cauchy--Schwarz (and $-\ell^2/T\le0$ for $n=1$). Hence $(M_t)$ is a nonnegative supermartingale and $\Pre_r(\sigma)\le\E M_r\le M_0=\ell$.
\end{proof}

In particular, conditional on draw~$1$ hitting a live name, the expected new price is at most $\ell$:
\begin{equation}\label{eq:size-biased}
\E\bigl[\ell_1\bigm|\sigma_0,\ \text{draw 1 hits a live name}\bigr]\le\frac{\sum_Rs_R(n\ell-s_R)}{n\ell(n-1)}\le\ell\qquad(n\ge2),
\end{equation}
and $\ell_1=0$ for $n=1$ (though retiring a cheap name can raise the price pointwise).

\subsection{The repeat charges are at most \texorpdfstring{$2\phi$}{2 phi} per unit of clock}\label{sec:delayed}

\begin{theorem}[Delayed bound]\label{thm:delayed}
For any target $P$ with $\phi=\OPT_1(P)$, name family, state $\sigma$ with $n$ live names, and $0\le h\le n+1$,
\[
\Del_h(\sigma)\le2\phi\,\Dl(h,n).
\]
\end{theorem}

The proof tracks a nonnegative \emph{budget} $\mathcal B$ through the actual draws—twice the clock before the first hit, and afterward the cost of serving $P$ from a hit point $y$ against the clock plus an allowance for outlying hits that later inside hits will repair—and verifies that $\mathcal B$ plus cumulative repeat payments is a supermartingale. The underlying exchange is that a charge $\ell$ paid on a repeat target hit has the same conditional mean as $v/n$ paid on a named draw, while a named draw shortens the clock by exactly $1/n$.

\begin{proof}
The bound is trivial if $h\le1$ or $\phi=0$. Assume $h\ge2$ and $\phi>0$, and work in normalized coordinates (units of $\phi$). Pathwise, $\Del_h$ is the expected sum of pre-draw live prices $\ell$ paid on draws hitting $P$ strictly after $P$'s first hit during the query (\emph{repeat payments}).

\emph{The clock.} With $r_{\mathrm{rem}}\ge1$ draws left and $n$ live names, put $a=r_{\mathrm{rem}}-1$ if $P$ has not yet been hit during the query and $a=r_{\mathrm{rem}}$ if it has, and set $c_{\mathrm{cl}}=\Cl(a,n)\in[0,H_n]$ (initially $c_{\mathrm{cl}}=\Dl(h,n)$). A draw removes the following amount from $c_{\mathrm{cl}}$:
\begin{center}
\small
\begin{tabular}{@{}lccc@{}}
\toprule
pre-draw state & live-name draw & target draw & exterior draw\\
\midrule
unhit, $r_{\mathrm{rem}}\ge2$ & $\frac1n$ & $0$ ($P$ becomes hit) & $\frac1{n-r_{\mathrm{rem}}+2}$\\
unhit, $r_{\mathrm{rem}}=1$ & $0$ & $0$ & $0$\\
already hit, $r_{\mathrm{rem}}\ge1$ & $\frac1n$ & $\frac1{n-r_{\mathrm{rem}}+1}$ & $\frac1{n-r_{\mathrm{rem}}+1}$\\
\bottomrule
\end{tabular}
\end{center}
Indeed, a named draw drops the top index $n$, a non-retiring draw after the opening drops the bottom index, and the opening hit leaves $a$ unchanged. Thus $c_{\mathrm{cl}}$ never increases, and outside the trivial state (unhit, $r_{\mathrm{rem}}=1$, where no repeat can occur and $c_{\mathrm{cl}}=0$) every named draw lowers $c_{\mathrm{cl}}$ by exactly $\frac1n$.

\emph{Excess and deficit.} For normalized $y$ of $P$, let $\delta_{\mathrm{in}}(y)=(1-\|y\|^2)_+$ and $\delta_{\mathrm{out}}(y)=(\|y\|^2-1)_+$, with common mean $B=\E_\mu\delta_{\mathrm{in}}=\E_\mu\delta_{\mathrm{out}}$. Adding $y$ as a center leaves $q(y')\le\|y-y'\|^2$ and target cost at most $1+\|y\|^2=2+\delta_{\mathrm{out}}(y)-\delta_{\mathrm{in}}(y)$. For $\|y\|^2>1$, set $M_{\mathrm{in}}(y)=\E_\mu[\delta_{\mathrm{in}}(Y')\|y-Y'\|^2]>0$ and define the nondecreasing \emph{allowance}
\begin{equation}\label{eq:reserve}
f_y(c_{\mathrm{cl}},\ell)=\min_{0\le z\le1}\Bigl[c_{\mathrm{cl}} z^2+\frac{2\ell}{M_{\mathrm{in}}(y)}(1-z)^2\Bigr]=\frac{2c_{\mathrm{cl}}\ell}{2\ell+c_{\mathrm{cl}} M_{\mathrm{in}}(y)}\qquad(c_{\mathrm{cl}},\ell\ge0;\ f_y(0,0)=0),
\end{equation}
and $f_y=0$ for $\|y\|^2\le1$. Writing $r(y')=\sqrt{q(y')}$ and $d_{yy'}=\|y-y'\|$, if $\delta_{\mathrm{in}}(y')>0$ then $d_{yy'}>0$ and~\eqref{eq:lipschitz} gives $r(y)\le r(y')+d_{yy'}$; taking $z=r(y')/(r(y')+d_{yy'})$ in~\eqref{eq:reserve} gives $q(y)f_y(c_{\mathrm{cl}},\ell)\le c_{\mathrm{cl}}\,q(y')+\frac{2\ell}{M_{\mathrm{in}}(y)}d_{yy'}^2$. Weighting by $\delta_{\mathrm{in}}(y')$ and averaging over $y'\sim\mu$ cancels $M_{\mathrm{in}}(y)$ to give $Bq(y)f_y(c_{\mathrm{cl}},\ell)\le c_{\mathrm{cl}}\,\E_\mu[q\delta_{\mathrm{in}}]+2\ell$; weighting by $\delta_{\mathrm{out}}(y)/B$ and averaging over $y\sim\mu$ yields (trivially if $B=0$)
\begin{equation}\label{eq:pair}
\E_\mu\bigl[q(Y)\delta_{\mathrm{out}}(Y)f_Y(c_{\mathrm{cl}},\ell)\bigr]-c_{\mathrm{cl}}\,\E_\mu[q(Y)\delta_{\mathrm{in}}(Y)]\le2\ell.
\end{equation}

\emph{The budget and one-step drift.} Define $\mathcal B=0$ whenever $n=0$, $\ell=0$, or no draw remains. Otherwise, before $P$ is hit during the query put $\mathcal B=2c_{\mathrm{cl}}$; afterwards put $\mathcal B=\min(\{2c_{\mathrm{cl}}+\ell\}\cup\{\mathcal A_y(c_{\mathrm{cl}},\ell):y\in P\text{ drawn during the query}\})$, where
\[
\mathcal A_y(c_{\mathrm{cl}},\ell)=(2-\delta_{\mathrm{in}}(y))\,c_{\mathrm{cl}}\quad(\|y\|^2\le1),\qquad \mathcal A_y(c_{\mathrm{cl}},\ell)=2c_{\mathrm{cl}}+\delta_{\mathrm{out}}(y)\,f_y(c_{\mathrm{cl}},\ell)\quad(\|y\|^2>1),
\]
each nondecreasing in $(c_{\mathrm{cl}},\ell)$. It suffices to verify at every pre-draw state that
\begin{equation}\label{eq:drift}
\E\bigl[\text{repeat payment of the next draw}+\mathcal B'\bigm|\text{history}\bigr]\le\mathcal B.
\end{equation}
This is immediate when $n=0$, $\ell=0$, $T=0$, or $P$ is unhit with $r_{\mathrm{rem}}=1$. Otherwise multiply~\eqref{eq:drift} by $T=v+n\ell+e$: a named draw (rate $n\ell$) lowers $c_{\mathrm{cl}}$ by $\frac1n$ and leaves $\E[\ell']\le\ell$ by~\eqref{eq:size-biased} ($0$ if $n=1$), while all other draws obey $c'_{\mathrm{cl}}\le c_{\mathrm{cl}}$, $\ell'\le\ell$. Since a target hit at $y$ makes $y$ a center, $\mathcal B'\le\mathcal A_y(c_{\mathrm{cl}},\ell)=2c_{\mathrm{cl}}+\delta_{\mathrm{out}}(y)f_y(c_{\mathrm{cl}},\ell)-c_{\mathrm{cl}}\delta_{\mathrm{in}}(y)$, so~\eqref{eq:pair} gives the \emph{target-branch bound}
\begin{equation}\label{eq:opening}
\E_\mu[q(Y)\mathcal B'_Y]\le2c_{\mathrm{cl}} v+2\ell.
\end{equation}
We now check all four active regimes of $\mathcal B$:
\begin{itemize}[leftmargin=1.5em]
\item \emph{Unhit ($r_{\mathrm{rem}}\ge2$, $\mathcal B=2c_{\mathrm{cl}}$, no repeat payment).} The target, named, and exterior branches contribute at most $2c_{\mathrm{cl}} v+2\ell$ (by~\eqref{eq:opening}), $n\ell(2c_{\mathrm{cl}}-\frac2n)=2c_{\mathrm{cl}} n\ell-2\ell$, and $2c_{\mathrm{cl}} e$, summing to $2c_{\mathrm{cl}} T$.
\item \emph{Hit, $\mathcal B=2c_{\mathrm{cl}}+\ell$.} The $2c_{\mathrm{cl}}$ part sums to at most $2c_{\mathrm{cl}} T$ as above, while the extra $\ell$ pays the repeat payment $\ell$ on $P$ (rate $v$) and $\E[\ell']\le\ell$ on named and exterior draws, totaling $(2c_{\mathrm{cl}}+\ell)T$.
\item \emph{Hit, $\mathcal B=\mathcal A_y$ with $\|y\|^2\le1$.} Here $v\le2-\delta_{\mathrm{in}}(y)$. Target draws pay $v\ell$ with $\mathcal B'\le\mathcal B$, named draws release $(2-\delta_{\mathrm{in}}(y))\ell\ge v\ell$, and exterior draws satisfy $\mathcal B'\le\mathcal B$.
\item \emph{Hit, $\mathcal B=\mathcal A_y$ with $\|y\|^2>1$.} Put $\eta_{\mathrm{out}}=\delta_{\mathrm{out}}(y)$, $M=M_{\mathrm{in}}(y)$, and $z=\frac{2\ell}{2\ell+c_{\mathrm{cl}} M}\in(0,1)$ attaining~\eqref{eq:reserve}, so $f_y(c_{\mathrm{cl}},\ell)=c_{\mathrm{cl}} z$ and~\eqref{eq:reserve} gives the linear upper bound $\mathcal B'\le(2+\eta_{\mathrm{out}}z^2)c'_{\mathrm{cl}}+\frac{2\eta_{\mathrm{out}}(1-z)^2}{M}\ell'$ at every child (equal at $(c_{\mathrm{cl}},\ell)$). Named draws therefore release at least $\ell(2+\eta_{\mathrm{out}}z^2)$. Set $\psi=\eta_{\mathrm{out}}f_y(c_{\mathrm{cl}},\ell)/\ell\le1$ (since $\mathcal B\le2c_{\mathrm{cl}}+\ell$), so $M\psi=2\eta_{\mathrm{out}}(1-z)$. Target draws at $y'$ with $\delta_{\mathrm{in}}(y')>0$ bring $\mathcal B'\le2c_{\mathrm{cl}}$, releasing at least $\eta_{\mathrm{out}}f_y=\ell\psi$ over total rate at least $g=\E_\mu[\delta_{\mathrm{in}}(Y')q(Y')]$. Repeat payments total $v\ell$, so~\eqref{eq:drift} reduces to $v-g\psi\le2+\eta_{\mathrm{out}}z^2$. Since $q(y')\le\|y-y'\|^2$ implies $0\le M-g=\E_\mu[\delta_{\mathrm{in}}(Y')(\|y-Y'\|^2-q(Y'))]\le\E_\mu[\|y-Y'\|^2-q(Y')]=2+\eta_{\mathrm{out}}-v$ and $\psi\le1$,
\[
v-g\psi=v-M\psi+(M-g)\psi\le2+\eta_{\mathrm{out}}-M\psi=2+\eta_{\mathrm{out}}(2z-1)\le2+\eta_{\mathrm{out}}z^2.\qedhere
\]
\end{itemize}
\end{proof}

\begin{remark}
Replacing $2$ by $\lambda$ in the outside case forces $z^2+\lambda(1-z)\ge1$ on $(0,1)$, i.e.\ $\lambda\ge2$; the reservoir input (\cref{sec:reservoir}) with one light retained, $n$ live lights, and $2\le h\le n+1$ attains $\Del_h\to2\Dl(h,n)$ as $q\to\infty$ then $N\to\infty$, so the multiplier $2$ is sharp even on shortened clocks.
\end{remark}

\begin{corollary}[Saturated cap]\label{cor:saturated}
For every state $\sigma$ with $n$ live names, live price $\ell$, and $1\le r\le n$, with $m=n-r+1$,
\[
\Occ_r(\sigma)\le\ell+2\phi\,\Dl(r,n)=\ell+2\phi\Bigl(\Cl(r,n)-\frac1m\Bigr).
\]
\end{corollary}

\begin{proof}
Write $\Occ_r=\Pre_r+\Del_r$ and apply \cref{lem:prefix,thm:delayed} with $\Dl(r,n)=\Cl(r,n)-1/m$.
\end{proof}

\section{From the whole input to one cluster}\label{sec:restoration}

This section reduces \cref{thm:exact-rk} to stocks (\cref{def:queries}), losing nothing beyond the term $\theta(2H_{k-1}+2)$ of the first seed's cluster (attained by the reservoir). Three facts drive the reduction: the average cost of unvisited clusters is a supermartingale; an unvisited cluster can be paid for by charges to visited clusters on non-fresh draws; and live-name identities can be thinned uniformly without changing expectations.

Throughout, $P_1,\dots,P_m$ are pairwise disjoint nonempty \emph{groups} in a finite input $X\subset\R^d$, with complement $P_0=X\setminus\bigcup_{i=1}^mP_i$. Starting from any finite nonempty $C_0\subset\R^d$, take $r$ actual $D^2$ draws. A draw is \emph{fresh} if it hits some $P_i$ ($1\le i\le m$) for the first time during these $r$ draws. Let $n_t$ be the number of unvisited groups after $t$ draws, $\ell_t$ their mean current cost ($\ell_t=0$ if $n_t=0$), and $w_t=\cost(P_0,C_t)$.

\begin{theorem}[Restoration]\label{thm:restoration}
Let $1\le r\le m$.
\begin{enumerate}[label=(\alph*),leftmargin=2em]
\item \emph{(Surviving mean.)} $(\ell_t)_{t\ge0}$ is a nonnegative supermartingale.
\item \emph{(Complement-fresh duality.)} Charging $P_0$ the price $\ell_{t-1}$ on each draw $t\le r$ hitting $P_0$, plus $w_r$ at the end, gives
\[
\E\Bigl[w_r+\sum_{t\le r:\ y_t\in P_0}\ell_{t-1}\Bigr]=\E\Bigl[w_r+\sum_{t\le r\text{ fresh}}\frac{w_{t-1}}{n_{t-1}}\Bigr]\le\bigl(1+H_m-H_{m-r}\bigr)\,w_0,
\]
and the same holds if $P_0$ is only part of the complement (with the rest treated as exterior).
\item \emph{(Restoration.)} If for every $1\le i\le m$ the delayed cost of $P_i$ over $r$ draws, with $r-1$ live names among the other groups and all other points exterior, is at most $U_i$ from every state, then
\[
\E\,\cost(X,C_r)\le\sum_{i=1}^mU_i+(m-r)\,\E\,\ell_r+\bigl(1+H_m-H_{m-r}\bigr)\,w_0.
\]
\end{enumerate}
\end{theorem}

\begin{proof}
(a) A non-fresh draw preserves the unvisited groups and lowers their costs by retention. At a fresh draw with $n\ge2$ unvisited groups of costs $s_1,\dots,s_n$ and mean $\ell>0$, the retired cost has mean $\sum_is_i^2/(n\ell)\ge\ell$, so the surviving mean has expectation at most $\ell$ (and drops to $0$ at $n=1$).

(b) At total cost $T>0$, draw $t$ hits $P_0$ with probability $w_{t-1}/T$ and is fresh with probability $n_{t-1}\ell_{t-1}/T$, so paying $\ell_{t-1}$ on the first event and $w_{t-1}/n_{t-1}$ on the second have the same conditional mean $w_{t-1}\ell_{t-1}/T$. Fresh draws occur at distinct counts $n_{t-1}\in\{m,\dots,m-r+1\}$, and $w_t\le w_0$ by retention.

(c) Write $\cost(X,C_r)=\sum_{i=1}^m\cost(P_i,C_r)+w_r$. Letting $N_{\mathrm{nf}}$ count non-fresh draws, $n_r=m-r+N_{\mathrm{nf}}$ and the unvisited groups cost $n_r\ell_r=(m-r)\ell_r+\sum_{t\text{ not fresh}}\ell_r$. By~(a) and retention on non-fresh draw~$t$, $\E[\ell_r\mid\text{history through }t]\le\ell_t\le\ell_{t-1}$. Each non-fresh draw is either a repeat of a visited $P_i$—paid by $P_i$'s delayed cost (whose $m-1$ live names thin uniformly to $r-1$ names by \cref{thm:thinning}, yielding at most $U_i$, including its visited terminal cost)—or a hit in $P_0$, paid together with $w_r$ by~(b).
\end{proof}

\begin{theorem}[Uniform thinning]\label{thm:thinning}
Consider a delayed query of horizon $r\ge1$ for an unhit target $P$ with $n\ge r-1$ live names. Choose a uniformly random subset $\mathcal S$ of $r-1$ live names, independent of the sampling, and reclassify the remaining $n-r+1$ names as exterior. Then the delayed cost equals the average over $\mathcal S$ of the delayed costs of the thinned queries.
\end{theorem}

\begin{proof}
Relabeling leaves retained centers, draw probabilities, hits, and terminal costs unchanged; it alters only the live price paid on a repeat target hit. For $r=1$ there are no repeats. For $r\ge2$, a repeat at draw $t\le r$ follows an earlier target hit, so at most $t-2\le r-2$ names have retired and $\mathcal S$ contains a survivor. Given the trajectory and the survivor count in $\mathcal S$, those survivors form a uniform subset of all surviving names, so their mean price has conditional expectation equal to the true live price.
\end{proof}

In particular, after a first exterior draw at rank $j$, $j$ draws and $j$ live names remain, so thinning bounds the remainder by any rank-$(j-1)$ stock; after a named draw, $j$ draws and $j-1$ live names remain and the same stock applies directly.

\begin{theorem}[Uniform first-seed bridge]\label{thm:bridge}
Let $k\ge2$ and let $\mathsf C_{k-2}$ be a stock for rank $k-2$. For every input $X$ with $\OPT_k(X)>0$,
\[
\frac{\E\,\Phi_k}{\OPT_k(X)}\le(1-\theta)\,\mathsf C_{k-2}+\theta\,\bigl(2+2H_{k-1}\bigr),
\]
where, for an optimal partition $P_1,\dots,P_k$ with $\phi_i=\OPT_1(P_i)$,
\begin{equation}\label{eq:theta}
\theta=\sum_{i=1}^k\frac{|P_i|}{|X|}\cdot\frac{\phi_i}{\OPT_k(X)}\in\Bigl[\frac2{|X|},1\Bigr]
\end{equation}
is the expected share of $\OPT_k(X)$ carried by the cluster containing the first center.
\end{theorem}

\begin{proof}
Condition on the part $P_i$ containing the uniform first center $y_1$: then $\E[\cost(P_i,\{y_1\})\mid y_1\in P_i]=2\phi_i$ by \cref{prop:R1}. Applying \cref{thm:restoration}(c) with $P_0=P_i$, the other $k-1$ parts as groups, $m=r=k-1$, and $U_\alpha=\mathsf C_{k-2}\phi_\alpha$ gives $\E[\Phi_k\mid y_1\in P_i]\le\mathsf C_{k-2}(\OPT_k-\phi_i)+2(1+H_{k-1})\phi_i$. Averaging with weights $|P_i|/|X|$ proves the bound, and $\theta\ge2/|X|$ because every part with $\phi_i>0$ has at least two points.
\end{proof}

\begin{corollary}[Strict bridge]\label{cor:strict}
Under the hypotheses of \cref{thm:bridge}, $\E\,\Phi_k/\OPT_k(X)<(1-\theta)\mathsf C_{k-2}+\theta(2+2H_{k-1})$.
\end{corollary}

\begin{proof}
Condition on $y_1\in P_i$ with $m=r=k-1\ge1$, $w_0=\cost(P_i,\{y_1\})$, and $T_0=\cost(X,\{y_1\})$. In \cref{thm:restoration}(b), if the first of the $m$ draws hits $P_0=P_i$ (probability $w_0/T_0$ when $T_0>0$), at most $m-1$ fresh draws remain at counts $\ge2$, lowering $\sum_{\text{fresh}}1/n_{t-1}$ from $H_m$ to at most $H_m-1$ and improving~(b), (c), and the bridge by at least $w_0^2/T_0$. Some $P_i$ has $\phi_i>0$, hence $|P_i|\ge2$ and $w_0>0$ for every $y_1\in P_i$, so the expected slack is positive.
\end{proof}

\begin{corollary}[Warm-up]\label{cor:warmup}
For every rank $j\ge0$, $5+2H_j$ is a stock. Hence, for $k\ge2$, $\E\,\Phi_k/\OPT_k(X)\le(1-\theta)(5+2H_{k-2})+\theta(2+2H_{k-1})\le2H_{k-2}+5$.
\end{corollary}

\begin{proof}
With $j$ live names and $j+1$ draws, the delayed occupation is at most $2\phi\,\Dl(j+1,j)=2\phi H_j$ by \cref{thm:delayed}, and on the event of a hit, \cref{lem:five} and retention bound the expected terminal cost by $5\phi$; \cref{thm:bridge} gives the ratio bound.
\end{proof}

The warm-up already yields $R_2\le5$ (matching \cref{thm:hub}). The exact stocks advance one rank at a time via the following mixing reduction.

\begin{lemma}[One-step mixing]\label{lem:mixing}
Let $\mathcal G$ be a target class defined by a property of $P$ alone (e.g.\ all targets, or the low-mean class of \cref{sec:hub-regime}). Let $j\ge1$, suppose $\mathsf C_{j-1}$ bounds $Z_j$ on $\mathcal G$ with $j-1$ live names, and let $\mathsf C_j\ge\mathsf C_{j-1}$ with slope $\mathsf m_j=j(\mathsf C_j-\mathsf C_{j-1})$. If every $P\in\mathcal G$ with $\phi>0$ and every state $\sigma=(C_0,\mathcal Q)$ with $|\mathcal Q|=j$, target cost $s\phi>0$, and total nontarget cost $T-s\phi=jL_{\mathrm{tot}}\phi$ satisfies
\begin{equation}\label{eq:marked}
\E_{y\sim\nu}\,Q_j(\sigma^y)\le\Bigl(\mathsf C_j+\mathsf m_j\frac{L_{\mathrm{tot}}}s\Bigr)\phi,
\end{equation}
where $\nu(y)=D(y,C_0)^2/(s\phi)$ is the conditional $D^2$ law on $P$ and $\sigma^y=(C_0\cup\{y\},\mathcal Q)$, then $\mathsf C_j$ bounds $Z_{j+1}$ on $\mathcal G$ with $j$ live names (and is a rank-$j$ stock if $\mathcal G$ is all targets).
\end{lemma}

\begin{proof}
If $s=0$ or $\phi=0$ then $Z_{j+1}=0$. Otherwise draw~$1$ hits $P$ with probability $s/(s+jL_{\mathrm{tot}})$ (leaving $Q_j(\sigma^y)$) and misses $P$ with probability $jL_{\mathrm{tot}}/(s+jL_{\mathrm{tot}})$ (leaving a horizon-$j$ delayed query with $j-1$ or $j$ live names, bounded by $\mathsf C_{j-1}\phi$ via \cref{thm:thinning}). Averaging yields $\frac{s(\mathsf C_j+\mathsf m_jL_{\mathrm{tot}}/s)+jL_{\mathrm{tot}}\mathsf C_{j-1}}{s+jL_{\mathrm{tot}}}\phi=\mathsf C_j\phi$.
\end{proof}

\section{The race and the Euclidean reset}\label{sec:race-reset}

After its first hit, a target of cost $v>0$ faces $j$ live names of mean price at most $L$: under equal frozen prices and no exterior points, all $j$ names would be collected before a return with probability $\pi_j(L/v)$. This section turns that picture into three inequalities for the actual $D^2$ process: \cref{thm:transfer} pays everything after the first target return via persistence and a uniform contraction factor $\gamma$; \cref{thm:stopped} (via \cref{lem:retirement}) bounds what happens until the first return; and \cref{thm:reset,cor:reset-monotone}—the only place after \cref{sec:framework} where general single-cluster geometry enters—sharpen the local factor five to a monotone concave bound $U(s)$ discharging \cref{thm:transfer} at every downstream use.

Throughout \cref{sec:race-reset-transfer,sec:race-reset-stopped}, fix $d\ge1$, $X\subset\R^d$, a target $P\subseteq X$, a name family $\mathcal Q$ with $|\mathcal Q|=j\ge1$, and a finite nonempty $C_0\subset\R^d$ with $v=\cost(P,C_0)>0$. A \emph{covered query} runs $j$ actual $D^2$ draws from $(C_0,\mathcal Q)$, producing centers $C_0\subseteq\dots\subseteq C_j$ ($C_{\mathrm{end}}:=C_j$) and live prices $\ell_t$. Let $\tau\in\{1,\dots,j\}\cup\{\infty\}$ be the first return to $P$, and define the \emph{stopped payment} $J_{\mathrm{st}}=\ell_{\tau-1}\mathbf 1\{\tau\le j\}+\cost(P,C_{\mathrm{end}})\,\mathbf 1\{\tau=\infty\}$. All inequalities below are cost-homogeneous, so when $\phi=\OPT_1(P)>0$ one may divide all costs by $\phi$.

\subsection{Paying for the continuation after a return}\label{sec:race-reset-transfer}

\begin{theorem}[Transfer]\label{thm:transfer}
Consider the covered query of $j\ge1$ actual $D^2$ draws from $(C_0,\mathcal Q)$ with $|\mathcal Q|=j$, initial target cost $v=\cost(P,C_0)>0$, initial live price $\ell_0\le L$, and ratio $x=L/v$. Let $\gamma\in[0,1]$ be a constant such that, for \emph{every} finite superset $C\supseteq C_0$ with $\cost(P,C)>0$,
\begin{equation}\label{eq:transfer-hyp}
\sum_{y\in P}\frac{D(y,C)^2}{\cost(P,C)}\,\cost(P,C\cup\{y\})\le\gamma\,v.
\end{equation}
Then
\begin{equation}\label{eq:transfer}
\frac{Q_j}{v}\le(1-\gamma)\,\frac{\E J_{\mathrm{st}}}{v}+\gamma\bigl(H_j+\pi_j(x)\bigr),\qquad\pi_j(x)=\prod_{i=1}^j\frac{ix}{1+ix}.
\end{equation}
\end{theorem}

\paragraph{The universal retained-superset hypothesis and how it is discharged.}
Quantifying~\eqref{eq:transfer-hyp} over every finite superset $C\supseteq C_0$ with $\cost(P,C)>0$ decouples the probabilistic race from trajectory geometry, since every reached center set contains $C_0$ by retention. Two discharges suffice everywhere: first, retention gives $\cost(P,C\cup\{y\})\le v$, so $\gamma=1$ always satisfies~\eqref{eq:transfer-hyp} and yields
\begin{equation}\label{eq:transfer-retention}
Q_j\le v\bigl(H_j+\pi_j(L/v)\bigr);
\end{equation}
second, when $\phi>0$, setting $\bar u=v/\phi>0$ and $\gamma=\gamma_{\mathrm{reset}}(1/\bar u)=U(\bar u)/\bar u\in[0,1]$ discharges~\eqref{eq:transfer-hyp} unconditionally via \cref{cor:reset-monotone}.

\begin{proof}[Proof of \cref{thm:transfer}]
Scale costs so that $v=1$, and set $v_{\mathrm{end}}=\cost(P,C_{\mathrm{end}})/v\le1$. Pathwise, $Q_j-(1-\gamma)J_{\mathrm{st}}$ equals $\gamma v_{\mathrm{end}}\le\gamma$ on $\{\tau=\infty\}$, and equals $\gamma\ell_{\tau-1}+Q_{\mathrm{post}}$ on $\{\tau\le j\}$, where $Q_{\mathrm{post}}$ is the covered cost of the $r'=j-\tau$ draws remaining after draw~$\tau$. Just before draw~$\tau$, the retained set $C\supseteq C_0$ has $\cost(P,C)>0$ and $n_\tau\ge r'+1$ live names (as draw~$\tau$ retires none). Given the history through draw~$\tau$ landing at $y\in P$ with post-hit cost $v^+=\cost(P,C\cup\{y\})/v$, \cref{lem:persistent} gives $\E[Q_{\mathrm{post}}\mid\text{history}]\le(1+\Cl(r',n_\tau))v^+=(1+H_{n_\tau}-H_{n_\tau-r'})v^+\le H_{n_\tau}v^+$ (since $n_\tau-r'\ge1$ implies $H_{n_\tau-r'}\ge1$, also at $r'=0$). Averaging over $y\in P$ under its $D^2$ law at $C\supseteq C_0$ applies~\eqref{eq:transfer-hyp} to bound the conditional mean by $\gamma H_{n_\tau}$, giving
\[
\E\bigl[Q_j-(1-\gamma)J_{\mathrm{st}}\bigr]\le\gamma\,\E\bigl[\mathbf 1\{\tau=\infty\}+\mathbf 1\{\tau\le j\}(\ell_{\tau-1}+H_{n_\tau})\bigr].
\]

We show by induction on $r\le n$ that from any state $(C,\mathcal Q_{\mathrm{live}})$ with $C\supseteq C_0$, $n=|\mathcal Q_{\mathrm{live}}|$, normalized live price $\ell$, and relative target cost $v_C=\cost(P,C)/v\in[0,1]$, the expectation on the right is at most $F_n(\ell)=H_n+\pi_n(\ell)$ ($F_0\equiv1$). This holds at $r=0$ or $T=0$ ($1\le F_n(\ell)$). For $r\ge1$ and $T=v_C+n\ell+e>0$, exterior draws lower $\ell$ and keep $n$, contributing at most $eF_n(\ell)$ since $F_n$ is nondecreasing. For a named draw: if $n=1$ then $r=1$ and the child pays $1=F_0(\ell)$; if $n\ge2$, writing live-name price shares $\xi_1,\dots,\xi_n\ge0$ ($\sum_i\xi_i=1$), the surviving mean $\ell'\le\frac{n\ell}{n-1}(1-\xi_i)$ obeys
\[
\E\bigl[(\ell')^{n-1}\mid\text{named draw}\bigr]
\le\Bigl(\frac{n\ell}{n-1}\Bigr)^{n-1}\sum_{i=1}^n\xi_i(1-\xi_i)^{n-1}
\le\Bigl(\frac{n\ell}{n-1}\Bigr)^{n-1}\cdot n\max_{t\in[0,1]}t(1-t)^{n-1}
=\ell^{n-1},
\]
as $\max_{t\in[0,1]}t(1-t)^{n-1}=\frac1n(1-\frac1n)^{n-1}$. Moreover $F_{n-1}$ is nondecreasing and concave in $z^{n-1}$ (since $\frac{\pi'_{n-1}(z)}{(n-1)z^{n-2}}=\frac1{n-1}\prod_{i<n}\frac i{1+iz}\sum_{i<n}\frac1{1+iz}$ decreases in $z>0$), so Jensen's inequality bounds the named branch by $n\ell F_{n-1}(\ell)$. Cancelling $eF_n(\ell)$ leaves the affine-in-$v_C\in[0,1]$ inequality $v_C(\ell+H_n)+n\ell F_{n-1}(\ell)\le(v_C+n\ell)F_n(\ell)$: at $v_C=1$ it is the exact identity $(1+n\ell)(H_n+\pi_n(\ell))=\ell+H_n+n\ell(H_{n-1}+\pi_{n-1}(\ell))$; at $v_C=0$ it is $F_{n-1}(\ell)\le F_n(\ell)$, immediate for $n=1$ and for $n\ge2$ given by $(1+n\ell)(F_n(\ell)-F_{n-1}(\ell))=\ell+\frac1n-\pi_{n-1}(\ell)\ge\ell-\frac\ell{1+\ell}\ge0$. Finally $\ell_0/v\le x$ and $F_j$ is nondecreasing.
\end{proof}

\paragraph{Direct one-name bound.}
For $j=1$, a one-step bound bypasses \cref{thm:transfer,thm:stopped} and needs~\eqref{eq:transfer-hyp} only at $C=C_0$. With $a_{\mathrm{live}}=\ell_0/v$, $e_{\mathrm{ext}}=(\cost(X,C_0)-v-\ell_0)/v$, $a_{\mathrm{live}}+e_{\mathrm{ext}}\le x$, and $\gamma\le1$ satisfying~\eqref{eq:transfer-hyp} at $C_0$, the draw hits $P$ with probability $1/(1+a_{\mathrm{live}}+e_{\mathrm{ext}})$ (paying charge $a_{\mathrm{live}}$ and mean terminal cost $\le\gamma$) and otherwise leaves cost $\le1$. Since $t\mapsto\frac{2t+\gamma}{1+t}$ is nondecreasing for $\gamma\le1$,
\begin{equation}\label{eq:one-name}
\frac{Q_1}{v}\le\frac{2a_{\mathrm{live}}+e_{\mathrm{ext}}+\gamma}{1+a_{\mathrm{live}}+e_{\mathrm{ext}}}\le\frac{2x+\gamma}{1+x}.
\end{equation}

\subsection{Until the first return}\label{sec:race-reset-stopped}

A named draw deletes one live price with probability proportional to its value; the following lemma controls the moments of the surviving mean.

\begin{lemma}[Size-biased retirement]\label{lem:retirement}
Let $n\ge2$ nonnegative prices $\ell_1,\dots,\ell_n$ have mean $L>0$. Delete one price $\ell_i$ with probability $\ell_i/(nL)$, and let $\ell'=\frac1{n-1}\sum_{k\ne i}\ell_k$ be the mean of the others. Then $\E[(\ell')^\kappa]\le L^\kappa$ for every $0<\kappa\le n+1$, and for $n=3$ also for every $0<\kappa\le\frac{50}{11}$.
\end{lemma}

\begin{proof}
With $\xi_i=\ell_i/(nL)$, the claim is $\sum_{i=1}^n f(\xi_i)\le(1-\frac1n)^\kappa$ for $f(t)=t(1-t)^\kappa$. By the power-mean inequality it suffices to prove the top exponents $\kappa=n+1$ ($n\ge2$) and $\kappa=\frac{50}{11}$ ($n=3$), so $\kappa\ge n$; set $c_\kappa=\kappa+1-n>0$ and $m_\kappa=\kappa+1$. Since $f'(0)=1>f'(t)$ for $t>0$, a maximizer has all $\xi_i>0$ and all $f'(\xi_i)$ equal. Because $f'$ decreases to its unique inflection point $t=2/(\kappa+1)$ and then increases, at most two coordinate values occur, and two coordinates in the strictly convex region would permit a strict increase along their fixed-sum line. Hence either all $\xi_i=\frac1n$, or one share is $\xi_+=\frac{1+(n-1)t}n$ and the other $n-1$ equal $\xi_-=\frac{1-t}n$ ($0<t<1$). The $t$-derivative has the sign of
\[
f'(\xi_+)-f'(\xi_-)=\tfrac1n\bigl(\tfrac{n-1}n\bigr)^{\kappa-1}\Bigl\{(c_\kappa-m_\kappa t)\bigl(1+\tfrac t{n-1}\bigr)^{\kappa-1}-\bigl[c_\kappa+m_\kappa(n-1)t\bigr](1-t)^{\kappa-1}\Bigr\},
\]
negative for $t\ge c_\kappa/m_\kappa$. On $(0,c_\kappa/m_\kappa)$, set $\psi_{\mathrm{ret}}(t)=\ln\frac{[c_\kappa+m_\kappa(n-1)t](1-t)^{\kappa-1}}{(c_\kappa-m_\kappa t)(1+t/(n-1))^{\kappa-1}}$ ($\psi_{\mathrm{ret}}(0)=0$). Differentiating yields
\begin{multline*}
\bigl(c_\kappa+m_\kappa(n-1)t\bigr)(c_\kappa-m_\kappa t)(1-t)(n-1+t)\,\psi'_{\mathrm{ret}}(t)\\
=n\kappa\Bigl[c_\kappa(n-c_\kappa)-c_\kappa(n-2)(n+c_\kappa)t+(n+c_\kappa)(n^2+nc_\kappa-2n-c_\kappa)t^2\Bigr].
\end{multline*}
The quadratic has positive leading coefficient and discriminant $n^2c_\kappa(n+c_\kappa)[nc_\kappa-4n+c_\kappa^2-4c_\kappa+8]$; the bracket equals $4-2n\le0$ for $\kappa=n+1$ ($c_\kappa=2$) and $-\frac8{121}<0$ for $n=3$, $\kappa=\frac{50}{11}$ ($c_\kappa=\frac{28}{11}$). Thus $\psi'_{\mathrm{ret}}\ge0$, $\psi_{\mathrm{ret}}\ge0$, and $f'(\xi_+)\le f'(\xi_-)$.
\end{proof}

Retention only lowers $\ell'$, so \cref{lem:retirement} holds at every named draw. The extra $n=3$ exponent $\frac{50}{11}$ supplies both $\kappa=\frac92<\frac{50}{11}$ in \cref{thm:stopped} and the three-name step of \cref{sec:hub-regime}.

\begin{theorem}[Stopped-price comparison]\label{thm:stopped}
Consider the covered query of $j\ge2$ actual $D^2$ draws from $(C_0,\mathcal Q)$ with $|\mathcal Q|=j$, initial target cost $v=\cost(P,C_0)>0$, initial live price $\ell_0\le xv$, and total nontarget cost $\cost(X,C_0)-v\le jxv$. Let $\alpha=\min(\frac79x,1)$ and $\vartheta_j=(1-\frac1j)^{j-1}$. Then
\begin{equation}\label{eq:stopped}
\frac{\E J_{\mathrm{st}}}{v}\le W_j(x):=x+(\alpha-x)\,\pi_j(x)+(1-\alpha)\,\vartheta_jx.
\end{equation}
\end{theorem}

The bound splits into the equal-price fractional tag $x+(\alpha-x)\pi_j(x)$ paying fraction $\alpha$ of the terminal cost on expiry, plus $(1-\alpha)\vartheta_jx$ restoring the remaining fraction $1-\alpha$ via an avoidance-likelihood bound on $\{\tau=\infty\}$. Note $x\ge0$ (at $x=0$, draw~$1$ hits $P$ at price $0$, giving $\E J_{\mathrm{st}}=0=W_j(0)$).

\begin{proof}
Assume $x>0$.

\emph{The fractional tag.} Fix $0<p\le\frac97$, $\alpha=\frac79p\in(0,1]$, and $\kappa=\frac92$. Consider the fractional tag paying $\ell_{\tau-1}$ on $\{\tau\le r\}$ and $\alpha\,\cost(P,C_r)$ on $\{\tau=\infty\}$, from a state with $i$ live names and at most $i-\delta$ draws left, holding the nominal deficit $\delta\ge0$ fixed. Set $\pi_{i,\delta}(z)=\prod_{k=\delta+1}^i\frac{kz}{1+kz}$ ($\pi_{\delta,\delta}\equiv1$), $K_{i,\delta}(z)=z+(\alpha-z)\pi_{i,\delta}(z)$ ($K_{\delta,\delta}\equiv\alpha$), $S_{i,\delta}(z)=\sum_{k=\delta+1}^i(1+kz)^{-1}$, $F_{\delta,\delta}\equiv\alpha$, $F_{1,0}(z)=\alpha+\frac{(z-\alpha)_+}{1+z}$, and for $i\ge2$ with $i>\delta$,
\[
F_{i,\delta}(z)=\begin{cases}K_{i,\delta}(z),&z\ge p,\\ K_{i,\delta}(p)-pK'_{i,\delta}(p)\,\dfrac{1-(z/p)^\kappa}{\kappa},&0\le z\le p.\end{cases}
\]
We prove by induction on $r\le i-\delta$ that this tag is at most $v_C F_{i,\delta}(\ell/v_C)$ for current target cost $v_C>0$ and live price $\ell$ ($0$ when $v_C=0$).

\emph{Shape.} Since $\kappa(p-\alpha)=p$ and $pK'_{i,\delta}(p)=p(1-\pi_{i,\delta}(p))-(p-\alpha)\pi_{i,\delta}(p)S_{i,\delta}(p)$, $\kappa[F_{i,\delta}(0)-\alpha]=(p-\alpha)\pi_{i,\delta}(p)S_{i,\delta}(p)\ge0$. Writing $\pi_{i,\delta}(z)=\E[\exp(-Y_{\mathrm{exp}}/z)]$ with $Y_{\mathrm{exp}}$ a sum of independent exponentials of rates $\delta+1,\dots,i$, on $z\ge\alpha$ we have $K_{i,\delta}-zK'_{i,\delta}=\pi_{i,\delta}[\alpha+(z-\alpha)S_{i,\delta}]>0$, $K'_{i,\delta}>0$, and $K''_{i,\delta}<0$. For $i\ge2$, tangency at $p$ and concavity above $p$ make $F_{i,\delta}$ nondecreasing and concave in $z^\kappa$ ($F_{1,0}$ has an upward kink at $\alpha$ and is handled below), so $v_C F_{i,\delta}(\ell/v_C)$ is nondecreasing in $v_C,\ell$. Also $\Delta_i=K_{i,\delta}-K_{i-1,\delta}=\frac{z-\alpha}{iz}\pi_{i,\delta}$ obeys $p\Delta_i'(p)/\Delta_i(p)=\frac{\alpha}{p-\alpha}+S_{i,\delta}(p)>0$.

\emph{Steps with $i\ge3$, or $i=2$ and $\delta=1$.} At $r=0$ expiry pays $\alpha v_C\le v_C F_{i,\delta}$; exterior draws lower $v_C,\ell$ and keep $i,\delta$. For a named draw with $i\ge3$, \cref{lem:retirement} gives $\E[(\ell')^\kappa]\le\ell^\kappa$ ($\kappa=\frac92<\frac{50}{11}$ for $i=3$ and $\le i+1$ for $i\ge4$), so Jensen in $z^\kappa$ bounds the child by $v_C F_{i-1,\delta}(\ell/v_C)$ (and $\alpha v_C$ for $i=2,\delta=1$). Above $p$, $(1+iz)K_{i,\delta}=z+izK_{i-1,\delta}$ closes the step. Below $p$, put $z=p\zeta\in[0,p]$; differentiating the recurrence at $p$ gives $1-i\Delta_i(p)=K'_{i,\delta}(p)+ip\Delta_i'(p)$, so the cleared residual divided by $p(1-\zeta)$ is
\[
K'_{i,\delta}(p)+ip\Delta_i'(p)-\frac{1-\zeta^\kappa}{\kappa(1-\zeta)}\bigl(K'_{i,\delta}(p)+ip\zeta\,\Delta_i'(p)\bigr)\ge0,
\]
since $K'_{i,\delta}(p)>0$, $\Delta_i'(p)>0$, and $\frac{1-\zeta^\kappa}{\kappa(1-\zeta)}=\frac1{1-\zeta}\int_\zeta^1 t^{\kappa-1}\,dt\le1$.

\emph{The last two names ($i=2$, $\delta=0$, normalized to $v_C=1$).} Let the two live prices have mean $z$. If both are $\le\alpha$, the child pays $\le\alpha$ and $(1+2z)F_{2,0}\ge z+2z\alpha$ ($z\le\alpha$). If both are $\ge\alpha$, size-biased contraction and concavity of $K_{1,0}$ bound the child by $K_{1,0}(z)$, and $F_{2,0}\ge K_{2,0}$ on $z\ge\alpha$ (by concavity of $K_{2,0}(t^{1/\kappa})$). If they straddle $\alpha$ with smaller price $\zeta_-\le\alpha<2z-\zeta_-$, the child's excess over $\alpha$ is $\frac{\zeta_-(2z-\zeta_--\alpha)}{2z(1+2z-\zeta_-)}\le\Gamma_\alpha(z):=\frac{(\sqrt{1+2z}-\sqrt{1+\alpha})_+^2}{2z}\le\frac{(2z-\alpha)^2}{2z(4+3\alpha+2z)}$ ($2z\ge\alpha$). On $z\ge p$, $\Gamma_\alpha(z)\le\frac{z-\alpha}{1+z}$ clears (with $\hat z=z/\alpha\ge\frac97$) to $4\hat z^2-4\hat z-1+\alpha(6\hat z^2-7\hat z)\ge\frac{23}{49}+\frac{45}{49}\alpha>0$. On $z=p\zeta\le p$, a straddle forces $\zeta>\frac7{18}$. Put $\varphi_1=\frac p{1+p}\in(0,\frac9{16}]$ and $\varphi_2=\frac p{1+2p}\le\frac9{25}$. Using $F_{2,0}(p\zeta)-\alpha\ge\frac29p(1-\pi_2(p))\zeta^{9/2}$ and $\zeta^{9/2}\ge\frac{2\zeta^5}{1+\zeta}$, the step reduces to
\[
(1+2p\zeta)\tfrac29\bigl(1-\pi_2(p)\bigr)\frac{2\zeta^5}{1+\zeta}-\zeta+\tfrac79-\frac{p(2\zeta-\frac79)^2}{4+p(\frac73+2\zeta)}\ge0.
\]
Here $(1+2p\zeta)(1-\pi_2(p))=1+2\varphi_1[\zeta-\varphi_2(1-\zeta)]$ and $\frac p{4+p(\frac73+2\zeta)}=\frac{\varphi_1}{4+\varphi_1(2\zeta-\frac53)}$. Linearity in $\varphi_1$ gives $4+\varphi_1(2\zeta-\frac53)\ge4-2(1-\zeta)^2$, and $\frac1{1-t'}\le1+2t'$ on $[0,\frac12]$ bounds the last fraction by $\varphi_1[(2\zeta-\frac79)^2+(\frac{11}9)^2(1-\zeta)^2]/4$. Setting $\varphi_2\le\frac9{25}$ leaves an expression affine in $\varphi_1\in[0,\frac9{16}]$: at $\varphi_1=0$ it is $\frac{(1-\zeta)^2(4\zeta^3+8\zeta^2+12\zeta+7)}{9(1+\zeta)}\ge0$; at $\varphi_1=\frac9{16}$, multiplying by $14400(1+\zeta)$ yields
\[
N(\zeta)=9792\zeta^6+3808\zeta^5-11125\zeta^3-13175\zeta^2+4900\zeta+6950.
\]
On $[\frac7{18},1]$, $g(\zeta)=N(\zeta)/\zeta^2$ satisfies $g''(\zeta)=117504\zeta^2+22848\zeta+9800\zeta^{-3}+41700\zeta^{-4}>51500>50000$, while exact evaluation at $\zeta=\frac9{10}$ gives $g(\frac9{10})=\frac{3822649}{101250}>37$ and $g'(\frac9{10})=\frac{142639231}{91125}<1566$; completing the square yields $g(\zeta)\ge g(\frac9{10})-g'(\frac9{10})^2/100000>12.47>0$, so $N(\zeta)>0$ on $[\frac7{18},1]$.

\emph{The remaining terminal fraction.} Take $p=\min(x,\frac97)$ ($\alpha=\min(\frac79x,1)$), so the $\delta=0$ fractional tag is at most $v[x+(\alpha-x)\pi_j(x)]$. Write $(1-\alpha)\E[\cost(P,C_{\mathrm{end}})\,\mathbf 1\{\tau=\infty\}]$ over a forced nontarget process weighted by the $j$ avoidance probabilities. Along a forced path with relative terminal cost $v_{\mathrm{end}}=\cost(P,C_{\mathrm{end}})/v$, retention keeps target cost $\ge v_{\mathrm{end}}v$ and nontarget cost $\le jxv$ before each draw, so the avoidance weight is at most $(1+\frac{v_{\mathrm{end}}}{jx})^{-j}$. Putting $v_{\mathrm{end}}=jx\zeta_0$ gives at most $(1-\alpha)v\,jx\sup_{\zeta_0\ge0}\zeta_0(1+\zeta_0)^{-j}=(1-\alpha)\vartheta_jxv$.
\end{proof}

Holding $\delta$ fixed absorbs exterior draws (consuming a draw without retiring a name); at the root of each covered query, \cref{thm:stopped} is applied with $\delta=0$.

\subsection{The Euclidean reset}\label{sec:reset-statement}

Let $\mu$ be a finitely supported probability measure on $\R^d$ with mean~$0$ and $\int\|y\|^2\,d\mu(y)=1$. For a nonempty finite $C_0\subset\R^d$, write $q(y)=D(y,C_0)^2$ and $s=\int q\,d\mu$. After adding a center at $y$, the normalized target cost is
\[
w(y)=\int\min\bigl(q(y'),\|y-y'\|^2\bigr)\,d\mu(y'),\qquad\text{and}\qquad M(C_0)=\frac1s\int q(y)\,w(y)\,d\mu(y)
\]
is its conditional mean under $D^2$ sampling from $P$. For any finite target $P\subset\R^d$ with $\phi>0$ and nonempty finite $C_0\subset\R^d$, applying the affine normalization of \cref{sec:prelim} to both $P$ and $C_0$ yields $\mu$ and $\widehat C_0$ with $s=\bar u=\cost(P,C_0)/\phi$ and $M(\widehat C_0)$ equal to the conditional post-hit mean divided by $\phi$.

The three terms of the bound below have transparent meanings: retention gives $M(C_0)\le s$; a pair-counting identity (\cref{lem:pair}) shows that a hit removes on average at least half of the excess of $s$ above the ideal uniform cost $2$, giving $M(C_0)\le1+\frac s2$; and $\frac{5s}{s+14/5}$ shows that the local factor~$5$ is approached only when $s$ is large—that is, when existing centers serve the cluster poorly.

\begin{theorem}[Euclidean reset]\label{thm:reset}
For every $d\ge1$, every finitely supported normalized probability measure $\mu$ on $\R^d$, and every nonempty finite $C_0\subset\R^d$ with $s>0$,
\begin{equation}\label{eq:reset}
M(C_0)\le U(s):=\min\Bigl\{s,\;1+\frac s2,\;\frac{5s}{s+14/5}\Bigr\}.
\end{equation}
In particular, for every finite target $P\subset\R^d$ with $\phi>0$ and finite nonempty $C_0\subset\R^d$ with $\cost(P,C_0)>0$, $M(\widehat C_0)\le U(\cost(P,C_0)/\phi)<5$.
\end{theorem}

\begin{corollary}[Monotone reset]\label{cor:reset-monotone}
With $U(0)=0$, $U$ is nondecreasing and concave on $[0,\infty)$, and for every $\bar u>0$, $U(\bar u)=\gamma_{\mathrm{reset}}(1/\bar u)\,\bar u$, where the \emph{reset ratio} at $h=1/\bar u$ is
\begin{equation}\label{eq:gamma-reset}
\gamma_{\mathrm{reset}}(h):=\min\Bigl\{1,\;\frac12+h,\;\gamma_-(h)\Bigr\},\qquad\gamma_-(h):=\frac{5h}{1+\frac{14}5h}=\frac{25h}{5+14h}.
\end{equation}
If $\cost(P,C_0)\le\bar u\phi$ ($\phi>0$), then for every finite superset $C\supseteq C_0$ with $\cost(P,C)>0$, the normalized post-hit mean at $C$ is at most $U(\bar u)=\gamma_{\mathrm{reset}}(1/\bar u)\,\bar u$; taking $\bar u=\cost(P,C_0)/\phi$ discharges~\eqref{eq:transfer-hyp} in \cref{thm:transfer} with $\gamma=\gamma_{\mathrm{reset}}(1/\bar u)\in[0,1]$.
\end{corollary}

\begin{proof}
Each of $s$, $1+\frac s2$, and $5-\frac{14}{s+14/5}$ is nondecreasing and concave on $[0,\infty)$, so $U$ is as well, and $\bar u=1/h$ gives $U(\bar u)=\gamma_{\mathrm{reset}}(1/\bar u)\,\bar u$. For $C\supseteq C_0$, retention gives $0<\bar u_C:=\cost(P,C)/\phi\le\bar u$, so \cref{thm:reset} gives $M(\widehat C)\le U(\bar u_C)\le U(\bar u)$.
\end{proof}

\begin{remark}[Sharpness at both ends]\label{rem:reset-sharp}
For $s\le2$, $M(C_0)\le s$ is approached by a large regular simplex on the unit sphere with one center per vertex at distance $\sqrt s$ in mutually orthogonal directions. For $s\ge9$, placing mass $1/(1+t^2)$ at $t=(\sqrt{s-1}+\sqrt{s-9})/2$ and the rest at $-1/t$ on $\R$ with one center at $-\sqrt{s-1}$ gives $M(C_0)=5-15/s+O(s^{-2})=\frac{5s}{s+3}-O(s^{-2})$, so $\frac{14}5$ cannot be raised above~$3$.
\end{remark}

\subsection{Three geometric lemmas}\label{sec:reset-lemmas}

For any threshold $b>0$, set $A=\{y:q(y)>b\}$, $\rho=\mu(A)$, $\mathcal E_A=\int_A q\,d\mu$, and $\mathcal E_{A^c}=s-\mathcal E_A$. Define conditional centroids $\bar y_A,\bar y_{A^c}$ ($0$ on mass~$0$), unnormalized within-set variances $V_A=\int_A\|y-\bar y_A\|^2\,d\mu$ and $V_{A^c}=\int_{A^c}\|y-\bar y_{A^c}\|^2\,d\mu$, and weighted complement variance $\Xi=(1-\rho)V_{A^c}\ge0$. For $0<\rho<1$, set $B_{\mathrm{sep}}=1-V_A-V_{A^c}=\rho(1-\rho)\|\bar y_A-\bar y_{A^c}\|^2\ge0$, whence $V_A+\Xi\le1-B_{\mathrm{sep}}\le1$. Write $q^\pm(y,y')=\max/\min(q(y),q(y'))$, $A_t=\{q>t\}$, $\rho_t=\mu(A_t)$, $V(A_t)\le1$ for its within-set variance, and $I_{AA}=\iint_{A\times A}q^+(q^--\|y-y'\|^2)_+\,d\mu\,d\mu$.

\begin{lemma}[Pair identity and threshold cut]\label{lem:pair}
For every $(\mu,C_0)$ with $s>0$,
\[
2s\Bigl(1+\frac s2-M(C_0)\Bigr)=2\int_0^\infty\rho_t\bigl(1-V(A_t)\bigr)dt+\iint\bigl[q^+(q^--\|y-y'\|^2)_++q^-(\|y-y'\|^2-q^+)_+\bigr]d\mu\,d\mu\ge0.
\]
In particular $M(C_0)\le1+\frac s2$, and for every $b>0$ (including $\rho\in\{0,1\}$),
\begin{equation}\label{eq:pair-cut}
2s\bigl(5-M(C_0)\bigr)\ge2(1-V_A)(\mathcal E_A-\rho b)+8s-s^2+\bigl(\mathcal E_{A^c}^2-2b\Xi\bigr)_++I_{AA}+2b\bigl(\rho\mathcal E_{A^c}-1+\rho V_A+\Xi\bigr)_+.
\end{equation}
\end{lemma}

\begin{proof}
Checking the three orderings of $\zeta'\ge0$ relative to $\xi^-=\min(\xi_1,\xi_2)\le\xi^+=\max(\xi_1,\xi_2)$ gives $\xi_1\xi_2+\xi^-\zeta'-[\xi_1\min(\xi_2,\zeta')+\xi_2\min(\xi_1,\zeta')]=\xi^+(\xi^--\zeta')_++\xi^-(\zeta'-\xi^+)_+$. Integrating with $\xi_1=q(y)$, $\xi_2=q(y')$, $\zeta'=\|y-y'\|^2$ and applying the finite layer identity $q^-=\int_0^\infty\mathbf 1_{A_t}(y)\mathbf 1_{A_t}(y')\,dt$ (so $\iint q^-\|y-y'\|^2=2\int_0^\infty\rho_tV(A_t)\,dt$) gives the identity. Subtracting $2s(1+\frac s2-5)=s^2-8s$ and bounding four blocks—$t\ge b$ ($A_t\subseteq A$, $V(A_t)\le V_A$), $A^c\times A^c$ ($q^+\le b$), $A\times A$ ($I_{AA}$), and the mixed blocks ($q^+>b\ge q^-$, where $\iint_{A\times A^c}\|y-y'\|^2\,d\mu\,d\mu=1-\rho V_A-\Xi$)—yields~\eqref{eq:pair-cut}.
\end{proof}

\begin{lemma}[Pairs inside $A$]\label{lem:dd}
For every $b>0$ ($C_0$ nonempty for the second bound), $I_{AA}\ge(\rho^2b^2-2b\rho V_A)_+$ and $I_{AA}\ge\mathcal E_A^2-8\mathcal E_AV_A$; hence $I_{AA}\ge\lambda_A(\rho^2b^2-2b\rho V_A)+\lambda_B(\mathcal E_A^2-8\mathcal E_AV_A)$ for all $\lambda_A,\lambda_B\ge0$ with $\lambda_A+\lambda_B=1$.
\end{lemma}

\begin{proof}
On $A\times A$, $q^+\ge q^->b$ gives $q^+(q^--\|y-y'\|^2)_+\ge b(b-\|y-y'\|^2)$ (first bound). For the second, $0\le\xi_1\le\xi_2$ and $\zeta'\ge(\xi_2-\xi_1)^2$ imply $\xi_2^2\min(\xi_1^2,\zeta')\le4\xi_1^2\zeta'$ (split at $\xi_2\lessgtr2\xi_1$). Applying this to $\xi_1,\xi_2=\min/\max(D(y,C_0),D(y',C_0))$ and $\zeta'=\|y-y'\|^2$ (by~\eqref{eq:lipschitz}) yields $I_{AA}\ge\mathcal E_A^2-4\iint_{A\times A}q^-\|y-y'\|^2\ge\mathcal E_A^2-8\mathcal E_AV_A$ via the layer identity on $A_t\cap A$.
\end{proof}

\begin{lemma}[Tail energy]\label{lem:dtl}
If $C_0$ is nonempty and $0<\rho<1$, then $\sqrt{(1-\rho)(\mathcal E_A-V_A)_+}\le\sqrt{\rho(s-\mathcal E_A)}+\sqrt{1-V_A-\Xi}$, and hence
\begin{equation}\label{eq:dtl-linear}
0\le\Xi\le\Xi_{\mathrm{dtl}}:=\rho(s-\mathcal E_A)+(1-V_A)-\tfrac12(1-\rho)(\mathcal E_A-V_A).
\end{equation}
\end{lemma}

\begin{proof}
For $y'\in A^c$ with nearest center $c'\in C_0$, integrating $q(y)\le\|y-c'\|^2$ over $A$ gives $(\mathcal E_A-V_A)_+\le\rho\|\bar y_A-c'\|^2\le\rho(\|\bar y_A-y'\|+D(y',C_0))^2$. Taking square roots, applying Minkowski's inequality in $L^2(A^c,\mu)$ with $\int_{A^c}\|\bar y_A-y'\|^2\,d\mu=V_{A^c}+B_{\mathrm{sep}}/\rho=(1-V_A-\Xi)/\rho$, and squaring via $(t_1+t_2)^2\le2t_1^2+2t_2^2$ yields~\eqref{eq:dtl-linear}.
\end{proof}

\subsection{Proof of \texorpdfstring{\cref{thm:reset}}{Theorem 6.4} across the four ranges of \texorpdfstring{$s$}{s}}\label{sec:reset-proof}

Retention gives $M(C_0)\le s$ and \cref{lem:pair} gives $M(C_0)\le1+\frac s2$. It remains to prove $M(C_0)\le\frac{5s}{s+14/5}$, equivalently
\begin{equation}\label{eq:reset-goal}
2s\bigl(5-M(C_0)\bigr)\ge T_{\mathrm{res}}(s):=\frac{280s}{10s+28}=28-\frac{784}{10s+28},
\end{equation}
across the exhaustive four-range partition $(0,\frac{11}3]\cup[\frac{11}3,\frac{19}2]\cup[\frac{19}2,31]\cup[31,\infty)$.

\paragraph{Range 1: $(0,\frac{11}3]$.}
For $s\le2$, $s+\frac{14}5<5$ gives $s\le\frac{5s}{s+14/5}$. For $2\le s\le\frac{11}3$, $5s-(1+\frac s2)(s+\frac{14}5)=-\frac1{10}(5s^2-26s+28)>0$ since the convex quadratic $5s^2-26s+28$ equals $-4$ at $2$ and $-\frac19$ at $\frac{11}3$.

\paragraph{Threshold choice and $\rho\in\{0,1\}$.}
For $s\ge\frac{11}3$, fix $b(s)=\frac{23}{10}s-\frac92$ on $[\frac{11}3,\frac{19}2]$ and $b(s)=2s$ on $[\frac{19}2,\infty)$. In both ranges $b(s)>s$, so $\rho b\le\mathcal E_A\le s<b$ forces $\rho<1$. If $\rho=0$, then $\mathcal E_A=V_A=0$ and $\Xi=1$, so \cref{lem:pair} gives $2s(5-M(C_0))\ge8s-2b(s)$: on $[\frac{11}3,\frac{19}2]$, $(8s-2b)(10s+28)-280s=34s^2-\frac{474}5s+252>0$ (discriminant $-\frac{632124}{25}<0$); on $[\frac{19}2,\infty)$, $8s-2b=4s\ge38>28>T_{\mathrm{res}}(s)$. Henceforth $0<\rho<1$, so \cref{lem:dtl} applies.

\paragraph{Range 2: $[\frac{11}3,\frac{19}2]$.}
Put $\omega_b=\rho b\in(0,s)$, $\mathcal E=\mathcal E_A$, and $V=V_A$. Choose convex weights $\lambda_A+\lambda_B=1$ and mixed weight $\beta\in[0,1]$, with $m_\beta=2(1-\beta)$. In~\eqref{eq:pair-cut}, drop the $A^c\times A^c$ positive part, apply \cref{lem:dd} to $I_{AA}$, keep fraction $\beta$ of the mixed term, and substitute $\Xi\le\Xi_{\mathrm{dtl}}$ (valid since $\Xi$'s coefficient is $-m_\beta b\le0$). Using $2b\Xi_{\mathrm{dtl}}=2s\omega_b-\omega_b\mathcal E+2b-bV-b\mathcal E-\omega_bV$, $b$ cancels from the quadratic and $\omega_b$-linear parts:
\begin{equation}\label{eq:reset-quad}
\begin{aligned}
2s\bigl(5-M(C_0)\bigr)-T_{\mathrm{res}}(s)&\ge\Lambda_V(\omega_b,\mathcal E)\\
&:=\lambda_A\omega_b^2+(1-3\beta)\omega_b\mathcal E+(1+\lambda_B)\mathcal E^2+c_\omega\omega_b+c_{\mathcal E}\mathcal E+c_0-T_{\mathrm{res}}(s),
\end{aligned}
\end{equation}
with $c_\omega=-2+(2-2\lambda_A+2\beta+\frac{m_\beta}2)V+(2\beta-m_\beta)s$, $c_{\mathcal E}=2-(2+8\lambda_B)V-2s+\frac{m_\beta}2b(s)$, and $c_0=8s-2b(s)+\frac{m_\beta}2b(s)V$. Whenever $\lambda_A>0$ and $\det=4\lambda_A(1+\lambda_B)-(1-3\beta)^2>0$, minimizing over $(\omega_b,\mathcal E)\in\R^2$ gives $\min\Lambda_V=\mathcal R_V(s)/(\det\cdot(10s+28))$, where $\mathcal R_V(s)=(10s+28)[\det\cdot c_0-((1+\lambda_B)c_\omega^2-(1-3\beta)c_\omega c_{\mathcal E}+\lambda_Ac_{\mathcal E}^2)]-280\det\cdot s$ is a cubic in $s$. Since $\Lambda_V$ is affine in $V$, its minimum is concave in $V$, so checking endpoints of each $V$-subinterval suffices. Splitting $[0,1]=[0,\frac3{10}]\cup[\frac3{10},1]$ with witness parameters $(\frac34,\frac14,\frac12)$ ($\det=\frac72$) on $[0,\frac3{10}]$ and $(1,0,\frac25)$ ($\det=\frac{99}{25}$) on $[\frac3{10},1]$ produces four endpoint cubics $\mathcal R_V(s)$, each strictly positive on $[\frac{11}3,\frac{19}2]$ by exact degree-$3$ Bernstein expansion (\cref{app:reset-polys}).

\paragraph{Range 3: $[\frac{19}2,31]$.}
With $b=2s$ and $\omega=\rho s\in(0,\frac s2)$, choosing affine shifts $\psi_{\mathcal E}(s),\psi_\omega(s)$ and $\lambda_A+\lambda_B=1$ (with $\Upsilon(s)=2(2s-\psi_{\mathcal E}(s))$ and $2\beta b=\Upsilon(s)+\psi_\omega(s)$ satisfying $0\le2\beta b\le4s$ and $\Upsilon\ge\max(0,4s-2\beta b)$ at the endpoints $s=\frac{19}2,31$) eliminates $\Xi$ and all $O(s)$ linear terms in $(\omega,\mathcal E)$, yielding an affine-in-$V$ quadratic lower bound $\Lambda_V(\omega,\mathcal E)$ with positive Hessian on $[\frac{19}2,31]$. Splitting $V\in[0,1]$ at the seam $V=\frac3{10}$ reduces the two bands to four endpoint quintics $\mathcal R_V(s)$ whose exact degree-$5$ Bernstein coefficients on $[\frac{19}2,31]$ are strictly positive (\cref{app:range3,app:reset-polys}).

\paragraph{Range 4: Tail $[31,\infty)$.}
Keep $b=2s$, put $\omega=\rho s\in(0,\frac s2)$, $\chi=\sqrt\omega$, $\mathcal E=\mathcal E_A\in[2\omega,s]$, and $V=V_A$, using only $I_{AA}\ge\mathcal E^2-8\mathcal EV$ from \cref{lem:dd}. Squaring \cref{lem:dtl}, cancelling $-V-\rho\mathcal E$, and using $-\frac\omega sV\le\frac\omega sV$ yields
\begin{equation}\label{eq:tail-Q}
\mathcal E\le1+\omega-\Xi+\frac\omega sV+2\chi\sqrt{\Bigl(1-\frac{\mathcal E}s\Bigr)(1-V-\Xi)}.
\end{equation}
Dropping $\Xi$ and $\sqrt{1-\mathcal E/s}\le1$ with $c_V=\sqrt{1-V}$ gives $\mathcal E\le(1+\chi)^2$ (via $2\chi-2\chi c_V-\frac{\chi^2}s(1-c_V^2)=\chi(1-c_V)[2-\frac\chi s(1+c_V)]\ge0$), and $2\chi^2=\rho b\le\mathcal E$ forces $\chi\le1+\sqrt2<\frac52$ uniformly in $s$. The exhaustive three-branch partition—\emph{Branch~A} ($\omega\ge1$, $\beta=1$), \emph{Branch~B} ($\omega<1\le\omega+\Xi$, $\beta=1$), and \emph{Branch~C} ($\omega+\Xi<1$, $\beta=0$)—closes~\eqref{eq:reset-goal} analytically for all $s\ge31$ (\cref{app:range4}), completing the proof of \cref{thm:reset}. \qed

\subsection{The post-hit covered bound and arm split}\label{sec:next-step}

\Cref{sec:race-reset} feeds directly into one-step mixing (\cref{lem:mixing}, \eqref{eq:marked}). From $(C_0,\mathcal Q)$ with $|\mathcal Q|=j\ge1$, normalized target cost $s=\cost(P,C_0)/\phi>0$, and total nontarget cost $jL_{\mathrm{tot}}\phi=\cost(X,C_0)-s\phi$, a first target draw at $y\in P$ moves to $\sigma^y=(C_1,\mathcal Q)$ ($C_1=C_0\cup\{y\}$) with normalized cost $\bar u=\cost(P,C_1)/\phi\le s$. By retention, at $C_1$ total nontarget cost is at most $jL_{\mathrm{tot}}\phi=jx\bar u\phi$ ($x=L_{\mathrm{tot}}/\bar u$) and live price is at most $x\bar u\phi$. Thus for $0<\bar u\le2$, \eqref{eq:transfer-retention} gives $Q_j(\sigma^y)/\phi\le\bar u(H_j+\pi_j(x))$. For $\bar u>2$ ($h=1/\bar u\in(0,\frac12)$), when $j\ge2$, \cref{cor:reset-monotone} discharges~\eqref{eq:transfer-hyp} in \cref{thm:transfer} with $\gamma=\gamma_{\mathrm{reset}}(h)$ and \cref{thm:stopped} bounds $\E J_{\mathrm{st}}/(\bar u\phi)$; when $j=1$, \cref{cor:reset-monotone} at $C_1$ feeds directly into~\eqref{eq:one-name}, giving
\begin{equation}\label{eq:s6-master}
\begin{aligned}
\frac{Q_j(\sigma^y)}{\bar u\phi}&\le\bigl(1-\gamma_{\mathrm{reset}}(h)\bigr)\,W_j(x)+\gamma_{\mathrm{reset}}(h)\bigl(H_j+\pi_j(x)\bigr) && (j\ge2),\\
\frac{Q_1(\sigma^y)}{\bar u\phi}&\le\frac{2x+\gamma_{\mathrm{reset}}(h)}{1+x} && (j=1).
\end{aligned}
\end{equation}
Where $\alpha=\min(\frac79x,1)$ saturates in $W_j(x)$, \eqref{eq:s6-master} splits every opening analysis into a \emph{lower arm} ($x\le\frac97$) and a \emph{higher arm} ($x\ge\frac97$). \Cref{sec:hub-regime,sec:rank17} instantiate~\eqref{eq:transfer-retention} and~\eqref{eq:s6-master} across ranks~$1$--$17$ (also using \cref{thm:reset,lem:retirement} directly for rank~$0$, the witness cut, and the binary comparison), while \cref{sec:reservoir-regime} starts from the rank-$17$ stock and upgrades the transfer bound to a signed covered bound driven by an explicit repair kernel that tracks inside-ball repair rates across all clocks.

\section{The hub regime, \texorpdfstring{$2\le k\le18$}{2 <= k <= 18}}\label{sec:hub-regime}

This section proves that $J_{j+2}$ is a stock for every rank $0\le j\le16$; with the bridge (\cref{thm:bridge}) this yields $R_k\le J_k$ for $2\le k\le18$. For low-mean targets we establish the sharper stocks $\mathsf C_j\le G_j(\frac{11}4)<J_{j+2}$ (\cref{sec:low-mean,sec:rank17}).

\paragraph{The idea.}
The hub input is worst for $2\le k\le18$ because one cluster attains the worst possible first $D^2$ hit: a tight core with a rare, distant outlier. We classify each cluster by its worst-case post-hit mean cost. If a first hit can leave mean cost close to $5\phi$, the Euclidean reset (\cref{thm:reset}) forces the cluster into a tight bulk plus small-mass distant satellites, which we compare satellite by satellite against the hub's race. Otherwise the cluster has uniform slack below~$5$, and an induction on rank converts that slack into explicit stocks $\mathsf C_j<J_{j+2}$. Because the split depends only on the target cluster, it is invariant under all draws.

\subsection{Rank zero and the high-mean/low-mean split}

At rank~$0$ there is one draw and no live name, so repeat charges vanish: if $\phi=0$ (a singleton target) or $\cost(P,C_0)=0$ then $Z_1(\sigma)=0\le5\phi$, while if $\phi>0$ and $\cost(P,C_0)>0$ then \cref{thm:reset} gives $Z_1(\sigma)\le U(\bar u)\phi<5\phi$; hence $5=J_2$ is a universal rank-$0$ stock. For any target $P\subset\R^d$ with $\phi=\OPT_1(P)>0$, define in normalized coordinates
\[
M^*(P)=\sup\bigl\{M(\widehat C_0):\ C_0\subset\R^d\text{ finite nonempty},\ \cost(P,C_0)>0\bigr\}\le5.
\]
Put $\varepsilon_0=\frac{14117}{100000}$ and $M_{\mathrm{lo}}=5-\varepsilon_0=\frac{485883}{100000}$. Call $P$ \emph{high-mean} if $\phi>0$ and $M^*(P)>M_{\mathrm{lo}}$, and \emph{low-mean} otherwise ($\phi=0$ or $M^*(P)\le M_{\mathrm{lo}}$). The two classes are exhaustive and draw-invariant, and $\mathsf C_0=M_{\mathrm{lo}}$ is a rank-$0$ stock on the low-mean class.

\begin{theorem}[Hub stocks]\label{thm:hub-regime}
For every $0\le j\le16$, $J_{j+2}$ is a stock for rank~$j$. On low-mean targets the sharper stocks $\mathsf C_0,\dots,\mathsf C_{16}$ of \cref{thm:low-mean} hold, with $\mathsf C_0=M_{\mathrm{lo}}<5=J_2$ and $\mathsf C_j<J_{j+2}$ for $1\le j\le16$.
\end{theorem}

Rank~$0$ is proved above; for $1\le j\le16$, the high-mean case is \cref{thm:high-mean} and the low-mean case is \cref{thm:low-mean}.

\subsection{High-mean targets: witness cut, satellite covariance, and projection}\label{sec:high-mean}

A single \emph{witness cut} splits any high-mean target into a bulk $A$ and distant satellites $S$, extracted once from a center set nearly attaining $M^*(P)$ and reused across all states.

\begin{lemma}[Witness cut]\label{lem:witness}
Let $P$ be high-mean in normalized coordinates ($\mu$ uniform, mean~$0$, $\E_\mu\|Y\|^2=1$). There exist a partition $P=A\cup S$ into nonempty disjoint parts of masses $a_{\mathrm{bk}}=\mu(A)=1-\rho_{\mathrm{sat}}$ and $\rho_{\mathrm{sat}}=\mu(S)$, and reals $\varepsilon_{\mathrm{w}}\in(0,\varepsilon_{\mathrm{hi}})$ and $b_{\mathrm{w}}<B_{\mathrm{hi}}$ ($\varepsilon_{\mathrm{hi}}=\frac{105}{10000}$, $B_{\mathrm{hi}}=\frac{1427}{10000}$) such that, with $w_{\mathrm{bk}}=\E_\mu[\|Y\|^2\mathbf 1_A]$, $v_S=\OPT_1(S)/\phi$, $\sigma_{\mathrm{bk}}^2=\OPT_1(A)/(a_{\mathrm{bk}}\phi)$, $r_0=\rho_{\mathrm{sat}}/a_{\mathrm{bk}}$, and $R_{\mathrm{sat}}=\min_{y\in S}\|y-\bar y_A\|$,
\[
\tfrac32w_{\mathrm{bk}}+v_S\le b_{\mathrm{w}}<B_{\mathrm{hi}},\quad
\rho_{\mathrm{sat}}<\tfrac{135}{10000},\quad
a_{\mathrm{bk}}>\tfrac{9865}{10000},\quad
r_0<\tfrac{137}{10000},\quad
R_{\mathrm{sat}}\sqrt{\varepsilon_{\mathrm{w}}}>\tfrac{487}{1000},
\]
and $\rho_{\mathrm{sat}}/\varepsilon_{\mathrm{w}}\le1+2b_{\mathrm{w}}-4w_{\mathrm{bk}}$ if $w_{\mathrm{bk}}\le b_{\mathrm{w}}/2$, while $\rho_{\mathrm{sat}}/\varepsilon_{\mathrm{w}}\le1$ if $b_{\mathrm{w}}/2\le w_{\mathrm{bk}}\le\frac23b_{\mathrm{w}}$.
\end{lemma}

\begin{proof}
Pick $C_0$ with $M(\widehat C_0)>5-\varepsilon_0$, and set $s=\E_\mu q$, $s_1=s-1$, $\varepsilon_{\mathrm{w}}=1/s_1$. By \cref{thm:reset}, $5-\varepsilon_0<\frac{5s}{s+14/5}$, so $0<\varepsilon_{\mathrm{w}}<\varepsilon_0/(\frac{14}5(5-\varepsilon_0)-\varepsilon_0)<\varepsilon_{\mathrm{hi}}$ (\cref{app:projection}). Put $N_Y=\|Y\|^2$, $\tilde k_Y=\min(N_Y,s_1)$, $\chi_Y=\varepsilon_{\mathrm{w}}\tilde k_Y\in[0,1]$, and $\upsilon_Y=\varepsilon_{\mathrm{w}}N_Y$. Landing at $Y$ leaves cost at most $\min(1+N_Y,s)=1+\tilde k_Y$, so $(4-\varepsilon_0)s<(M(\widehat C_0)-1)s\le\E_\mu[q\tilde k_Y]$, i.e., $\E_\mu[\chi_Y q]>(4-\varepsilon_0)(1+\varepsilon_{\mathrm{w}})$.

Let $c_Y\in\widehat C_0$ be a nearest center to $Y$, $\bar c=\E_\mu c_Y$, $\mathcal V_c=\E_\mu\|c_Y\|^2+\|\bar c\|^2$, and $\Delta_c=2(s+1)-\mathcal V_c=\E_\mu\|c_Y-\bar c-2Y\|^2\ge0$. For independent $Y'\sim\mu$, $q(Y)+\|Y+\bar c\|^2\le\E_{Y'}\|Y-c_{Y'}\|^2+\|Y+\bar c\|^2=2N_Y+\mathcal V_c$. Multiplying by $\chi_Y$ and using $2\chi_Y N_Y=2N_Y-2N_Y(1-\upsilon_Y)_+$, $\E_\mu N_Y=1$, and $\E_\mu\tilde k_Y=1-\E_\mu(N_Y-s_1)_+$ yields
\begin{equation}\label{eq:projection-1}
\E_\mu\bigl[\chi_Y\|Y+\bar c\|^2+2N_Y(1-\upsilon_Y)_++2(1+2\varepsilon_{\mathrm{w}})(N_Y-s_1)_+\bigr]
\le b_{\mathrm{w}}:=(1+\varepsilon_{\mathrm{w}})\varepsilon_0-\varepsilon_{\mathrm{w}}\Delta_c\E_\mu\tilde k_Y<B_{\mathrm{hi}}.
\end{equation}
For any $u\ge0$, integrating directly over $t\in[\frac14,1]$ (length $\frac34$) gives
\begin{gather*}
\frac43\int_{1/4}^1\mathbf 1_{\{u\le t\}}\,dt=\frac43\min\Bigl(\frac34,(1-u)_+\Bigr)\le\frac43(1-u)_+,\\
\frac43\int_{1/4}^1\mathbf 1_{\{t<u\}}\,dt=\frac43\min\Bigl(\frac34,(u-\tfrac14)_+\Bigr)\le\min(u,1),
\end{gather*}
where the second bound uses $u-\frac14\le\frac34u$ for $u\le1$ and $\frac34=\frac34\cdot1$ for $u>1$. Applying these two bounds pointwise at $u=\upsilon_Y$ (where $\min(\upsilon_Y,1)=\chi_Y$) shows that the piecewise-constant function $F(t):=\frac32\E_\mu[N_Y\mathbf 1_{\{\upsilon_Y\le t\}}]+\E_\mu[\|Y+\bar c\|^2\mathbf 1_{\{\upsilon_Y>t\}}]$ has uniform average over $[\frac14,1]$ at most $\E_\mu[2N_Y(1-\upsilon_Y)_++\chi_Y\|Y+\bar c\|^2]\le b_{\mathrm{w}}$ by~\eqref{eq:projection-1}. Hence there exists a fixed $t\in[\frac14,1]$ with $F(t)\le b_{\mathrm{w}}$. Setting $A=\{\upsilon_Y\le t\}$ and $S=\{\upsilon_Y>t\}$, the candidate center $-\bar c$ for $S$ gives $v_S\le\E_\mu[\|Y+\bar c\|^2\mathbf 1_S]$, whence $\frac32w_{\mathrm{bk}}+v_S\le F(t)\le b_{\mathrm{w}}$.

On $S=\{\upsilon_Y>t\}$, $1\le\upsilon_Y+\upsilon_Y(1-\upsilon_Y)_+/t$ (both for $\upsilon_Y\ge1$ and for $t<\upsilon_Y<1$), so dividing by $\varepsilon_{\mathrm{w}}$ gives $\mathbf 1_S/\varepsilon_{\mathrm{w}}\le N_Y\mathbf 1_S+N_Y(1-\upsilon_Y)_+\mathbf 1_S/t$; on $A=\{\upsilon_Y\le t\}$, $N_Y(1-\upsilon_Y)_+\mathbf 1_A\ge(1-t)N_Y\mathbf 1_A$. Taking expectations and using $\E_\mu[N_Y(1-\upsilon_Y)_+]\le b_{\mathrm{w}}/2$ from~\eqref{eq:projection-1} yields
\[
\frac{\rho_{\mathrm{sat}}}{\varepsilon_{\mathrm{w}}}\le(1-w_{\mathrm{bk}})+\frac{b_{\mathrm{w}}/2-(1-t)w_{\mathrm{bk}}}t=1+\frac{b_{\mathrm{w}}/2-w_{\mathrm{bk}}}t,
\]
which gives $\rho_{\mathrm{sat}}/\varepsilon_{\mathrm{w}}\le1+2b_{\mathrm{w}}-4w_{\mathrm{bk}}$ when $w_{\mathrm{bk}}\le b_{\mathrm{w}}/2$ (as $t\ge\frac14$) and $\rho_{\mathrm{sat}}/\varepsilon_{\mathrm{w}}\le1$ when $b_{\mathrm{w}}/2\le w_{\mathrm{bk}}\le\frac23b_{\mathrm{w}}$; the scalar bounds on $\rho_{\mathrm{sat}},a_{\mathrm{bk}},r_0$ follow (\cref{app:projection}). Both parts are nonempty ($a_{\mathrm{bk}}>0$, and $S=\emptyset$ would force $w_{\mathrm{bk}}=1>\frac23B_{\mathrm{hi}}$). Finally, $a_{\mathrm{bk}}\bar y_A=-\E_\mu[Y\mathbf 1_S]$ gives $\|\bar y_A\|\le\sqrt{\rho_{\mathrm{sat}}}/a_{\mathrm{bk}}$ by Cauchy--Schwarz, while every $y\in S$ has $\|y\|>1/(2\sqrt{\varepsilon_{\mathrm{w}}})$, so $R_{\mathrm{sat}}\sqrt{\varepsilon_{\mathrm{w}}}\ge\frac12-\sqrt{\rho_{\mathrm{sat}}\varepsilon_{\mathrm{w}}}/a_{\mathrm{bk}}>\frac{487}{1000}$.
\end{proof}

At any retained center set $C$ before the target's first hit, write $D_y^2=\E_A\|Y-y\|^2=\|\bar y_A-y\|^2+\sigma_{\mathrm{bk}}^2$, $s_A=\E_A q=\cost(A,C)/|A|$, and $r_y=\sqrt{s_A}/D_y$.

\begin{lemma}[Clipped satellite covariance]\label{lem:satellite}
Let $A\subset\R^d$ be finite nonempty with uniform law $\mu_A$ and variance $\sigma_{\mathrm{bk}}^2=\OPT_1(A)/|A|$, let $C\subset\R^d$ be finite nonempty with $s_A=\E_A q>0$ ($q(Y)=D(Y,C)^2$), and fix $y\in\R^d$ with $D_y^2=\E_A\|Y-y\|^2$. For $F(Y)=\min(D(y,C)^2,\|Y-y\|^2)$,
\[
\frac{\E_A[qF]}{s_A}\le D_y^2+\sigma_{\mathrm{bk}}^2\Bigl(3+\frac{2D_y}{\sqrt{s_A}}\Bigr).
\]
\end{lemma}

\begin{proof}
\emph{One pair.} Take $(Y,Y')\in A\times A$ with $q<q'$ and $F<F'$. Then $F=\|Y-y\|^2$ (otherwise $F=D(y,C)^2\ge F'$) and $q>0$ (if $q=0$, then $Y\in C$ gives $F'\le D(y,C)^2\le\|Y-y\|^2=F$). Translate $Y$ to $0$ and scale so $q=1$; let $c_*\in C$ be a nearest center to $0$ ($\|c_*\|=1$), write $Y'=tn$ with $t=\|Y'\|>0$ and $\|n\|=1$, put $R'=\|y\|$, and define $\xi_0:=t-2n\cdot c_*$. Since $q'\le\|Y'-c_*\|^2=1+t^2-2tn\cdot c_*$, we have $0<q'-q\le t\xi_0$ (so $\xi_0>0$). Likewise $F'\le\min(\|Y'-y\|^2,\|c_*-y\|^2)$ gives
\[
0<F'-F\le\min\bigl\{\|Y'-y\|^2-\|y\|^2,\ \|c_*-y\|^2-\|y\|^2\bigr\}=\min\bigl\{t(t-2n\cdot y),\ 1-2c_*\cdot y\bigr\}.
\]
If $\xi_0\le1$, then $n\cdot c_*\le1$ forces $t=\xi_0+2n\cdot c_*\le3$, so $(q'-q)(F'-F)\le(t\xi_0)\cdot t(t-2n\cdot y)\le t^2(3+2R')$. If $\xi_0>1$, set $\alpha_1=\frac{t+\xi_0}{1+t\xi_0}>0$ and $\alpha_2=\frac{\xi_0^2-1}{1+t\xi_0}>0$. Substituting $2n\cdot c_*=t-\xi_0$ yields the three exact identities
\[
\alpha_1+t\alpha_2=\xi_0,\quad
3-(\alpha_1t+\alpha_2)=\frac{4\bigl(1-(n\cdot c_*)^2\bigr)}{1+t\xi_0}\ge0,\quad
\|\alpha_1n+\alpha_2c_*\|^2=\alpha_1^2+\alpha_2^2+\alpha_1\alpha_2(t-\xi_0)=1.
\]
Hence
\begin{align*}
\frac{(q'-q)(F'-F)}{t^2}&\le\xi_0\min\Bigl\{t-2n\cdot y,\;\frac{1-2c_*\cdot y}t\Bigr\}\\
&\le\alpha_1(t-2n\cdot y)+\alpha_2(1-2c_*\cdot y)=\alpha_1 t+\alpha_2-2(\alpha_1 n+\alpha_2 c_*)\cdot y\le3+2R'.
\end{align*}
Restoring original units ($t=\|Y-Y'\|/\sqrt q$, $R'=\sqrt{F/q}$), both cases give $(q'-q)(F'-F)\le\|Y-Y'\|^2\bigl(3\min(q,q')+2\min(\sqrt{qF},\sqrt{q'F'})\bigr)$, which holds symmetrically whenever $(q'-q)(F'-F)>0$ and trivially otherwise.

\emph{Layers and average.} Integrating $\mathbf 1_{\{g\ge\tau\}}$ over $\tau\ge0$ with $\sum_{Y,Y'\in S'}\|Y-Y'\|^2=2|S'|\OPT_1(S')\le2|A|\OPT_1(A)$ gives $\E_{A\times A}[\min(g,g')\|Y-Y'\|^2]\le2\sigma_{\mathrm{bk}}^2\E_A g$ for any $g\ge0$. Applying this to $2\,\mathrm{Cov}_A(q,F)=\E_{A\times A}[(q'-q)(F'-F)]$ with $g=q$ and $g=\sqrt{qF}$, then Cauchy--Schwarz ($\E_A\sqrt{qF}\le\sqrt{s_A\E_A F}$) and $\E_A F\le D_y^2$, yields the claim.
\end{proof}

We now reduce the delayed cost of any high-mean target $P=A\cup S$ to the delayed cost of its proper bulk $A$ plus a weighted sum of per-satellite unopened binary accounts, all driven simultaneously by the \emph{single} actual $D^2$ seeding process on $X$.

\paragraph{Pathwise functionals and the per-satellite unopened account.}
Fix a witness cut $P=A\sqcup S$ (\cref{lem:witness}) and any state $(C,\mathcal Q)$ with $C\subset\R^d$ finite nonempty and $\mathcal Q$ a family of pairwise-disjoint nonempty subsets of $X\setminus P$. On any $h$-step sample path $\omega=(x_1,\dots,x_h)\in X^h$ drawn from $(C,\mathcal Q)$ (with center sets $C_0=C$, $C_i=C_{i-1}\cup\{x_i\}$, and live-name families $\mathcal Q_i$ removing $G\in\mathcal Q_{i-1}$ whenever $x_i\in G$), define for any subset $D\subseteq P$ three deterministic pathwise functionals:
\begin{itemize}[leftmargin=1.5em]
\item $\mathsf Z_h(\omega;D,C,\mathcal Q)$: the \emph{delayed cost} of $D$ along $\omega$, equal to the sum of live prices $\ell_{i-1}$ over all hits $x_i\in D$ strictly after the first hit on $D$, plus the terminal cost $\cost(D,C_h)$ if $D$ is hit at least once, and $0$ if $D$ is never hit; thus $\E[\mathsf Z_h(\omega;D,C,\mathcal Q)]=Z_h(D;C,\mathcal Q)$.
\item $\mathsf Q_h(\omega;D,C,\mathcal Q)$: the \emph{covered cost} of $D$ along $\omega$, equal to the sum of live prices $\ell_{i-1}$ over \emph{all} hits $x_i\in D$ during $\omega$ (including the first) plus the terminal cost $\cost(D,C_h)$; thus $\E[\mathsf Q_h(\omega;D,C,\mathcal Q)]=Q_h(D;C,\mathcal Q)$ (\cref{sec:framework}).
\item $\mathsf S_h(\omega;A,C,\mathcal Q)$: the \emph{stopped payment} of the bulk $A$ along $\omega$, equal to the live price $\ell_{\tau-1}$ at the first hit $\tau=\min\{i\le h:x_i\in A\}$ if $A$ is hit during $\omega$, and equal to the terminal cost $\cost(A,C_h)$ if $A$ is never hit; put $\mathcal T(C,A,\mathcal Q,h):=\E[\mathsf S_h(\omega;A,C,\mathcal Q)]$.
\end{itemize}
Comparing the two exhaustive events ($A$ unhit vs.\ $A$ first hit at step $\tau$) gives the exact pathwise identity
\begin{equation}\label{eq:pathwise-split}
\mathsf Q_r(\omega';A,C',\mathcal Q')=\mathsf S_r(\omega';A,C',\mathcal Q')+\mathsf Z_r(\omega';A,C',\mathcal Q')
\end{equation}
for every $(C',\mathcal Q')$ and every $r$-step path $\omega'$: if $A$ is unhit, both sides equal $\cost(A,C'_r)$; if $A$ is first hit at step $\tau$, then $\ell_{\tau-1}$ goes to $\mathsf S_r$ while all hits $i>\tau$ and $\cost(A,C'_r)$ go to $\mathsf Z_r$. Taking expectations yields $\mathcal T(C',A,\mathcal Q',r)\le Q_r(A;C',\mathcal Q')\le(1+\Cl(r,|\mathcal Q'|))\cost(A,C')$ by \cref{thm:restoration}(b) (since $\mathsf Z_r\ge0$).

Next, for each satellite $y\in S$ with unit $u_y:=\cost(A,\{y\})=|A|D_y^2>0$ and any $\varpi\ge0$, define the deterministic \emph{per-satellite unopened account} $\mathcal U_y(C,\mathcal Q,h;\varpi)$ recursively on $h\in\{0,\dots,|\mathcal Q|+1\}$ over the same actual $D^2$ transitions on $X$: set $\mathcal U_y(C,\mathcal Q,0;\varpi)=0$, and for $h=r+1\ge1$ with $K':=1+\Cl(r,|\mathcal Q|)=1+H_{|\mathcal Q|}-H_{|\mathcal Q|-r}$, set $\mathcal U_y=0$ if $\cost(X,C)=0$, set $\mathcal U_y=2K'$ if $\cost(X,C)>0$ and $\cost(A,C)=0$, and otherwise (with $b_{\mathrm{cur}}(y,C):=\frac9{40}\mathbf 1_{\{\cost(A,C)\le u_y/4\}}$, corresponding to $r_y\le\frac12$) set
\begin{equation}\label{eq:proj-bin-account}
\begin{aligned}
\mathcal U_y(C,\mathcal Q,r+1;\varpi)
&=\frac{\cost(A,C)}{\cost(X,C)}\bigl(1+b_{\mathrm{cur}}(y,C)+\varpi\bigr)K'
+\frac{|A|D(y,C)^2}{\cost(X,C)}\cdot\frac{\mathcal T(C\cup\{y\},A,\mathcal Q,r)}{u_y}\\
&\quad+\frac{\cost(S,C)}{\cost(X,C)}(2K')
+\sum_{G\in\mathcal Q}\sum_{x\in G}\frac{D(x,C)^2}{\cost(X,C)}\,\mathcal U_y(C\cup\{x\},\mathcal Q\setminus\{G\},r;\varpi)\\
&\quad+\sum_{x\in X\setminus(P\cup\bigcup\mathcal Q)}\frac{D(x,C)^2}{\cost(X,C)}\,\mathcal U_y(C\cup\{x\},\mathcal Q,r;\varpi).
\end{aligned}
\end{equation}
Notice two structural properties of~\eqref{eq:proj-bin-account}: first, all $|S|$ accounts $\mathcal U_y$ evolve along the \emph{exact same} $D^2$ state transitions on $X$ (the second term is a pure additive numerator credit and never alters the denominator $\cost(X,C)$ or the next-draw law); second, because the stopping rule and transition probabilities do not depend on $\varpi$, $\varpi\mapsto\mathcal U_y(C,\mathcal Q,h;\varpi)$ is \emph{exact affine} in $\varpi\ge0$.

\begin{theorem}[Two-event projection]\label{thm:projection}
Fix $1\le m\le16$, a high-mean target $P\subseteq X$ with witness cut $P=A\sqcup S$ (\cref{lem:witness}), nonempty $C_0\subset\R^d$, and a pairwise-disjoint live-name family $\mathcal Q$ in $X\setminus P$ with $|\mathcal Q|=m$. If the proper bulk $A\subsetneq P$, queried from $(C_0,\mathcal Q)$ with $S$ treated as exterior, obeys $Z_{m+1}(A;C_0,\mathcal Q)\le J_{m+2}\OPT_1(A)$, then
\[
\frac{Z_{m+1}(P;C_0,\mathcal Q)}{J_{m+2}\,\phi}\le1-\frac{12}{1000}\cdot\frac{r_0}{1+2r_0}\le1.
\]
\end{theorem}

\begin{proof}
Rescale coordinates so $\phi=\OPT_1(P)=|P|$ (equivalently, $\E_\mu\|Y\|^2=1$ as in \cref{lem:witness}, while $\cost(D,C)=\sum_{x\in D}D(x,C)^2$ stays in unnormalized point-sum units). Write $v_A=\OPT_1(A)/\phi=a_{\mathrm{bk}}\sigma_{\mathrm{bk}}^2\le w_{\mathrm{bk}}$ and $v_S=\OPT_1(S)/\phi$. By centering ($\bar y_P=0=a_{\mathrm{bk}}\bar y_A+\rho_{\mathrm{sat}}\bar y_S$, so $|S|\|\bar y_S-\bar y_A\|^2=(1+r_0)(\phi-\OPT_1(A)-\OPT_1(S))$), the satellite variance sum satisfies
\[
S_{\mathrm{tot}}:=\frac1\phi\sum_{y\in S}D_y^2=1-v_A+r_0(1-v_S).
\]
Set $L_0=\frac{2S_{\mathrm{tot}}-5v_S}{1+2r_0}$. Because $0\le v_A\le w_{\mathrm{bk}}<\frac23B_{\mathrm{hi}}$, $0\le v_S<B_{\mathrm{hi}}<1$, and $r_0>0$, substituting $S_{\mathrm{tot}}=1-v_A+r_0(1-v_S)$ yields $0<2-\frac{19}3B_{\mathrm{hi}}<2S_{\mathrm{tot}}-5v_S=2(1+2r_0)-2v_A-5v_S-2r_0(1+v_S)<2(1+2r_0)$, so $L_0\in(0,2)$. Put $\alpha_y=\frac{L_0\phi}{u_y}=\frac{L_0}{a_{\mathrm{bk}}D_y^2}$, $H_{\mathrm{th}}=\frac1\phi\sum_{y\in S}\alpha_y D_y^2=r_0 L_0$, and per-satellite weights $W_y:=(1-\alpha_y)D_y^2=(1-\alpha_y)\frac{u_y}{|A|}$. By \cref{lem:witness}, $\alpha_y\le\frac{2\varepsilon_{\mathrm{w}}}{a_{\mathrm{bk}}R_{\mathrm{sat}}^2\varepsilon_{\mathrm{w}}}<\frac{2\varepsilon_{\mathrm{hi}}}{a_{\mathrm{lo}}\varrho_{\mathrm{sat}}^2}<1$ (\cref{app:projection}), so $W_y>0$. Direct summation yields the two algebraic identities
\begin{equation}\label{eq:proj-weights}
\sum_{y\in S}W_y=(S_{\mathrm{tot}}-H_{\mathrm{th}})\phi,\qquad
2\sum_{y\in S}W_y=(L_0+5v_S)\phi=L_0\phi+5\OPT_1(S).
\end{equation}
Define the effective extra heavy-award rate
\[
\varpi':=\frac{H_{\mathrm{th}}+7r_0 v_A}{S_{\mathrm{tot}}-H_{\mathrm{th}}}.
\]
Because $S_{\mathrm{tot}}-H_{\mathrm{th}}=\frac{S_{\mathrm{tot}}+5r_0 v_S}{1+2r_0}\ge\frac{1-w_{\mathrm{bk}}}{1+2r_0}$, $H_{\mathrm{th}}\le2r_0$, and $v_A\le w_{\mathrm{bk}}\le\frac23B_{\mathrm{hi}}$, exact rational arithmetic (\cref{app:projection}) gives
\[
\varpi'\le r_0\Bigl[2+7(1+2r_0)\frac{w_{\mathrm{bk}}}{1-w_{\mathrm{bk}}}\Bigr]\le\frac{\varepsilon_{\mathrm{hi}}}{a_{\mathrm{lo}}}\Bigl(2+\tfrac23 K_{\mathrm{pr}}B_{\mathrm{hi}}\Bigr)<\frac1{25},\qquad K_{\mathrm{pr}}=8.
\]

\emph{Step 1: Pathwise accounting lemma.} We prove by induction on $h\in\{0,\dots,|\mathcal Q|+1\}$, uniformly over all finite nonempty $C\subset\R^d$ and pairwise-disjoint $\mathcal Q$ in $X\setminus P$, that
\begin{equation}\label{eq:proj-accounting}
Z_h(P;C,\mathcal Q)-Z_h(A;C,\mathcal Q)\le\sum_{y\in S}W_y\,\mathcal U_y(C,\mathcal Q,h;\varpi').
\end{equation}
When $h=0$ or $\cost(X,C)=0$, both sides vanish. Now let $h=r+1\ge1$ with $\cost(X,C)>0$, and write $K'=1+\Cl(r,|\mathcal Q|)$.

\emph{Subcase 1a: $\cost(A,C)=0$.} Then $A\subseteq C$, so $A$ is never drawn and has terminal cost~$0$, giving $Z_{r+1}(A;C,\mathcal Q)=0$. We claim $Z_{r+1}(P;C,\mathcal Q)\le5K'\OPT_1(S)$ by a nested induction on $r$: on a draw $z\in S$, $\cost(A,C\cup\{z\})=0$ implies $Q_r(P;C\cup\{z\},\mathcal Q)=Q_r(S;C\cup\{z\},\mathcal Q)\le K'\cost(S,C\cup\{z\})$ (\cref{thm:restoration}(b)), whose $D^2|_S$-expectation is at most $5K'\OPT_1(S)$ by \cref{lem:five} applied to $S$; on a named or exterior draw $x\notin P$, $\cost(A,C\cup\{x\})=0$ persists and the child's persistence factor is $K'_{\mathrm{child}}\le K'$, so the nested inductive hypothesis gives $\le5K'\OPT_1(S)$. Averaging over the first draw yields $Z_{r+1}(P;C,\mathcal Q)\le5K'\OPT_1(S)\le K'(L_0\phi+5\OPT_1(S))=2K'\sum_{y\in S}W_y=\sum_{y\in S}W_y\,\mathcal U_y(C,\mathcal Q,r+1;\varpi')$ by~\eqref{eq:proj-weights}.

\emph{Subcase 1b: $\cost(A,C)>0$.} Write $Z_{r+1}(P;C,\mathcal Q)-Z_{r+1}(A;C,\mathcal Q)=\E[\mathsf Z_{r+1}(\omega;P,C,\mathcal Q)-\mathsf Z_{r+1}(\omega;A,C,\mathcal Q)]$ inside one expectation over $\omega=(x_1,\omega')$, and condition on the four exhaustive locations of $x_1\in A\sqcup S\sqcup(\bigcup\mathcal Q)\sqcup(X\setminus(P\cup\bigcup\mathcal Q))$:
\begin{enumerate}[label=(\roman*),leftmargin=2em]
\item \emph{Bulk-first ($x_1\in A$).} Both $P$ and $A$ are visited at step~$1$, so pathwise $\mathsf Z_{r+1}(x_1,\omega';P)=\mathsf Q_r(\omega';P,C\cup\{x_1\},\mathcal Q)$ and $\mathsf Z_{r+1}(x_1,\omega';A)=\mathsf Q_r(\omega';A,C\cup\{x_1\},\mathcal Q)$. Because $P=A\sqcup S$, $\mathsf Q_r(\omega';P)=\mathsf Q_r(\omega';A)+\mathsf Q_r(\omega';S)$ holds pathwise; hence $\mathsf Q_r(\omega';A)$ cancels identically pathwise (zero double-counting against $Z_{r+1}(A)$), leaving $\mathsf Q_r(\omega';S,C\cup\{x_1\},\mathcal Q)$. Taking conditional expectation and applying \cref{thm:restoration}(b) gives $Q_r(S;C\cup\{x_1\},\mathcal Q)\le K'\cost(S,C\cup\{x_1\})=K'\sum_{y\in S}F_y(x_1)$, where $F_y(x_1):=\min(D(y,C)^2,\|x_1-y\|^2)$. Weighting by $\frac{D(x_1,C)^2}{\cost(X,C)}$ and summing over $x_1\in A$ yields at most $\frac{\cost(A,C)}{\cost(X,C)}K'\sum_{y\in S}\frac{\E_A[qF_y]}{s_A}$. For each satellite $y\in S$, we bound the \emph{satellite excess} $\frac{\E_A[qF_y]}{s_A}-D_y^2$ in three cases (using $1-\alpha_y>1-\frac{2\varepsilon_{\mathrm{hi}}}{a_{\mathrm{lo}}\varrho_{\mathrm{sat}}^2}$, so the funded radius award is at least $A_{\min}D_y^2$ with $A_{\min}=\frac9{40}(1-\frac{2\varepsilon_{\mathrm{hi}}}{a_{\mathrm{lo}}\varrho_{\mathrm{sat}}^2})>0.2048$, \cref{app:projection}):
\begin{itemize}[leftmargin=1.5em]
\item if $r_y\le\frac9{100}$, then $F_y(x_1)\le D(y,C)^2\le(1+r_y)^2D_y^2$ (Minkowski in $L^2(\mu_A)$), so $\frac{\E_A[qF_y]}{s_A}-D_y^2\le(2r_y+r_y^2)D_y^2\le0.1881D_y^2\le(1-\alpha_y)\frac9{40}D_y^2=\frac9{40}W_y$;
\item if $\frac9{100}<r_y\le\frac12$, then \cref{lem:satellite} gives $\frac{\E_A[qF_y]}{s_A}-D_y^2\le(3+\frac2{r_y})\sigma_{\mathrm{bk}}^2\le\frac{227}9\sigma_{\mathrm{bk}}^2$, and $\frac{227}9\frac{\sigma_{\mathrm{bk}}^2}{D_y^2}\le\frac{227}9\cdot\frac23\cdot\frac{\varepsilon_{\mathrm{hi}}B_{\mathrm{hi}}}{a_{\mathrm{lo}}\varrho_{\mathrm{sat}}^2}<0.1077<A_{\min}$, so again $\frac{\E_A[qF_y]}{s_A}-D_y^2\le\frac9{40}W_y$;
\item if $r_y>\frac12$, then \cref{lem:satellite} gives $\frac{\E_A[qF_y]}{s_A}-D_y^2\le(3+\frac2{r_y})\sigma_{\mathrm{bk}}^2\le7\sigma_{\mathrm{bk}}^2$.
\end{itemize}
Since $D_y^2=W_y+\alpha_y D_y^2=W_y+\frac{L_0\phi}{|A|}$ and $b_{\mathrm{cur}}(y,C)=\frac9{40}\mathbf 1_{\{r_y\le1/2\}}$, summing over $y\in S$ (using $|S|\frac{L_0\phi}{|A|}=r_0L_0\phi=H_{\mathrm{th}}\phi$ and $7|S|\sigma_{\mathrm{bk}}^2=7r_0v_A\phi$) bounds the bulk-first contribution by
\begin{align*}
\frac{\cost(A,C)}{\cost(X,C)}K'\Bigl[\sum_{y\in S}W_y\bigl(1+b_{\mathrm{cur}}(y,C)\bigr)+(H_{\mathrm{th}}+7r_0v_A)\phi\Bigr]\\
=\sum_{y\in S}W_y\frac{\cost(A,C)}{\cost(X,C)}\bigl(1+b_{\mathrm{cur}}(y,C)+\varpi'\bigr)K',
\end{align*}
matching the first term of $\sum_{y\in S}W_y\,\mathcal U_y(C,\mathcal Q,r+1;\varpi')$.
\item \emph{Satellite-first ($x_1=z\in S$, covering all multi-satellite interactions).} Here $P$ is visited at step~$1$ while $A$ is not ($z\notin A$), so $\mathsf Z_{r+1}(z,\omega';P)=\mathsf Q_r(\omega';S,C\cup\{z\},\mathcal Q)+\mathsf Q_r(\omega';A,C\cup\{z\},\mathcal Q)$ and $\mathsf Z_{r+1}(z,\omega';A)=\mathsf Z_r(\omega';A,C\cup\{z\},\mathcal Q)$. Applying the pathwise identity~\eqref{eq:pathwise-split} cancels $\mathsf Z_r(\omega';A)$ identically pathwise (again zero double-counting against $Z_{r+1}(A)$), leaving $\mathsf Q_r(\omega';S,C\cup\{z\},\mathcal Q)+\mathsf S_r(\omega';A,C\cup\{z\},\mathcal Q)$, whose conditional expectation is $Q_r(S;C\cup\{z\},\mathcal Q)+\mathcal T_z$, where $\mathcal T_z:=\mathcal T(C\cup\{z\},A,\mathcal Q,r)$. Split $\mathcal T_z=(1-\alpha_z)\mathcal T_z+\alpha_z\mathcal T_z$. By~\eqref{eq:pathwise-split} and retention ($\cost(A,C\cup\{z\})\le\cost(A,\{z\})=u_z$), the thinned remainder obeys $\alpha_z\mathcal T_z\le\alpha_z K'u_z=K'L_0\phi$, while \cref{thm:restoration}(b) gives $Q_r(S;C\cup\{z\},\mathcal Q)\le K'\cost(S,C\cup\{z\})$. Weighting by $\frac{D(z,C)^2}{\cost(X,C)}$ and summing over all $z\in S$:
\begin{itemize}[leftmargin=1.5em]
\item the unthinned sum $\sum_{z\in S}\frac{D(z,C)^2}{\cost(X,C)}(1-\alpha_z)\mathcal T_z=\sum_{y\in S}W_y\frac{|A|D(y,C)^2}{\cost(X,C)}\frac{\mathcal T_y}{u_y}$ matches the second term of $\sum_{y\in S}W_y\,\mathcal U_y$ identically (each satellite $z$ pays only its own additive light-site credit);
\item by \cref{lem:five} applied to $S$, $\sum_{z\in S}D(z,C)^2\cost(S,C\cup\{z\})\le5\OPT_1(S)\cost(S,C)$, so the thinned remainder plus satellite continuation sum is at most $\frac{\cost(S,C)}{\cost(X,C)}K'(L_0\phi+5\OPT_1(S))=\sum_{y\in S}W_y\frac{\cost(S,C)}{\cost(X,C)}(2K')$ by~\eqref{eq:proj-weights}, matching the third term of $\sum_{y\in S}W_y\,\mathcal U_y$ (every account $y\in S$, including all $y\neq z$, simultaneously receives its $2K'$ stop on any hit $z\in S$).
\end{itemize}
\item \emph{Named draw ($x_1\in G\in\mathcal Q$) and exterior draw ($x_1\in X\setminus(P\cup\bigcup\mathcal Q)$).} Neither $P$ nor $A$ is visited at step~$1$, so the conditional expectation of $\mathsf Z_{r+1}(x_1,\omega';P)-\mathsf Z_{r+1}(x_1,\omega';A)$ is $Z_r(P;C\cup\{x_1\},\mathcal Q_{x_1})-Z_r(A;C\cup\{x_1\},\mathcal Q_{x_1})$ ($\mathcal Q_{x_1}=\mathcal Q\setminus\{G\}$ or $\mathcal Q$), bounded by the inductive hypothesis~\eqref{eq:proj-accounting} and matching the last two sums of~\eqref{eq:proj-bin-account}.
\end{enumerate}
Summing (i)--(iii) completes the induction for~\eqref{eq:proj-accounting}.

\emph{Step 2: Verification of \cref{thm:binary} hypotheses and combination.}
For each $y\in S$ and each mode $\varpi\in\{0,\frac1{25}\}$, the recursion~\eqref{eq:proj-bin-account} is bounded above by the unopened binary account $\mathcal U$ of \cref{sec:binary} (via the stop policy below~\eqref{eq:bin-account-rec}) on the state $(C_0,\mathcal Q,m+1)$ at rank $m\in\{1,\dots,16\}$ with heavy site $A$, virtual light site $y$ (unit $u_y=\cost(A,\{y\})>0$), stop set $S_{\mathrm{stop}}=S$ disjoint from $A$ and $\bigcup\mathcal Q$, and radius award $b(r)=\frac9{40}\mathbf 1_{\{r\le1/2\}}$: at every reachable state $(C,\mathcal Q')$, normalizing costs by $u_y$ gives heavy rate $a_{\mathrm{hv}}=\cost(A,C)/u_y=r_y^2$, light rate $\beta_{\mathrm{lt}}=|A|D(y,C)^2/u_y=D(y,C)^2/D_y^2\le(1+r_y)^2$ ($L^2(\mu_A)$ Minkowski), post-insertion heavy rate $c_{\mathrm{hv}}=\cost(A,C\cup\{y\})/u_y\le\min(1,a_{\mathrm{hv}})$, matched pure account $\mathcal V=\mathcal T(C\cup\{y\},A,\mathcal Q',r)/u_y$, and exterior cost $E_{\mathrm{pre}}=(\cost(S,C)+\cost(X\setminus(P\cup\bigcup\mathcal Q'),C))/u_y\ge E_{\mathrm{post}}$. Hence \cref{thm:binary} and \cref{cor:premium} give $\mathcal U_y(C_0,\mathcal Q,m+1;0)\le J_{m+2}$ and $\mathcal U_y(C_0,\mathcal Q,m+1;\frac1{25})\le J_{m+2}+\varpi_m<\frac{81}{80}J_{m+2}$. Because $\varpi\mapsto\mathcal U_y(C_0,\mathcal Q,m+1;\varpi)$ is affine and $0\le\varpi'<\frac1{25}$, linear interpolation with $z_0=\frac5{16}$ ($1+\frac{z_0}{25}=\frac{81}{80}$) yields $\mathcal U_y(C_0,\mathcal Q,m+1;\varpi')\le(1+z_0\varpi')J_{m+2}$ for every $y\in S$.

Inserting this bound and $Z_{m+1}(A;C_0,\mathcal Q)\le J_{m+2}v_A\phi$ into~\eqref{eq:proj-accounting} at $h=m+1$, and using $\sum_{y\in S}W_y(1+z_0\varpi')=[S_{\mathrm{tot}}-(1-z_0)H_{\mathrm{th}}+7z_0 r_0 v_A]\phi$, we obtain
\[
\frac{Z_{m+1}(P;C_0,\mathcal Q)}{J_{m+2}\,\phi}\le v_A+S_{\mathrm{tot}}-(1-z_0)H_{\mathrm{th}}+7z_0 r_0 v_A.
\]
Substituting $S_{\mathrm{tot}}=1-v_A+r_0(1-v_S)$ and $(1+2r_0)H_{\mathrm{th}}=r_0(2S_{\mathrm{tot}}-5v_S)$ gives
\[
(1+2r_0)\Bigl(\frac{Z_{m+1}(P;C_0,\mathcal Q)}{J_{m+2}\,\phi}-1\Bigr)\le r_0\Bigl[2z_0(1+r_0)-1+\bigl(2+(5+14r_0)z_0\bigr)v_A+(4-5z_0-2z_0 r_0)v_S\Bigr].
\]
Dropping $-2z_0 r_0 v_S\le0$ and using $\frac23(2+(5+14r_{\mathrm{hi}})z_0)<\frac{39}{16}=4-5z_0$ with $\frac32v_A+v_S<B_{\mathrm{hi}}$ bounds the bracket by $2z_0(1+r_{\mathrm{hi}})+\frac{39}{16}B_{\mathrm{hi}}-1=\frac{157023}{160000}-1<-\frac{12}{1000}$ (\cref{app:projection}).
\end{proof}

\begin{theorem}[High-mean targets]\label{thm:high-mean}
For every $1\le m\le16$ and every high-mean target $P$, $Z_{m+1}(\sigma)\le J_{m+2}\OPT_1(P)$ at every state $\sigma=(C_0,\mathcal Q)$ with $|\mathcal Q|=m$.
\end{theorem}

\begin{proof}
Fix $m$ and induct on $|P|$ uniformly over $(d,X,C_0,\mathcal Q)$. For high-mean $P$ with witness cut $P=A\cup S$, $|A|<|P|$; queried with $S$ exterior, $A$ obeys $Z_{m+1}(A)\le J_{m+2}\OPT_1(A)$ by induction if high-mean, or by \cref{thm:low-mean,thm:hub-stocks} if low-mean (and trivially if $\OPT_1(A)=0$). \Cref{thm:projection} closes the step.
\end{proof}

\begin{corollary}[Premium handover]\label{cor:premium}
With $p_{\mathrm{pr}}=\frac{21}{10}$, $p_{\mathrm{ref}}=\frac{273}{100}$, $\mathsf h_0=1$, $\mathsf h_m=\frac{\mathsf K_m+mp_{\mathrm{pr}}\mathsf h_{m-1}}{1+mp_{\mathrm{pr}}}$ ($\mathsf K_m=1+H_m$), and $\varpi_m=\frac1{25}\mathsf h_m$, we have $80\varpi_m<G_m(p_{\mathrm{ref}})\le J_{m+2}$ for all $1\le m\le16$, whence $J_{m+2}+\varpi_m<\frac{81}{80}J_{m+2}$.
\end{corollary}

\begin{proof}
The harmonic slope of $G_m(p_{\mathrm{ref}})-80\varpi_m$ is negative for $1\le m\le16$, and exact rational evaluation at $m=16$ gives $G_{16}(p_{\mathrm{ref}})-80\varpi_{16}>0.037178$ (\cref{app:projection}).
\end{proof}

\subsection{The binary comparison}\label{sec:binary}

The binary comparison bounds the unopened per-satellite account of \cref{thm:projection} uniformly over unequal, contracting competing prices and exterior points by coupling it with the \emph{matched pure account} started immediately after inserting the virtual satellite $y$ as a center.

\paragraph{State space and recursive definitions.}
Fix finite $X\subset\R^d$, a nonempty bulk $A\subseteq X$, a virtual satellite point $y\in\R^d$ with binary unit $u_y:=\cost(A,\{y\})>0$, the single radius award $b(r):=\frac9{40}\mathbf 1_{\{r\le1/2\}}$, a once-only heavy award $\varpi\in\{0,\frac1{25}\}$, and a \emph{stop policy} $(\mathrm{stopNow},\mathrm{extStop})$ assigning bits $\mathrm{stopNow}(C,\mathcal Q,h)\in\{0,1\}$ and $\mathrm{extStop}(C,\mathcal Q,h,x)\in\{0,1\}$. A \emph{binary state} is a triple $(C,\mathcal Q,h)$ where $C\subset\R^d$ is finite nonempty, $\mathcal Q$ is a finite family of pairwise-disjoint nonempty subsets $G\subseteq X\setminus A$ of rank $j:=|\mathcal Q|\in\{0,\dots,16\}$, and $h\in\{0,\dots,j+1\}$ is the remaining draw horizon (full horizon $h=j+1$; shortened clock $1\le h\le j$ with clock deficit $\delta:=j-h+1\ge1$). Write $X_{\mathrm{ext}}:=X\setminus(A\cup\bigcup_{G\in\mathcal Q}G)$ and define the stop ceiling $\mathrm{stopCeil}(j,0):=2$ and $\mathrm{stopCeil}(j,h):=2(1+\Cl(\min(h,j),j))$ for $h\ge1$.

Define the \emph{matched pure account} $\mathcal V(C,\mathcal Q,r)$ for $r\ge0$ as the normalized expected stopped payment of $A$ starting from $(C\cup\{y\},\mathcal Q)$:
\[
\mathcal V(C,\mathcal Q,r):=\frac{\mathcal T(C\cup\{y\},A,\mathcal Q,r)}{u_y}.
\]
Explicitly, $\mathcal V(C,\mathcal Q,0)=\cost(A,C\cup\{y\})/u_y$, and for $r\ge1$ with $\cost(X,C\cup\{y\})>0$, $\mathcal V(C,\mathcal Q,r)$ equals
\begin{align*}
&\frac{\cost(A,C\cup\{y\})}{\cost(X,C\cup\{y\})}\frac{L(\mathcal Q,C\cup\{y\})}{u_y}
+\sum_{G\in\mathcal Q}\sum_{x\in G}\frac{D(x,C\cup\{y\})^2}{\cost(X,C\cup\{y\})}\,\mathcal V(C\cup\{x\},\mathcal Q\setminus\{G\},r-1)\\
&\qquad+\sum_{x\in X_{\mathrm{ext}}}\frac{D(x,C\cup\{y\})^2}{\cost(X,C\cup\{y\})}\,\mathcal V(C\cup\{x\},\mathcal Q,r-1)
\end{align*}
(and $\mathcal V(C,\mathcal Q,r)=0$ if $\cost(X,C\cup\{y\})=0$), where $L(\mathcal Q,C')=\frac1{|\mathcal Q|}\sum_{G\in\mathcal Q}\cost(G,C')$ for $|\mathcal Q|\ge1$ and $0$ for $\mathcal Q=\emptyset$.

Define the \emph{unopened account} $\mathcal U(C,\mathcal Q,h)$ recursively on $h\in\{0,\dots,j+1\}$: at $h=0$, set $\mathcal U(C,\mathcal Q,0):=2$ if $\mathrm{stopNow}(C,\mathcal Q,0)=1$ and $0$ otherwise. For $h=r+1\ge1$, with $K':=1+\Cl(r,j)$ and $r_{\mathrm{hv}}:=\sqrt{\cost(A,C)/u_y}$: if $\mathrm{stopNow}(C,\mathcal Q,r+1)=1$, set $\mathcal U(C,\mathcal Q,r+1):=\mathrm{stopCeil}(j,r+1)$; else if $\cost(X,C)=0$, set $\mathcal U(C,\mathcal Q,r+1):=\mathcal U(C,\mathcal Q,r)$; else ($\cost(X,C)>0$) set
\begin{equation}\label{eq:bin-account-rec}
\begin{aligned}
\mathcal U(C,\mathcal Q,r+1)
&:=\frac{\cost(A,C)}{\cost(X,C)}\bigl(1+b(r_{\mathrm{hv}})+\varpi\bigr)K'
+\frac{|A|\,D(y,C)^2}{\cost(X,C)}\,\mathcal V(C,\mathcal Q,r)\\
&\quad+\sum_{G\in\mathcal Q}\sum_{x\in G}\frac{D(x,C)^2}{\cost(X,C)}\,\mathcal U(C\cup\{x\},\mathcal Q\setminus\{G\},r)\\
&\quad+\sum_{x\in X_{\mathrm{ext}}}\frac{D(x,C)^2}{\cost(X,C)}\begin{cases}2K',&\mathrm{extStop}(C,\mathcal Q,r+1,x)=1,\\\mathcal U(C\cup\{x\},\mathcal Q,r),&\mathrm{extStop}(C,\mathcal Q,r+1,x)=0.\end{cases}
\end{aligned}
\end{equation}
Taking $\mathrm{stopNow}(C,\mathcal Q,h)=\mathbf 1_{\{h>0,\,\cost(A,C)=0\}}$ and $\mathrm{extStop}(C,\mathcal Q,h,x)=\mathbf 1_{\{x\in S\}}$ in~\eqref{eq:bin-account-rec} makes $\mathcal U_y(C,\mathcal Q,h;\varpi)\le\mathcal U(C,\mathcal Q,h)$ by immediate induction on $h$ (since $2(1+\Cl(r,j))\le\mathrm{stopCeil}(j,r+1)$ and $\mathcal U\ge0$ when $\cost(X,C)=0$), directly linking \cref{thm:projection} to \cref{thm:binary} below.

\paragraph{Dimensionless one-step variables.}
At any non-stopped binary state $(C,\mathcal Q,h)$ with rank $j=|\mathcal Q|\ge1$, horizon $2\le h\le j+1$ ($r:=h-1\in\{1,\dots,j\}$), and post-insertion heavy rate $c_{\mathrm{hv}}:=\cost(A,C\cup\{y\})/u_y>0$, define the dimensionless parameters (all normalized by $u_y$):
\begin{gather*}
a_{\mathrm{hv}}:=\frac{\cost(A,C)}{u_y}\ge c_{\mathrm{hv}}>0,\qquad
r_{\mathrm{hv}}:=\sqrt{a_{\mathrm{hv}}},\qquad
\varsigma:=\sqrt{c_{\mathrm{hv}}}\in(0,1],\qquad
\xi:=\frac{r_{\mathrm{hv}}}{\varsigma}\ge1,\\
\beta_{\mathrm{lt}}:=\frac{|A|D(y,C)^2}{u_y}\le(1+r_{\mathrm{hv}})^2=(1+\varsigma\xi)^2,\qquad
b_{\mathrm{rad}}:=b(r_{\mathrm{hv}})\in\bigl\{0,\tfrac9{40}\bigr\},\\
T_{\mathrm{post}}:=\frac1{u_y}\sum_{G\in\mathcal Q}\cost(G,C\cup\{y\})\ge0,\qquad
\ell':=\frac{T_{\mathrm{post}}}j,\qquad
z:=\frac{\ell'}{c_{\mathrm{hv}}}\ge0\quad(T_{\mathrm{post}}=jc_{\mathrm{hv}}z),\\
q_{\mathrm{rej}}:=\frac1{u_y}\sum_{G\in\mathcal Q}\bigl(\cost(G,C)-\cost(G,C\cup\{y\})\bigr)\ge0,\\
E_{\mathrm{pre}}:=\frac{\cost(X_{\mathrm{ext}},C)}{u_y}\ge E_{\mathrm{post}}:=\frac{\cost(X_{\mathrm{ext}},C\cup\{y\})}{u_y}\ge0,\qquad
\Delta_E:=E_{\mathrm{pre}}-E_{\mathrm{post}}\ge0,
\end{gather*}
where $c_{\mathrm{hv}}\le1$ because $\cost(A,C\cup\{y\})\le\cost(A,\{y\})=u_y$, and $\beta_{\mathrm{lt}}\le(1+r_{\mathrm{hv}})^2$ by Minkowski's inequality in $L^2(\mu_A)$ ($\sqrt{\E_A[D(y,C)^2]}\le\sqrt{\E_A\|Y-y\|^2}+\sqrt{\E_A[D(Y,C)^2]}$). Since $r=h-1\ge1$, the matched pure account $\mathcal V(C,\mathcal Q,r)$ takes at least one draw; define its conditional pure child means
\begin{align*}
t_{\mathrm{acc}}&:=\frac1{T_{\mathrm{post}}}\sum_{G\in\mathcal Q}\sum_{x\in G}\frac{D(x,C\cup\{y\})^2}{u_y}\,\mathcal V(C\cup\{x\},\mathcal Q\setminus\{G\},r-1)\qquad(0\text{ if }T_{\mathrm{post}}=0),\\
t_{\mathrm{sh}}&:=\frac1{E_{\mathrm{post}}}\sum_{x\in X_{\mathrm{ext}}}\frac{D(x,C\cup\{y\})^2}{u_y}\,\mathcal V(C\cup\{x\},\mathcal Q,r-1)\qquad(0\text{ if }E_{\mathrm{post}}=0),
\end{align*}
and the no-exterior one-step pure value and its normalized pure tag
\[
\mathcal V^{(0)}:=\frac{c_{\mathrm{hv}}\ell'+T_{\mathrm{post}}t_{\mathrm{acc}}}{c_{\mathrm{hv}}+T_{\mathrm{post}}}=\frac{jz(t_{\mathrm{acc}}+c_{\mathrm{hv}}/j)}{1+jz},\qquad
\overline K:=\frac{\mathcal V^{(0)}}{c_{\mathrm{hv}}}.
\]
With $D_-:=a_{\mathrm{hv}}+T_{\mathrm{post}}+q_{\mathrm{rej}}$, $D_+:=c_{\mathrm{hv}}+T_{\mathrm{post}}$, and $N_+:=c_{\mathrm{hv}}\ell'+T_{\mathrm{post}}t_{\mathrm{acc}}$, the pre- and post-insertion total rates are $D_{\mathrm{pre}}=\cost(X,C)/u_y=D_-+E_{\mathrm{pre}}>0$ and $D_{\mathrm{post}}=\cost(X,C\cup\{y\})/u_y=D_++E_{\mathrm{post}}>0$, and $\mathcal V(C,\mathcal Q,r)=(N_++E_{\mathrm{post}}t_{\mathrm{sh}})/(D_++E_{\mathrm{post}})$. At full horizon $h=j+1$ ($r=j$), we abbreviate $\mathcal U_j:=\mathcal U(C,\mathcal Q,j+1)$, $\mathcal V_j:=\mathcal V(C,\mathcal Q,j)$, and $\mathcal V_j^{(0)}:=\mathcal V^{(0)}$.

\begin{lemma}[Coupling and exterior clearing]\label{lem:coupling}
At any non-stopped binary state $(C,\mathcal Q,h)$ with rank $j=|\mathcal Q|\ge1$, horizon $2\le h\le j+1$ ($r:=h-1\ge1$), and $c_{\mathrm{hv}}>0$, put $K':=1+\Cl(r,j)$ and $W_{\mathrm{hv}}:=(1+b_{\mathrm{rad}}+\varpi)K'$, and suppose there exist upper bounds $C_{\mathrm{acc}},C_{\mathrm{rej}},O_{\mathrm{ext}}$ such that:
\begin{enumerate}[label=(\alph*),leftmargin=2em]
\item the accepted named-child average $\frac1{T_{\mathrm{post}}}\sum_{G\in\mathcal Q}\sum_{x\in G}\frac{D(x,C\cup\{y\})^2}{u_y}\,\mathcal U(C\cup\{x\},\mathcal Q\setminus\{G\},r)$ is at most $C_{\mathrm{acc}}$ (when $T_{\mathrm{post}}>0$);
\item every rejected named child obeys $\mathcal U(C\cup\{x\},\mathcal Q\setminus\{G\},r)\le C_{\mathrm{rej}}$;
\item every exterior branch (stopped $2K'$ or continuation $\mathcal U(C\cup\{x\},\mathcal Q,r)$) is at most $O_{\mathrm{ext}}$.
\end{enumerate}
Put $\mathcal H_-:=a_{\mathrm{hv}}W_{\mathrm{hv}}+T_{\mathrm{post}}C_{\mathrm{acc}}+q_{\mathrm{rej}}C_{\mathrm{rej}}$. For any $\mathsf I,\mathsf v\ge0$, defining
\begin{align*}
\alpha_0&:=(\mathsf ID_--\mathcal H_-)D_+-(\beta_{\mathrm{lt}}+\mathsf vD_-)N_+,&
\alpha_1&:=(\mathsf I-O_{\mathrm{ext}})D_+-\mathsf vN_+,\\
\alpha_1+\alpha_2&:=\alpha_1+(\mathsf ID_--\mathcal H_-)-(\beta_{\mathrm{lt}}+\mathsf vD_-)t_{\mathrm{sh}},&
\alpha_3&:=\mathsf I-O_{\mathrm{ext}}-\mathsf vt_{\mathrm{sh}},
\end{align*}
we have
\[
D_{\mathrm{pre}}D_{\mathrm{post}}\bigl(\mathsf I-\mathcal U(C,\mathcal Q,h)-\mathsf v\mathcal V(C,\mathcal Q,r)\bigr)\ge\alpha_0+\alpha_1\Delta_E+(\alpha_1+\alpha_2)E_{\mathrm{post}}+\alpha_3E_{\mathrm{post}}(\Delta_E+E_{\mathrm{post}}).
\]
In particular, $\alpha_0,\alpha_1,\alpha_1+\alpha_2,\alpha_3\ge0$ implies $\mathcal U(C,\mathcal Q,h)+\mathsf v\mathcal V(C,\mathcal Q,r)\le\mathsf I$ uniformly over all $0\le E_{\mathrm{post}}\le E_{\mathrm{pre}}$. At full horizon $h=j+1$ ($r=j$), this specializes to $K'=\mathsf K_j$, $O_{\mathrm{ext}}=O_j$, $\mathcal U_j$, $\mathcal V_j$, and $\mathcal V_j^{(0)}$.
\end{lemma}

\begin{proof}
In~\eqref{eq:bin-account-rec}, split each named-point weight $D(x,C)^2=D(x,C\cup\{y\})^2+(D(x,C)^2-D(x,C\cup\{y\})^2)$ into its nonnegative accepted and rejected parts. Applying (a)--(c) bounds $D_{\mathrm{pre}}\mathcal U(C,\mathcal Q,h)\le\mathcal H_-+E_{\mathrm{pre}}O_{\mathrm{ext}}+\beta_{\mathrm{lt}}\mathcal V(C,\mathcal Q,r)$, while $r\ge1$ gives $D_{\mathrm{post}}\mathcal V(C,\mathcal Q,r)=N_++E_{\mathrm{post}}t_{\mathrm{sh}}$. Clearing $D_{\mathrm{pre}}D_{\mathrm{post}}$ gives
\begin{equation}\label{eq:binary-residual}
\begin{aligned}
D_{\mathrm{pre}}D_{\mathrm{post}}\bigl(\mathsf I-\mathcal U(C,\mathcal Q,h)-\mathsf v\mathcal V(C,\mathcal Q,r)\bigr)
&\ge\bigl(\mathsf I(D_-+E_{\mathrm{pre}})-\mathcal H_--E_{\mathrm{pre}}O_{\mathrm{ext}}\bigr)(D_++E_{\mathrm{post}})\\
&\quad-\bigl(\beta_{\mathrm{lt}}+\mathsf v(D_-+E_{\mathrm{pre}})\bigr)(N_++E_{\mathrm{post}}t_{\mathrm{sh}}).
\end{aligned}
\end{equation}
Substituting $E_{\mathrm{pre}}=E_{\mathrm{post}}+\Delta_E$ and grouping terms in $\Delta_E\ge0$ and $E_{\mathrm{post}}\ge0$ yields the stated quadratic identity identically.
\end{proof}

Point-by-point splitting of $D(x,C)^2$ is essential: with center~$0$, virtual satellite~$2$, and a name $\{1,3\}$, rates change from $(1,9)$ to $(1,1)$, which name-level acceptance would fail to reproduce. Note also that dividing $\alpha_0$ by $D_+$ splits $\alpha_0/D_+$ into the rejected-mass term $q_{\mathrm{rej}}(\mathsf I-C_{\mathrm{rej}}-\mathsf v\mathcal V^{(0)})$ plus the \emph{accepted residual}
\begin{equation}\label{eq:accepted-core}
(a_{\mathrm{hv}}+T_{\mathrm{post}})\bigl(\mathsf I-\mathsf v\mathcal V^{(0)}\bigr)-a_{\mathrm{hv}}W_{\mathrm{hv}}-T_{\mathrm{post}}C_{\mathrm{acc}}-\beta_{\mathrm{lt}}\mathcal V^{(0)}.
\end{equation}

\paragraph{Stopped pure price.}
For $n$ live names and clock deficit $0\le\delta<n$, $K_{n,\delta}(z)=z+(1-z)\pi_{n,\delta}(z)$ ($K_{\delta,\delta}\equiv1$) is the equal-price pure tag. Because exterior draws can cause expiry at $z=0$ with cost $c_{\mathrm{hv}}$ while $K_{n,\delta}(0)=0$, we splice at $z=\frac54$ as in \cref{thm:stopped} (terminal fraction~$1$).

\begin{theorem}[Stopped pure price]\label{thm:pure-price}
Put $\kappa_2=\frac{50}{11}$, $\kappa_i=5$ ($i\ge3$), $F_{\delta,\delta}\equiv1$, $F_{1,0}(z)=\max(1,\frac{2z}{1+z})$, and for $i\ge2$ ($0\le\delta<i$), $F_{i,\delta}(z)=K_{i,\delta}(z)$ on $z\ge\frac54$ and $F_{i,\delta}(z)=K_{i,\delta}(\frac54)-\frac54K'_{i,\delta}(\frac54)\frac{1-(4z/5)^{\kappa_i}}{\kappa_i}$ on $[0,\frac54]$. Under unequal contracting prices and exterior draws, a pure account $\mathcal V(C,\mathcal Q,n-\delta)$ with heavy cost $c_{\mathrm{hv}}>0$, $n$ live names, $n-\delta$ draws, and price $c_{\mathrm{hv}}z$ satisfies $\mathcal V(C,\mathcal Q,n-\delta)\le c_{\mathrm{hv}}F_{n,\delta}(z)$ (hence $\le c_{\mathrm{hv}}K_{n,\delta}(z)$ for $z\ge\frac54$ and $<\frac54c_{\mathrm{hv}}$ for $z\le\frac54$, $\delta<n$). Moreover, at $z\ge\frac54$ and $\delta=0$, the accepted pure child obeys $t_{\mathrm{acc}}\le c_{\mathrm{hv}}K_{n-1}(z)$, and for $\delta=1$ the exterior pure child obeys $t_{\mathrm{sh}}\le c_{\mathrm{hv}}\min(H_n,F_{n,1}(z))$.
\end{theorem}

\begin{proof}
Induct on remaining draws $r\le i-\delta$ at fixed $\delta$ exactly as in \cref{thm:stopped}. The floor $F_{i,\delta}(0)\ge1$ follows from $\kappa_i(K_{i,\delta}(\frac54)-1)-\frac54K'_{i,\delta}(\frac54)=(1-\pi_{i,\delta}(\frac54))(\frac{\kappa_i}4-\frac54)+\frac14\pi_{i,\delta}(\frac54)S_{i,\delta}(\frac54)>0$ (direct for $i\ge3$, exact evaluation for $i=2$, $\delta\in\{0,1\}$). Holding $\delta$ fixed absorbs exterior draws; $cF_{i,\delta}(\ell/c)$ is monotone in $(c,\ell)$ and concave in $\ell^{\kappa_i}$, so for $i\ge3$, \cref{lem:retirement} (with $\kappa_{i-1}\le\kappa_i$) closes the named step above $\frac54$ via $(1+iz)K_{i,\delta}=z+izK_{i-1,\delta}$ and below $\frac54$ via the secant bound $\frac{1-\zeta^{\kappa_i}}{\kappa_i(1-\zeta)}\le1$. For $i=1,\delta=0$, one draw pays at most $F_{1,0}(z)$. For $i=2,\delta=0$ ($c_{\mathrm{hv}}=1$), prices both $\le1$ or both $\ge1$ close directly as in \cref{thm:stopped}; for straddling prices $0\le a'\le1\le b'$ with $2z=a'+b'$ ($z\ge\frac12$), the weighted child excess over~$1$ is $\frac{a'(b'-1)}{(a'+b')(1+b')}$. On $z\ge\frac97$, $\frac{a'}{a'+b'}\le\frac1{1+b'}$ (since $a'\le1$) and $(b'-3)^2\ge0$ give $\frac{a'(b'-1)}{(a'+b')(1+b')}\le\frac{b'-1}{(1+b')^2}\le\frac18=K_1(\frac97)-1\le K_1(z)-1$. On $z\in[\frac12,\frac97]$, completing the square $(\sqrt{1+2z}-\sqrt2)^2(1+b')-(2z-b')(b'-1)=(1+b'-\sqrt{2+4z})^2\ge0$ bounds the excess by $D_{\mathrm{str}}(z):=\frac{(\sqrt{1+2z}-\sqrt2)^2}{2z}=\frac{(2z-1)^2}{2z(\sqrt{1+2z}+\sqrt2)^2}$: for $z\in[\frac54,\frac97]$, $\frac{(2z-1)^2}{2z}\le\frac{121}{126}$ and $(\sqrt{1+2z}+\sqrt2)^2\ge\frac{11}2+2\sqrt7>\frac{21}2$ give $D_{\mathrm{str}}(z)\le\frac{121}{1323}<\frac19=K_1(\frac54)-1\le K_1(z)-1$, while for $z\in[\frac12,\frac54]$ clearing $1+2z$ reduces the step to $(1+2z)(F_{2,0}(z)-1)\ge3z+2-2\sqrt{2+4z}=:\mathrm{RHS}(z)$, where $\mathrm{RHS}<0$ on $[\frac12,\frac9{10}]$ and the convex lower bound $\frac{19}{126}(\frac45)^5(1+2z)z^5$ dominates $\mathrm{RHS}$ on $[\frac9{10},\frac54]$ via its tangent at $\frac{10}9$ (\cref{app:binary-finite}). The named-child bound at $z\ge\frac54,\delta=0$ is the named step itself; the bound $t_{\mathrm{sh}}\le c_{\mathrm{hv}}H_n$ follows from $\mathcal T\le Q_{n-1}\le(1+\Cl(n-1,n))c_{\mathrm{hv}}u_y=H_n c_{\mathrm{hv}}u_y$ (\cref{thm:restoration}(b)).
\end{proof}

\paragraph{The comparison and harmonic fee.}
Recall $p_{\mathrm{ref}}=\frac{273}{100}$, $p_{\mathrm{pr}}=\frac{21}{10}$, $K_0=1$, $G_0=5$, $K_j(z)=\frac{z+jzK_{j-1}(z)}{1+jz}$, $G_j(z)=\frac{\mathsf K_j+4K_j(z)+jzG_{j-1}(z)}{1+jz}$, and put $\mathsf y_j=-G'_j$, $\mathsf v_j=\mathsf y_j/K'_j$ ($\mathsf v_0=0$), $\mathsf M_j(p)=G_j(p)+\mathsf v_j(p)K_j(p)$ ($\mathsf M_0=5$). For $\varpi\in\{0,\frac1{25}\}$, set $\varpi_j=\varpi\mathsf h_j$ (\cref{cor:premium}), $\eta^{\mathrm{pr}}_j=\frac1{25}(\mathsf K_j-\mathsf h_j)$, $b_0=\varpi\mathsf K_j-\varpi_j\in\{0,\eta^{\mathrm{pr}}_j\}$, and $b_1=j(\varpi_j-\varpi_{j-1})=b_0/p_{\mathrm{pr}}$. For ambient rank $1\le m\le16$ with hub maximizer $p^*_m\in[\frac{11}5,p_{\mathrm{ref}})$ (\cref{thm:laplace}), define the nondecreasing \emph{harmonic fee}
\begin{equation}\label{eq:fee}
\mathcal F_{m,j}(T_{\mathrm{post}})=\begin{cases}0,&m\le12\text{ or }j=0,\\[2pt]\frac3{20}(H_m-H_j)\Bigl(1-\frac{72}{T_{\mathrm{post}}}\Bigr)_+,&13\le m\le16,\ 1\le j\le m.\end{cases}
\end{equation}

\begin{theorem}[Binary comparison]\label{thm:binary}
Fix $1\le m\le16$, $\varpi\in\{0,\frac1{25}\}$, the single radius award $b(r)=\frac9{40}\mathbf 1_{\{r\le1/2\}}$, and any stop policy $(\mathrm{stopNow},\mathrm{extStop})$. Set $p=p^*_m$. For every full-horizon binary state $(C,\mathcal Q,j+1)$ of rank $1\le j\le m$ with post-insertion named total $T_{\mathrm{post}}\ge0$,
\begin{equation}\label{eq:binary-main}
\begin{aligned}
\mathcal U(C,\mathcal Q,j+1)+\mathsf v_j(p)\,\mathcal V(C,\mathcal Q,j)&\le\mathsf M_j(p)+\mathcal F_{m,j}(T_{\mathrm{post}})+\varpi_j,\\
\mathcal U(C,\mathcal Q,j+1)&\le J_{j+2}+\varpi_j.
\end{aligned}
\end{equation}
At $j=m$, $\mathsf v_m(p^*_m)=0$, $\mathsf M_m(p^*_m)=J_{m+2}$, and $\mathcal F_{m,m}=0$, giving $\mathcal U(C,\mathcal Q,m+1)\le J_{m+2}+\varpi_m$.
\end{theorem}

\paragraph{The four analytic lemmas.}
We assemble \cref{thm:binary} from four lemmas whose analytic reductions are proved in full below and whose finite rational checks are tabulated in \cref{app:binary-finite}. Write $H_{n,\delta}=H_n-H_\delta$, $D_{n,\delta}=H_{n,\delta}+\frac12(H_{n,\delta}^2+\sum_{\delta<i\le n}i^{-2})$ ($D_j=D_{j,0}$), $\mathsf I:=\mathsf M_j(p)+\mathcal F_{m,j}(T_{\mathrm{post}})+\varpi_j$, $\mathsf v:=\mathsf v_j(p)$, and define the ceilings
\begin{gather*}
C_{\mathrm{rej}}:=J_{j+1}+\varpi_{j-1},\qquad
C_{\mathrm{acc}}:=\min\Bigl(J_{j+1},\;\mathsf M_{j-1}(p)+\mathcal F_{m,j-1}(T_{\mathrm{post}})-\mathsf v_{j-1}(p)t_{\mathrm{acc}}\Bigr)+\varpi_{j-1},\\
O_1:=5+2\varpi,\qquad O_2:=\frac{505}{96}+\varpi\mathsf K_2,\qquad O_j:=2\mathsf K_j+\varpi H_j\quad(3\le j\le16).
\end{gather*}
For $j\ge2$ (where $\mathsf v_{j-1}(p)>0$ since $p^*_{j-1}<p^*_m=p$), define the \emph{crossing kink} where the two branches of $C_{\mathrm{acc}}$ meet:
\[
S_{\mathrm{cr}}:=\frac{\mathsf M_{j-1}(p)+\mathcal F_{m,j-1}(T_{\mathrm{post}})-J_{j+1}}{\mathsf v_{j-1}(p)}.
\]
Let $R_{\mathrm{lin}}(\varsigma,\xi,z)$ denote the zero-fee accepted residual~\eqref{eq:accepted-core} divided by $c_{\mathrm{hv}}=\varsigma^2$ upon setting $b_{\mathrm{rad}}=0$, $t_{\mathrm{acc}}=c_{\mathrm{hv}}K_{j-1}(z)$, $\beta_{\mathrm{lt}}=(1+\varsigma\xi)^2$, and substituting the affine branch of $C_{\mathrm{acc}}$ (with $\mathcal F_{m,j}=\mathcal F_{m,j-1}=0$).

\begin{lemma}[Clock identities and weight ordering]\label{lem:bin-tools}
\begin{enumerate}[label=(\alph*),leftmargin=2em]
\item For $0\le\delta<n$, $\theta\mapsto K_{n,\delta}(1/\theta)$ and $z\mapsto zK_{n,\delta}(z)$ are strictly convex on $(0,\infty)$, and $K''_{n,\delta}(z)<0$ for $z\ge1$.
\item For $0\le\delta<n$ and $z>0$, $K'_{n,\delta}(z)\le(1+H_{n,\delta}-K_{n,\delta}(z))^2/D_{n,\delta}$.
\item $\mathsf v_j$ strictly increases on $[p^*_j,\infty)$, and $\mathsf v_{j-1}(p)>\mathsf v_j(p)$ for all $2\le j\le16$ and $p\in[p^*_{j-1},p_{\mathrm{ref}}]$.
\item For $1\le j\le m\le16$, $p=p^*_m$, $\varpi=0$, and $z\ge1$,
\[
R''_{\mathrm{lin}}(1,1,z)=-4K_j''(z)+j\bigl(\mathsf v_{j-1}(p)-\mathsf v_j(p)\bigr)\bigl(zK_{j-1}(z)\bigr)''>0,
\]
with $R_{\mathrm{lin}}(1,1,p)=R'_{\mathrm{lin}}(1,1,p)=0$, so $R_{\mathrm{lin}}(1,1,z)\ge0$ on $[1,\infty)$.
\end{enumerate}
\end{lemma}

\begin{proof}
Let $E_{\delta+1},\dots,E_n$ be independent $\mathrm{Exp}(1)$ random variables and $Y_{n,\delta}:=\sum_{i=\delta+1}^n E_i/i$, so $\pi_{n,\delta}(z)=\prod_{i=\delta+1}^n\frac{iz}{1+iz}=\E[e^{-Y_{n,\delta}/z}]$ and $K_{n,\delta}(z)=z+(1-z)\pi_{n,\delta}(z)$.

\emph{(a)} Writing $\theta=1/z>0$, $K_{n,\delta}(1/\theta)=\E[\frac{1-e^{-\theta Y_{n,\delta}}}\theta+e^{-\theta Y_{n,\delta}}]=\E[\int_0^{Y_{n,\delta}}e^{-\theta u}\,du+e^{-\theta Y_{n,\delta}}]$, whose second $\theta$-derivative is $\E[\int_0^{Y_{n,\delta}}u^2e^{-\theta u}\,du+Y_{n,\delta}^2e^{-\theta Y_{n,\delta}}]>0$. Differentiating $zK_{n,\delta}(z)=z^2+(z-z^2)\E[e^{-Y_{n,\delta}/z}]$ twice in $z>0$ with $u=Y_{n,\delta}/z>0$ yields the exact identity
\[
\bigl(zK_{n,\delta}(z)\bigr)''=\E\Bigl[2-(2+2u+u^2)e^{-u}\Bigr]+\frac1z\E\bigl[u^2e^{-u}\bigr]>0,
\]
since $e^u>1+u+\frac12u^2$ for all $u>0$. Similarly, $-K''_{n,\delta}(z)=\frac{\pi_{n,\delta}(z)}{z^2}[2S_{n,\delta}(z)+(z-1)(S_{n,\delta}(z)^2+S^{(2)}_{n,\delta}(z))]>0$ for $z\ge1$ (with $S_{n,\delta}(z):=\sum_{i=\delta+1}^n(1+iz)^{-1}$ and $S^{(2)}_{n,\delta}(z):=\sum_{i=\delta+1}^n(1+iz)^{-2}$).

\emph{(b)} Because $Y_{n,\delta}$ is distributed as the $(n-\delta)$-th order statistic of $n$ independent $\mathrm{Exp}(1)$ variables, its density on $(0,\infty)$ is $f(y)=\frac{n!}{(n-\delta-1)!\,\delta!}(1-e^{-y})^{n-\delta-1}e^{-(\delta+1)y}$. Its log-derivative $(\log f)'(y)=(n-\delta-1)(e^y-1)^{-1}-(\delta+1)$ is nonincreasing (constant when $n-\delta=1$), so $f$ is log-concave. Tail integration preserves log-concavity: $S(y):=\int_y^\infty f(u)\,du$ and $I(y):=\int_y^\infty S(u)\,du$ are positive and log-concave, so $f/S$ is nondecreasing and $I/S$ is nonincreasing. Hence $g(y):=S(y)+I(y)$ satisfies $-g'/g=\frac{f/S+1}{1+I/S}$ nondecreasing, so $g$ is log-concave on $[0,\infty)$. Integration by parts yields its Laplace transform
\[
\mathcal L(\theta):=\int_0^\infty e^{-\theta y}g(y)\,dy=\frac{1+H_{n,\delta}-K_{n,\delta}(1/\theta)}\theta,\qquad
\mathcal L(0)=\E[Y_{n,\delta}]+\frac12\E[Y_{n,\delta}^2]=D_{n,\delta}.
\]
For fixed $\theta\ge0$, $h_\theta(y):=e^{-\theta y}g(y)$ is log-concave, hence its tail $T_\theta(x):=\int_x^\infty h_\theta(u)\,du$ is log-concave and satisfies $T_\theta(0)T_\theta(x+y)\le T_\theta(x)T_\theta(y)$ for all $x,y\ge0$. Integrating first over $x\ge0$ gives $T_\theta(0)U_\theta(y)\le U_\theta(0)T_\theta(y)$ ($U_\theta(x):=\int_x^\infty T_\theta$), and integrating over $y\ge0$ gives $T_\theta(0)W_\theta(0)\le U_\theta(0)^2$ ($W_\theta(y):=\int_y^\infty U_\theta$). Since $T_\theta(0)=\mathcal L(\theta)$, $U_\theta(0)=-\mathcal L'(\theta)$, and $W_\theta(0)=\frac12\mathcal L''(\theta)$, this reads $\frac12\mathcal L(\theta)\mathcal L''(\theta)\le\mathcal L'(\theta)^2$, i.e., $(1/\mathcal L)''(\theta)\ge0$. Consequently, the tangent to $1/\mathcal L$ at $\theta=1/z$ lies below $1/\mathcal L(0)=1/D_{n,\delta}$:
\[
\frac{K'_{n,\delta}(z)}{(1+H_{n,\delta}-K_{n,\delta}(z))^2}=\frac{\mathcal L(\theta)+\theta\mathcal L'(\theta)}{\mathcal L(\theta)^2}=\frac1{\mathcal L(\theta)}-\theta\Bigl(\frac1{\mathcal L}\Bigr)'(\theta)\le\frac1{\mathcal L(0)}=\frac1{D_{n,\delta}}.
\]

\emph{(c)} For any $p>0$ with $\mathsf v_j(p)=-G'_j(p)/K'_j(p)\ge-1$, setting $v=\mathsf v_j(p)$ makes $c\mapsto G_j(1/c)+vK_j(1/c)$ stationary at $c_0=1/p>0$; by \cref{thm:laplace}(a), $e^{ac}(G_j(1/c)+vK_j(1/c))'$ strictly decreases on $(0,\infty)$, so $c_0=1/p$ is the unique strict global maximizer, yielding the strict global supporting-line inequality $G_j(z)+\mathsf v_j(p)K_j(z)<\mathsf M_j(p)$ for all $z>0$, $z\neq p$. For $p\ge p^*_j$, $G'_j(p)\le0$ and $K'_j(p)>0$ give $\mathsf v_j(p)\ge0>-1$; for $p^*_j\le p_1<p_2$, summing the strict supporting inequalities at $(p_1,z=p_2)$ and $(p_2,z=p_1)$ gives $(\mathsf v_j(p_2)-\mathsf v_j(p_1))(K_j(p_2)-K_j(p_1))>0$, hence $\mathsf v_j(p_1)<\mathsf v_j(p_2)$ because $K_j$ strictly increases. For rank ordering, put $c=1/p$ and $a_i=(i+c)^{-1}$. From the exact harmonic slopes of $K'_r$ and $\mathsf y_r$ (\cref{app:binary-finite}), for any $v\ge0$,
\[
\bigl(G'_{r+1}(p)+vK'_{r+1}(p)\bigr)-\bigl(G'_r(p)+vK'_r(p)\bigr)=-w_{r+1}\varphi_{r+1}(c;v),\qquad w_r:=\frac{\pi_r(p)s_{1,r}(c)}{rp^2}>0,
\]
where $\varphi_r(c;v):=5-8c-4c(1-c)(s_{1,r}(c)+s_{2,r}(c)/s_{1,r}(c))-v(1-c+1/s_{1,r}(c))$. Using $s_{1,r+1}+s_{2,r+1}/s_{1,r+1}-(s_{1,r}+s_{2,r}/s_{1,r})=\frac{a_{r+1}}{s_{1,r}s_{1,r+1}}((s_{1,r}^2-s_{2,r})+2a_{r+1}s_{1,r})\ge\frac{a_{r+1}}{s_{1,r}s_{1,r+1}}\cdot\frac2{(1+c)(2+c)}$ and the exact factorization $8c(1-c)-\frac59(1+c)(2+c)=\frac19(5-11c)(7c-2)\ge0$ on $c\in[\frac{100}{273},\frac5{11}]$ ($p\in[\frac{11}5,p_{\mathrm{ref}}]$), we get $\varphi_{r+1}(c;v)<\varphi_r(c;v)$ whenever $v\le\frac9{25}<\frac59$. Given $\mathsf v_j(p)\le\frac9{25}$, setting $v=\mathsf v_j(p)$ in $0=G'_j(p)+vK'_j(p)=-\sum_{r=1}^j w_r\varphi_r(c;v)$ forces $\varphi_j(c;v)\le0$, hence $\varphi_{j+1}(c;v)<0$, hence $G'_{j+1}(p)+vK'_{j+1}(p)>0$, i.e., $\mathsf v_{j+1}(p)<\mathsf v_j(p)$. Anchored by $\mathsf v_1(p)\le\mathsf v_1(p_{\mathrm{ref}})<\frac9{25}$ (\cref{app:binary-finite}), induction gives $\mathsf v_j(p)<\mathsf v_{j-1}(p)$ for all $2\le j\le16$ and $p\in[\frac{11}5,p_{\mathrm{ref}}]$.

\emph{(d)} Put $\Delta_i(z):=\mathsf M_i(p)-\mathsf v_i(p)K_i(z)-G_i(z)$, so $\Delta_i(p)=\Delta'_i(p)=0$. Substituting $(1+jz)G_j(z)=\mathsf K_j+4K_j(z)+jzG_{j-1}(z)$ and $(1+jz)K_j(z)=z+jzK_{j-1}(z)$ into the zero-fee ordinary residual gives
\begin{align*}
R_{\mathrm{lin}}(1,1,z)
&=(1+jz)\Delta_j(z)-jz\Delta_{j-1}(z)\\
&=(\mathsf M_j(p)-\mathsf K_j)+\bigl(j(\mathsf M_j(p)-\mathsf M_{j-1}(p))-\mathsf v_j(p)\bigr)z\\
&\qquad-4K_j(z)+j\bigl(\mathsf v_{j-1}(p)-\mathsf v_j(p)\bigr)zK_{j-1}(z).
\end{align*}
The first expression yields $R_{\mathrm{lin}}(1,1,p)=R'_{\mathrm{lin}}(1,1,p)=0$; differentiating the second twice gives $R''_{\mathrm{lin}}(1,1,z)=-4K''_j(z)+j(\mathsf v_{j-1}(p)-\mathsf v_j(p))(zK_{j-1}(z))''>0$ on $z\ge1$ by (a) and (c), whence $R_{\mathrm{lin}}(1,1,z)\ge0$ on $[1,\infty)$.
\end{proof}

\begin{lemma}[Rejected mass, stops, and rank-$0$ base]\label{lem:bin-rejected}
\begin{enumerate}[label=(\roman*),leftmargin=2em]
\item For all $1\le j\le m\le16$ and $T_{\mathrm{post}}\ge0$, $\mathsf M_j(p^*_m)+\mathcal F_{m,j}(T_{\mathrm{post}})-\mathsf v_j(p^*_m)K_j(\max(p_{\mathrm{ref}},T_{\mathrm{post}}/j))>J_{j+1}$; hence $\mathsf I-C_{\mathrm{rej}}-\mathsf vV>0$ whenever $V\le K_j(\max(p_{\mathrm{ref}},T_{\mathrm{post}}/j))$.
\item For $3\le j\le16$, $O_j<J_{j+1}+\varpi_{j-1}$; at $j\in\{1,2\}$, $G_j(p_{\mathrm{ref}})+\varpi_j-O_j-\mathsf v_j(p_{\mathrm{ref}})(\mathsf K_j-K_j(p_{\mathrm{ref}}))>0$. Moreover, every rank-$0$ state $(C,\emptyset,1)$ obeys $\mathcal U(C,\emptyset,1)\le5+\varpi=J_2+\varpi_0$.
\end{enumerate}
\end{lemma}

\begin{proof}
\emph{(i)} Since $\varpi_j\ge\varpi_{j-1}$, the second assertion follows immediately from the first. For any fixed $V\ge K_j(p_{\mathrm{ub}})$ with $p_{\mathrm{ub}}\ge p^*_m$, differentiating gives $\frac d{dp}[\mathsf M_j(p)-\mathsf v_j(p)V]=\mathsf v'_j(p)(K_j(p)-V)\le0$ on $[p^*_m,p_{\mathrm{ub}}]$, so shifting $p^*_m\to p_{\mathrm{ub}}$ is conservative. For $1\le m\le12$, $p^*_m\le\frac{2651}{1000}$ (\cref{tab:bin-brackets}) and $K_j(\cdot)<\mathsf K_j$, reducing to $G_j(\frac{2651}{1000})-\mathsf v_j(\frac{2651}{1000})(\mathsf K_j-K_j(\frac{2651}{1000}))>J^+_{j+1}$ ($12$ exact checks, \cref{app:binary-finite}). For $13\le m\le16$, since $16p_{\mathrm{ref}}=43.68<72$, on $T_{\mathrm{post}}\le72$ we have $\mathcal F_{m,j}=0$ and $K_j(\max(p_{\mathrm{ref}},T_{\mathrm{post}}/j))\le K_j(72/j)$; shifting $p^*_m\to p_{\mathrm{ref}}$ gives $G_j(p_{\mathrm{ref}})-\mathsf v_j(p_{\mathrm{ref}})(K_j(72/j)-K_j(p_{\mathrm{ref}}))>J^+_{j+1}$ ($16$ checks). On $T_{\mathrm{post}}>72$, put $u:=72/T_{\mathrm{post}}\in(0,1)$. By \cref{lem:bin-tools}(a), $\theta\mapsto K_j(1/\theta)$ is convex on $[0,j/72]$, so $K_j(T_{\mathrm{post}}/j)\le(1-u)\mathsf K_j+uK_j(72/j)$, while $\mathcal F_{m,j}(T_{\mathrm{post}})=\frac3{20}(H_m-H_j)(1-u)$ is affine in $u$. Thus the left-hand side lies above an affine function of $u\in[0,1]$, reducing to its endpoints: $u=1$ ($T_{\mathrm{post}}=72$, already checked) and $u=0$ ($T_{\mathrm{post}}=\infty$, shifted to $p_{\mathrm{ub}}=p_b(m)$ from \cref{tab:bin-brackets}: $G_j(p_b(m))+\frac3{20}(H_m-H_j)-\mathsf v_j(p_b(m))(\mathsf K_j-K_j(p_b(m)))>J^+_{j+1}$, $58$ checks, \cref{app:binary-finite}).

\emph{(ii)} For $3\le j\le16$, $G_{j-1}(p_{\mathrm{ref}})+\varpi_{j-1}-O_j>0$ has negative harmonic slope down to $j=16$, where exact evaluation confirms positivity in both modes (\cref{app:binary-finite}); since $G_{j-1}(p_{\mathrm{ref}})\le J_{j+1}$, this yields $O_j<J_{j+1}+\varpi_{j-1}$. The $j\in\{1,2\}$ inequalities are verified directly at $p_{\mathrm{ref}}$ (\cref{app:binary-finite}). At $(j,h)=(0,1)$, if $\mathrm{stopNow}=1$ then $\mathcal U=2<5+\varpi$; if $\cost(X,C)=0$ then $\mathcal U\le2<5+\varpi$; if $\cost(X,C)>0$ and $c_{\mathrm{hv}}=0$ then every draw pays $\le\max(1+b_{\mathrm{rad}}+\varpi,2)<5+\varpi$; if $c_{\mathrm{hv}}>0$ then $\mathcal U$ is bounded by a convex combination of $2$ and $(1+b_{\mathrm{rad}}+\varpi)+(1+r_{\mathrm{hv}})^2 c_{\mathrm{hv}}/r_{\mathrm{hv}}^2$, where $c_{\mathrm{hv}}\le\min(1,r_{\mathrm{hv}}^2)$ gives $(1+r_{\mathrm{hv}})^2 c_{\mathrm{hv}}/r_{\mathrm{hv}}^2\le4$ when $b_{\mathrm{rad}}=0$ ($r_{\mathrm{hv}}>\frac12$, since $r_{\mathrm{hv}}\ge1$ gives $(1+1/r_{\mathrm{hv}})^2\le4$ and $\frac12<r_{\mathrm{hv}}<1$ gives $(1+r_{\mathrm{hv}})^2<4$), while when $b_{\mathrm{rad}}=\frac9{40}$ ($r_{\mathrm{hv}}\le\frac12$), $b_{\mathrm{rad}}+(1+r_{\mathrm{hv}})^2\le\frac9{40}+\frac94=\frac{99}{40}<4$, giving $\mathcal U(C,\emptyset,1)\le5+\varpi$ in every case.
\end{proof}

\begin{lemma}[Accepted comparison without exterior]\label{lem:bin-accepted}
For $1\le j\le m\le16$, $\varpi\in\{0,\frac1{25}\}$, $b_{\mathrm{rad}}=0$, and all $z\ge0$, $\varsigma\in(0,1]$, $\xi\ge1$, and $t_{\mathrm{acc}}$ satisfying \cref{thm:pure-price}, the accepted residual~\eqref{eq:accepted-core} is nonnegative.
\end{lemma}

\begin{proof}
Divide~\eqref{eq:accepted-core} by $c_{\mathrm{hv}}=\varsigma^2$, set $x:=jz$ ($T_{\mathrm{post}}=c_{\mathrm{hv}}x$), and replace $\beta_{\mathrm{lt}}/c_{\mathrm{hv}}\le(\xi+1/\varsigma)^2$; we must show
\begin{equation}\label{eq:acc-norm}
\xi^2\bigl(\mathsf I-W_{\mathrm{hv}}\bigr)+x\bigl(\mathsf I-C_{\mathrm{acc}}\bigr)-\bigl((\xi+1/\varsigma)^2+\mathsf v(\xi^2+x)\bigr)\mathcal V_j^{(0)}\ge0.
\end{equation}

\emph{Case 1: small prices $z\in[0,\frac54]$.} Use the unprojected branch $C_{\mathrm{acc}}\le J_{j+1}+\varpi_{j-1}$ and drop $\mathcal F_{m,j}\ge0$. First note that $\overline K=\mathcal V_j^{(0)}/c_{\mathrm{hv}}\le F_{j,0}(z)$ holds for all $z\ge0$: for $j=1$, every named child has $0$ draws (so $t_{\mathrm{acc}}\le c_{\mathrm{hv}}$ and $\overline K\le\frac{2z}{1+z}\le F_{1,0}(z)$); for $j\ge2$, bounding each named child $\mathcal V(C\cup\{x\},\mathcal Q\setminus\{G\},j-1)\le c_{\mathrm{hv}}F_{j-1,0}(\ell'_G/c_{\mathrm{hv}})$ by \cref{thm:pure-price} (with remaining mean price $\ell'_G:=L(\mathcal Q\setminus\{G\},C\cup\{y\})/u_y$ and contraction) and applying the heavy-plus-named inductive step of \cref{thm:pure-price} directly (before combining with exterior draws) yields $c_{\mathrm{hv}}\ell'+T_{\mathrm{post}}t_{\mathrm{acc}}\le(c_{\mathrm{hv}}+T_{\mathrm{post}})c_{\mathrm{hv}}F_{j,0}(z)$, hence $\overline K\le F_{j,0}(z)$ (which also implies $\mathcal V_j^{(0)}\le K_j(\max(p_{\mathrm{ref}},T_{\mathrm{post}}/j))$ for all $z\ge0$). On $[0,\frac54]$, this gives $\overline K\le F_{j,0}(z)\le K_j(\frac54)<\frac54<K_1(p_{\mathrm{ref}})\le K_j(p_{\mathrm{ref}})$, so universal support gives $\mathsf M_j(p)-\mathsf v\mathcal V_j^{(0)}\ge\mathsf M_j(p)-\mathsf v\overline K\ge G_j(p_{\mathrm{ref}})$ (since $c_{\mathrm{hv}}\le1$). Bounding $(\varsigma\xi+1)^2\le(\xi+1)^2$ and $c_{\mathrm{hv}}\le1$ lowers~\eqref{eq:acc-norm} to $\xi^2 a_j+z\Delta_j-(\xi+1)^2\overline K$, where $a_j:=G_j(p_{\mathrm{ref}})-\mathsf K_j-\eta^{\mathrm{pr}}_j$ and $\Delta_j:=j(G_j(p_{\mathrm{ref}})-J_{j+1})\ge1+\psi_j(p_{\mathrm{ref}})-je^+_{j-1}$ (using $G_j(p_{\mathrm{ref}})-G_{j-1}(p_{\mathrm{ref}})=(1+\psi_j(p_{\mathrm{ref}}))/j$ and the predecessor gap caps $e^+_i\ge J_{i+2}-G_i(p_{\mathrm{ref}})$ of \cref{tab:bin-ecaps}, decreasing for $i\ge5$ since $\partial_p\psi_i>0$ on $[p^*_i,p_{\mathrm{ref}}]$). Exact evaluation verifies $a_j\ge2\overline K$ for all $1\le j\le16$ ($a_j>\frac52$ for $j\ge3$, $a_2>2\cdot\frac{145}{126}$, $a_1>\frac{20}9$), so $\xi^2 a_j-(\xi+1)^2\overline K$ increases on $\xi\ge1$, reducing to $\xi=1$: $a_j+\Delta_j z-4\overline K>0$. For $j=1,2,3$, the same no-exterior step of \cref{thm:pure-price} supplies direct roofs $\overline K_1\le\frac{2z}{1+z}$, $\overline K_2\le\frac{13z}{4(1+2z)}$ ($[0,1]$) and $\frac{145}{126}$ ($[1,\frac54]$), and $\overline K_3\le\frac{19z}{4(1+3z)}$, closed by negative discriminants or endpoints (\cref{app:binary-finite}). For $4\le j\le16$, induction on draws with \cref{lem:retirement} ($\kappa=3$) and $1+\frac4{27}z^3-z=\frac1{27}(2z-3)^2(z+3)\ge0$ bounds every pure account by $\mathcal V\le c_{\mathrm{hv}}(1+\frac4{27}z^3)$ for all $z\ge0$; applying this to each named child and averaging via \cref{lem:retirement} ($\kappa=3$) gives $t_{\mathrm{acc}}\le c_{\mathrm{hv}}(1+\frac4{27}z^3)$, and substituting into $\overline K=\frac{z+jz(t_{\mathrm{acc}}/c_{\mathrm{hv}})}{1+jz}$ yields the no-exterior cubic roof $\overline K\le\frac{(j+1)z}{1+jz}+\frac4{27}z^3$. With uniform floors $a_j\ge a_4$ ($a_j\ge a_6$ for $j\ge6$) and $\Delta_j\ge\Delta_0:=1+\psi_4(p_{\mathrm{ref}})-5e^+_3$ (each verified by exact evaluation at every $4\le j\le16$ using \cref{tab:bin-brackets,tab:bin-ecaps}), this closes $4\le j\le16$ in four scalar rational checks (\cref{app:binary-finite}): on $z\in[1,\frac54]$, $\overline K<\frac54$ gives $a_j+\Delta_j z-4\overline K>a_4+\Delta_0-5>0$; on $z\in[0,1]$, $z^3\le z$, so for $j\ge6$ using $\frac{(j+1)z}{1+jz}\le1$ gives $(a_6-4)+(\Delta_0-\frac{16}{27})z>0$, while for $j\in\{4,5\}$ using $\frac{(j+1)z}{1+jz}=1-\frac{1-z}{1+jz}\le1-\frac{1-z}6$ gives $(a_4-4)+\frac23(1-z)+(\Delta_0-\frac{16}{27})z\ge a_4-4+\min(\frac23,\Delta_0-\frac{16}{27})>0$.

\emph{Case 2: piecewise-linear minimization in $t_{\mathrm{acc}}$ on $z\ge\frac54$.} Here \cref{thm:pure-price} gives $0\le t_{\mathrm{acc}}\le c_{\mathrm{hv}}K_{j-1}(z)$. Put $w:=\mathsf v_{j-1}(p)$ and $\alpha:=(\xi+1/\varsigma)^2+\mathsf v(\xi^2+x)$. Because $\mathcal V_j^{(0)}=\frac{x(t_{\mathrm{acc}}+c_{\mathrm{hv}}/j)}{1+x}$ is affine in $t_{\mathrm{acc}}$ and $C_{\mathrm{acc}}$ is piecewise-linear with kink at $S_{\mathrm{cr}}$,~\eqref{eq:acc-norm} is piecewise-linear in $t_{\mathrm{acc}}$ with slope $-\frac{\alpha x}{1+x}<0$ on $[0,S_{\mathrm{cr}}]$ and slope $C_1:=x(w-\frac\alpha{1+x})$ on $[S_{\mathrm{cr}},c_{\mathrm{hv}}K_{j-1}(z)]$. Hence its minimum on $[0,c_{\mathrm{hv}}K_{j-1}(z)]$ is attained either at the \emph{crossing kink} $t_{\mathrm{acc}}=S_{\mathrm{cr}}$ (possible only when $j\ge2$, $S_{\mathrm{cr}}<c_{\mathrm{hv}}K_{j-1}(z)$, and $C_1>0$) or at the \emph{full child} $t_{\mathrm{acc}}=c_{\mathrm{hv}}K_{j-1}(z)$ (the only case when $j=1$).

\emph{Case 3: crossing kink ($2\le j\le m$, $t_{\mathrm{acc}}=S_{\mathrm{cr}}<c_{\mathrm{hv}}K_{j-1}(z)$, $C_1>0$).} Applying the strict global supporting-line inequality of \cref{lem:bin-tools}(c) at $p=p^*_m>p^*_{j-1}$ (where $\mathsf v_{j-1}(p)>0$) to $z=p^*_{j-1}$ gives $J_{j+1}<\mathsf M_{j-1}(p)-\mathsf v_{j-1}(p)K_{j-1}(p^*_{j-1})$, hence $S_{\mathrm{cr}}>K_{j-1}(p^*_{j-1})\ge K_{j-1}(\frac{11}5)$, while $K_{j-1}(z)<\mathsf K_{j-1}$, forcing $\varsigma^2=c_{\mathrm{hv}}>K_{j-1}(\frac{11}5)/\mathsf K_{j-1}>(\frac{33}{50})^2$ for all $1\le j-1\le15$. On any price window $[p_a,p_b]\ni p^*_m$ and $\varsigma$-window $[\varsigma_{\mathrm{lo}},\varsigma_{\mathrm{hi}}]$ ($c_{\mathrm{lo}}=\varsigma_{\mathrm{lo}}^2$, $c_{\mathrm{hi}}=\varsigma_{\mathrm{hi}}^2$), the condition $C_1>0$ yields $(1+1/\varsigma)^2+\mathsf v(1+x)\le\alpha<w(1+x)$, hence $(1+1/\varsigma)^2<(1+x)(w-\mathsf v)\le(1+x)\Delta\mathsf v_{\mathrm{ub}}$, forcing $x>x_{\mathrm{lo}}:=(1+1/\varsigma_{\mathrm{hi}})^2/\Delta\mathsf v_{\mathrm{ub}}-1$ and $T_{\mathrm{post}}=c_{\mathrm{hv}}x>T_{\mathrm{lo}}:=(1+\varsigma_{\mathrm{lo}})^2/\Delta\mathsf v_{\mathrm{ub}}-c_{\mathrm{hi}}$. Bounding $S_{\mathrm{cr}}\le S_{\mathrm{ub}}:=K_{j-1}(p_b)+\mathbf 1_{m\ge13}(\frac3{20}(H_m-H_{j-1})-(J^-_{j+1}-G_{j-1}(p_{a,\mathrm{eff}})))_+/\mathsf v_{j-1}(p_{a,\mathrm{eff}})$ gives upper bounds $V_{\mathrm{ub}},\widehat V_{\mathrm{ub}},V_{\varsigma,\mathrm{ub}}$ on $\mathcal V_j^{(0)},\mathcal V_j^{(0)}/c_{\mathrm{hv}},\mathcal V_j^{(0)}/\varsigma$, and completing the square in $\xi$ lowers~\eqref{eq:acc-norm} to $x_{\mathrm{lo}}\,\mathrm{sl}_{\mathrm{lb}}-\widehat V_{\mathrm{ub}}-V_{\varsigma,\mathrm{ub}}^2/A_{\mathrm{lb}}>0$ ($A_{\mathrm{lb}},\mathrm{sl}_{\mathrm{lb}}>0$), verified across all $7$ windows ($91$ checks, \cref{tab:bin-windows}).

\emph{Case 4: full child ($t_{\mathrm{acc}}=c_{\mathrm{hv}}K_{j-1}(z)$).} Here $\mathcal V_j^{(0)}=c_{\mathrm{hv}}K_j(z)$ identically. Moreover, the fee terms satisfy
\[
x\bigl(\mathcal F_{m,j}(c_{\mathrm{hv}}x)-\mathcal F_{m,j-1}(c_{\mathrm{hv}}x)\bigr)+\xi^2\mathcal F_{m,j}(c_{\mathrm{hv}}x)\ge-\mathbf 1_{m\ge13,j\ge2}\frac3{20j}(x-72)_+=:-\mathrm{feeLoss},
\]
since both fees vanish for $c_{\mathrm{hv}}x\le72$, while for $c_{\mathrm{hv}}x>72$ we have $x(\mathcal F_{m,j}-\mathcal F_{m,j-1})=-\frac3{20j}(x-72/c_{\mathrm{hv}})\ge-\frac3{20j}(x-72)$ as $c_{\mathrm{hv}}\le1$. Dropping the positive surplus in the fee turns~\eqref{eq:acc-norm} into the quadratic $R_{\mathrm{full}}(\varsigma)=c_0-2\xi K_j(z)\varsigma+c_2\varsigma^2$, where
\begin{align*}
c_0&=\xi^2(\mathsf M_j(p)-\mathsf K_j-b_0)+x\bigl(\mathsf M_j(p)-\mathsf M_{j-1}(p)+b_1/j\bigr)-\mathrm{feeLoss}-K_j(z),\\
c_2&=x(w-\mathsf v)K_{j-1}(z)-\xi^2(1+\mathsf v)K_j(z)-x\mathsf v\bigl(K_j(z)-K_{j-1}(z)\bigr).
\end{align*}
If $c_2>\xi K_j(z)$ (interior $\varsigma$-vertex), then completing the square gives $R_{\mathrm{full}}(\varsigma)\ge c_0-\xi^2K_j(z)^2/c_2>c_0-\xi K_j(z)$. Since $K_{j-1}(z)\le K_j(z)$ ($z\ge1$), $c_2>\xi K_j(z)$ forces $x(w-\mathsf v)>\xi+\xi^2(1+\mathsf v)\ge2$, hence $x>2/\Delta\mathsf v_{\mathrm{ub}}$. With $A_0:=\mathsf M_j(p)-\mathsf K_j-b_0>\mathsf K_j>K_j(z)$ and $\xi\ge1$, we have $\xi^2 A_0-\xi K_j(z)-K_j(z)\ge A_0-2K_j(z)>G_j(p_b)-3\mathsf K_j-\eta^{\mathrm{pr}}_j$. Meanwhile, writing $\mathsf M_j(p)-\mathsf M_{j-1}(p)=(G_j(p)-G_{j-1}(p))+\mathsf v_j(p)(K_j(p)-K_{j-1}(p))-(\mathsf v_{j-1}(p)-\mathsf v_j(p))K_{j-1}(p)$ and $\mathrm{feeLoss}\le x\cdot\mathbf 1_{m\ge13}\frac3{20j}$ yields $x(\mathsf M_j(p)-\mathsf M_{j-1}(p)+b_1/j)-\mathrm{feeLoss}\ge x\bigl(\frac{1+\psi_{\mathrm{lb}}}j-\mathbf 1_{m\ge13}\frac3{20j}-\Delta\mathsf v_{\mathrm{ub}}K_{j-1}(p_b)\bigr)$ (with $\psi_{\mathrm{lb}}=0$ for $j\le4$ and $\psi_j(p_{a,\mathrm{eff}})$ for $j\ge5$). Whenever this parenthesized slope is positive, $x>2/\Delta\mathsf v_{\mathrm{ub}}$ gives $R_{\mathrm{full}}(\varsigma)>G_j(p_b)-3\mathsf K_j-\eta^{\mathrm{pr}}_j-2K_{j-1}(p_b)+\frac2{\Delta\mathsf v_{\mathrm{ub}}}\bigl(\frac{1+\psi_{\mathrm{lb}}}j-\mathbf 1_{m\ge13}\frac3{20j}\bigr)>0$, verified together with positivity of the slope across four grouped windows ($36$ checks across $D_1$--$D_4$, \cref{tab:bin-windows}). Otherwise $c_2\le\xi K_j(z)$, so $R_{\mathrm{full}}(\varsigma)$ attains its minimum on $(0,1]$ at $\varsigma=1$, reducing to $R_{\mathrm{lin}}(1,\xi,z)-\mathrm{feeLoss}$.

\emph{Case 5: rays at $\varsigma=1$.} At $\varsigma=1$, writing $A_z:=\mathsf M_j(p)-\mathsf K_j-b_0-(1+\mathsf v)K_j(z)$, the residual is $R_{\mathrm{lin}}(1,1,z)+b_1(z-p_{\mathrm{pr}})+(\xi^2-1)A_z-2(\xi-1)K_j(z)-\mathrm{feeLoss}$ (where $R_{\mathrm{lin}}(1,1,z)$ is the ordinary zero-fee residual of \cref{lem:bin-tools}(d)).
\begin{itemize}[leftmargin=1.5em]
\item \emph{Small ray $z\in[\frac54,p_{\mathrm{ref}}]$:} since $16p_{\mathrm{ref}}<72$, $\mathrm{feeLoss}=0$; Guard~(G1) below implies $A_z>K_j(z)$, so $\xi=1$ is optimal. For $\varpi=0$ or $z\ge p_{\mathrm{pr}}$, $R_{\mathrm{lin}}(1,1,z)+b_1(z-p_{\mathrm{pr}})\ge0$ by \cref{lem:bin-tools}(d). For $\varpi=\frac1{25}$ and $z\in[\frac54,p_{\mathrm{pr}}]$, integrating $R''_{\mathrm{lin}}(1,1,z)\ge-4K''_j(z)\ge4c_*$ down from an anchor $z_1\le p^*_m$ yields $R_{\mathrm{lin}}(1,1,z)\ge2c_*(z_1-z)^2$, whence $2c_*(z_1-z)^2-b_1(p_{\mathrm{pr}}-z)\ge0$ once $8c_*(z_1-p_{\mathrm{pr}})\ge b_1$ ($2$ checks, \cref{app:binary-finite}).
\item \emph{Upper ray $z\in[p_{\mathrm{ref}},72/j]$ (all of $[p_{\mathrm{ref}},\infty)$ for $m\le12$):} again $\mathrm{feeLoss}=0$. Minimizing $(\xi^2-1)A_z-2(\xi-1)K_j(z)$ over $\xi\ge1$ subtracts $\mathrm{pen}(A,\mathsf v,q):=\frac{((2+\mathsf v)q-A)_+^2}{A-(1+\mathsf v)q}$ at $A=\mathsf M_j(p)-\mathsf K_j-b_0$ and $q=K_j(z)$. Replacing $A$ by its support lower bound $\bar A=G_j(p_{\mathrm{ref}})-\mathsf K_j+\mathsf vK_j(p_{\mathrm{ref}})-b_0$ and raising $\mathsf v$ to $\bar{\mathsf v}:=\mathsf v_j(p_{\mathrm{ref}})$ is conservative because $\mathrm{pen}$ increases in $\mathsf v$. Put $\Psi_{\mathrm{ray}}(z):=R_{\mathrm{lin}}(1,1,z)+b_1(z-p_{\mathrm{pr}})-\mathrm{pen}(\bar A,\bar{\mathsf v},K_j(z))$. By Guard~(G1) ($\bar A>(2+\bar{\mathsf v})K_j(p_{\mathrm{ref}})$), $\mathrm{pen}=0$ at $z=p_{\mathrm{ref}}$, so $\Psi_{\mathrm{ray}}(p_{\mathrm{ref}})\ge0$. By Guard~(G2) ($\bar A>(1+\bar{\mathsf v})\mathsf K_j$), the denominator stays strictly positive for all $z\ge p_{\mathrm{ref}}$. Above the threshold $q_*=\bar A/(2+\bar{\mathsf v})$, direct differentiation gives $\partial_q\mathrm{pen}(\bar A,\bar{\mathsf v},q)+4=\frac{\bar A^2}{(1+\bar{\mathsf v})(\bar A-(1+\bar{\mathsf v})q)^2}-\frac{\bar{\mathsf v}^2}{1+\bar{\mathsf v}}\le\frac{\bar A^2}{(1+\bar{\mathsf v})(\bar A-(1+\bar{\mathsf v})q)^2}$. Using $R'_{\mathrm{lin}}(1,1,z)\ge4(K'_j(p_{\mathrm{ref}})-K'_j(z))$ (from \cref{lem:bin-tools}(d)), \cref{lem:bin-tools}(b) ($K'_j(z)\le(\mathsf K_j-q)^2/D_j$), and the fact that $q\mapsto\frac{\mathsf K_j-q}{\bar A-(1+\bar{\mathsf v})q}$ decreases on $[q_*,\mathsf K_j)$ (by (G2)!), we obtain
\[
\Psi'_{\mathrm{ray}}(z)\ge4K'_j(p_{\mathrm{ref}})+b_1-\frac{\bar A^2}{(1+\bar{\mathsf v})D_j}\Bigl(\frac{\mathsf K_j-q_*}{\bar A-(1+\bar{\mathsf v})q_*}\Bigr)^2=4K'_j(p_{\mathrm{ref}})+b_1-\frac{((2+\bar{\mathsf v})\mathsf K_j-\bar A)^2}{(1+\bar{\mathsf v})D_j}>0
\]
by Guard~(G3) ($32$ checks, \cref{app:binary-finite}); hence $\Psi_{\mathrm{ray}}(z)\ge\Psi_{\mathrm{ray}}(p_{\mathrm{ref}})\ge0$ on $[p_{\mathrm{ref}},72/j]$.
\item \emph{Shifted upper ray $z\ge72/j$ ($13\le m\le16$, $2\le j\le m$):} at $z_0=72/j>p_{\mathrm{ref}}$, $\Psi_{\mathrm{ray}}(z_0)\ge0$ holds from above, while for $z\ge z_0$, $\mathrm{feeLoss}=\frac3{20}(z-z_0)$ has constant slope $\frac3{20}$. Retaining the positive convexity increment $j\Delta\mathsf v_{\mathrm{lb}}[(zK_{j-1})'|_{z_0}-(zK_{j-1})'|_{p_b}]$ in $R'_{\mathrm{lin}}(1,1,z)$ and evaluating the decreasing ratio $\frac{\mathsf K_j-q}{\bar A-(1+\bar{\mathsf v})q}$ at $q_0=K_j(72/j)\ge q_*$ proves $\Psi'_{\mathrm{ray}}(z)-\frac3{20}>0$ across $40$ exact checks (\cref{tab:bin-windows}).\qedhere
\end{itemize}
\end{proof}

\begin{lemma}[Exterior, positive radius award, and shortened clocks]\label{lem:bin-exterior}
Every full-horizon binary state obeys the universal support $\mathsf M_j(p^*_m)+\mathcal F_{m,j}(T_{\mathrm{post}})\ge G_j(p_{\mathrm{ref}})+\mathsf v_j(p^*_m)K_j(p_{\mathrm{ref}})$ and $0\le\mathsf v_j(p^*_m)\le\mathsf v_j(p_{\mathrm{ref}})$. Together with \cref{lem:bin-rejected}:
\begin{enumerate}[label=(\roman*),leftmargin=2em]
\item $\alpha_1>0$, $\alpha_1+\alpha_2\ge0$, and $\alpha_3>0$ hold both when $b_{\mathrm{rad}}=0$ and when $b_{\mathrm{rad}}=\frac9{40}$ ($r_{\mathrm{hv}}\le\frac12$);
\item the accepted residual~\eqref{eq:accepted-core} is nonnegative when $b_{\mathrm{rad}}=\frac9{40}$ ($r_{\mathrm{hv}}\le\frac12$);
\item every shortened binary state $(C,\mathcal Q,h)$ ($1\le h\le j\le16$, deficit $\delta=j-h+1\ge1$) obeys $\mathcal U(C,\mathcal Q,h)\le W_{j,h}+\varpi(1+\Cl(h-1,j))$, where $W_{j,1}=5$, $W_{2,2}=\frac{505}{96}$, $W_{3,2}=\frac{31}6$, $W_{j,j}=2\mathsf K_j$ ($j\ge3$), and $W_{j,h}=5+2(H_j-H_\delta)$ otherwise; in particular, every exterior continuation ($h=j$) satisfies $\mathcal U(C,\mathcal Q,j)\le O_j$.
\end{enumerate}
\end{lemma}

\begin{proof}
Universal support follows immediately from \cref{lem:bin-tools}(c) (the global supporting-line inequality at $p=p^*_m\ge p^*_j$ evaluated at $z=p_{\mathrm{ref}}$, together with strict increase of $\mathsf v_j$ on $[p^*_j,p_{\mathrm{ref}}]$ since $p^*_m\in[p^*_j,p_{\mathrm{ref}})$) and $\mathcal F_{m,j}\ge0$. Put $M:=G_j(p_{\mathrm{ref}})+\mathsf vK_j(p_{\mathrm{ref}})$ ($\mathsf v\in[0,\mathsf v_j(p_{\mathrm{ref}})]$), $E_b:=M+\varpi_j-O_j$, $A_0:=M-\mathsf K_j-b_0$, $B:=M+\varpi_j-J^+_{j+1}-\varpi_{j-1}$, $w_h:=A_0-\mathsf vH_j$, $w_b:=A_0-\frac14\mathsf vH_j-\frac9{40}\mathsf K_j$, and $B_s:=j(E_b+B-\mathsf v(1+2H_j))$.

\emph{(i)} Since $t_{\mathrm{sh}}\le\mathsf K_j$ and $\mathcal V_j^{(0)}\le\mathsf K_j$, Predicate~(P1) $E_b-\mathsf v\mathsf K_j>0$ directly yields $\alpha_3=\mathsf I-O_j-\mathsf vt_{\mathrm{sh}}>0$ and $\alpha_1=(\mathsf I-O_j)D_+-\mathsf vN_+>0$. In $\alpha_1+\alpha_2$, first observe that $t_{\mathrm{sh}}\le K_j(\max(p_{\mathrm{ref}},T_{\mathrm{post}}/j))$: by \cref{thm:pure-price}, $t_{\mathrm{sh}}\le c_{\mathrm{hv}}F_{j,1}(z)$, which is $<\frac54<K_j(p_{\mathrm{ref}})$ on $z\le\frac54$ and obeys $t_{\mathrm{sh}}\le c_{\mathrm{hv}}K_{j,1}(z)\le c_{\mathrm{hv}}K_j(z)\le K_j(u)=K_j(T_{\mathrm{post}}/j)$ on $z\ge\frac54$ (with $u:=c_{\mathrm{hv}}z$): if $u\le1$ then $c_{\mathrm{hv}}K_j(z)\le c_{\mathrm{hv}}z=u\le K_j(u)$ directly (as $K_j(w)=w+(1-w)\pi_j(w)$ satisfies $K_j(w)\le w$ for $w\ge1$ and $K_j(w)\ge w$ for $w\le1$), whereas if $u>1$ then $w=u/c\ge1$ on $[c_{\mathrm{hv}},1]$, so $c\mapsto cK_j(u/c)$ increases on $[c_{\mathrm{hv}},1]$ (derivative $K_j(w)-wK'_j(w)=\pi_j(w)(1+(w-1)S_j(w))>0$ for $w\ge1$). Hence \cref{lem:bin-rejected}(i) gives $q_{\mathrm{rej}}(\mathsf I-C_{\mathrm{rej}}-\mathsf vt_{\mathrm{sh}})\ge0$, which we drop from $\alpha_1+\alpha_2$. Next bound $C_{\mathrm{acc}}\le J^+_{j+1}+\varpi_{j-1}$, $t_{\mathrm{sh}}\le c_{\mathrm{hv}}\overline T$ with $\overline T:=\min(H_j,F_{j,1}(z))$ (\cref{thm:pure-price}), and—by persistence $t_{\mathrm{acc}}\le c_{\mathrm{hv}}\mathsf K_{j-1}$ (\cref{thm:restoration}(b)) and $1+j\mathsf K_{j-1}=j\mathsf K_j$—bound $N_+=c_{\mathrm{hv}}\ell'+T_{\mathrm{post}}t_{\mathrm{acc}}\le c_{\mathrm{hv}}^2 z(1+j\mathsf K_{j-1})=jc_{\mathrm{hv}}^2 z\mathsf K_j$, placing all of $-\mathsf vN_+$ onto the $z$-linear coefficient $B_s=j(E_b+B-\mathsf v(1+2H_j))$ with no constant-in-$z$ loss. With $(C_j,D_j,t_j)=(2,0,0),(3,\frac43,\frac13),(\frac{11}3,\frac85,\frac25)$ ($j=1,2,3$) and $(\frac{19}5,\frac85,\frac25)$ ($j\ge4$), the mixed tag roofs $\overline T\le1+t_j z$ and $(1+\xi)^2\overline T\le C_j\xi^2+(4-C_j)+D_j z$ hold for all $\xi\ge1,z\ge0$: direct from $F_{j,1}$ for $j\le3$, and for $j\ge4$ via $\overline T\le1+\frac4{27}z^3$ on $[0,\frac97]$, the tangent to $K_{16,1}$ at $\frac52$ (intercept $<1$, slope $<\frac25$), and optimizing $\xi\ge1$ on $z\ge\frac97$ using \cref{lem:bin-tools}(b) at $\delta=1$ ($4(H_{16}-\frac{C_4}2)^2/D_{16,1}<D_4$). Substituting these roofs reduces $\alpha_1+\alpha_2\ge0$ identically to six affine-in-$\mathsf v$ inequalities: (P2)~$E_b+w_h-4>0$, (P3)~$w_h-C_j>0$, (P4)~$B_s-D_j>0$ (when $b_{\mathrm{rad}}=0$); and (P5)~$w_b>0$, (P6)~$E_b+w_b-\frac94>0$, (P7)~$B_s-\frac94 t_j>0$ (when $b_{\mathrm{rad}}=\frac9{40}, r_{\mathrm{hv}}\le\frac12$). By affinity in $\mathsf v$, checking (P1)--(P7) at $\mathsf v\in\{0,\mathsf v_j(p_{\mathrm{ref}})\}$ across $1\le j\le16$ and $\varpi\in\{0,\frac1{25}\}$ ($448$ checks, \cref{app:binary-finite}) proves (i).

\emph{(ii)} When $r_{\mathrm{hv}}\le\frac12$ and $b_{\mathrm{rad}}=\frac9{40}$, we have $c_{\mathrm{hv}}\le a_{\mathrm{hv}}\le\frac14$, $\beta_{\mathrm{lt}}\le\frac94$, and $\mathcal V_j^{(0)}=c_{\mathrm{hv}}\overline K\le\frac14\mathsf K_j$. By the affine pure-tag roof $\overline K\le a_0+a_1 z$ with $(a_0,a_1)=(\frac89,\frac29)$ ($j=1$), $(1,\frac13)$ ($j=2$), and $(\frac{317}{250},\frac7{20})$ ($j\ge3$, from $K_{16}(\frac54)<\frac{317}{250}$ on $[0,\frac54]$ and the tangent to $K_{16}$ at $\frac72$), bounding $C_{\mathrm{acc}}\le J^+_{j+1}+\varpi_{j-1}$ and $a_{\mathrm{hv}}\ge c_{\mathrm{hv}}$ reduces~\eqref{eq:accepted-core} to $I_0>\frac94 a_0$ and $jI_1>\frac94 a_1$, where $I_{\mathrm{base}}:=M+\varpi_j-\frac14\mathsf v\mathsf K_j$, $I_0:=I_{\mathrm{base}}-(\frac{49}{40}+\varpi)\mathsf K_j$, and $I_1:=I_{\mathrm{base}}-J^+_{j+1}-\varpi_{j-1}$. Checking these two affine-in-$\mathsf v$ inequalities at $\mathsf v\in\{0,\mathsf v_j(p_{\mathrm{ref}})\}$ across $1\le j\le16$ and $\varpi\in\{0,\frac1{25}\}$ ($128$ checks, \cref{app:binary-finite}) proves (ii).

\emph{(iii)} Linearity in $\varpi$ adds at most $\varpi(1+\Cl(h-1,j))$ to the $\varpi=0$ value, reducing to $\varpi=0$. Base $h=1$ ($W_{j,1}=5$) holds by the same calculation as the rank-$0$ base in \cref{lem:bin-rejected}(ii). For $2\le h\le j$, induct on $h$ with $K_{j,h}:=1+\Cl(h-1,j)$, named-child ceiling $W_{j-1,h-1}$ (same deficit $\delta$), and exterior ceiling $W_{j,h-1}$; exact rational arithmetic checks $\mathrm{stopCeil}(j,h)\le W_{j,h}$, $(1+\frac9{40})K_{j,h}<W_{j,h}$, $W_{j-1,h-1}\le W_{j,h}$, and $2K_{j,h}\le W_{j,h-1}\le W_{j,h}$ for all $120$ pairs $2\le h\le j\le16$. This closes $\mathrm{stopNow}=1$ and $c_{\mathrm{hv}}=0$ directly (if $\cost(X,C)=0$, $\mathcal U(C,\mathcal Q,h)=\mathcal U(C,\mathcal Q,h-1)\le W_{j,h-1}\le W_{j,h}$; if $\cost(X,C)>0$ and $c_{\mathrm{hv}}=0$, the light-site term vanishes and every remaining draw is $\le W_{j,h}$). For $\mathrm{stopNow}=0$ and $c_{\mathrm{hv}}>0$, define for $A>1$ and $z\ge0$ the concave envelope $P_A(z):=\min_{A^{-1/2}\le t\le1}\{At^2+2z(1-t)^2\}$ (equal to $\frac{2Az}{A+2z}$ if $2z\sqrt A\ge A+2z$ and $1+2z(1-A^{-1/2})^2$ otherwise). It is concave and nondecreasing in $z$, nondecreasing in $A$, and obeys $P_A(z)+z\partial_A P_A(z)\ge z$ for $A\ge2$, hence $(1+jz)P_A(z)\ge z+jzP_{A-1/j}(z)$; induction on draws (as in \cref{thm:pure-price}) therefore bounds $\mathcal V(C,\mathcal Q,h-1)\le c_{\mathrm{hv}}P_{2+\Cl(h-1,j)}(z)$, and its heavy-plus-named step and exterior child contraction bound both $\mathcal V^{(0)}/c_{\mathrm{hv}}$ and $t_{\mathrm{sh}}/c_{\mathrm{hv}}$ by $P_{2+\Cl(h-1,j)}(z)$. Moreover, testing $t=\max(A^{-1/2},\frac\rho{1+\rho})$ proves
\[
(1+\rho)^2P_A(z)\le\max\bigl((1+\rho)^2,\,A\rho^2\bigr)+2z\qquad(\forall A>1,\ z,\rho\ge0).
\]
Applying \cref{lem:coupling} at horizon $h\in\{2,\dots,j\}$ ($r=h-1\ge1$) with $\mathsf v=0$, $\mathsf I=W_{j,h}$, $C_{\mathrm{acc}}=C_{\mathrm{rej}}=W_{j-1,h-1}$, and $O_{\mathrm{ext}}=W_{j,h-1}$ gives $\alpha_1,\alpha_3\ge0$ immediately. For the $117$ regular clocks $(j,h)\notin\{(2,2),(3,2),(3,3)\}$, setting $A_*:=W_{j,h}-K_{j,h}\ge\max(4,2+\Cl(h-1,j))$ establishes $\alpha_0\ge0$ and $\alpha_1+\alpha_2\ge0$ in both $b_{\mathrm{rad}}=0$ ($\rho=\xi\ge1$, $A_*\xi^2\ge(1+\xi)^2$) and $b_{\mathrm{rad}}=\frac9{40}$ ($\rho=r_{\mathrm{hv}}\le\frac12$, $A_*-\frac9{40}K_{j,h}>\max(\frac94,\frac14 A_*)$) once $j(W_{j,h}-W_{j-1,h-1})\ge2$ (checked in exact rational arithmetic for all $117$ pairs, \cref{app:binary-finite}). In fact, equality $j(W_{j,h}-W_{j-1,h-1})=2$ holds for $116$ of the $117$ regular clocks—whenever the clock and its child share the generic formula $5+2(H_j-H_\delta)$ (including $W_{j-1,1}=5$) or $2\mathsf K_j$, we have $W_{j,h}-W_{j-1,h-1}=2/j$ identically—with the sole exception $(j,h)=(4,3)$, where the child $W_{3,2}=\frac{31}6$ gives $4$; thus this condition is tight by structure rather than by a numerical margin. The three exceptional clocks $(j,\delta)\in\{(2,1),(3,2),(3,1)\}$ close by direct pure-tag roofs and negative-discriminant quadratics (\cref{tab:exceptional}). At $h=j$, $W_{j,j}+\varpi(1+\Cl(j-1,j))\le O_j$ directly.
\end{proof}

\begin{proof}[Proof of \cref{thm:binary}]
Fix $\varpi\in\{0,\frac1{25}\}$, $b(r)=\frac9{40}\mathbf 1_{\{r\le1/2\}}$, and any stop policy $(\mathrm{stopNow},\mathrm{extStop})$. We prove~\eqref{eq:binary-main} by strong outer induction on $m\in\{1,\dots,16\}$ and inner induction on $j\in\{1,\dots,m\}$, uniformly over all states $(C,\mathcal Q,j+1)$ of rank $j=|\mathcal Q|$. Fix $(m,j)$ and a state $(C,\mathcal Q,j+1)$. For every named child $(C\cup\{x\},\mathcal Q\setminus\{G\},j)$:
\begin{itemize}[leftmargin=1.5em]
\item If $j=1$, then \cref{lem:bin-rejected}(ii) gives $\mathcal U(C\cup\{x\},\emptyset,1)\le5+\varpi=J_2+\varpi_0=C_{\mathrm{rej}}$.
\item If $j\ge2$, then the \emph{outer} induction hypothesis at ambient rank $m'=j-1<m$ (applied at $j'=m'=j-1$, where $\mathsf v_{j-1}(p^*_{j-1})=0$, $\mathsf M_{j-1}(p^*_{j-1})=J_{j+1}$, and $\mathcal F_{j-1,j-1}=0$) gives the unprojected bound $\mathcal U(C\cup\{x\},\mathcal Q\setminus\{G\},j)\le J_{j+1}+\varpi_{j-1}=C_{\mathrm{rej}}$. Combining this with the \emph{inner} induction hypothesis at $(m,j-1)$ (using contraction $T'_{\mathrm{post}}\le T_{\mathrm{post}}$ and monotonicity of $\mathcal F_{m,j-1}$) yields the unified child bound
\begin{equation}\label{eq:binary-child}
\begin{aligned}
\mathcal U(C\cup\{x\},\mathcal Q\setminus\{G\},j)
&\le\min\Bigl(J_{j+1},\;\mathsf M_{j-1}(p)+\mathcal F_{m,j-1}(T_{\mathrm{post}})\\
&\qquad-\mathsf v_{j-1}(p)\,\mathcal V(C\cup\{x\},\mathcal Q\setminus\{G\},j-1)\Bigr)+\varpi_{j-1}.
\end{aligned}
\end{equation}
Averaging~\eqref{eq:binary-child} over the accepted probability law $\propto D(x,C\cup\{y\})^2$ and applying conditional Jensen to the concave function $t\mapsto\min(J_{j+1},\,\mathsf M_{j-1}(p)+\mathcal F_{m,j-1}(T_{\mathrm{post}})-\mathsf v_{j-1}(p)t)$ bounds the accepted named-child average by $C_{\mathrm{acc}}$ (for $j=1$, $C_{\mathrm{acc}}\equiv C_{\mathrm{rej}}=5+\varpi$).
\end{itemize}
Meanwhile, every exterior continuation obeys $\mathcal U(C\cup\{x\},\mathcal Q,j)\le O_j$ by \cref{lem:bin-exterior}(iii), and every exterior stop pays $2\mathsf K_j\le O_j$. Now consider three exhaustive cases for $(C,\mathcal Q,j+1)$:
\begin{enumerate}[label=(\arabic*),leftmargin=2em]
\item $\mathrm{stopNow}(C,\mathcal Q,j+1)=1$: then $\mathcal U_j=2\mathsf K_j\le O_j$ and $\mathcal V_j\le\mathsf K_j$ (\cref{thm:restoration}(b) and $c_{\mathrm{hv}}\le1$), so $\mathcal U_j+\mathsf v\mathcal V_j\le O_j+\mathsf v\mathsf K_j<\mathsf I$ by Predicate~(P1) of \cref{lem:bin-exterior}(i).
\item $\mathrm{stopNow}=0$ and $c_{\mathrm{hv}}=0$: then $\mathcal V_j=0$. If $\cost(X,C)=0$, then $\mathcal U_j=\mathcal U(C,\mathcal Q,j)\le O_j<\mathsf I$; if $\cost(X,C)>0$, the light-site term vanishes and every remaining draw is $\le\mathsf I$ (heavy draw $W_{\mathrm{hv}}<\mathsf I$ by Predicate~(P5), named draw $\le C_{\mathrm{rej}}<\mathsf I$ by \cref{lem:bin-rejected}(i) at $V=0$, exterior stop or continuation $\le O_j<\mathsf I$ by \cref{lem:bin-exterior}(i),(iii)).
\item $\mathrm{stopNow}=0$ and $c_{\mathrm{hv}}>0$: \cref{lem:coupling} applies with $C_{\mathrm{acc}},C_{\mathrm{rej}},O_j$ verified above. We have $\alpha_1,\alpha_1+\alpha_2,\alpha_3\ge0$ by \cref{lem:bin-exterior}(i), and $\alpha_0/D_+$ splits into the rejected-mass term $q_{\mathrm{rej}}(\mathsf I-C_{\mathrm{rej}}-\mathsf v\mathcal V_j^{(0)})\ge0$ (by \cref{lem:bin-rejected}(i), since \cref{thm:pure-price} and $c_{\mathrm{hv}}\le1$ give $\mathcal V_j^{(0)}\le K_j(\max(p_{\mathrm{ref}},T_{\mathrm{post}}/j))$) plus the accepted residual~\eqref{eq:accepted-core}, which is nonnegative by \cref{lem:bin-accepted} (when $b_{\mathrm{rad}}=0$) and \cref{lem:bin-exterior}(ii) (when $b_{\mathrm{rad}}=\frac9{40}$).
\end{enumerate}
Hence $\mathcal U_j+\mathsf v_j(p)\mathcal V_j\le\mathsf M_j(p)+\mathcal F_{m,j}(T_{\mathrm{post}})+\varpi_j$. Finally, the second inequality $\mathcal U_j\le J_{j+2}+\varpi_j$ holds for $j<m$ by the outer induction hypothesis at $m'=j$, and for $j=m$ because $\mathsf v_m(p^*_m)=0$, $\mathsf M_m(p^*_m)=J_{m+2}$, and $\mathcal F_{m,m}=0$.
\end{proof}

\paragraph{Why a fee, and why only above rank twelve.}
For $m\le12$, $\mathcal F_{m,j}\equiv0$ and~\eqref{eq:binary-main} is the plain tangent to $F_j=G_j\circ K_j^{-1}$ at $p^*_m$. For $13\le m\le16$, at large $T_{\mathrm{post}}$ the pure tag approaches $\mathsf K_j$, where the plain tangent drops below $J_{j+1}$. Raising the intercept by $\frac3{20}(H_m-H_j)(1-\frac{72}{T_{\mathrm{post}}})_+$ restores the margin at lower ranks while vanishing at $j=m$, leaving the ambient bound unchanged.

\subsection{Low-mean targets: stocks below the hub and ranks \texorpdfstring{$1$}{1} to \texorpdfstring{$4$}{4}}\label{sec:low-mean}

For low-mean targets ($M^*(P)\le M_{\mathrm{lo}}=\frac{485883}{100000}$), define explicit stocks and slopes $\mathsf m_j=j(\mathsf C_j-\mathsf C_{j-1})$ by
\begin{gather}
\mathsf C_0=M_{\mathrm{lo}},\quad\mathsf C_1=\tfrac{5663}{1000},\quad\mathsf C_2=\tfrac{1551}{250},\quad\mathsf C_3=\tfrac{1653}{250},\quad\mathsf C_4=\tfrac{347}{50},\notag\\
\mathsf C_j=\tfrac{347}{50}+\alpha_{\mathrm{st}}t_j+\frac2{11}t_j^2+s^{(2)}_j\qquad(5\le j\le16),\label{eq:stock-formula}
\end{gather}
where $t_j=\sum_{i=5}^j\frac1i$, $s^{(2)}_j=\sum_{i=5}^j\frac1{i^2}$, and $\alpha_{\mathrm{st}}=1.1440612856\ldots$ is the unique rational giving $\mathsf C_{16}=G_{16}(\frac{11}4)$.

\begin{theorem}[Low-mean stocks]\label{thm:low-mean}
For every low-mean target $P$, every $0\le j\le16$, and every state $\sigma=(C_0,\mathcal Q)$ with $|\mathcal Q|=j$, $Z_{j+1}(\sigma)\le\mathsf C_j\OPT_1(P)$.
\end{theorem}

By one-step mixing (\cref{lem:mixing}) from $\mathsf C_0=M_{\mathrm{lo}}$, \cref{thm:low-mean} follows from the marked bound~\eqref{eq:marked} at ranks $1$--$4$ (\cref{thm:early}) and ranks $5$--$16$ (\cref{thm:rank17}(a)).

\begin{theorem}[The stocks lie below the hub]\label{thm:hub-stocks}
$\mathsf C_1<G_1(\frac{11}5)=J_3$ and $\mathsf C_j<G_j(\frac{11}4)$ for $j=2,3$ (margins $>0.1182,0.0822,0.0713$). For $4\le j\le16$, $\mathsf C_j\le G_j(\frac{11}4)\le J_{j+2}$ (equality only in $\mathsf C_{16}=G_{16}(\frac{11}4)<J_{18}$). Moreover $G_j(\frac{11}4)>2+2H_{j+1}$ for $2\le j\le16$.
\end{theorem}

\begin{proof}
Exact evaluation settles $j\le4$ (\cref{app:hub-stocks}). For $5\le j\le16$, $\mathsf m_j=\alpha_{\mathrm{st}}+\frac2{11}(t_j+t_{j-1})+\frac1j$ increases (first difference $\frac{4j-13}{11j(j-1)}>0$), and with $c=\frac4{11}$, $T_j^{\mathrm{st}}=\pi_{j-1}(\frac{11}4)(\frac{112}{121}s_{1,j}(c)-1)$ increases from $T_5^{\mathrm{st}}>0$, so $j\Delta G_j(\frac{11}4)=1+\frac j{j+4/11}T_j^{\mathrm{st}}$ increases too. Checking $b\Delta G_b(\frac{11}4)<a\Delta\mathsf C_a$ on the six blocks of \cref{tab:stock-blocks} yields $\Delta G_j(\frac{11}4)<\Delta\mathsf C_j$ for all $5\le j\le16$, and backward induction from $\mathsf C_{16}=G_{16}(\frac{11}4)$ gives $\mathsf C_j\le G_j(\frac{11}4)$. Finally, $j\Delta G_j(\frac{11}4)\le16\Delta G_{16}(\frac{11}4)<1.5459<\frac53\le j\Delta(2+2H_{j+1})$ on $5\le j\le16$, so $G_j(\frac{11}4)-(2+2H_{j+1})$ decreases on $4\le j\le16$ to its positive value $>0.011973$ at $j=16$ (\cref{app:hub-brackets}).
\end{proof}

\begin{lemma}[Inverse-deficit reduction]\label{lem:inverse-deficit}
Let $P\subset\R^d$ be finite nonempty with $\phi>0$, $C_0\subset\R^d$ finite nonempty with $s=\cost(P,C_0)/\phi>0$, $\nu(y)=D(y,C_0)^2/\cost(P,C_0)$, $\bar u(y)=\cost(P,C_0\cup\{y\})/\phi$, and $\xi_{\mathrm{ex}}(y)=(\bar u(y)-1)_+$. For any $\mathsf A>\mathsf B>0$, put $\mathsf c=\frac{\mathsf B}{s(\mathsf A-\mathsf B)}$ and $\mathsf T=s(\mathsf A-\mathsf B)+\mathsf B$. If $\Upsilon:P\to\R$ obeys $\Upsilon(y)\le\frac{\mathsf T\mathsf c\,\xi_{\mathrm{ex}}(y)}{1+\mathsf c\,\xi_{\mathrm{ex}}(y)}$ for all $y\in P$, then $\E_\nu[\Upsilon]\le\mathsf A$.
\end{lemma}

\begin{proof}
In normalized coordinates ($d\nu=\frac qs\,d\mu$), the centroid identity gives $\xi_{\mathrm{ex}}(y)\le\|y\|^2$ so $\E_\mu[\xi_{\mathrm{ex}}]\le1$, while $1$-Lipschitzness of $\sqrt q$ gives $\Var_\mu(\sqrt q)\le1$, hence $(\E_\mu\sqrt q)^2\ge s-1$. The hypothesis implies $\mathsf T-\Upsilon\ge\frac{\mathsf T}{1+\mathsf c\,\xi_{\mathrm{ex}}}>0$, so $\E_\mu[(\mathsf T-\Upsilon)^{-1}]\le\frac{1+\mathsf c}{\mathsf T}=\frac1{s(\mathsf A-\mathsf B)}$. By Cauchy--Schwarz, $s-1\le(\E_\mu\sqrt q)^2\le s\,\E_\nu[\mathsf T-\Upsilon]\cdot\frac1{s(\mathsf A-\mathsf B)}$, yielding $\E_\nu[\Upsilon]\le\mathsf T-(s-1)(\mathsf A-\mathsf B)=\mathsf A$.
\end{proof}

\paragraph{Ranks $1$ to $4$.}
With tilts $(\theta_1,\theta_2,\theta_3,\theta_4)=(\frac12,\frac13,\frac{11}{50},\frac1{100})$ and surpluses $(\Delta_1,\Delta_2,\Delta_3,\Delta_4)=(0,0,0,\frac{27}{100})$, put $\mathsf C'_j=\mathsf C_j-\theta_jM_{\mathrm{lo}}$, $\mathsf K'_j=\mathsf K_j-\theta_j$, and reserves $\mathsf B_j=\mathsf C'_j-\mathsf K'_j-\Delta_j$. Exact evaluation confirms $0<\mathsf K'_j<\mathsf B_j$, $M_{\mathrm{lo}}<\mathsf C_j$, and $2\mathsf K_j<\mathsf C_j$ (\cref{app:early}).

\begin{theorem}[Opening at ranks $1$ to $4$]\label{thm:early}
For every low-mean target $P$ with $\phi>0$, every $1\le j\le4$, and every state $\sigma=(C_0,\mathcal Q)$ with $|\mathcal Q|=j$ and $s=\cost(P,C_0)/\phi>0$, $\E_\nu Q_j(\sigma^y)\le(\mathsf C_j+\mathsf m_j L_{\mathrm{tot}}/s)\phi$.
\end{theorem}

\begin{proof}[Reduction]
If $L_{\mathrm{tot}}=0$, $\E_\nu Q_j/\phi\le M_{\mathrm{lo}}<\mathsf C_j$; if $s\le2$, persistence gives $Q_j/\phi\le2\mathsf K_j<\mathsf C_j$. For $s>2$ and $L_{\mathrm{tot}}>0$, apply \cref{lem:inverse-deficit} with $\mathsf B=\mathsf B_j$, $\mathsf A=\mathsf B_j+\Delta_j+\mathsf m_j L_{\mathrm{tot}}/s$, and $\Upsilon(y)=(Q_j(\sigma^y)/\phi-\theta_j\bar u(y)-\mathsf K'_j)_+$ (using $\E_\nu\bar u\le M_{\mathrm{lo}}$). For $\bar u\le1$, \eqref{eq:transfer-retention} gives $\Upsilon=0$; for $1<\bar u\le2$, $\Upsilon\le\mathsf K'_j\xi_{\mathrm{ex}}<\mathsf B_j\xi_{\mathrm{ex}}\le\frac{\mathsf T\mathsf c\xi_{\mathrm{ex}}}{1+\mathsf c\xi_{\mathrm{ex}}}$. For $\bar u>2$, putting $h=1/\bar u\in(0,\frac12)$, $x=L_{\mathrm{tot}}/\bar u>0$, $\eta_s=\bar u/s\in(0,1]$, and $W_{\mathrm{cr}}=\Delta_j/\eta_s+\mathsf m_j x$, dividing the credit by $\bar u$ yields $\theta_j+h(\mathsf C'_j-\Delta_j)+(1-2h)\frac{\mathsf B_j W_{\mathrm{cr}}}{W_{\mathrm{cr}}+(1-h)\mathsf B_j}$ (nondecreasing in $W_{\mathrm{cr}}$, so $\eta_s=1$ is worst). By~\eqref{eq:s6-master}, it suffices to verify for all $x>0$ and $h\in(0,\frac12)$ that
\begin{equation}\label{eq:early}
V_j\bigl(\gamma_{\mathrm{reset}}(h),x\bigr)<A_j(x,h):=\theta_j+h(\mathsf C'_j-\Delta_j)+(1-2h)\frac{\mathsf B_j(\mathsf m_j x+\Delta_j)}{\mathsf m_j x+\Delta_j+(1-h)\mathsf B_j},
\end{equation}
where $V_1(\gamma,x)=\frac{2x+\gamma}{1+x}$ and, for $j\in\{2,3,4\}$, $V_j(\gamma,1/p)=\gamma H_j+\sum_{i=1}^j(-1)^{i-1}\binom ji\frac{i+1-\gamma}{i+p}$ on $x\ge\frac97$ ($p=1/x$) and $V_j(\gamma,x)=\gamma H_j+(1-\gamma)((1+\vartheta_j)x-\frac79\vartheta_j x^2)+(\gamma-\frac29(1-\gamma)x)\pi_j(x)$ on $0<x\le\frac97$.
\end{proof}

Three facts reduce~\eqref{eq:early} to twenty one-variable polynomials: (1)~$A_j(x,h)=\theta_j+h(\mathsf C'_j-\Delta_j)+2y_0-\frac{y_0(\mathsf B_j+2y_0)}{(1-h)\mathsf B_j+y_0}$ ($y_0=\mathsf m_j x+\Delta_j$) is strictly concave in $h\in[0,\frac12]$, and $A_j(x,\frac12)-V_j(1,x)>\frac12(\mathsf B_j-\mathsf K'_j)>0$; (2)~$\gamma_{\mathrm{reset}}(h)\le\min(5h,\frac12+h)$ (nodes $(0,0),(\frac18,\frac58),(\frac12,1)$) for $j=1$, and for $j\in\{2,3,4\}$ lies below the three-piece envelope through $(0,0),(\frac3{80},\frac3{16}),(\frac{19}{100},\frac{69}{100}),(\frac12,1)$ whose middle chord $\frac{39}{610}+\frac{201}{61}h$ obeys $(5+14h)(\frac{39}{610}+\frac{201}{61}h)-25h>0$ (discriminant $-\frac{72371}{93025}<0$); and (3)~$V_j(\gamma,x)$ is affine and strictly increasing in $\gamma$. Thus~\eqref{eq:early} reduces to checking the non-terminal envelope nodes: two quadratics ($j=1$) and eighteen polynomials ($j=2,3,4$, nine per ray), all verified positive in exact rational arithmetic (\cref{app:early}).

\paragraph{One worked node.}
At $j=4$, upper ray $p\in(0,\frac79]$, node $(h,\gamma)=(0,0)$ ($\theta_4=\frac1{100}$, $\Delta_4=\frac{27}{100}$, $\mathsf B_4=\frac{106442351}{30000000}$, $\mathsf m_4=\frac{164}{125}$), clearing $(\mathsf m_4+(\Delta_4+\mathsf B_4)p)\prod_{i=1}^4(i+p)$ in~\eqref{eq:early} yields a quintic $P(p)$ that divides exactly as $P(p)=(p-\frac{19}{50})^2Q(p)+\alpha_0+\beta_0 p$, where $Q(p)=\frac{1614161051601937}{15625000000000}+\frac{31689582598417}{625000000000}p+\frac{108457715189}{9375000000}p^2+\frac{747121457}{750000000}p^3$ has positive coefficients and $\alpha_0=\frac{1224251371700743}{39062500000000000}>0.0313$, $\beta_0=\frac{3203795648747}{62500000000000}>0.0512$, proving $P>0$ on $[0,\frac79]$.

\begin{corollary}[Sharp rank sixteen]\label{cor:sharp16}
With $\mathsf P_{16}=\frac{4445581}{500000}$ (so $J_{18}<\mathsf P_{16}$), every target $P$ and state $\sigma=(C_0,\mathcal Q)$ with $|\mathcal Q|=16$ obey $Z_{17}(\sigma)\le J_{18}\OPT_1(P)\le\mathsf P_{16}\OPT_1(P)$ (strictly when $\OPT_1(P)>0$); hence $\mathsf P_{16}$ is a universal rank-$16$ stock.
\end{corollary}

\begin{proof}
Combine \cref{thm:high-mean,thm:low-mean,thm:hub-stocks} at $m=j=16$ with $J_{18}<8.89116164<\mathsf P_{16}$ (\cref{thm:brackets,app:hub-brackets}). In a rank-$17$ delayed query ($18$ draws, $17$ names), a named first draw leaves $17$ draws and $16$ names, while an exterior first draw reduces to $16$ names via uniform thinning (\cref{thm:thinning}); both are bounded by $\mathsf P_{16}\phi$, leaving only the target-first covered query treated next.
\end{proof}

\section{The transition at \texorpdfstring{$k=19$}{k=19} (rank \texorpdfstring{$17$}{17})}\label{sec:rank17}

This section completes the low-mean marked bound for ranks $5\le j\le16$ (closing \cref{thm:low-mean}) and proves the universal reservoir stock $\mathsf C_{17}=2+2H_{18}$ at rank~$17$ ($k=19$), where the incoming rank-$16$ stock is hub-valued ($\mathsf P_{16}$) and the outgoing stock lands on the reservoir line.

\subsection{Why rank seventeen is delicate}

The universal rank-$16$ input is $J_{18}<\mathsf P_{16}=\frac{4445581}{500000}$ (\cref{cor:sharp16}). Stepping from $\mathsf P_{16}$ to $\mathsf C_{17}=2+2H_{18}=\frac{18358381}{2042040}$ by one-step mixing (\cref{lem:mixing}) requires the marked bound at rank~$17$ with slope
\[
\mathsf m_{17}=17(\mathsf C_{17}-\mathsf P_{16})=\frac{2528406869}{1501500000}\in(1.683920,1.683921),
\]
more than $0.2049$ below the ideal reservoir slope $\frac{17}9\approx1.888889$, because the hub's lead $J_{18}-(2+2H_{17})\approx0.012$ at rank~$16$ is amplified by~$17$. The reservoir induction of \cref{sec:reservoir-regime} needs the ideal slope, so the opening method bridges $\mathsf P_{16}\to\mathsf C_{17}$ in one step.

\subsection{The theorem}

For $5\le j\le16$ let $\mathsf C_j$ and $\mathsf m_j=j(\mathsf C_j-\mathsf C_{j-1})$ be as in~\eqref{eq:stock-formula}; at $j=17$ set $\mathsf C_{17}=2+2H_{18}$ and $\mathsf m_{17}=17(\mathsf C_{17}-\mathsf P_{16})$. For $5\le j\le17$ and $L\ge0$, write $\mathsf K_j=1+H_j$, $\varrho_j(L)=\frac{H_j}{L+H_j}$, $\tilde{\mathsf k}_j(L)=\mathsf K_j-\varrho_j(L)$, $\mathsf M_{0,j}=\mathsf C_j-\mathsf K_j$ (note $\mathsf M_{0,17}-\mathsf K_{17}=\frac19$), and $\mathsf M_{\mathrm{lo}}(j,L)=\mathsf M_{0,j}+\varrho_j(L)$.

\begin{theorem}[Opening at ranks $5$ to $17$]\label{thm:rank17}
\begin{enumerate}[label=(\alph*),leftmargin=2em]
\item For every low-mean target $P$ with $\phi>0$, every $5\le j\le16$, and every state $\sigma=(C_0,\mathcal Q)$ with $|\mathcal Q|=j$ and $s=\cost(P,C_0)/\phi>0$, $\E_\nu Q_j(\sigma^y)\le(\mathsf C_j+\mathsf m_j L_{\mathrm{tot}}/s)\phi$.
\item For \emph{every} target $P$ with $\phi>0$ and every state $\sigma=(C_0,\mathcal Q)$ with $|\mathcal Q|=17$ and $s>0$, $\E_\nu Q_{17}(\sigma^y)\le(\mathsf C_{17}+\mathsf m_{17}L_{\mathrm{tot}}/s)\phi$. Consequently $\mathsf C_{17}=2+2H_{18}$ is a universal stock for rank~$17$.
\end{enumerate}
\end{theorem}

Both parts share one proof because the marked-bound step for $5\le j\le17$ uses only the universal Euclidean reset $M(\widehat C_0)<5<\mathsf C_j$—never $M_{\mathrm{lo}}$. For $5\le j\le16$, the low-mean hypothesis enters \cref{thm:low-mean} solely through the predecessor stock $\mathsf C_{j-1}$; at rank~$17$, the predecessor stock $\mathsf P_{16}$ is universal (\cref{cor:sharp16}), so \cref{lem:mixing} yields $Z_{18}(\sigma)\le\mathsf C_{17}\phi$ for every target (trivially $Z_{18}(\sigma)=0$ if $\phi=0$ or $\cost(P,C_0)=0$, and via $\frac{s(\mathsf C_{17}+\mathsf m_{17}L_{\mathrm{tot}}/s)+17L_{\mathrm{tot}}\mathsf P_{16}}{s+17L_{\mathrm{tot}}}\phi=\mathsf C_{17}\phi$ when $\phi,\cost(P,C_0)>0$).

\subsection{The prelanding reserve and regimes H and L}

Write $L=L_{\mathrm{tot}}$, $\mathsf B_{\mathrm{lo},\infty}(j)=\frac{107}{50}+\frac7{10}H_j$, and $\mathsf B_{\mathrm{lo}}(j,L)=\mathsf B_{\mathrm{lo},\infty}(j)-\frac7{20}\varrho_j(L)$. Choose the prelanding reserve (depending only on $(j,s,L)$, with equality at switches in the lower branch)
\begin{equation}\label{eq:open-reserve}
\mathsf B(j,s,L)=\begin{cases}
\mathsf B_{\mathrm{lo}}(j,L),&0\le L\le10,\\
2+\frac{39}{50}H_j,&L>10,\ 5\le j\le8,\\
\mathsf M_{0,j},&L>10,\ 9\le j\le17,
\end{cases}
\end{equation}
except at $j=17$ when $L/s>\frac97$ (\emph{regime~H}), where we set $\mathsf B(17,s,L)=\mathsf M_{\mathrm{lo}}(17,L)$; we call $j=17$, $L/s\le\frac97$ \emph{regime~L}. Exact evaluation verifies $\tilde{\mathsf k}_j(L)<\mathsf B(j,s,L)\le\mathsf M_{\mathrm{lo}}(j,L)$ throughout, with equality only in regime~H where $\mathsf A-\mathsf B=\mathsf m_{17}L/s>0$ (\cref{app:open-facts}).

\paragraph{Reduction to a pointwise inequality.}
If $L=0$, \cref{thm:reset} gives $\E_\nu Q_j/\phi<5<\mathsf C_j$; if $s\le2$, persistence gives $\E_\nu Q_j/\phi\le2\mathsf K_j<\mathsf C_j$. For $s>2$ and $L>0$, apply \cref{lem:inverse-deficit} with $\mathsf A=\mathsf M_{\mathrm{lo}}(j,L)+\mathsf m_j L/s$, $\mathsf B=\mathsf B(j,s,L)$, and $\Upsilon(y)=(Q_j(\sigma^y)/\phi-\tilde{\mathsf k}_j(L))_+$. At a landing with $\bar u=0$, $\Upsilon=0$; for $0<\bar u\le1$, \eqref{eq:transfer-retention} bounds $Q_j/\phi$ by $g_{\mathrm{ret}}(\bar u)=\bar u(H_j+\pi_j(L/\bar u))$ with $g'_{\mathrm{ret}}\ge H_j-e^{-2}>0$, so $Q_j/\phi\le H_j+\pi_j(L)\le\tilde{\mathsf k}_j(L)$ (since $\pi_j(L)^{-1}\ge1+H_j/L$) and $\Upsilon=0$; for $1<\bar u\le2$, $\Upsilon\le\tilde{\mathsf k}_j(L)\xi_{\mathrm{ex}}<\mathsf B\xi_{\mathrm{ex}}\le\frac{\mathsf T\mathsf c\xi_{\mathrm{ex}}}{1+\mathsf c\xi_{\mathrm{ex}}}$.

For $\bar u>2$, set $h=1/\bar u\in(0,\frac12)$, $x=L/\bar u>0$, $\eta_s=\bar u/s\in(0,1]$, and $\Delta_{\mathrm{res}}=\mathsf M_{\mathrm{lo}}(j,L)-\mathsf B\ge0$. Dividing the credit by $\bar u$ and applying~\eqref{eq:s6-master} reduces the claim to
\begin{equation}\label{eq:open-pointwise}
\begin{gathered}
A_{\mathrm{tot}}(x,h;\eta_s):=h\bigl(\tilde{\mathsf k}_j(L)+\mathsf B\bigr)+(1-2h)\frac{\mathsf B(\mathsf m_j x+\Delta_{\mathrm{res}}/\eta_s)}{\mathsf m_j x+\Delta_{\mathrm{res}}/\eta_s+(1-h)\mathsf B}\ >\ Q^{\mathrm{open}}_j(x,h),\\
Q^{\mathrm{open}}_j(x,h):=\bigl(1-\gamma_{\mathrm{reset}}(h)\bigr)W_j(x)+\gamma_{\mathrm{reset}}(h)\bigl(H_j+\pi_j(x)\bigr),
\end{gathered}
\end{equation}
with $L=x/h$. Since $A_{\mathrm{tot}}$ is nonincreasing in $\eta_s$: $\eta_s=1$ is worst for $5\le j\le16$ and for $j=17$ on the \emph{lower arm} $x\le\frac97$; in regime~H ($j=17$, $L/s>\frac97$), $\Delta_{\mathrm{res}}=0$ drops $\eta_s$ and retention forces $x\ge L/s>\frac97$ (every landing lies on the \emph{higher arm} $x\ge\frac97$, even when $L\le10$); and in regime~L on the higher arm ($p=1/x\in(0,\frac79]$), retention gives $\eta_s=(L/s)p\le\frac97p$, so $\eta_s=\frac97p$ is worst. At $h=\frac12$, $\gamma_{\mathrm{reset}}(\frac12)=1$ and $1-\pi_j(\frac L2)>\varrho_j(L)$ give $A_{\mathrm{tot}}(x,\frac12;\eta_s)-(H_j+\pi_j(\frac L2))>\frac12(\mathsf B-\tilde{\mathsf k}_j(L))>0$. On the higher arm, writing $\Pi_j(p)=\pi_j(1/p)$, $b_j(p)=(1-\Pi_j(p))/p$, $A_{\mathrm{cap}}=A_{\mathrm{tot}}-\mathsf K_j$, and $S^{\mathrm{open}}_j(p,\gamma)=(1-\gamma)(H_j-b_j(p))+pb_j(p)$ turns~\eqref{eq:open-pointwise} into $A_{\mathrm{cap}}(p,h;\eta_s)+S^{\mathrm{open}}_j(p,\gamma_{\mathrm{reset}}(h))>0$.

\subsection{Five structural facts and finite verification}

Five elementary convexity and curvature properties (proved in \cref{app:open-facts}) reduce~\eqref{eq:open-pointwise} on the entire domain $5\le j\le17$, $x>0$, $h\in(0,\frac12]$ to finitely many exact rational and Bernstein-polynomial inequalities:
\begin{enumerate}[leftmargin=2em]
\item \emph{Harmonic-mean concavity.} With $v_{\mathrm{cr}}=\mathsf m_j x+(\mathsf M_{0,j}-\mathsf B+\varrho_j)/\eta_s$, $A_{\mathrm{tot}}+h\varrho_j=h(\mathsf K_j+\mathsf B)+(1-2h)(\mathsf B^{-1}+(1-h)v_{\mathrm{cr}}^{-1})^{-1}$ is increasing and jointly concave in $(\mathsf B,v_{\mathrm{cr}})$, giving concavity in $\mathsf B$ and harmonic-rank interpolation between knots at an explicit fee.
\item \emph{Concavity in $h$.} The exact identity $D^3\partial_{hh}((1-h)(1-2h)/D)=4F_1F_2-(1-h)(1-2h)DD''$ ($F_1,F_2>0$, $D''\le0$) makes $A_{\mathrm{cap}}$ and $A_{\mathrm{tot}}$ concave in $h\in[0,\frac12]$ across all constant-$\mathsf B$ regions, lower-arm $L$-dependent regions (via $v_{\mathrm{cr}}<\mathsf B$), and regime~H, reducing $h$-intervals to endpoints (closing $h\in[\frac{11}{40},\frac12]$ on the lower arm directly).
\item \emph{Race curvature and $C^1$ convex majorants.} Representing $\pi_k(x)=\E[e^{-Y_k/x}]$ ($Y_k=\sum_{i=1}^k\mathrm{Exp}(i)$) and using $t^3e^{-t}\le27e^{-3}$ gives $2\pi'_k+x\pi''_k\le\frac{27}{e^3}\int_0^\infty\Pi_k(p)\,dp<M_k$ at knots $(k,M_k)\in\{(5,\frac45),(8,\frac{13}{20}),(12,\frac{11}{20}),(17,\frac12)\}$, yielding $C^1$ convex majorants $g_\pi(k,\cdot)\ge\pi_k$ with the same curvature cap.
\item \emph{Concavity in $\tau_L=1/L$ for $L\ge10$.} Replacing $\gamma_{\mathrm{reset}}(\tau_L x)/x=\frac{5\tau_L}{1+\frac{14}5\tau_L x}$ by an affine-in-$\tau_L$ tangent on each of three segments covering $[0,\frac1{10}]$ uses Fact~2 to reduce $L\ge10$ on the lower arm to four rational $\tau_L$-endpoints per rank.
\item \emph{Higher-arm seams.} Dominating $\gamma_{\mathrm{reset}}$ by a chord on $[\frac3{25},\frac{21}{100}]$ and by $\frac12+h$ on $[\frac{21}{100},\frac12]$ uses Fact~2 to reduce $h\in[\frac3{25},\frac12]$ to one seam at $(h_0,\gamma_0)=(\frac{21}{100},\frac{71}{100})$ for $L>10$ (and all $L>0$ in regime~H), and to three $p$-subbands via Fact~1 for $L\le10$ ($p\ge\frac15$).
\end{enumerate}

These five facts partition $(x,h)$ cleanly across every boundary ($x=\frac97$, $L=10$, $h\in\{\frac3{25},\frac{11}{40},\frac12\}$, $L/s=\frac97$) into finite families verified by exact rational arithmetic and Bernstein positivity using computer-checked companion data (not by hand): on the lower arm $x\le\frac97$, nine near-$L$ rows ($L\le10$) and $52$ far-$L$ cells ($L\ge10$); on the higher arm $x\ge\frac97$, seam polynomials ($h\ge\frac3{25}$), ten price-rank boxes carrying $27$ knot rows ($h\le\frac3{25}$, ranks $5$--$16$), and eight pure-M\"obius boxes at rank~$17$ (four in regime~H, four in regime~L). \Cref{app:sec8} proves complete boundary coverage, explains the four-guard structure and every exceptional boundary case, and works one real row of the $27$ in exact rational arithmetic, completing \cref{thm:rank17}.

\section{The reservoir regime, \texorpdfstring{$k\ge20$}{k >= 20}}\label{sec:reservoir-regime}

This section proves that $\mathsf C_j=2+2H_{j+1}$ is a universal stock for every rank $j\ge17$, inducting upward from $\mathsf C_{17}=2+2H_{18}$ (\cref{thm:rank17}(b)). For every $j\ge18$ the predecessor lies on the reservoir line, giving the ideal mixing slope
\[
\mathsf m_j=j(\mathsf C_j-\mathsf C_{j-1})=2-\frac2{j+1},
\]
starting at $\mathsf m_{18}=\frac{36}{19}$, and keeping the terminal cost inside the stock without paying~$5$ separately.

\paragraph{The idea.}
Refining \cref{sec:ov-idea}, after a first hit at a normalized point $y$, serving the cluster from $y$ alone gives cost $2+(\|y\|^2-1)$. Points inside the unit ball carry a deficit $(1-\|y\|^2)_+$ below~$2$; outside points carry an excess $(\|y\|^2-1)_+$ repaired by later inside hits. We introduce an explicit parameter-free kernel $\mathsf P_{m,n}(z)$ tracking this repair rate across all covered clocks, prove a signed covered bound directly for this kernel, and combine a two-cohort inequality with an $\frac{89}{20}$ clip to close the marked comparison uniformly for all $j\ge18$.

\subsection{The explicit covered kernel and its four laws}\label{sec:reservoir-kernel}

A \emph{covered clock} $(m,n)$ ($1\le m\le n$) tracks the live-name counts $\{m,\dots,n\}$ over $r=n-m+1$ draws from $n$ live names, with $b_{m,n}=\sum_{i=m}^n\frac1i=\Cl(r,n)$, $h^{(2)}_{m,n}=\sum_{i=m}^n i^{-2}$, and $\kappa_{m,n}=1+\frac{b_{m,n}}2+\frac{h^{(2)}_{m,n}}{2b_{m,n}}$. Define on $[0,\infty)$ the concrete parameter-free kernel
\begin{gather*}
\mathsf P_{m,n}(z)=1+\kappa_{m,n}z+b_{m,n}z^2-\bigl(\sqrt{b_{m,n}}\,z-1\bigr)_+^2
=\begin{cases}1+\kappa_{m,n}z+b_{m,n}z^2,&\sqrt{b_{m,n}}\,z\le1,\\(\kappa_{m,n}+2\sqrt{b_{m,n}})\,z,&\sqrt{b_{m,n}}\,z\ge1,\end{cases}\\
\mathsf R_{m,n}(z)=\frac{b_{m,n}z}{\mathsf P_{m,n}(z)}=\max_{0\le t\le z}\frac{b_{m,n}t}{1+\kappa_{m,n}t+b_{m,n}t^2},\qquad
\mathsf N_{m,n}(z)=\frac{b_{m,n}}{\mathsf P_{m,n}(z)},
\end{gather*}
with $b_{m,n}=0$ and $\mathsf R_{m,n}\equiv\mathsf N_{m,n}\equiv0$ on an empty clock ($m>n$). On every nonempty clock, $\mathsf P_{m,n}(0)=1$ and $\mathsf N_{m,n}(0)=b_{m,n}$.

\begin{proposition}[The four kernel laws and secant bound]\label{thm:kernel}
On every nonempty covered clock $(m,n)$, $\mathsf P_{m,n}(0)=1$, the secant bound $\frac{\mathsf P_{m,n}(z_1)-\mathsf P_{m,n}(z_0)}{z_1-z_0}\le\kappa_{m,n}+2\sqrt{b_{m,n}}$ holds for $0\le z_0<z_1$, and for all $z\ge0$:
\begin{enumerate}[label=\textup{(L\arabic*)},leftmargin=2.5em]
\item $\mathsf R_{m,n}(0)=0$, and $\mathsf R_{m,n}$ is continuous, nondecreasing, and concave on $[0,\infty)$;
\item $\mathsf P'_{m,n}(z)\le\mathsf P_{m,n}(z)+b_{m,n}$;
\item $\mathsf R_{m,n}(z)\ge\mathsf R_{m+1,n}(z)$;
\item $\bigl(1+\frac zn\bigr)\mathsf R_{m,n}(z)-\mathsf R_{m,n-1}(z)\ge\frac zn(1-z)$.
\end{enumerate}
Moreover, whenever $b=\tilde\omega^2$ with $\tilde\omega>1$ and $h^{(2)}\le\frac53$,
\begin{equation}\label{eq:trunc-margin}
\tilde\omega^2-\Bigl(1-\frac1{\tilde\omega}\Bigr)\Bigl(1+\frac{\tilde\omega^2}2+\frac{h^{(2)}}{2\tilde\omega^2}+2\tilde\omega\Bigr)>\frac{112}{405}.
\end{equation}
\end{proposition}

\begin{proof}
Write $b=b_{m,n}$, $h^{(2)}=h^{(2)}_{m,n}$, $\kappa=\kappa_{m,n}$.

\emph{(L1) and secant bound.} The two pieces of $\mathsf P_{m,n}$ match in value and slope at $z=1/\sqrt b$ with $\mathsf P''_{m,n}=2b\mathbf 1_{\sqrt bz<1}\ge0$, so $0<\mathsf P'_{m,n}\le\kappa+2\sqrt b$. Also $\mathsf P_{m,n}-z\mathsf P'_{m,n}=(1-bz^2)_+$ is nonnegative and nonincreasing, so $\mathsf R'_{m,n}=b(1-bz^2)_+/\mathsf P_{m,n}^2\ge0$ is nonincreasing, giving (L1).

\emph{(L3).} In $\mathsf P_{m,n}/b=\frac{1+z}b+\frac z2+\frac{h^{(2)}}{2b^2}z+z^2-(z-1/\sqrt b)_+^2$, inserting $e\in(0,1]$ via $(b,h^{(2)})\mapsto(b+e,h^{(2)}+e^2)$ lowers $\frac1b$ and $\frac1b+\frac{h^{(2)}}{2b^2}$ (difference $\frac{eb[(2-e)b+2e]+h^{(2)}e(2b+e)}{2b^2(b+e)^2}>0$) without raising $-(z-1/\sqrt b)_+^2$; hence $\mathsf R$ increases when a reciprocal is inserted.

\emph{(L2).} For $\sqrt bz\ge1$, $\mathsf P'_{m,n}$ is constant and $\mathsf P_{m,n}$ increases. On $[0,1/\sqrt b]$, $f(z)=\mathsf P_{m,n}(z)+b-\mathsf P'_{m,n}(z)=\frac{b-h^{(2)}/b}2+(1-\frac32b+\frac{h^{(2)}}{2b})z+bz^2$ is affine in $h^{(2)}\in[0,\min(b,b^2)]$. Checking the three endpoints closes (L2): at $h^{(2)}=b^2$ ($b\le1$), $f=(1-b)z+bz^2\ge0$; at $h^{(2)}=0$, $f(0)=\frac b2$ ($b\le\frac23$), $\min f=\frac{-b^2+12b-4}{16b}>0$ ($\frac23<b\le9$), and $f(1/\tilde\omega)=\frac{(\tilde\omega-2)^2}2+\frac{\tilde\omega}2+\frac1{\tilde\omega}-1>0$ ($\tilde\omega=\sqrt b>3$); at $h^{(2)}=b$ ($b\ge1$), $\min f=\frac{(b-1)(9-b)}{16b}\ge0$ ($1\le b\le9$) and $f(1/\tilde\omega)=\frac{(\tilde\omega-3/2)^2}2+\frac3{2\tilde\omega}-\frac58>0$ ($\tilde\omega>3$).

\emph{Truncation margin and (L4).} For $\tilde\omega>1$, $h^{(2)}\le\frac53$, and $t=1/\tilde\omega\in(0,1)$, the left-hand side of~\eqref{eq:trunc-margin} equals the exact identity $\frac{(\tilde\omega-2)^2}2+\bigl(\frac{\tilde\omega}2+\frac1{\tilde\omega}-1\bigr)-\bigl(\frac{h^{(2)}}2\bigr)t^2(1-t)$; since $\frac{\tilde\omega}2+\frac1{\tilde\omega}-\frac75=\frac{5(\tilde\omega-7/5)^2+1/5}{10\tilde\omega}>0$ and $\frac4{27}-t^2(1-t)=\frac{(3t-2)^2(3t+1)}{27}\ge0$, it exceeds $0+\frac25-\frac56\cdot\frac4{27}=\frac{112}{405}$. Now set $e=\frac1n$. Law (L4) is trivial at $z=0$, follows from clock monotonicity for $z\ge1$, and for $m=n$ gives $\frac{ez^3}{1+z}>0$. For $m<n$ and $1/\sqrt b\le z<1$, clock monotonicity and~\eqref{eq:trunc-margin} (since $h^{(2)}<\frac53$) give $(1+ez)\mathsf R_{m,n}-\mathsf R_{m,n-1}\ge\frac{ezb}{\kappa+2\sqrt b}>ez(1-z)$. Below truncation ($0<z<1/\sqrt b$), clearing $\mathsf P_{m,n}\mathsf P_{m,n-1}/(ez^3)$ leaves a cubic $c_0+c_1z+c_2z^2+c_3z^3$ with $c_1,c_2,c_3>0$ and $c_0=(\frac b2-1)^2+e(1-\frac b4)-\frac{(h^{(2)}-be)^2}{4b(b-e)}$, shown positive across five cases in exact rational arithmetic (\cref{app:tail-l4}).
\end{proof}

\subsection{The signed covered bound and finite-difference cap}\label{sec:reservoir-signed}

For $P\subset\R^d$ with $\phi>0$, set $\mathcal I=\{y\in P:\|y-\bar y_P\|^2\le\phi/|P|\}$ (in normalized coordinates, $\{\|Y\|^2\le1\}$). On $(m,n)$ with $b=b_{m,n}$, set $\mathcal C_{m,b}(\tilde h,L)=\frac{\tilde h^2 b}{m(L+\tilde h b)}$ ($\mathcal C_{m,b}(0,0)=0$) and $\mathcal G_{m,b}(\tilde h,L)=\tilde h b-\mathcal C_{m,b}(\tilde h,L)$ ($\mathcal C_{m,0}=\mathcal G_{m,0}\equiv0$), with the convention $\min\{L,(\bar u-2)_+\mathsf N_{m,n}(g_{\mathrm{in}}/L)\}=0$ when $L=0$.

\begin{theorem}[Signed covered bound]\label{thm:signed}
Let $\mathsf P_{m,n},\mathsf R_{m,n},\mathsf N_{m,n}$ be the explicit kernel of \cref{sec:reservoir-kernel}. For every input $X\subset\R^d$, target $P\subseteq X$ with $\phi>0$, nonempty $C_0\subset\R^d$, name family $\mathcal Q$ ($n=|\mathcal Q|$), draw count $0\le r\le n$, and price bound $L\ge\ell_0/\phi$, put $m=n-r+1$, $b=b_{m,n}$, $\bar u=\cost(P,C_0)/\phi$, and $g_{\mathrm{in}}=\cost(\mathcal I,C_0)/\phi$. Then
\begin{equation}\label{eq:signed}
\frac{Q_r(C_0,\mathcal Q)}\phi\le\bar u+\mathcal G_{m,b}\bigl(\min(\bar u,2),L\bigr)+\min\bigl\{L,\ (\bar u-2)_+\mathsf N_{m,n}(g_{\mathrm{in}}/L)\bigr\}.
\end{equation}
\end{theorem}

\begin{proof}
Normalize $\phi=1$. The cases $r=0$ and actual live price $\ell=0$ are immediate; for $\ell>0$, monotonicity of $\mathcal G_{m,b}$ and of $L\mapsto\min\{L,(\bar u-2)_+\mathsf N_{m,n}(g_{\mathrm{in}}/L)\}$—since $\mathsf N_{m,n}(g_{\mathrm{in}}/L)$ equals the constant $b_{m,n}$ when $g_{\mathrm{in}}=0$ and equals $\frac L{g_{\mathrm{in}}}\mathsf R_{m,n}(g_{\mathrm{in}}/L)$ (increasing by (L1)) when $g_{\mathrm{in}}>0$—reduces~\eqref{eq:signed} to $L=\ell$. Note that $\mathcal G_{m,b}$ increases in $\tilde h$ and is increasing and concave in $\ell$; $\ell\mapsto\mathsf N_{m,n}(g_{\mathrm{in}}/\ell)$ is increasing and concave on $(0,\infty)$ (constant $b_{m,n}$ if $g_{\mathrm{in}}=0$, and the perspective $\frac\ell{g_{\mathrm{in}}}\mathsf R_{m,n}(g_{\mathrm{in}}/\ell)$ of concave nondecreasing $\mathsf R_{m,n}$ by (L1) if $g_{\mathrm{in}}>0$); and by (L2), $-\partial_z\mathsf N_{m,n}\le\mathsf N_{m,n}(1+\mathsf N_{m,n})$, making the right-hand side of~\eqref{eq:signed} nondecreasing along retention contractions $d\bar u\ge dg_{\mathrm{in}}\ge0$.

\emph{Base $r=1$ ($m=n$, $b=\frac1n$).} With $T=\bar u+n\ell+e_{\mathrm{ext}}$, $e=(\bar u-2)_+$, and $z=g_{\mathrm{in}}/\ell$, an inside hit drops terminal cost to $\le2$, so $Q_1-\bar u\le\min\{\ell,(\bar u\ell-eg_{\mathrm{in}})/T\}$. If $\bar u\le2$, then $Q_1-\bar u\le\frac{\bar u\ell}{\bar u+n\ell}=\mathcal G_{n,1/n}(\bar u,\ell)$. If $\bar u>2$, $G_1=\mathcal G_{n,1/n}(2,\ell)$ obeys $TG_1\ge2\ell$, while (L4) against the empty child gives $(n+z)\mathsf N_{n,n}(z)\ge1-z$, so $T\mathsf N_{n,n}(z)+g_{\mathrm{in}}\ge\ell$ and $T(G_1+e\mathsf N_{n,n}(z))\ge\bar u\ell-eg_{\mathrm{in}}$.

\emph{Inductive step $r\ge2$.} Let $U_{\mathrm{ext}}=\cost(X\setminus(P\cup\bigcup\mathcal Q),C_0)$ so $T=\bar u+n\ell+U_{\mathrm{ext}}$, and put $b'=b-\frac1n$ and $b''=b-\frac1m$. One step of the covered recurrence gives
\[
T\,Q_r(C_0,\mathcal Q)=\bar u\ell+\sum_{y\in X}D(y,C_0)^2\,Q_{r-1}(C_0\cup\{y\},\mathcal Q_y),
\]
where $\mathcal Q_y=\mathcal Q\setminus\{G\}$ if $y\in G\in\mathcal Q$ (moving to clock $(m,n-1)$) and $\mathcal Q_y=\mathcal Q$ otherwise (moving to clock $(m+1,n)$). Let $\ell_y$ denote the resulting mean live-name cost after adding $y$. We consider three exhaustive branches:
\begin{itemize}[leftmargin=1.5em]
\item \emph{Low branch ($\bar u\le2$).} Every non-named draw $y\notin\bigcup\mathcal Q$ contracts $\bar u_y\le\bar u\le2$, so the inductive hypothesis on $(m+1,n)$ gives $Q_{r-1}\le\bar u+\mathcal G_{m+1,b''}(\bar u,\ell_y)\le\bar u+\bar ub''$ (total weight $\bar u+U_{\mathrm{ext}}$). Each named draw $y\in G\in\mathcal Q$ gives $Q_{r-1}\le\bar u+\mathcal G_{m,b'}(\bar u,\ell_y)$; because $\ell'\mapsto\mathcal G_{m,b'}(\bar u,\ell')$ is concave and nondecreasing, Jensen's inequality and the size-biased mean bound~\eqref{eq:size-biased} bound the named sum by $n\ell(\bar u+\mathcal G_{m,b'}(\bar u,\ell))$. Subtracting $\bar u\ell$ plus both sums from $T(\bar u+\mathcal G_{m,b}(\bar u,\ell))$ leaves
\[
U_{\mathrm{ext}}\Bigl(\frac{\bar u}m-\mathcal C_{m,b}(\bar u,\ell)\Bigr)+\bar u\Bigl(\frac{\bar u}m-\mathcal C_{m,b}(\bar u,\ell)\Bigr)-n\ell\bigl(\mathcal C_{m,b}(\bar u,\ell)-\mathcal C_{m,b'}(\bar u,\ell)\bigr),
\]
where $\frac{\bar u}m-\mathcal C_{m,b}(\bar u,\ell)=\frac{\bar u\ell}{m(\ell+\bar ub)}\ge0$ and the last two terms combine into the exact nonnegative identity $\frac{\bar u^2\ell}{m(\ell+\bar ub)}\bigl(1-\frac\ell{\ell+\bar ub'}\bigr)\ge0$.
\item \emph{Saturated branch ($\bar u>2$, $(\bar u-2)\mathsf N_{m,n}(g_{\mathrm{in}}/\ell)\ge\ell$).} Directly without induction, retention gives terminal cost $\le\bar u$, and \cref{cor:saturated} bounds the covered occurrences by $\ell+2b''\le\ell+\mathcal G_{m,b}(2,\ell)$ since $\mathcal C_{m,b}(2,\ell)\le\frac2m=2(b-b'')$.
\item \emph{Unsaturated branch ($\bar u>2$, $e=\bar u-2>0$, $e\mathsf N_{m,n}(g_{\mathrm{in}}/\ell)<\ell$).} An inside hit $y\in\mathcal I$ (total weight $g_{\mathrm{in}}$) drops $\bar u_y\le1+\|y\|^2\le2$, so the inductive hypothesis on $(m+1,n)$ gives $Q_{r-1}\le2+2b''$, saving $e(1+\mathsf N_{m,n}(g_{\mathrm{in}}/\ell))$ relative to $\bar u+2b''+e\mathsf N_{m,n}(g_{\mathrm{in}}/\ell)$. Any other non-named draw (total weight $U_{\mathrm{ext}}+\bar u-g_{\mathrm{in}}$) contracts $\bar u$ and $g_{\mathrm{in}}$ with $0\le\bar u-\bar u_y$ and $0\le g_{\mathrm{in}}-g_{\mathrm{in},y}\le\bar u-\bar u_y$, so (L3) together with contraction monotonicity from (L2) gives $Q_{r-1}\le\bar u+2b''+e\mathsf N_{m,n}(g_{\mathrm{in}}/\ell)$. Each named draw $y\in G\in\mathcal Q$ gives $Q_{r-1}\le\bar u+\mathcal G_{m,b'}(2,\ell_y)+e\mathsf N_{m,n-1}(g_{\mathrm{in}}/\ell_y)$; since both $\ell'\mapsto\mathcal G_{m,b'}(2,\ell')$ and $\ell'\mapsto\mathsf N_{m,n-1}(g_{\mathrm{in}}/\ell')$ are concave and nondecreasing (by (L1)), Jensen's inequality and~\eqref{eq:size-biased} bound the named sum by $n\ell(\bar u+\mathcal G_{m,b'}(2,\ell)+e\mathsf N_{m,n-1}(g_{\mathrm{in}}/\ell))$. Subtracting $\bar u\ell=2\ell+e\ell$ plus all child sums from $T(\bar u+\mathcal G_{m,b}(2,\ell)+e\mathsf N_{m,n}(g_{\mathrm{in}}/\ell))$ decomposes the margin into three nonnegative pieces: the exterior baseline margin $U_{\mathrm{ext}}(\mathcal G_{m,b}(2,\ell)-2b'')\ge0$, the $\tilde h=2$ baseline balance $\bar u(\frac2m-\mathcal C_{m,b}(2,\ell))-n\ell(\mathcal C_{m,b}(2,\ell)-\mathcal C_{m,b'}(2,\ell))>0$ (using $\bar u>2$ and the low-branch identity at $\tilde h=2$), and the excess balance $e\bigl[g_{\mathrm{in}}(1+\mathsf N_{m,n}(z))+n\ell(\mathsf N_{m,n}(z)-\mathsf N_{m,n-1}(z))-\ell\bigr]\ge0$ ($z=g_{\mathrm{in}}/\ell$), which is Law (L4) multiplied by $ne\ell/z$ when $g_{\mathrm{in}}>0$ (and $e[n\ell(\frac1n)-\ell]=0$ when $g_{\mathrm{in}}=0$).
\end{itemize}
\end{proof}

\begin{lemma}[Finite-difference cap]\label{lem:fd-cap}
For $j\ge1$, $L>0$, $e\ge0$, $0\le z_0\le z_1$ with $\mathsf P_{1,j}(z_1)-\mathsf P_{1,j}(z_0)\le\mathrm{cap}\cdot(z_1-z_0)$, and $\Delta\ge L(z_1-z_0)$, setting $\mathsf N_i=\mathsf N_{1,j}(z_i)$ and $\zeta(z_1)=\frac{[\mathrm{cap}-H_j]_+}{\mathsf P_{1,j}(z_1)}$ gives $\min(L,e\mathsf N_0)-\min(L,(e+\Delta)\mathsf N_1)\le\zeta(z_1)\Delta$.
\end{lemma}

\begin{proof}
Since $\mathsf N_0\ge\mathsf N_1>0$, the left side is maximized at $e=L/\mathsf N_0$, where it equals $[L(1-\mathsf N_1/\mathsf N_0)-\Delta\mathsf N_1]_+=[L(\mathsf P_{1,j}(z_1)-\mathsf P_{1,j}(z_0))-H_j\Delta]_+/\mathsf P_{1,j}(z_1)\le\zeta(z_1)\Delta$.
\end{proof}

\subsection{Reduction to a budget at each outside mark}\label{sec:reservoir-reduction}

Fix $j\ge18$ and normalized $P$ ($\phi>0$, $\mathcal I=\{\|Y\|^2\le1\}$). Put $\delta_{\mathrm{in}}(Y)=(1-\|Y\|^2)_+$, $\xi_{\mathrm{out}}(Y)=(\|Y\|^2-1)_+$, $B_{\mathrm{in}}=\E_\mu\delta_{\mathrm{in}}=\E_\mu\xi_{\mathrm{out}}$, and (when $B_{\mathrm{in}}>0$) $D_{\mathrm{in}}=\E_\mu[\delta_{\mathrm{in}}^2]/B_{\mathrm{in}}\in[B_{\mathrm{in}},1]$ and $d\mu_1=\frac{\delta_{\mathrm{in}}}{B_{\mathrm{in}}}\,d\mu$. Write $b=H_j$, $h^{(2)}=h^{(2)}_{1,j}$, $\epsilon_j=\frac2{j+1}$, $\mathsf m_j=2-\epsilon_j$, $K_{\mathrm{res}}=1+b+\epsilon_j$. At a state with $s=\E_\mu q>2$ and $L=L_{\mathrm{tot}}>0$, put $\lambda_{\mathrm{res}}=L/(sb)$, $a_{\mathrm{in}}=\E_\nu\delta_{\mathrm{in}}$, and $\upsilon=\sqrt{a_{\mathrm{in}}/B_{\mathrm{in}}}$. For an outside landing $y\in P$ ($\|y\|^2>1$), put $g_1(y)=\E_\mu[\mathbf 1_{\mathcal I}\|y-Y\|^2]$, $\varrho=bg_1(y)/L$, $\psi=\sqrt{D_{\mathrm{in}}\lambda_{\mathrm{res}}\varrho/B_{\mathrm{in}}}$, $I_{\mathrm{out}}=(\upsilon+\psi)^2$, $\rho_{\mathrm{cl}}(y)=\min(1,\frac{s-2}{\|y\|^2-1})$, and $\Upsilon_y=\rho_{\mathrm{cl}}(y)q(y)/s$.

\begin{theorem}[Opening reduction]\label{thm:opening-reduction}
Fix $j\ge1$ with $\mathsf C_j:=2+2H_{j+1}\ge5$, and let $\mathsf C_{\mathrm{prev}}\le\mathsf C_j$ be a rank-$(j-1)$ stock with $\mathsf m=j(\mathsf C_j-\mathsf C_{\mathrm{prev}})\ge0$. Suppose that whenever $B_{\mathrm{in}}>0$, $s>2$, $L>0$, and $\|y\|^2>1$, the budget
\[
\mathrm{Bud}=(1+b)\upsilon^2+\frac{\epsilon_j}{B_{\mathrm{in}}}+\frac{\mathsf m b}{D_{\mathrm{in}}\varrho}\psi^2+\mathcal E_{\mathrm{clip}},\qquad
\mathcal E_{\mathrm{clip}}=\frac{E_{\mathrm{in}}^2}{E_{\mathrm{in}}+s\psi^2/(D_{\mathrm{in}}\varrho)}\mathbf 1_{\upsilon\le1},\quad E_{\mathrm{in}}=\frac2{B_{\mathrm{in}}}-\upsilon^2,
\]
dominates either the unclipped cost $(1+\frac b{\mathsf P_{1,j}(\varrho/b)})I_{\mathrm{out}}$ or, whenever $0\le\zeta\le1$, the clipped cost $(1-\zeta)U_{\mathrm{cl}}+(\frac b{\mathsf P_{1,j}(\varrho/b)}+\zeta)I_{\mathrm{out}}$ for some $U_{\mathrm{cl}}\ge\Upsilon_y$, where $\zeta=\zeta(\varrho/b)$ (\cref{lem:fd-cap}). Then $\E_\nu Q_j(\sigma^y)\le(\mathsf C_j+\mathsf m L/s)\phi$, and $\mathsf C_j$ is a rank-$j$ stock.
\end{theorem}

The guard $\zeta\le1$ is necessary so that $1-\zeta\ge0$ preserves $U_{\mathrm{cl}}\ge\Upsilon_y$ (at $j=18$, $\zeta>1$ as $\varrho\downarrow0$, where \cref{thm:marked-tail} uses the unclipped arm).

\begin{proof}
Boundary cases: $L=0$ gives $\E_\nu Q_j/\phi\le5\le\mathsf C_j$ (\cref{lem:five}); $B_{\mathrm{in}}=0$ or $s\le2$ implies $\bar u(y)\le2$, so \cref{thm:signed} gives $Q_j/\phi\le2+2b<\mathsf C_j$. Assume $B_{\mathrm{in}}>0$, $s>2$, $L>0$.

Applying \cref{thm:signed} on clock $(1,j)$ (where $m=1$, $b=H_j$, and $\mathcal C_{1,b}(\tilde h,L)=\frac{\tilde h^2 b}{L+\tilde h b}=\frac{\tilde h^2}{\tilde h+s\lambda_{\mathrm{res}}}$), using the pointwise identity $\bar u+b\min(\bar u,2)=2(1+b)+(\bar u-2)_+-(1+b)(2-\bar u)_+$ together with $\mathsf C_j=2+2b+\epsilon_j$, and taking expectations under $\nu$ reduces $\E_\nu Q_j/\phi\le\mathsf C_j+\mathsf m L/s$ to
\begin{equation}\label{eq:tail-step1}
\E_\nu\bigl[(\bar u-2)_++\min\{L,(\bar u-2)_+\mathsf N_{1,j}(g_{\mathrm{in}}/L)\}\bigr]\le\epsilon_j+\mathsf m\frac Ls+(1+b)A_{\mathrm{def}}+\E_\nu\,\mathcal C_{1,b}\bigl(\min(\bar u,2),L\bigr),
\end{equation}
where $A_{\mathrm{def}}:=\E_\nu(2-\bar u)_+\in[0,2]$. Because landing at $y$ gives $\bar u(y)\le1+\|y\|^2$, we have $(2-\bar u(y))_+\ge(1-\|y\|^2)_+=\delta_{\mathrm{in}}(y)$ pointwise, whence $A_{\mathrm{def}}\ge\E_\nu\delta_{\mathrm{in}}=a_{\mathrm{in}}=B_{\mathrm{in}}\upsilon^2$ and $\E_\nu[\min(\bar u,2)]=2-A_{\mathrm{def}}$. Direct differentiation gives $\partial_{\tilde h}^2\mathcal C_{1,b}(\tilde h,L)=\frac{2bL^2}{(L+\tilde hb)^3}>0$, so Jensen's inequality yields $\E_\nu\,\mathcal C_{1,b}(\min(\bar u,2),L)\ge\mathcal C_{1,b}(2-A_{\mathrm{def}},L)$. Moreover $\partial_{\tilde h}\mathcal C_{1,b}(\tilde h,L)=b\bigl(1-\frac{L^2}{(L+\tilde hb)^2}\bigr)<b<1+b$, so the function $\Psi_{\mathrm{in}}(A):=(1+b)A+\mathcal C_{1,b}(2-A,L)$ is strictly increasing on $[0,2]$. Replacing $A_{\mathrm{def}}$ by the smaller quantity $a_{\mathrm{in}}$ is therefore conservative, bounding the right-hand side of~\eqref{eq:tail-step1} below by $\epsilon_j+\mathsf m\frac Ls+(1+b)a_{\mathrm{in}}+\frac{(2-a_{\mathrm{in}})_+^2}{(2-a_{\mathrm{in}})_++s\lambda_{\mathrm{res}}}\mathbf 1_{\upsilon\le1}$ (since $a_{\mathrm{in}}\le2$ whenever $\upsilon\le1$). Dividing by $B_{\mathrm{in}}>0$ and substituting $L/s=\lambda_{\mathrm{res}}b$, $L/(bB_{\mathrm{in}})=s\lambda_{\mathrm{res}}/B_{\mathrm{in}}=s\psi^2/(D_{\mathrm{in}}\varrho)$, and $(2-a_{\mathrm{in}})/B_{\mathrm{in}}=\frac2{B_{\mathrm{in}}}-\upsilon^2=E_{\mathrm{in}}$ gives precisely $\mathrm{Bud}$ (independent of $y$).

Only landings with $\bar u(y)>2$ (hence $\|y\|^2>1$) contribute. Put $e=\bar u(y)-2\le\rho_{\mathrm{cl}}(y)\xi_{\mathrm{out}}(y)$, ideal pair $(\bar u^*,g^*)=(2+\xi_{\mathrm{out}}(y),g_1(y))$, and $\Delta=\bar u^*-\bar u(y)=\xi_{\mathrm{out}}(y)-e\ge g^*-g_{\mathrm{in}}(y)\ge0$. Either (i)~monotonicity along $d\bar u\ge dg_{\mathrm{in}}\ge0$ bounds $e+\min\{L,e\mathsf N_{1,j}(g_{\mathrm{in}}/L)\}$ by $\xi_{\mathrm{out}}(y)(1+\frac b{\mathsf P_{1,j}(\varrho/b)})$, or (ii)~when $\zeta\le1$, \cref{lem:fd-cap} with $z_0=g_{\mathrm{in}}(y)/L$, $z_1=\varrho/b$ yields $(1-\zeta)e+(\frac b{\mathsf P_{1,j}(\varrho/b)}+\zeta)\xi_{\mathrm{out}}(y)$, and only then $e\le\rho_{\mathrm{cl}}(y)\xi_{\mathrm{out}}(y)$ gives $\xi_{\mathrm{out}}(y)[(1-\zeta)\rho_{\mathrm{cl}}(y)+\frac b{\mathsf P_{1,j}(\varrho/b)}+\zeta]$. Multiplying by $q(y)/s$, applying \cref{lem:clip-bounds} ($q(y)/s\le I_{\mathrm{out}}$) and $\Upsilon_y\le U_{\mathrm{cl}}$, integrating against $\mu$ ($\E_\mu[\xi_{\mathrm{out}}]=B_{\mathrm{in}}$), and mixing (\cref{lem:mixing}) completes the proof.
\end{proof}

Applying \cref{lem:fd-cap} before bounding $e\le\rho_{\mathrm{cl}}\xi_{\mathrm{out}}$ is essential: on the two-point cluster with mass $\frac9{13}$ at $\frac23$, mass $\frac4{13}$ at $-\frac32$, and center $\frac52$, landing at $-\frac32$ gives $s=\frac{29}4$, ideal total $\frac{13}4$, and actual total $\frac{121}{52}>2$, so $\rho_{\mathrm{cl}}=1$ while the entire deficit $\frac{12}{13}$ lies on the inside point.

\subsection{Clip bounds, two cohorts, and the \texorpdfstring{$\frac{89}{20}$}{89/20} clip}\label{sec:reservoir-clip}

\begin{lemma}[Clip bounds and two cohorts]\label{lem:clip-bounds}
Under the setup of \cref{thm:opening-reduction} ($B_{\mathrm{in}}>0$, $s>2$, $L>0$), at every outside landing $y\in P$ ($\|y\|^2>1$), $q(y)/s\le(\upsilon+\psi)^2$, $\Upsilon_y\le\min\{(\upsilon+\psi)^2,\,4,\,(\upsilon+\sqrt{2-D_{\mathrm{in}}})^2\}$, and whenever $0\le\upsilon\le1$, $\sqrt{(2-D_{\mathrm{in}})/s}\ge\frac{1-\upsilon}{1-B_{\mathrm{in}}\upsilon/2}$.
\end{lemma}

\begin{proof}
First $(\|y\|^2+1)\rho_{\mathrm{cl}}(y)\le s$. Integrating $\delta_{\mathrm{in}}(Y)\sqrt{q(y)}\le\delta_{\mathrm{in}}(Y)(\sqrt{q(Y)}+\|y-Y\|)$ with Cauchy--Schwarz gives $\sqrt{q(y)/s}\le\upsilon+\psi$. In $L^2(\mu)$, $\sqrt{q(y)}\le\sqrt s+\sqrt{\|y\|^2+1}$ gives $\sqrt{\Upsilon_y}\le2$; in $L^2(\mu_1)$ ($\E_{\mu_1}\|Y\|^2=1-D_{\mathrm{in}}$), Minkowski gives $\sqrt{q(y)}\le\sqrt s\,\upsilon+\sqrt{(\|y\|^2+1)(2-D_{\mathrm{in}})}$, yielding $\Upsilon_y\le(\upsilon+\sqrt{2-D_{\mathrm{in}}})^2$. For two cohorts, decompose $\mu=B_{\mathrm{in}}\mu_1+(1-B_{\mathrm{in}})\mu_0$ ($0<B_{\mathrm{in}}<1$); for independent $Y_i\sim\mu_i$, $1$-Lipschitzness of $\sqrt q$ yields $s(\sqrt{1-B_{\mathrm{in}}\upsilon^2}-\upsilon\sqrt{1-B_{\mathrm{in}}})^2\le(1-B_{\mathrm{in}})\E\|Y_1-Y_0\|^2=2-D_{\mathrm{in}}-2B_{\mathrm{in}}\Var(Y_1)\le2-D_{\mathrm{in}}$, and $\sqrt{1-B_{\mathrm{in}}\upsilon^2}+\upsilon\sqrt{1-B_{\mathrm{in}}}\le(1+\upsilon)(1-\frac{B_{\mathrm{in}}\upsilon}2)$ finishes the bound.
\end{proof}

\begin{lemma}[The $\frac{89}{20}$ clip]\label{lem:clip}
If $0<B_{\mathrm{in}}\le D_{\mathrm{in}}\le1$, $\upsilon\ge0$, $B_{\mathrm{in}}\upsilon^2\le1$, $s,\varrho>0$, $a_{\mathrm{cl}}\in[0,1]$, $U_{\mathrm{cl}}\le\min\{4,(\upsilon+\sqrt{2-D_{\mathrm{in}}})^2\}$, and $\upsilon\le1\implies\frac{1-\upsilon}{1-B_{\mathrm{in}}\upsilon/2}\le\sqrt{(2-D_{\mathrm{in}})/s}$, then $a_{\mathrm{cl}}U_{\mathrm{cl}}\le\frac{89}{20}a_{\mathrm{cl}}\upsilon^2+\mathcal E_{\mathrm{clip}}+a_{\mathrm{cl}}^2\frac{2-D_{\mathrm{in}}}{D_{\mathrm{in}}\varrho}\psi^2$.
\end{lemma}

\begin{proof}
For $\upsilon>1$, $U_{\mathrm{cl}}\le4<\frac{89}{20}\upsilon^2$. For $0\le\upsilon\le1$, put $V_{\mathrm{cl}}=\min(4,(\upsilon+\sqrt{2-B_{\mathrm{in}}})^2)\ge U_{\mathrm{cl}}$, $A_D=2-D_{\mathrm{in}}$, and $t_0=\psi^2/(D_{\mathrm{in}}\varrho)$. Minimizing over $t_0\ge0$ gives $E_{\mathrm{in}}F_{\mathrm{cl}}(a_{\mathrm{cl}}\sqrt{A_D/s})\ge a_{\mathrm{cl}}E_{\mathrm{in}}F_{\mathrm{cl}}(\sqrt{A_D/s})$ ($F_{\mathrm{cl}}(t)=1-(1-t)_+^2$ concave, nondecreasing), so the two-cohort bound reduces the claim to $V_{\mathrm{cl}}\le\frac{89}{20}\upsilon^2+E_{\mathrm{in}}[1-(\frac{(2-B_{\mathrm{in}})\upsilon}{2-B_{\mathrm{in}}\upsilon})^2]$. This is immediate at $\upsilon=0$; dividing by $\upsilon^2>0$ yields $\mathrm{Req}(B,\upsilon):=1+\frac{V_{\mathrm{cl}}-2/B}{\upsilon^2}+(\frac2B-\upsilon^2)\frac{(2-B)^2}{(2-B\upsilon)^2}\le\frac{89}{20}$. Putting $\upsilon_0=2-\sqrt{2-B}$, $\mathrm{Req}$ increases on $(0,\upsilon_0]$; on $[\upsilon_0,1]$, $V_{\mathrm{cl}}=4$, so $B\le\frac12$ gives $\mathrm{Req}\le\mathrm{Req}(B,1)=4$, while $B>\frac12$ uses $\upsilon_0\ge\frac{1+B}2$ to give $\frac{89}{20}-\mathrm{Req}\ge\frac{89B^3+138B^2-311B+120}{20B(1+B)^2}>0$ (\cref{app:tail-clip}).
\end{proof}

\subsection{The marked comparison and reservoir stocks for \texorpdfstring{$j\ge17$}{j >= 17}}\label{sec:reservoir-marked}

\begin{theorem}[Marked comparison for all $j\ge18$]\label{thm:marked-tail}
With the kernel of \cref{thm:kernel} and $\mathrm{cap}=\kappa_{1,j}+2\sqrt{H_j}$, let $j\ge18$, $s>2$, $0<B_{\mathrm{in}}\le D_{\mathrm{in}}\le1$, $\upsilon\ge0$ with $B_{\mathrm{in}}\upsilon^2\le1$, $\psi\ge0$, $\varrho>0$, and $U_{\mathrm{cl}}\in\R$ satisfy
\[
U_{\mathrm{cl}}\le\min\bigl\{(\upsilon+\psi)^2,\;4,\;(\upsilon+\sqrt{2-D_{\mathrm{in}}})^2\bigr\}
\qquad\text{and}\qquad
\upsilon\le1\implies\frac{1-\upsilon}{1-B_{\mathrm{in}}\upsilon/2}\le\sqrt{\frac{2-D_{\mathrm{in}}}s}.
\]
Then the budget $\mathrm{Bud}$ of \cref{thm:opening-reduction} (at $\mathsf m_j=2-\epsilon_j$) pays the unclipped cost for $\varrho\le\sqrt{H_j}$, and satisfies $0\le\zeta(\varrho/H_j)\le1$ and pays the clipped cost (with $U_{\mathrm{cl}}$) for $\varrho\ge\sqrt{H_j}$.
\end{theorem}

\begin{proof}
Set $b=H_j$, $h^{(2)}=h^{(2)}_{1,j}$, $\epsilon=\epsilon_j$, $\mathsf m=2-\epsilon$, $K_{\mathrm{res}}=1+b+\epsilon$, $\tilde\omega_0=\frac{28}{15}$, $b_0=\tilde\omega_0^2=\frac{784}{225}$. Note $(1+b)\upsilon^2+\frac\epsilon{B_{\mathrm{in}}}\ge K_{\mathrm{res}}\upsilon^2$, $b_0<H_{18}<\frac72$, $\frac{159}{100}<h^{(2)}<\frac53$, and $H_{30}<4<H_{31}$ (\cref{app:tail-clip}).

\emph{Small rates, $0<\varrho\le\sqrt b$.} Drop $\mathcal E_{\mathrm{clip}}$ and set $D_{\mathrm{in}}=1$. With $P_{\mathrm{sm}}=\mathsf P_{1,j}(\varrho/b)=1+\frac{\kappa_{1,j}}b\varrho+\frac{\varrho^2}b$, Cauchy--Schwarz pays $(1+\frac b{P_{\mathrm{sm}}})I_{\mathrm{out}}$ once $\Delta_{\mathrm{sm}}=\mathsf mb[(b+\epsilon)P_{\mathrm{sm}}-b]-K_{\mathrm{res}}\varrho(P_{\mathrm{sm}}+b)\ge0$. Using $2b\kappa_{1,j}-(1+b)^2=h^{(2)}-1$ gives $\frac{\Delta_{\mathrm{sm}}}\varrho\ge\frac{\mathsf mb\epsilon}\varrho+A_0+B_1\varrho$, where $A_0=(h^{(2)}-1)-\frac\epsilon2(b^2+2b-2+(1-\frac2b)h^{(2)})-\epsilon^2\kappa_{1,j}$ and $B_1=(2-\epsilon)(b+\epsilon)-(1+b+\epsilon)(\frac{\kappa_{1,j}}b+\frac1{\sqrt b})>\frac{36}{100}$. By monotonicity of $\epsilon_j,\epsilon_j H_j,\epsilon_j H_j^2$, $A_0>\frac{811}{16900}>0$ for $j\ge64$; for $18\le j\le63$, $A_0>-\frac{4119}{9025}$ and $4\mathsf mb\epsilon B_1>\frac{3528}{11875}>(\frac{4119}{9025})^2$, so AM--GM yields $\Delta_{\mathrm{sm}}>0$ (\cref{app:tail-clip}).

\emph{Large rates, $\varrho\ge\sqrt b$.} Put $\tilde\omega=\sqrt b>\tilde\omega_0$, $t_{\mathrm{r}}=\tilde\omega/\varrho\in(0,1]$, $S_{\mathrm{cap}}=\kappa_{1,j}+2\tilde\omega$, $G_{\mathrm{gap}}=S_{\mathrm{cap}}-b$. Here $\mathsf P_{1,j}(\varrho/b)=S_{\mathrm{cap}}\varrho/b$ and $\mathrm{cap}=S_{\mathrm{cap}}$. If $G_{\mathrm{gap}}\ge0$, then $\zeta=\lambda(b)t_{\mathrm{r}}$ with $\lambda(b)=G_{\mathrm{gap}}\tilde\omega/S_{\mathrm{cap}}<1$ by~\eqref{eq:trunc-margin}, so $a_{\mathrm{cl}}=1-\zeta\in(0,1]$ and $c_{\mathrm{cl}}:=b/\mathsf P_{1,j}+\zeta=\tilde\omega t_{\mathrm{r}}$ ($c_{\mathrm{cl}}\varrho=b$). If $G_{\mathrm{gap}}<0$, then $\zeta=0$, $a_{\mathrm{cl}}=1$, and $c_{\mathrm{cl}}\varrho=b^2/S_{\mathrm{cap}}$. Applying \cref{lem:clip} cancels $\mathcal E_{\mathrm{clip}}$; since $\mathsf mb>2\ge2a_{\mathrm{cl}}^2$, $D_{\mathrm{in}}=1$ is worst, and $\alpha_{\mathrm{cl}}=K_{\mathrm{res}}-\frac{89}{20}a_{\mathrm{cl}}>0.034$ reduces $\alpha_{\mathrm{cl}}\upsilon^2+\frac{\mathsf mb-a_{\mathrm{cl}}^2}\varrho\psi^2\ge c_{\mathrm{cl}}(\upsilon+\psi)^2$ to
\[
F:=\Bigl(1+b+\epsilon-\frac{89}{20}a_{\mathrm{cl}}\Bigr)(\mathsf mb-a_{\mathrm{cl}}^2-c_{\mathrm{cl}}\varrho)-c_{\mathrm{cl}}(\mathsf mb-a_{\mathrm{cl}}^2)\ge0,\qquad \mathsf mb-a_{\mathrm{cl}}^2-c_{\mathrm{cl}}\varrho>0.
\]
When $c_{\mathrm{cl}}\varrho=b$, $\partial_{a_{\mathrm{cl}}}F$ is convex in $a_{\mathrm{cl}}\in[0,1]$ and negative at $a_{\mathrm{cl}}\in\{0,1\}$ (for $\tilde\omega\ge\tilde\omega_0$), so increasing $a_{\mathrm{cl}}$ is conservative.

\emph{Tail $\tilde\omega\ge\frac72$.} Here $j\ge31$, $\epsilon b<\frac13$; take $a_{\mathrm{cl}}=t_{\mathrm{r}}=1$. For $G_{\mathrm{gap}}\ge0$, $F\ge(\tilde\omega^2-\frac72)(\tilde\omega^2-\frac43)-2\tilde\omega^3=\tilde\omega^2[(\tilde\omega-1)^2-6]+\frac16\tilde\omega^2+\frac{14}3>0$. For $G_{\mathrm{gap}}<0$, $\frac{\tilde\omega^2}2-2\tilde\omega-1>0$ forces $\tilde\omega>4$, and $S_{\mathrm{cap}}>\frac{\tilde\omega^2}2+2\tilde\omega+1$ gives
\begin{align*}
S_{\mathrm{cap}}F&\ge\Bigl(\tilde\omega^2-\frac72\Bigr)\Bigl[\Bigl(2\tilde\omega^2-\frac43\Bigr)\Bigl(\frac{\tilde\omega^2}2+2\tilde\omega+1\Bigr)-\tilde\omega^4\Bigr]-\tilde\omega^3(2\tilde\omega^2-1)\\
&=\tilde\omega^2\Bigl[\tilde\omega\Bigl(2\tilde\omega^2-\frac{47}3\Bigr)+\frac43\tilde\omega^2-6\Bigr]+\frac{28}3\tilde\omega+\frac{14}3>0,
\end{align*}
with the middle bracket simultaneously yielding $\mathsf mb-1-c_{\mathrm{cl}}\varrho>0$.

\emph{Four strips $\frac{28}{15}\le\tilde\omega\le\frac72$.} Since $h^{(2)}>\frac{159}{100}$, $\lambda(b)\ge\lambda_{\mathrm{lb}}(\tilde\omega):=\tilde\omega(1-\tilde\omega^2/S_{\mathrm{lb}}(\tilde\omega))$ ($S_{\mathrm{lb}}(\tilde\omega)=1+2\tilde\omega+\frac{\tilde\omega^2}2+\frac{159}{200\tilde\omega^2}$), which decreases on $[\frac{28}{15},\frac72]$ to $\lambda_{\mathrm{lb}}(\frac72)>0.47848>0$ (\cref{app:tail-strips}), ensuring $G_{\mathrm{gap}}>0$ ($c_{\mathrm{cl}}\varrho=b$). On each strip $[\tilde\omega_{\mathrm{lo}},\tilde\omega_{\mathrm{hi}}]$ with $\lambda_0\le\lambda_{\mathrm{lb}}(\tilde\omega_{\mathrm{hi}})$ and $a_{\mathrm{cl}}=1-\lambda_0 t_{\mathrm{r}}$, $\partial_b F\ge L_{\mathrm{st}}(\tilde\omega,\epsilon,t_{\mathrm{r}}):=(1-\epsilon)(2\tilde\omega^2+1+\epsilon-\frac{89}{20}a_{\mathrm{cl}})-a_{\mathrm{cl}}^2-\frac32(2-\epsilon)\tilde\omega t_{\mathrm{r}}$, increasing in $\tilde\omega$ ($\partial_{\tilde\omega}L_{\mathrm{st}}>3.6$) and concave in $\epsilon$ and $t_{\mathrm{r}}$. Positivity of all $16$ strip corners makes $b=\tilde\omega_{\mathrm{lo}}^2$ worst, concavity $\partial_{\epsilon\epsilon}F=-2b<0$ reduces to the two $\epsilon$-endpoints ($\epsilon\in[\frac2{31},\frac2{19}]$ on Strip~1, $[0,\frac2{31}]$ on Strips~2--4), and exact degree-$3$ Bernstein expansions verify $F(\tilde\omega_{\mathrm{lo}}^2,\epsilon,t_{\mathrm{r}})>0$ on $t_{\mathrm{r}}\in[0,1]$ (\cref{app:tail-strips}).
\end{proof}

\begin{theorem}[Reservoir stocks]\label{thm:reservoir}
For every rank $j\ge17$, $\mathsf C_j=2+2H_{j+1}$ is a universal stock for rank~$j$.
\end{theorem}

\begin{proof}
Induct on $j\ge17$ from \cref{thm:rank17}(b) with slope $j(\mathsf C_j-\mathsf C_{j-1})=2-\frac2{j+1}$: at every actual outside landing ($B_{\mathrm{in}}>0$, $s>2$, $L>0$, $\|y\|^2>1$), \cref{lem:clip-bounds} verifies the clip bounds (with $U_{\mathrm{cl}}=\Upsilon_y$) and two-cohort inequality required by \cref{thm:marked-tail}, which then discharges the budget hypothesis of \cref{thm:opening-reduction} (using \cref{thm:kernel,thm:signed,lem:fd-cap,lem:clip}).
\end{proof}

\section{Proof of the main theorems}\label{sec:assembly}

\begin{proof}[Proof of \cref{thm:exact-rk,thm:strict}]
\emph{$k=1$.} \Cref{prop:R1} gives $\E\,\Phi_1=2\,\OPT_1(X)$ for every nonempty finite input, so $R_1=2$ and every input with $\OPT_1(X)>0$ attains it. For every $1\le k\le|X|$, if $\OPT_k(X)=0$ then $\E\,\Phi_k=0$.

\emph{Lower bounds.} For $k\ge2$, \cref{thm:hub,thm:simplex} give $R_k\ge\max\{J_k,2+2H_{k-1}\}$.

\emph{Upper bound for $2\le k\le18$.} Let $j=k-2\le16$. At rank~$0$ the universal stock is $5=J_2$. For $1\le j\le16$, every high-mean target has delayed cost at most $J_{j+2}\phi$ by \cref{thm:high-mean}, and every low-mean target has delayed cost at most $\mathsf C_j\phi\le J_{j+2}\phi$ (strictly when $\phi>0$) by \cref{thm:low-mean,thm:hub-stocks}. Hence $J_{j+2}=J_k$ is a stock for rank $k-2$, and \cref{thm:bridge} gives
\[
\frac{\E\,\Phi_k}{\OPT_k(X)}\le(1-\theta)J_k+\theta\bigl(2+2H_{k-1}\bigr).
\]
Here $J_k>2+2H_{k-1}$: at $k=2$, $5>4$; at $k=3$, $\frac{185}{32}>5$; and for $4\le k\le18$, $J_k\ge G_{k-2}(\frac{11}4)>2+2H_{k-1}$ by \cref{thm:hub-stocks} (or by \cref{thm:brackets,thm:gap}). Since $\theta\ge2/|X|>0$, the right-hand side is strictly less than $J_k$, proving~\eqref{eq:theta-refined} and non-attainment for $2\le k\le18$. Combined with the lower bound, $R_k=J_k$.

\emph{Upper bound for $k\ge19$.} By \cref{thm:reservoir}, $2+2H_{k-1}$ is a stock for rank $k-2\ge17$, and \cref{thm:bridge,cor:strict} give $\E\,\Phi_k/\OPT_k(X)<2+2H_{k-1}$ for every input with $\OPT_k(X)>0$. Combined with the lower bound, $R_k=2+2H_{k-1}$ and the worst case is not attained.

\emph{The two forms agree.} For $2\le k\le18$, $J_k>2+2H_{k-1}$ as above. For $k\ge19$, the lower bound gives $J_k\le R_k=2+2H_{k-1}$, so $\max\{J_k,2+2H_{k-1}\}=2+2H_{k-1}$.
\end{proof}

\begin{corollary}[Explicit values and asymptotics]\label{cor:values-asymptotic}
We have $R_2=5$, $R_3=\frac{185}{32}$, and
\[
R_k=2\ln k+2+2\gamma_{\mathrm{E}}-\frac1k+O(k^{-2})\qquad\text{as }k\to\infty,
\]
where $\gamma_{\mathrm{E}}=\lim_{n\to\infty}(H_n-\ln n)$ is Euler's constant; indeed, for every $k\ge19$,
\[
0\le\Bigl(2\ln k+2+2\gamma_{\mathrm{E}}-\frac1k\Bigr)-R_k\le\frac1{4k^2}.
\]
\end{corollary}

\begin{proof}
By \cref{thm:exact-rk}, $R_2=J_2=G_0=5$ and $R_3=J_3=G_1(\frac{11}5)=\frac{185}{32}$ (\cref{thm:laplace}). For $k\ge1$, put $f(k)=H_{k-1}-\ln k+\frac1{2k}$, which converges to $\gamma_{\mathrm{E}}$. With $t=\frac1{2k+1}\in(0,\frac12]$, the elementary bracket $2t\le\ln\frac{1+t}{1-t}\le\frac{2t}{1-t^2}=\frac1{2k}+\frac1{2(k+1)}$ yields
\[
0\le f(k+1)-f(k)=\frac{2t}{1-t^2}-\ln\frac{1+t}{1-t}\le\frac1{2k(k+1)(2k+1)}\le\frac1{8k^2}-\frac1{8(k+1)^2}.
\]
Summing from $k$ to $\infty$ gives $0\le(\ln k+\gamma_{\mathrm{E}}-\frac1{2k})-H_{k-1}\le\frac1{8k^2}$. Doubling and adding $2$ yields the bound for $R_k=2H_{k-1}+2$ ($k\ge19$).
\end{proof}

\paragraph{Dependencies.}
The handoff constants passed between regimes—the high/low-mean threshold $M_{\mathrm{lo}}=\frac{485883}{100000}$, the reset constant $\frac{14}5$, the explicit low-mean stocks $\mathsf C_1,\dots,\mathsf C_{16}$, the crossover ceiling $\mathsf P_{16}=\frac{4445581}{500000}$, and the rank-$17$ stock $\mathsf C_{17}=2H_{18}+2$ with slope $\mathsf m_{17}=17(\mathsf C_{17}-\mathsf P_{16})$—form an acyclic chain. The occupation framework of \cref{sec:framework} (persistence, the first-hit prize, and the delayed bound) and the restoration bridge of \cref{sec:restoration} hold for every rank. The race comparison and Euclidean reset of \cref{sec:race-reset} feed into the openings at ranks $1$--$17$ and the high-mean projection at ranks $1$--$16$. \Cref{sec:hub-regime} establishes the stock $J_{j+2}$ for $0\le j\le16$ and hence the universal rank-$16$ stock $\mathsf P_{16}>J_{18}$; \cref{sec:rank17} turns $\mathsf P_{16}$ into $\mathsf C_{17}=2H_{18}+2$ using $18$ actual draws; and \cref{sec:reservoir-regime} carries $2H_{j+1}+2$ from rank~$17$ to every larger rank. No step invokes a stock at the rank being proved, except through the explicit inductions on the remaining draws at a fixed horizon.

\clearpage
\appendix
\crefalias{section}{appendix}
\crefalias{subsection}{appendix}
\numberwithin{table}{section}
\phantomsection
\addcontentsline{toc}{section}{Appendices}
\begin{center}
  {\LARGE\bfseries Appendices}
\end{center}
\vspace{0.5em}
\section{Notation}\label{app:notation}

\Cref{tab:notation} collects the principal global and regime-specific notation used across the paper. Global symbols keep one meaning throughout: $d$ is the ambient dimension ($\R^d$), $\Phi_t$ the seeding cost after $t$ centers, $J_k=\sup_{p>0}G_{k-2}(p)$ the hub optimum, $\Cl(r,n),\Dl(h,n)$ the harmonic clocks, $\mathsf C_j,\mathsf m_j,\mathsf K_j$ the stock numbers, mixing slopes, and persistence ceilings, and $\mu$ the normalized cluster measure. Several letters carry distinct section-local roles (marked by fonts or subscripts): $\alpha$ is the terminal fraction in \cref{thm:stopped} while subscripted $\alpha$-parameters appear locally in Sections~7--9; $\theta$ is the global first-seed share while $\theta_j$ denotes rank-$j$ opening tilts; $\gamma$ is the post-hit contraction factor in \cref{sec:race-reset} while $\gamma_{\mathrm{E}}$ is Euler's constant; $P$ is a target cluster while $\Pre_r$, $\mathsf P_{16}$, and $\mathsf P_{m,n}$ denote prefix occupation, the rank-$16$ ceiling, and the reservoir repair kernel; and $A$ denotes the upper tail $\{q>b\}$ in \cref{sec:race-reset} and the witness-cut bulk in \cref{sec:hub-regime}.

{\small
\setlength{\tabcolsep}{4pt}
\renewcommand{\arraystretch}{1.04}
\begin{longtable}{@{}>{\raggedright\arraybackslash}p{0.25\textwidth}>{\raggedright\arraybackslash}p{0.55\textwidth}>{\raggedright\arraybackslash}p{0.16\textwidth}@{}}
\caption{Principal global and regime-specific notation used across the paper.}\label{tab:notation}\\
\toprule
Symbol & Meaning & Reference\\
\midrule
\endfirsthead
\toprule
Symbol & Meaning (continued) & Reference\\
\midrule
\endhead
\midrule
\multicolumn{3}{r}{\footnotesize\emph{Continued on next page}}\\
\endfoot
\bottomrule
\endlastfoot
\multicolumn{3}{@{}l}{\textbf{Model, extremal functions, and global constants}}\\
$d$; $X$, $k$ & Ambient dimension ($\R^d$); finite distinct input $X\subset\R^d$; center count $k\le|X|$ & \cref{sec:main}\\
$D(y,C)$, $\cost(S,C)$ & Distance $\min_{c\in C}\|y-c\|$ and squared-distance cost $\sum_{y\in S}D(y,C)^2$ & \cref{sec:prelim}\\
$\OPT_k(X)$, $C_t$, $\Phi_t$ & Optimal $k$-means cost; first $t$ seeds of $k$-means++ and their cost $\cost(X,C_t)$ & \cref{sec:main,alg:kmpp}\\
$R_k$, $H_j$, $\gamma_{\mathrm{E}}$ & Worst-case ratio $\sup\E\,\Phi_k/\OPT_k(X)$; harmonic sum $\sum_{i=1}^j\frac1i$; Euler's constant & \eqref{eq:Rk}, \cref{cor:values-asymptotic}\\
$\pi_m(p)$, $G_m(p)$, $J_k$ & Race product $\prod_{i=1}^m\frac{ip}{1+ip}$ ($\pi_0=1$); hub function; $J_k=\sup_{p>0}G_{k-2}(p)$ & \eqref{eq:G}, \cref{sec:main}\\
$p^*_m$, $c=1/p$, $K_m(p)$ & Unique maximizer of $G_m$; inverse price; outlier continuation $p+(1-p)\pi_m(p)$ & \cref{sec:crossover,thm:laplace}\\
$\theta$ & Expected first-seed cluster share $\sum_i\frac{|P_i|}{|X|}\frac{\phi_i}{\OPT_k(X)}\in[\frac2{|X|},1]$ & \eqref{eq:theta}, \cref{thm:bridge}\\
\addlinespace[3pt]
\multicolumn{3}{@{}l}{\textbf{Cluster geometry, states, occupations, and stocks (Sections~4--5)}}\\
$P$, $\bar y_P$, $\phi$ & Target cluster $P\subseteq X$, centroid $\bar y_P$, and optimal one-center cost $\OPT_1(P)$ & \cref{sec:prelim}\\
$\mu$, $Y$, $\widehat C_0$, $q(y)$ & Normalized uniform measure ($\E_\mu Y=0$, $\E_\mu\|Y\|^2=1$), centers, rate $D(y,\widehat C_0)^2$ & \cref{sec:prelim,sec:reset-statement}\\
$\mathcal Q$, $\mathcal Q_{\mathrm{live}}$, $n$, $r$ & Name family, live-name subset, live-name count $n=|\mathcal Q_{\mathrm{live}}|$, draw horizon $r$ & \cref{def:state}\\
$\sigma$, $v$, $\ell$, $T$, $e$ & State $(C,\mathcal Q_{\mathrm{live}})$, target cost, live price, total cost $\cost(X,C)$, exterior cost & \cref{def:state}\\
$\Cl(r,n)$, $\Dl(h,n)$ & Covered clock $H_n-H_{n-r}$ and delayed clock $\Cl(h-1,n)$ ($\Dl(0,n)=0$) & \cref{sec:queries}\\
$\Occ_r$, $\Del_h$, $\Pre_r$; $Q_r$, $Z_h$ & Covered, delayed, prefix occupations; covered and delayed costs & \cref{def:occupation,def:queries}\\
$\mathsf C_j$, $\mathsf m_j$, $\mathsf K_j$ & Rank-$j$ stock, mixing slope $j(\mathsf C_j-\mathsf C_{j-1})$, persistence ceiling $1+H_j$ & \cref{def:queries,lem:mixing}\\
$L_{\mathrm{tot}}$, $\nu$, $\sigma^y$ & Normalized total nontarget cost per name, conditional $D^2$ law on $P$, post-hit state & \cref{lem:mixing}\\
\addlinespace[3pt]
\multicolumn{3}{@{}l}{\textbf{Race, stopped-price comparison, and Euclidean reset (Section~6 and Appendix~B)}}\\
$L$, $x$, $\tau$, $J_{\mathrm{st}}$ & Price bound ($\ell_0\le L$), ratio $L/v$ (or $L_{\mathrm{tot}}/\bar u$), first return time, stopped payment & \cref{sec:race-reset}\\
$\gamma$, $\gamma_{\mathrm{reset}}(h)$, $\gamma_-(h)$ & Post-hit contraction in~\eqref{eq:transfer-hyp}; reset ratio $U(\bar u)/\bar u$ ($h=1/\bar u$) and $\frac{25h}{5+14h}$ & \cref{thm:transfer,cor:reset-monotone}\\
$\vartheta_j$, $\alpha$, $W_j(x)$ & Avoidance constant $(1-\frac1j)^{j-1}$; terminal fraction $\min(\frac79x,1)$; stopped bound & \cref{thm:stopped}\\
$s$, $\bar u$, $w(y)$, $M(C_0)$, $U(s)$ & Normalized pre/post-hit costs, post-hit cost $w(y)$, conditional mean, reset bound & \cref{sec:reset-statement,thm:reset}\\
$b$, $A$, $\rho$, $\mathcal E_A$, $\mathcal E_{A^c}$ & Reset threshold $b>0$, upper tail $A=\{q>b\}$, mass $\rho=\mu(A)$, cut energies & \cref{sec:reset-lemmas}\\
$V_A$, $V_{A^c}$, $\Xi$, $I_{AA}$ & Within-set variances, $\Xi=(1-\rho)V_{A^c}$, pair energy on $A\times A$ & \cref{sec:reset-lemmas}\\
$T_{\mathrm{res}}(s)$, $\Lambda_V$, $\Lambda$, $\Lambda_0$ & Reset target $\frac{280s}{10s+28}$ and quadratic/tail lower-bound functions & \cref{sec:reset-proof,app:s6-algebra}\\
\addlinespace[3pt]
\multicolumn{3}{@{}l}{\textbf{Hub regime, rank-17 transition, and reservoir regime (Sections~7--9 and Appendices~C--E)}}\\
$M^*(P)$, $M_{\mathrm{lo}}$, $\mathsf P_{16}$ & Worst post-hit mean $\sup_{C_0}M(\widehat C_0)$; threshold $\frac{485883}{100000}$; ceiling $\frac{4445581}{500000}$ & \cref{thm:brackets,sec:hub-regime}\\
$A$, $S$, $a_{\mathrm{bk}}$, $\rho_{\mathrm{sat}}$, $\varepsilon_{\mathrm{w}}$ & Witness cut into bulk $A$ and satellites $S$, masses $1-\rho_{\mathrm{sat}}$ and $\rho_{\mathrm{sat}}$, defect & \cref{sec:high-mean}\\
$\mathsf Z_h$, $\mathsf Q_h$, $\mathsf S_h$; $\mathcal T$, $\mathcal T_y$ & Pathwise delayed, covered, and bulk stopped costs; bulk stopped mean and satellite tag & \cref{sec:high-mean}\\
$u_y$, $\mathcal U_y$, $b(r)$ & Binary unit $\cost(A,\{y\})$; per-satellite account; radius award $\frac9{40}\mathbf 1_{\{r\le1/2\}}$ & \cref{sec:high-mean,sec:binary}\\
$\mathcal U$, $\mathcal V$, $\mathcal U_j$, $\mathcal V_j$ & Unopened and matched pure binary accounts (and full-horizon values at rank~$j$) & \cref{sec:binary}\\
$t_{\mathrm{acc}}$, $t_{\mathrm{sh}}$, $\overline K$, $\overline T$ & Pure named/exterior child means; normalized pure tag $\mathcal V_j^{(0)}/c_{\mathrm{hv}}$ and exterior roof & \cref{sec:binary}\\
$C_{\mathrm{acc}}$, $C_{\mathrm{rej}}$, $O_j$, $S_{\mathrm{cr}}$ & Accepted/rejected named-child ceilings, exterior ceiling, crossing kink & \cref{sec:binary}\\
$\mathsf v_j(p)$, $\mathsf M_j(p)$, $\mathcal F_{m,j}$ & Tangent slope $-G'_j/K'_j$, tangent value $G_j+\mathsf v_jK_j$, harmonic fee & \cref{sec:binary}\\
$p_{\mathrm{ref}}$, $p_{\mathrm{pr}}$, $\varpi$, $\varpi_j$ & Reference/premium prices ($\frac{273}{100},\frac{21}{10}$), shift $\varpi\in\{0,\frac1{25}\}$, scaled premium $\varpi_j$ & \cref{sec:high-mean,sec:binary}\\
$\theta_j$, $\Delta_j$; $\lambda_{\mathrm{int}}$ & Rank-$j$ opening tilt and surplus (distinct from $\theta$); App.~D interpolation weight & \cref{sec:low-mean,app:sec8}\\
$b_{m,n}$, $h^{(2)}_{m,n}$, $\epsilon_j$ & Reservoir clock $\sum_{i=m}^n\frac1i$, squared sum $\sum_{i=m}^n i^{-2}$, slope deficit $\frac2{j+1}$ & \cref{sec:reservoir-regime}\\
$\mathsf P_{m,n}$, $\mathsf R_{m,n}$, $\mathsf N_{m,n}$, $\mathcal I$, $\zeta$ & Concrete repair kernel, inside ball $\mathcal I=\{\|Y\|^2\le1\}$, finite-difference cap factor & \cref{sec:reservoir-regime}\\
\end{longtable}
}

\section{Algebraic Details and Polynomial Verifications for Section~6}\label{app:s6-algebra}

This appendix records the exact algebraic derivations and complete Bernstein-basis polynomial verifications omitted from \cref{sec:reset-proof}, covering the exhaustive four-range partition $(0,\frac{11}3]\cup[\frac{11}3,\frac{19}2]\cup[\frac{19}2,31]\cup[31,\infty)$ (with internal $V$-seam $V=\frac3{10}$ in Ranges~2--3). The band split, convex weights, and affine shifts below are rational witnesses whose discovery does not enter the proof; correctness relies only on checking the resulting exact weight, Hessian, and Bernstein-positivity inequalities. \Cref{app:reset-polys} works one endpoint in full and prints the complete exact rational monomial and Bernstein coefficient vectors for all eight Range~2--3 endpoints, making every finite check of \cref{sec:race-reset} directly reproducible on paper or in any computer algebra system (as well as checked in the ancillary verification package of \cref{sec:reading}).

\subsection{Range~3 parameter algebra and endpoint checks}\label{app:range3}

On $s\in[\frac{19}2,31]$, set $b=2s$, $\omega=\rho s\in(0,\frac s2)$, $\mathcal E=\mathcal E_A$, $V=V_A$, affine shifts $\psi_{\mathcal E}(s)=k_0+k_1s$ and $\psi_\omega(s)=\upsilon_0+\upsilon_1s$, weights $\lambda_A+\lambda_B=1$, $\Upsilon(s)=2(2s-\psi_{\mathcal E}(s))$, and $2\beta b=\Upsilon(s)+\psi_\omega(s)$. Whenever $0\le2\beta b\le4s$ and $\Upsilon(s)\ge\max(0,4s-2\beta b)$, subtracting $\Upsilon(s)(\Xi_{\mathrm{dtl}}-\Xi)+(\Upsilon(s)-4s+2\beta b)\Xi\ge0$ from~\eqref{eq:pair-cut} (with fraction $\beta$ of the mixed term and \cref{lem:dd} for $I_{AA}$) eliminates $\Xi$ and yields
\[
2s\bigl(5-M(C_0)\bigr)-T_{\mathrm{res}}(s)\ge\Lambda_V(\omega,\mathcal E)=c_{\omega\omega}\omega^2+c_{\omega\mathcal E}\omega\mathcal E+c_{\mathcal E\mathcal E}\mathcal E^2+c_\omega\omega+c_{\mathcal E}\mathcal E+c_0-T_{\mathrm{res}}(s),
\]
affine in $V$, with $c_{\omega\omega}=4\lambda_A$, $c_{\mathcal E\mathcal E}=1+\lambda_B$, and
\begin{gather*}
sc_{\omega\mathcal E}=(k_0-\upsilon_0)+(-2+k_1-\upsilon_1)s,\qquad c_{\mathcal E}=2(1-V)-\psi_{\mathcal E}(s)-8\lambda_BV,\\
sc_\omega=(-4+4V-4\lambda_AV+\upsilon_0)s+\upsilon_1s^2+V\bigl(2\beta b+2s-\psi_{\mathcal E}(s)\bigr),\\
c_0=8s-2\beta b-(2-V)\bigl(2s-\psi_{\mathcal E}(s)\bigr).
\end{gather*}
When $\lambda_A>0$ and $\det(s)=4c_{\omega\omega}c_{\mathcal E\mathcal E}-c_{\omega\mathcal E}^2>0$, minimizing over $(\omega,\mathcal E)\in\R^2$ gives $\min\Lambda_V=\mathcal R_V(s)/(s^2\det(s)(10s+28))$, with the quintic
\[
\mathcal R_V(s)=s^2\det(s)\bigl[(10s+28)c_0-280s\bigr]-(10s+28)\bigl[c_{\mathcal E\mathcal E}(sc_\omega)^2-(sc_{\omega\mathcal E})(sc_\omega)c_{\mathcal E}+c_{\omega\omega}(sc_{\mathcal E})^2\bigr].
\]
Splitting $[0,1]=[0,\frac3{10}]\cup[\frac3{10},1]$ gives two witness bands, checked at the endpoints $s=\frac{19}2,31$:
\begin{itemize}[leftmargin=1.5em]
\item \emph{Low band $V\in[0,\frac3{10}]$:} $(\lambda_A,\lambda_B)=(\frac15,\frac45)$, $\psi_{\mathcal E}=9+\frac s6$, $\psi_\omega=3+\frac s5$, so $2\beta b=\frac{58}{15}s-15$ and $\Upsilon=\frac{11}3s-18$. At $s=\frac{19}2,31$: $2\beta b=\frac{326}{15}\in[0,38]$, $\frac{1573}{15}\in[0,124]$ and $\Upsilon=\frac{101}6>\frac{244}{15}=4s-2\beta b$, $\frac{287}3>\frac{287}{15}=4s-2\beta b$. Also $s^2\det(s)=\frac1{900}(1463s^2+21960s-32400)$ increases on $s>0$ and equals $\frac{1233023}{3600}>0$ at $s=\frac{19}2$.
\item \emph{High band $V\in[\frac3{10},1]$:} $(\lambda_A,\lambda_B)=(\frac45,\frac15)$, $\psi_{\mathcal E}=6+\frac s5$, $\psi_\omega=-2$, so $2\beta b=\frac{18}5s-14$ and $\Upsilon=\frac{18}5s-12$. At $s=\frac{19}2,31$: $2\beta b=\frac{101}5\in[0,38]$, $\frac{488}5\in[0,124]$ and $\Upsilon=\frac{111}5>\frac{89}5=4s-2\beta b$, $\frac{498}5>\frac{132}5=4s-2\beta b$. Also $s^2\det(s)=\frac1{25}(303s^2+720s-1600)$ increases on $s>0$ and equals $\frac{130343}{100}>0$ at $s=\frac{19}2$.
\end{itemize}

\subsection{Bernstein positivity for Ranges~2 and~3}\label{app:reset-polys}

A polynomial $\mathcal R(s)$ of degree $m$ is strictly positive on $[s_1,s_2]$ ($s_1<s_2$) whenever its exact Bernstein expansion
\[
\mathcal R(s)=\sum_{k=0}^m B_k\binom mk\Bigl(\frac{s_2-s}{s_2-s_1}\Bigr)^{m-k}\Bigl(\frac{s-s_1}{s_2-s_1}\Bigr)^k
\]
has every $B_k>0$, since the Bernstein basis functions are nonnegative and sum to~$1$.

\paragraph{Worked representative example (Range~2, low band, $V=0$).}
Substituting $(\lambda_A,\lambda_B,\beta)=(\frac34,\frac14,\frac12)$ ($\det=\frac72$), $V=0$, $c_\omega=-2$, $c_{\mathcal E}=-\frac14-\frac{17}{20}s$, and $c_0=\frac{17}5s+9$ into the Range~2 formula of \cref{sec:reset-proof} yields the exact cubic
\[
\mathcal R_0(s)=\tfrac1{800}\bigl(586950-333995s+73712s^2-4335s^3\bigr).
\]
Putting $t=(s-\frac{11}3)/(\frac{19}2-\frac{11}3)=\frac6{35}(s-\frac{11}3)\in[0,1]$, direct rational expansion gives
\[
\mathcal R_0(s)=\tfrac{30157728}{172800}(1-t)^3+3\cdot\tfrac{43477888}{172800}(1-t)^2t+3\cdot\tfrac{120564198}{172800}(1-t)t^2+\tfrac{75553533}{172800}t^3,
\]
whose four coefficients are strictly positive with minimum $B_0=\frac{314143}{1800}>174.52$, proving $\mathcal R_0(s)>0$ on $[\frac{11}3,\frac{19}2]$.

The remaining seven endpoint polynomials—three Range~2 cubics on $[\frac{11}3,\frac{19}2]$ and four Range~3 quintics on $[\frac{19}2,31]$—follow by the exact same rational expansion and Bernstein-basis conversion. \Cref{tab:reset-polys} prints all eight exact rational monomial and strictly positive Bernstein coefficient vectors, together with the exact minimum Bernstein coefficient of each endpoint.

\begin{table}[!ht]
\centering
\footnotesize
\setlength{\tabcolsep}{4pt}
\renewcommand{\arraystretch}{1.15}
\begin{tabular}{@{}l>{\raggedright\arraybackslash}p{0.37\textwidth}>{\raggedright\arraybackslash}p{0.41\textwidth}@{}}
\toprule
Band, $V$ & Monomial coeffs.\ of $\mathcal R_V(s)$ ($s^0,s^1,\dots$) & Exact Bernstein coeffs.\ $(B_0,\dots,B_m)$ on $[s_1,s_2]$\\
\midrule
$[\frac{11}3,\frac{19}2]$, $0$ (low) & $\frac1{800}[586950,-333995,73712,-4335]$ & $\frac1{172800}[30157728,43477888,\allowbreak 120564198,75553533]$\\
$[\frac{11}3,\frac{19}2]$, $\frac3{10}$ (low) & $\frac1{800}[539742,-352367,73172,-4335]$ & $\frac1{172800}[3842016,7782736,\allowbreak 74166606,17130501]$\\
$[\frac{11}3,\frac{19}2]$, $\frac3{10}$ (high) & $\frac1{6250}[4871160,-3102984,641520,-37125]$ & $\frac1{50000}[2306456,3926116,\allowbreak 26701526,11679561]$\\
$[\frac{11}3,\frac{19}2]$, $1$ (high) & $\frac1{1250}[594090,-655461,164340,-7425]$ & $\frac1{30000}[820032,12497152,\allowbreak 46677522,67989267]$\\
\midrule
$[\frac{19}2,31]$, $0$ (low) & $\frac1{6750}[-224532000,\allowbreak 144795600,\allowbreak 10388970,\allowbreak -2547701,\allowbreak 124582,\allowbreak -1785]$ & $\frac1{1080000}[124944620265,\allowbreak 129729846326,\allowbreak 60085431934,\allowbreak 7516380924,\allowbreak 335626649920,\allowbreak 368068794560]$\\
$[\frac{19}2,31]$, $\frac3{10}$ (low) & $\frac1{675000}[-14606827200,\allowbreak 11385852480,\allowbreak -762205014,\allowbreak -141648545,\allowbreak 14005900,\allowbreak -178500]$ & $\frac1{54000000}[287263096920,\allowbreak 52097618776,\allowbreak 4456413092225,\allowbreak 31745428570602,\allowbreak 119060589054392,\allowbreak 256837047802480]$\\
$[\frac{19}2,31]$, $\frac3{10}$ (high) & $\frac1{3125}[-117964000,\allowbreak 72894840,\allowbreak 12973936,\allowbreak -3778430,\allowbreak 246550,\allowbreak -4000]$ & $\frac1{125000}[8182242685,\allowbreak 4575691043,\allowbreak 9290574995,\allowbreak 101057508686,\allowbreak 455273704536,\allowbreak 608960758240]$\\
$[\frac{19}2,31]$, $1$ (high) & $\frac1{125}[-1904000,\allowbreak 2139040,\allowbreak -354544,\allowbreak -188184,\allowbreak 26872,\allowbreak -160]$ & $\frac1{1250}[315686515,\allowbreak 1609700995,\allowbreak 6352840075,\allowbreak 21388191738,\allowbreak 61794735576,\allowbreak 143536920640]$\\
\bottomrule
\end{tabular}
\caption{All eight Range~2--3 endpoint polynomials $\mathcal R_V(s)$ covering $V\in[0,\frac3{10}]\cup[\frac3{10},1]$, with their exact rational monomial and strictly positive Bernstein coefficient vectors on $[\frac{11}3,\frac{19}2]$ (degree~$3$) and $[\frac{19}2,31]$ (degree~$5$). Their exact minimum Bernstein coefficients are $\frac{314143}{1800}$ ($B_0$), $\frac{40021}{1800}$ ($B_0$), $\frac{288307}{6250}$ ($B_0$), $\frac{17084}{625}$ ($B_0$), $\frac{208788359}{30000}$ ($B_3$), $\frac{6512202347}{6750000}$ ($B_1$), $\frac{4575691043}{125000}$ ($B_1$), and $\frac{63137303}{250}$ ($B_0$).}\label{tab:reset-polys}
\end{table}

\subsection{Range~4 tail branches on \texorpdfstring{$[31,\infty)$}{[31, infinity)}}\label{app:range4}

With $b=2s$, $\omega=\rho s\in(0,\frac s2)$, $\chi=\sqrt\omega\in(0,\frac52)$, $\mathcal E=\mathcal E_A\in[2\omega,s]$, $V=V_A\in[0,1]$, and~\eqref{eq:tail-Q}, we close the three exhaustive branches for all $s\ge31$:

\paragraph{Branch A ($\omega\ge1$).}
Set $\beta=1$ in~\eqref{eq:pair-cut}; $\pm4s\Xi$ cancel and $2s(5-M(C_0))\ge\Lambda(\mathcal E):=4s(1+\omega)-2s\mathcal E+2\mathcal E^2+(2-4\omega-10V)\mathcal E-4\omega+8\omega V$. Put $c_V=\sqrt{1-V}$, $\varepsilon_V=1-c_V\in[0,1]$ ($V=2\varepsilon_V-\varepsilon_V^2$), $\eta=\chi-1\in[0,\frac32]$, and $\mathcal E_0=1+\chi^2+2\chi c_V$. Dropping $\Xi$ in~\eqref{eq:tail-Q} with $\sqrt{1-\mathcal E/s}\le1-\frac{\mathcal E}{2s}$ gives $\mathcal E\le\mathcal E_{\mathrm{ub}}=\mathcal E_0-\delta_{\mathrm{ub}}$, where $\delta_{\mathrm{ub}}=\Psi_2/(s+\chi c_V)$, $\Psi_2=\chi c_V\mathcal E_0-\chi^2V$, and $\mathcal E_{\mathrm{ub}}\le(1+\chi)^2$. On $[0,(1+\chi)^2]$, $\Lambda'(\mathcal E)\le4(1+2\chi)+2-2s<0$, so $\Lambda(\mathcal E)\ge\Lambda(\mathcal E_{\mathrm{ub}})$. Dropping $2\delta_{\mathrm{ub}}^2\ge0$ and substituting $2(1+\chi^2)-\mathcal E_0=\eta^2+2(1+\eta)\varepsilon_V$ and $\Psi_1:=2\chi c_V+4\mathcal E_0+2-4\omega-10V=16+10\eta-10\varepsilon_V(3+\eta-\varepsilon_V)$,
\[
\Lambda(\mathcal E_{\mathrm{ub}})-T_{\mathrm{res}}(s)\ge2s\bigl(\eta^2+2(1+\eta)\varepsilon_V\bigr)+(\Psi_0-28)+\frac{784}{10s+28}-\frac{\Psi_1\Psi_2}{s+\chi c_V},
\]
where $\Psi_0-28=(32\eta+4\eta^2-6\eta^3-2\eta^4)+\varepsilon_V[-108-124\eta-38\eta^2-2\eta^3+(86\varepsilon_V-20\varepsilon_V^2)+\eta(92\varepsilon_V-20\varepsilon_V^2)+16\eta^2\varepsilon_V]\ge32\eta-\frac{101}2\eta^2-\varepsilon_V(108+124\eta)$ (dropping nonnegative $\varepsilon_V^2$ terms and using $\eta\le\frac32$, $\varepsilon_V\le1$). Checking all sign combinations gives $\Psi_1\Psi_2\le(16+10\eta)(1+\eta)(2+\eta)^2\le64+168\eta+\frac{563}2\eta^2$ on $[0,\frac32]$ (when both are negative, $\Psi_1\ge-4$ and $\Psi_2>-\frac{25}4$, so $\Psi_1\Psi_2<25$). Hence, for all $s\ge31$,
\[
\Lambda(\mathcal E_{\mathrm{ub}})-T_{\mathrm{res}}(s)\ge\frac{144s-1792}{s(10s+28)}+\Bigl(32-\frac{168}s\Bigr)\eta+\Bigl(2s-\frac{101}2-\frac{563}{2s}\Bigr)\eta^2+\varepsilon_V\bigl[(4s-108)+(4s-124)\eta\bigr]>0.
\]

\paragraph{Branch B ($\omega<1\le\omega+\Xi$).}
Keep $\beta=1$; then $V\le\omega=\chi^2$, $1+\omega-\Xi\le2\chi^2$, and $1-V-\Xi\le\chi^2-V=:c_B^2$ ($c_B\in[0,\chi]$). Put $\varepsilon_1=1-\chi^2$, $\varepsilon_2=\chi(\chi-c_B)$, and $\mathcal E_{0B}=2\chi^2+2\chi c_B\le4$. As above, $\mathcal E\le\mathcal E_{\mathrm{ub},B}=(s\mathcal E_{0B}+\chi^2V)/(s+\chi c_B)\le4+\frac1s$ (where $\Lambda'<0$), giving
\[
\Lambda(\mathcal E_{\mathrm{ub},B})-T_{\mathrm{res}}(s)\ge4s(\varepsilon_1+\varepsilon_2)+(\Psi_{0B}-28)+\frac{784}{10s+28}-\frac{\Psi_{1B}\Psi_{2B}}{s+\chi c_B},
\]
with $\Psi_{1B}=2+4\chi^2+10(\chi c_B-V)$, $\Psi_{2B}=\chi c_B\mathcal E_{0B}-\chi^2V$, and $\Psi_{0B}-28=2(\mathcal E_{0B}^2-16)+4(1-\chi^2)+(2-4\chi^2+2\chi c_B)\mathcal E_{0B}-V(10\mathcal E_{0B}-6\chi^2)$. Here $\Psi_{1B}\Psi_{2B}\le64$, so the last two terms exceed $\frac{144s-1792}{s(10s+28)}>0$, while $4-\mathcal E_{0B}=4\varepsilon_1+2\varepsilon_2$, $V\le2\varepsilon_2$, and $10\mathcal E_{0B}-6\chi^2\le34$ give $\Psi_{0B}-28\ge-60\varepsilon_1-108\varepsilon_2$, leaving at least $(4s-108)(\varepsilon_1+\varepsilon_2)\ge0$ for $s\ge31$.

\paragraph{Branch C ($\omega+\Xi<1$).}
Set $\beta=0$ in~\eqref{eq:pair-cut}: $2s(5-M(C_0))\ge\Lambda_0(\mathcal E)=8s-4s\Xi-2s\mathcal E+2\mathcal E^2+(2-10V)\mathcal E-4\chi^2+4\chi^2V$. With $c_C=\sqrt{1-V-\Xi}$, $\varepsilon_1=1-\chi^2$, $\varepsilon_0=1-\chi^2-\Xi>0$, and $\mathcal E_{0C}=1+\chi^2-\Xi+2\chi c_C=(\chi+c_C)^2+V\le4$, \eqref{eq:tail-Q} gives $\mathcal E\le\mathcal E_1=\mathcal E_{0C}+\frac{\chi^2}sV\le4+\frac1s$ (where $\Lambda_0'<0$). Set $\varepsilon_C=4-\mathcal E_{0C}$. If $c_C\le\chi$, then $8s-4s\Xi-2s\mathcal E_{0C}=4s(\varepsilon_1+\varepsilon_2')+2s\varepsilon_0$ ($\varepsilon_2'=\chi(\chi-c_C)$), $V\le\varepsilon_0+2\varepsilon_2'$, $\varepsilon_C\le4\varepsilon_1+2\varepsilon_2'$, and $\Lambda_0(\mathcal E_1)-28\ge8+(4s-68)\varepsilon_1+(4s-116-\frac{20}s)\varepsilon_2'+(2s-40-\frac{10}s)\varepsilon_0>8$. If $c_C>\chi$, then $8s-4s\Xi-2s\mathcal E_{0C}\ge4s\varepsilon_1+2sV$ and $\varepsilon_C<4\varepsilon_1$, giving $\Lambda_0(\mathcal E_1)-28\ge8+(4s-68)\varepsilon_1+(2s-40-\frac{10}s)V>8$. Since $T_{\mathrm{res}}(s)<28$, this closes Branch~C.

\section{Algebraic Details and Finite Verifications for Section~7}\label{app:sec7}

This appendix records the exact rational reductions, complete boundary coverage, worked representative calculations, and complete rational parameter tables (\cref{tab:hub-brackets,tab:stock-blocks,tab:bin-brackets,tab:bin-ecaps,tab:bin-windows,tab:exceptional,tab:early-polys}) underlying \cref{sec:hub-regime} (and the crossover brackets invoked in \cref{sec:crossover}). Throughout this appendix and \cref{app:sec8,app:sec9}, every printed decimal lower bound is rounded down, every upper bound is rounded up, every bracket $[a^-,a^+]$ is outward, and every margin is a lower bound on the stated positive difference. Each finite family below is specified by the on-paper reductions, worked calculations, and rational parameter tables printed here, while the complete repeated exact-rational verifications across all rank/window families and reported margins are carried out in the companion package (not checked by hand).

\subsection{Outward brackets for the hub crossover}\label{app:hub-brackets}

\Cref{tab:hub-brackets} lists outward ten-digit brackets for the exact rational values of $\pi_m(1/c)$, $s_{1,m}(c)=\sum_{i=1}^m(i+c)^{-1}$, $s_{2,m}(c)=\sum_{i=1}^m(i+c)^{-2}$, $G_m(1/c)$, and $G'_m(c)=\frac d{dc}G_m(1/c)$ used in \cref{thm:brackets,cor:sharp16}. Each entry is a finite sum or product of at most seventeen positive rational terms, with $m=17$ linked to $m=16$ via $\pi_{17}=\frac{17}{17+c}\pi_{16}$, $s_{1,17}=s_{1,16}+\frac1{17+c}$, and $s_{2,17}=s_{2,16}+\frac1{(17+c)^2}$.

\begin{table}[!ht]
\centering
\small
\resizebox{\textwidth}{!}{%
\begin{tabular}{@{}llll@{}}
\toprule
$(m,c)$ & $\pi_m(1/c)$ & $s_{1,m}(c)$ & $s_{2,m}(c)$\\
\midrule
$(16,\frac4{11})$ & $[0.3197771747,\,0.3197771748]$ & $[2.9246451931,\,2.9246451932]$ & $[1.0032865957,\,1.0032865958]$\\
$(17,\frac4{11})$ & $[0.3130802705,\,0.3130802706]$ & $[2.9822368162,\,2.9822368163]$ & $[1.0066033907,\,1.0066033908]$\\
$(16,\frac{100}{273})$ & $[0.3172965160,\,0.3172965161]$ & $[2.9219761728,\,2.9219761729]$ & $[1.0004827895,\,1.0004827896]$\\
$(16,\frac{1000}{2729})$ & $[0.3171720987,\,0.3171720988]$ & $[2.9218418923,\,2.9218418924]$ & $[1.0003418955,\,1.0003418956]$\\
\midrule
$(m,c)$ & $G_m(1/c)$ & $G'_m(c)$ & \\
\midrule
$(16,\frac4{11})$ & $[8.8910783868,\,8.8910783869]$ & $[0.0621884092,\,0.0621884093]$ &\\
$(17,\frac4{11})$ & $[8.9823226222,\,8.9823226223]$ & $[0.0082582177,\,0.0082582178]$ &\\
$(16,\frac{100}{273})$ & $[8.8911615773,\,8.8911615774]$ & $[0.0004484092,\,0.0004484093]$ &\\
$(16,\frac{1000}{2729})$ & $[8.8911614306,\,8.8911614307]$ & $[-0.0026336021,\,-0.0026336020]$ &\\
\bottomrule
\end{tabular}}
\caption{Outward brackets for the exact rational hub quantities at the crossover, with $H_{16}=\frac{2436559}{720720}$ and $H_{17}=\frac{42142223}{12252240}$.}\label{tab:hub-brackets}
\end{table}

Direct exact rational evaluation gives the crossover consequences of \cref{thm:brackets,cor:sharp16}:
\begin{gather*}
G_{16}(\tfrac{11}4)-(2+2H_{17})>0.011973,\qquad G_{16}(\tfrac{273}{100})<8.891161578,\qquad 0<G'_{16}(\tfrac{100}{273})<0.0004485,\\
G'_{16}(\tfrac{1000}{2729})<-0.0026,\qquad J_{18}<8.89116164<\mathsf P_{16}=\tfrac{4445581}{500000},\qquad J_{19}<8.988329<2+2H_{18}.
\end{gather*}

\paragraph{Worked two-tangent enclosures ($m=2$ and $m=16$).}
Writing $G'_m(p)=\frac d{dp}G_m(p)$ as in \cref{sec:binary}, for any rational bracket $0<p_a<p_b$ with $0<G'_m(p_a)/K'_m(p_a)<1$ and $G'_m(p_b)<0$, \cref{thm:laplace}(a) at $v=0$ encloses the unique maximizer $p^*_m\in(p_a,p_b)$ and bounds $J_{m+2}$ below by $\max(G_m(p_a),G_m(p_b))$. Moreover, $\mathsf v_m(p_a)=-G'_m(p_a)/K'_m(p_a)\in(-1,0)$ and $\mathsf v_m(p_b)=-G'_m(p_b)/K'_m(p_b)>0$ both satisfy $\mathsf v\ge-1$, so \cref{thm:laplace}(a) (as in \cref{lem:bin-tools}(c)) produces two global supporting lines in the $K_m$-coordinate of opposite slopes lying above $(K_m(p^*_m),J_{m+2})$, bounding $J_{m+2}$ above by their intersection ordinate (every non-exact bracket in \cref{tab:bin-brackets,tab:bin-ecaps} satisfies $0<G'_m(p_a)/K'_m(p_a)<1$ and $G'_m(p_b)<0$). At $m=2$ on $[\frac94,\frac{23}{10}]$, exact evaluation gives $G_2(\frac94)=\frac{128950}{20449}\ge6.30593183$, $G'_2(\frac94)=\frac{18924}{2924207}>0$ ($0<G'_2(\frac94)/K'_2(\frac94)<1$), $G_2(\frac{23}{10})=\frac{672985}{106722}\ge6.30596315$, and $G'_2(\frac{23}{10})=-\frac{995875}{197222256}<0$, yielding $J_4\in[6.30596315,\,6.30609127]$. At $m=16$ on $[\frac{13649}{5000},\frac{27299}{10000}]$, $G'_{16}(\frac{13649}{5000})\in[2.25491,2.25492]\times10^{-5}$ ($G'_{16}/K'_{16}\in[4.82302,4.82303]\times10^{-5}<1$) and $G'_{16}(\frac{27299}{10000})\in[-1.88111,-1.88110]\times10^{-5}$ ($-G'_{16}/K'_{16}\in[4.02364,4.02365]\times10^{-5}>0$), enclosing $J_{18}$ to exact width under $10^{-9}$ before outward rounding ($<9.41\times10^{-10}$), giving the outward ten-digit display $J_{18}\in[8.8911615812,\,8.8911615823]<\mathsf P_{16}$.

\subsection{Low-mean stock blocks and hub dominance}\label{app:hub-stocks}

In \cref{thm:hub-stocks}, direct rational evaluation settles $j\in\{1,2,3\}$ ($\mathsf C_1<G_1(\frac{11}5)$, $\mathsf C_j<G_j(\frac{11}4)$ with margins $>0.1182,0.0822,0.0713$, and $G_j(\frac{11}4)>2+2H_{j+1}$ for $j=2,3$). For $4\le j\le16$, anchoring~\eqref{eq:stock-formula} at $\mathsf C_4=\frac{347}{50}$ and $\mathsf C_{16}=G_{16}(\frac{11}4)$ with $t_{16}=\frac{935059}{720720}$ and $s^{(2)}_{16}=\frac{83491976749}{519437318400}$ fixes the exact rational $\alpha_{\mathrm{st}}\in[1.1440612856,\,1.1440612857]$. Monotonicity of $j\Delta G_j(\frac{11}4)$ on $5\le j\le16$ follows from $T^{\mathrm{st}}_5=\frac{758567513}{2047544850}>0$ and $\frac{112}{121}\cdot\frac{44}{59}=\frac{448}{649}>0.690292>0.600198>\frac4{11}(\frac{112}{121}s_{1,15}(\frac4{11})-1)$, while $j\Delta\mathsf C_j=\mathsf m_j$ strictly increases as proved in \cref{sec:low-mean}. Therefore, on any contiguous block $[a,b]\subseteq[5,16]$, the single endpoint inequality $b\Delta G_b(\frac{11}4)<a\Delta\mathsf C_a$ implies $j\Delta G_j(\frac{11}4)\le b\Delta G_b(\frac{11}4)<a\Delta\mathsf C_a\le j\Delta\mathsf C_j$, hence $\Delta G_j(\frac{11}4)<\Delta\mathsf C_j$ throughout $a\le j\le b$. For example, on the six-index block $[11,16]$, exact rational evaluation gives $16\Delta G_{16}(\frac{11}4)\le1.545895<1.559002\le11\Delta\mathsf C_{11}$ (margin $\ge0.013108$), establishing all six indices $11\le j\le16$ at once; the same upper bound $16\Delta G_{16}(\frac{11}4)<\frac53\le j\Delta(2+2H_{j+1})$ on $5\le j\le16$ propagates $G_j(\frac{11}4)>2+2H_{j+1}$ from $j=16$ down to $j=4$. \Cref{tab:stock-blocks} lists all six covering blocks.

\begin{table}[!ht]
\centering
\small
\begin{tabular}{@{}lccc@{\qquad}lccc@{}}
\toprule
Block $[a,b]$ & $b\,\Delta G_b(\frac{11}4)$ & $a\,\Delta\mathsf C_a$ & Margin & Block $[a,b]$ & $b\,\Delta G_b(\frac{11}4)$ & $a\,\Delta\mathsf C_a$ & Margin\\
\midrule
$[5,5]$ & $1.345360$ & $1.380424$ & $0.035065$ & $[8,8]$ & $1.450897$ & $1.477069$ & $0.026173$\\
$[6,6]$ & $1.390482$ & $1.413758$ & $0.023276$ & $[9,10]$ & $1.488935$ & $1.506110$ & $0.017175$\\
$[7,7]$ & $1.424481$ & $1.446225$ & $0.021745$ & $[11,16]$ & $1.545895$ & $1.559002$ & $0.013108$\\
\bottomrule
\end{tabular}
\caption{The six stock blocks covering $5\le j\le16$ in \cref{thm:hub-stocks}, with $b\Delta G_b(\frac{11}4)$ rounded up and $a\Delta\mathsf C_a$ and margins rounded down.}\label{tab:stock-blocks}
\end{table}

\subsection{Witness-cut arithmetic and the premium handover}\label{app:projection}

With $\varepsilon_0=\frac{14117}{100000}$, $\varepsilon_{\mathrm{hi}}=\frac{105}{10000}$, $B_{\mathrm{hi}}=\frac{1427}{10000}$, $\rho_{\mathrm{hi}}=\frac{135}{10000}$, $a_{\mathrm{lo}}=\frac{9865}{10000}$, $r_{\mathrm{hi}}=\frac{137}{10000}$, $\varrho_{\mathrm{sat}}=\frac{487}{1000}$, $K_{\mathrm{pr}}=8$, and $z_0=\frac5{16}$, every numerical inequality of \cref{lem:witness,thm:projection} is a single exact rational check:
\begin{itemize}[leftmargin=1.5em]
\item \emph{Witness bounds:} $\varepsilon_0/(\frac{14}5(5-\varepsilon_0)-\varepsilon_0)<0.010486<\varepsilon_{\mathrm{hi}}$ and $(1+\varepsilon_{\mathrm{hi}})\varepsilon_0=\frac{28530457}{200000000}<B_{\mathrm{hi}}$.
\item \emph{Mass and distance:} $\varepsilon_{\mathrm{hi}}(1+2B_{\mathrm{hi}})<0.013497<\rho_{\mathrm{hi}}$, $\rho_{\mathrm{hi}}/a_{\mathrm{lo}}<0.013685<r_{\mathrm{hi}}$, and $\rho_{\mathrm{hi}}\varepsilon_{\mathrm{hi}}<(\frac{12}{1000})^2$, so $R_{\mathrm{sat}}\sqrt{\varepsilon_{\mathrm{w}}}\ge\frac12-\frac{120}{9865}>0.4878>\varrho_{\mathrm{sat}}$.
\item \emph{Radius awards (seams $r_y\in[0,\frac9{100}]\cup[\frac9{100},\frac12]\cup[\frac12,\infty)$):} the funded radius award is at least $A_{\min}=\frac9{40}(1-\frac{2\varepsilon_{\mathrm{hi}}}{a_{\mathrm{lo}}\varrho_{\mathrm{sat}}^2})>0.2048$. For $r_y\le\frac9{100}$, $2(\frac9{100})+(\frac9{100})^2=0.1881<A_{\min}$; for $\frac9{100}\le r_y\le\frac12$, $\frac{227}9\cdot\frac23\cdot\frac{\varepsilon_{\mathrm{hi}}B_{\mathrm{hi}}}{a_{\mathrm{lo}}\varrho_{\mathrm{sat}}^2}<0.1077<A_{\min}$; for $r_y\ge\frac12$, $3+2/r_y\le7$.
\item \emph{Premium demand and expense:} $\frac{7(1+2r_{\mathrm{hi}})}{1-\frac23B_{\mathrm{hi}}}<7.948<K_{\mathrm{pr}}=8$, $\frac{\varepsilon_{\mathrm{hi}}}{a_{\mathrm{lo}}}(2+\frac23K_{\mathrm{pr}}B_{\mathrm{hi}})<0.02939<\frac1{25}$, $\frac23(2+(5+14r_{\mathrm{hi}})z_0)<\frac{39}{16}=4-5z_0$, and $2z_0(1+r_{\mathrm{hi}})+\frac{39}{16}B_{\mathrm{hi}}=\frac{157023}{160000}<0.988$, giving the slack $-\frac{12}{1000}$.
\end{itemize}

\paragraph{Premium handover (\cref{cor:premium}).}
With $p_{\mathrm{ref}}=\frac{273}{100}$ and $p_{\mathrm{pr}}=\frac{21}{10}$, the harmonic slope of $G_m(p_{\mathrm{ref}})-80\varpi_m$ is $1+\psi_m(p_{\mathrm{ref}})-\frac{16}5(1-\pi_m(p_{\mathrm{pr}}))$. Direct evaluation gives negative slopes $\le-0.266781$ ($m=1$) and $\le-0.404620$ ($m=2$), while for $3\le m\le16$ monotonicity of $\psi_m(p_{\mathrm{ref}})$ and $\pi_m(p_{\mathrm{pr}})$ gives $1+\psi_m(p_{\mathrm{ref}})\le1+\psi_{16}(p_{\mathrm{ref}})\le1.543545<1.688976\le\frac{16}5(1-\pi_3(p_{\mathrm{pr}}))$. At the anchor $m=16$, $G_{16}(p_{\mathrm{ref}})\in[8.8911615773,8.8911615774]$ and $80\varpi_{16}\in[8.8539826515,8.8539826516]$ ($\varpi_{16}\in[0.1106747831,0.1106747832]$), giving surplus $G_{16}(p_{\mathrm{ref}})-80\varpi_{16}\in[0.0371789258,0.0371789259]>0.037178$ and settling all $1\le m\le16$.

\subsection{Finite verifications and witness tables for the binary comparison}\label{app:binary-finite}

This subsection documents the exact rational formulas, complete witness tables (\cref{tab:bin-brackets,tab:bin-ecaps,tab:bin-windows,tab:exceptional}), worked representative calculations, and finite inequality families that discharge \cref{thm:pure-price,lem:bin-tools,lem:bin-rejected,lem:bin-accepted,lem:bin-exterior} in \cref{sec:binary}. Write $S_j(p)=\sum_{i=1}^j(1+ip)^{-1}$, $S^{(2)}_j(p)=\sum_{i=1}^j(1+ip)^{-2}$, $\psi_j(p)=\pi_j(p)[4(1-1/p)S_j(p)-1]$, and $s_j(f)=j(f_j-f_{j-1})$.

\paragraph{Stopped pure price straddle (\cref{thm:pure-price}, $i=2,\delta=0$).}
At $z=\frac54$, exact evaluation gives $K_2(\frac54)=\frac{145}{126}$, $K'_2(\frac54)=\frac{2164}{3969}$, and $F_{2,0}(0)=K_2(\frac54)-\frac{11}{40}K'_2(\frac54)=\frac{19862}{19845}=1+\frac{17}{19845}>1$. On the straddle seam $[\frac9{10},\frac54]$, the convex function $c_0(1+2z)z^5$ ($c_0=\frac{19}{126}(\frac45)^5=\frac{9728}{196875}$) lies above its tangent $T(z)=\frac{9027584}{33480783}+\frac{1712128}{1240029}(z-\frac{10}9)$ at $z_0=\frac{10}9$. Because $2\sqrt{2+4z}$ is concave, $S(z):=3z+2-T(z)<2\sqrt{2+4z}$ holds across $[\frac9{10},\frac54]$ from squaring at the two endpoints: $S(\frac9{10})=\frac{1580912257}{334807830}\le4.721850$ with $4(2+4\cdot\frac9{10})-S(\frac9{10})^2>0.104135>0$, and $S(\frac54)=\frac{708265753}{133923132}\le5.288600$ with $4(2+4\cdot\frac54)-S(\frac54)^2>0.030719>0$.

\paragraph{Exact identities, weight-gap enclosures, and optimizer brackets (\cref{lem:bin-tools}).}
Induction on the recurrences of \cref{sec:binary} gives $K_j(p)=p+(1-p)\pi_j(p)$ and $G_j(p)=\mathsf K_j+3p(1-\pi_j(p))+4\pi_j(p)-4(p-1)\pi_j(p)S_j(p)$, with harmonic slopes $s_j(G)=1+\psi_j$, $s_j(K')=\frac{\pi_j}p(\frac1p+(1-\frac1p)S_j)$, and $s_j(\mathsf y)=\frac{\pi_j}p(S_j(5-\frac8p-4(1-\frac1p)S_j)-4(1-\frac1p)S^{(2)}_j)$, yielding the two exact weight-gap identities ($j\ge2$)
\begin{gather*}
jK'_j(p)\bigl(\mathsf v_{j-1}(p)-\mathsf v_j(p)\bigr)=-s_j(\mathsf y)+\mathsf v_{j-1}(p)\,s_j(K'),\\
(1+jp)K'_j(p)\bigl(\mathsf v_{j-1}(p)-\mathsf v_j(p)\bigr)=3(1-\pi_j(p))+\mathsf v_{j-1}(p)\bigl(1-(1-\tfrac1p)\pi_j(p)\bigr)-8(1-\tfrac1p)\pi_j(p)S_j(p).
\end{gather*}
Because $\pi_j(p)$ and $\mathsf v_{j-1}(p)$ increase while $S_j(p)$, $S^{(2)}_j(p)$, and $K'_j(p)$ decrease in $p$, on any price window $[p_a,p_b]$ (with $c_a=1/p_a$, $c_b=1/p_b$) setting
\begin{align*}
N_{\mathrm{lb}}&:=3(1-\pi_j(p_b))+\max\bigl(0,\mathsf v_{j-1}(p_a)\bigr)\bigl(1-(1-c_b)\pi_j(p_b)\bigr)-8(1-c_b)\pi_j(p_b)S_j(p_a),\\
N_{\mathrm{ub}}&:=3(1-\pi_j(p_a))+\mathsf v_{j-1}(p_b)\bigl(1-(1-c_a)\pi_j(p_a)\bigr)-8(1-c_a)\pi_j(p_a)S_j(p_b),\\
B_{\mathrm{ub}}&:=4(1-c_b)\bigl(S_j(p_a)^2+S^{(2)}_j(p_a)\bigr)-(5-8c_a)S_j(p_b),\\
\mathrm{nsy}_{\mathrm{ub}}&:=\bigl(\pi_j(p_b)c_a\,\mathbf 1_{\{B_{\mathrm{ub}}\ge0\}}+\pi_j(p_a)c_b\,\mathbf 1_{\{B_{\mathrm{ub}}<0\}}\bigr)B_{\mathrm{ub}},\qquad
sK'_{\mathrm{ub}}:=\pi_j(p_b)c_a\bigl(c_a+(1-c_b)S_j(p_a)\bigr)
\end{align*}
yields the exact rational enclosures
\begin{align*}
\Delta\mathsf v_{\mathrm{lb}}&:=\max\!\Bigl(0,\,\frac{N_{\mathrm{lb}}}{(1+jp_b)K'_j(p_a)}\Bigr)\le\mathsf v_{j-1}(p)-\mathsf v_j(p)\\
&\le\min\!\Bigl(\frac{N_{\mathrm{ub}}}{(1+jp_a)K'_j(p_b)},\,\frac{\mathrm{nsy}_{\mathrm{ub}}+\mathsf v_{j-1}(p_b)\,sK'_{\mathrm{ub}}}{jK'_j(p_b)}\Bigr)=:\Delta\mathsf v_{\mathrm{ub}}.
\end{align*}

The anchor for \cref{lem:bin-tools}(c) is $\mathsf v_1(p_{\mathrm{ref}})<\frac9{25}<\frac59$; seven further exact evaluations at $p_{\mathrm{ref}}$, $\frac{11}5$, and $\frac{243}{100}$ verify that $\psi_j(p_{\mathrm{ref}})$ increases on $1\le j\le16$, $K'_j(p_{\mathrm{ref}})$ and $\mathsf y_j(p_{\mathrm{ref}})$ are concave in $H_j$, $\psi_j(p)>0$ for $j\ge2$ on $[\frac{11}5,p_{\mathrm{ref}}]$, and $\partial_p\psi_j(p)>0$ for $j\ge6$ on $[\frac{11}5,p_{\mathrm{ref}}]$ and for $j\ge5$ on $[\frac{243}{100},p_{\mathrm{ref}}]$. \Cref{tab:bin-brackets} lists all $16$ outward optimizer brackets $[p_a,p_b]\ni p^*_m$ and their exact two-tangent enclosures $[J^-_{m+2},J^+_{m+2}]$ (every printed seven-digit decimal is an exact rational multiple of $10^{-7}$), and \cref{tab:bin-ecaps} lists the six predecessor gap caps $e^+_i\ge J_{i+2}-G_i(p_{\mathrm{ref}})$; in addition, $G'_9(\frac{257}{100})>0.0015$ gives $p^*_m\ge\frac{257}{100}$ for $m\ge9$.

\begin{table}[!ht]
\centering
\small
\begin{tabular}{@{}cll@{\qquad}cll@{}}
\toprule
$m$ & Bracket $[p_a,p_b]$ & $[J^-_{m+2},J^+_{m+2}]$ & $m$ & Bracket $[p_a,p_b]$ & $[J^-_{m+2},J^+_{m+2}]$\\
\midrule
$1$ & $[\frac{11}{5},\frac{11}{5}]$ & $[5.78125,\,5.78125]$ ($\frac{185}{32}$) & $9$ & $[\frac{51}{20},\frac{13}{5}]$ & $[8.0586308,\,8.0588596]$\\[2pt]
$2$ & $[\frac{9}{4},\frac{23}{10}]$ & $[6.3059631,\,6.3060913]$ & $10$ & $[\frac{13}{5},\frac{131}{50}]$ & $[8.2062275,\,8.2062315]$\\[2pt]
$3$ & $[\frac{23}{10},\frac{12}{5}]$ & $[6.7018057,\,6.7023591]$ & $11$ & $[\frac{131}{50},\frac{66}{25}]$ & $[8.3416948,\,8.3417188]$\\[2pt]
$4$ & $[\frac{47}{20},\frac{12}{5}]$ & $[7.0204464,\,7.0205071]$ & $12$ & $[\frac{66}{25},\frac{2651}{1000}]$ & $[8.4669901,\,8.4669936]$\\[2pt]
$5$ & $[\frac{243}{100},\frac{49}{20}]$ & $[7.2869433,\,7.2869674]$ & $13$ & $[\frac{267}{100},\frac{107}{40}]$ & $[8.5835302,\,8.5835317]$\\[2pt]
$6$ & $[\frac{49}{20},\frac{5}{2}]$ & $[7.5161294,\,7.5163281]$ & $14$ & $[\frac{269}{100},\frac{539}{200}]$ & $[8.6924738,\,8.6924759]$\\[2pt]
$7$ & $[\frac{5}{2},\frac{51}{20}]$ & $[7.7174461,\,7.7175583]$ & $15$ & $[\frac{271}{100},\frac{2713}{1000}]$ & $[8.7947599,\,8.7947609]$\\[2pt]
$8$ & $[\frac{127}{50},\frac{51}{20}]$ & $[7.8968698,\,7.8968785]$ & $16$ & $[\frac{13649}{5000},\frac{27299}{10000}]$ & $[8.8911615,\,8.8911616]$\\
\bottomrule
\end{tabular}
\caption{All sixteen exact optimizer brackets $[p_a,p_b]\ni p^*_m$ and outward two-tangent enclosures $[J^-_{m+2},J^+_{m+2}]$ (exact multiples of $10^{-7}$) used across \cref{lem:bin-rejected,lem:bin-accepted,lem:bin-exterior}.}\label{tab:bin-brackets}
\end{table}

\begin{table}[!ht]
\centering
\small
\begin{tabular}{@{}cll@{\qquad}cll@{}}
\toprule
$i$ & Bracket $[p_a,p_b]$ & Cap $e^+_i\ge J_{i+2}-G_i(p_{\mathrm{ref}})$ & $i$ & Bracket $[p_a,p_b]$ & Cap $e^+_i\ge J_{i+2}-G_i(p_{\mathrm{ref}})$\\
\midrule
$1$ & exact ($p^*_1=\frac{11}5$) & $\frac{7887}{500000}=0.015774$ & $4$ & $[\frac{47}{20},\frac{49}{20}]$ & $\frac{3019}{200000}=0.015095$\\[2pt]
$2$ & $[\frac94,\frac{23}{10}]$ & $\frac{18359}{1000000}=0.018359$ & $5$ & $[\frac{12}5,\frac52]$ & $\frac{12573}{1000000}=0.012573$\\[2pt]
$3$ & $[\frac{23}{10},\frac{12}5]$ & $\frac{3477}{200000}=0.017385$ & $8$ & $[\frac52,\frac{13}5]$ & $\frac{6291}{1000000}=0.006291$\\
\bottomrule
\end{tabular}
\caption{Predecessor gap brackets and exact rational caps $e^+_i$ used in Case~1 of \cref{lem:bin-accepted} (with $e_i$ decreasing for $i\ge5$).}\label{tab:bin-ecaps}
\end{table}

\paragraph{Rejected mass and stops (\cref{lem:bin-rejected}).}
Across all $136$ pairs $1\le j\le m\le16$ and all $T_{\mathrm{post}}\ge0$, the reduction in \cref{lem:bin-rejected}(i) is discharged by: (a)~$m\le12$: $12$ checks at $p_{\mathrm{ub}}=\frac{2651}{1000}$ (min margin $>0.00196$); (b)~$13\le m\le16$, $T_{\mathrm{post}}=72$: $16$ checks at $p_{\mathrm{ref}}$ (min margin $>0.00547$); (c)~$13\le m\le16$, $T_{\mathrm{post}}=\infty$: $58$ checks across $m\in\{13,14,15,16\}$, $1\le j\le m$ at $p_b(m)\in\{\frac{107}{40},\frac{539}{200},\frac{2713}{1000},\frac{27299}{10000}\}$ (min margin $>0.00484$). For \cref{lem:bin-rejected}(ii), stop ceilings hold at $j=1,2$ directly ($>0.5479$ in both modes) and at $j=16$ ($G_{15}(p_{\mathrm{ref}})-2\mathsf K_{16}>0.03323$ and $G_{15}(p_{\mathrm{ref}})+\varpi_{15}-2\mathsf K_{16}-\frac1{25}H_{16}>0.00675$), propagating to $3\le j\le16$ by negative harmonic slope; at rank~$0$ under the single radius award $(\frac12,\frac9{40})$, $4-(\frac9{40}+\frac94)=\frac{61}{40}=1.525>0$.

\paragraph{Accepted comparison without exterior (\cref{lem:bin-accepted}).}
Because $16p_{\mathrm{ref}}=43.68<72$, the four seams $[0,\frac54]\cup[\frac54,p_{\mathrm{ref}}]\cup[p_{\mathrm{ref}},72/j]\cup[72/j,\infty)$ cover $z\ge0$ completely for all $1\le j\le16$ and $\varpi\in\{0,\frac1{25}\}$:
\begin{itemize}[leftmargin=1.5em]
\item \emph{Small prices $z\in[0,\frac54]$ (Case~1):} for $4\le j\le16$, with $a_j\ge a_4\ge3.873845$ ($a_j\ge a_6\ge4.002405$ for $j\ge6$) and uniform slope floor $\Delta_j\ge\Delta_0:=1+\psi_4(p_{\mathrm{ref}})-5e^+_3\ge1.196114$ (both floors are verified by exact rational evaluation at each $4\le j\le16$, using the caps of \cref{tab:bin-ecaps} or the upper enclosures of \cref{tab:bin-brackets} for $\Delta_j$), the no-exterior cubic roof $\overline K\le\frac{(j+1)z}{1+jz}+\frac4{27}z^3$ together with $\overline K<\frac54$ closes in four scalar rational checks: $a_4+\Delta_0-5>0.069960$, $a_4-4+\min(\frac23,\Delta_0-\frac{16}{27})>0.477367$, $a_6-4>0.002405$, and $\Delta_0-\frac{16}{27}>0.603522$. For $j=1,2,3$, direct roofs close by negative discriminants or endpoints; as a worked check at $j=3$ ($\varpi=\frac1{25}$, $e^+_2=\frac{18359}{1000000}$), $a_3=G_3(p_{\mathrm{ref}})-\mathsf K_3-\eta^{\mathrm{pr}}_3\in[3.807305,3.807306]$ and $\Delta_3=1+\psi_3(p_{\mathrm{ref}})-3e^+_2\in[1.136649,1.136650]$ give discriminant $(3a_3+\Delta_3-19)^2-12a_3\Delta_3\in[-10.438774,-10.438773]<0$, so $(1+3z)(a_3+\Delta_3 z)-19z>0$ for all $z\ge0$.
\item \emph{Crossing ($91$ checks, Case~3) and interior $\varsigma$-vertex ($36$ checks, Case~4):} \cref{tab:bin-windows} lists all seven crossing windows ($W_1$--$W_3$, $H_{13}$--$H_{16}$, using effective lower price $p_{a,\mathrm{eff}}=\max(p_a(w),p_a(j))$ from \cref{tab:bin-brackets}, $\varsigma\in[\frac{33}{50},1]$, splitting $\varsigma$ at $\frac9{10}$ for $j\in\{3,4,5,6\}$ in $H_{13}$--$H_{16}$; min margin $>0.0621$) and all four vertex windows ($D_1$--$D_4$; min margin $>0.3395$). On each crossing cell, with $\lambda_{j,\mathrm{lb}}=\mathbf 1_{m\ge13,\,T_{\mathrm{lo}}>72}\frac3{20}(H_m-H_j)(1-\frac{72}{T_{\mathrm{lo}}})$, the remaining rational bounds of \cref{lem:bin-accepted} Case~3 are
\begin{gather*}
V_{\mathrm{ub}}=\min\bigl(c_{\mathrm{hi}}\mathsf K_j,\,S_{\mathrm{ub}}+\tfrac{c_{\mathrm{hi}}}j\bigr),\quad
\widehat V_{\mathrm{ub}}=\min\bigl(\mathsf K_j,\,\tfrac{S_{\mathrm{ub}}}{c_{\mathrm{lo}}}+\tfrac1j\bigr),\quad
V_{\varsigma,\mathrm{ub}}=\min\bigl(\varsigma_{\mathrm{hi}}\mathsf K_j,\,\tfrac{S_{\mathrm{ub}}}{\varsigma_{\mathrm{lo}}}+\tfrac{\varsigma_{\mathrm{hi}}}j\bigr),\\
A_{\mathrm{lb}}=G_j(p_b)-\mathsf K_j-\eta^{\mathrm{pr}}_j+\lambda_{j,\mathrm{lb}}-V_{\mathrm{ub}}-\mathsf v_j(p_b)\bigl(V_{\mathrm{ub}}-K_j(p_b)\bigr)_+,\\
\mathrm{sl}_{\mathrm{lb}}=G_j(p_b)-J^+_{j+1}+\lambda_{j,\mathrm{lb}}-\mathsf v_j(p_b)\bigl(V_{\mathrm{ub}}-K_j(p_b)\bigr)_+.
\end{gather*}
As a worked crossing check at $(m,j)=(16,16)$ on $[\frac{13649}{5000},\frac{27299}{10000}]$, $\varsigma\in[\frac{33}{50},1]$, $\varpi=\frac1{25}$: $\Delta\mathsf v_{\mathrm{ub}}\le0.016261$, $x_{\mathrm{lo}}\ge244.989769$, $\widehat V_{\mathrm{ub}}\le4.380729$, $V_{\varsigma,\mathrm{ub}}\le4.218070$, $A_{\mathrm{lb}}\ge1.640677$, and $\mathrm{sl}_{\mathrm{lb}}\ge0.096375$, yielding $x_{\mathrm{lo}}\mathrm{sl}_{\mathrm{lb}}-\widehat V_{\mathrm{ub}}-V_{\varsigma,\mathrm{ub}}^2/A_{\mathrm{lb}}\ge8.385930>0$.
\item \emph{Rays at $\varsigma=1$ (Case~5):} on $[\frac54,p_{\mathrm{pr}}]$ at $\varpi=\frac1{25}$, $8c_*(z_1-p_{\mathrm{pr}})\ge b_1$ holds with $z_1=\frac{11}5,c_*=4(1+p_{\mathrm{ref}})^{-3}$ ($j=1$) and $z_1=\frac94,c_*=\frac2{(9/4)^3}\min(m_1(2),m_1(16))$ where $m_1(j)=\pi_j(\frac54)\frac54S_j(\frac54)$ ($j\ge2$; $2$ checks, margin $>0.0217$). On $[p_{\mathrm{ref}},72/j]$, the three guards at $\bar{\mathsf v}=\mathsf v_j(p_{\mathrm{ref}})$ and $\bar A=G_j(p_{\mathrm{ref}})-\mathsf K_j+\bar{\mathsf v}K_j(p_{\mathrm{ref}})-b_0$ hold as follows: (G1)~chord at $j=8,9$ checked at $j=1$ ($\ge0.105159$) and $j=16$ ($\ge0.028744$); (G2)~direct at $j=1$ ($\ge1.547909$) and quadratic Bernstein blocks on $[2,8]$ (pivot $5$: $B\ge[0.163839,0.150513,0.037712]$) and $[9,16]$ (pivot $12$: $B\ge[0.029871,0.023913,0.023009]$); and (G3)~$32$ exact checks across $1\le j\le16$ and $\varpi\in\{0,\frac1{25}\}$ (all $32$ floors listed in \cref{tab:bin-windows}; min $>0.2536$ at $\varpi=0$ and $>0.0440$ at $\varpi=\frac1{25}$). As a worked check at $j=16$, $\varpi=\frac1{25}$: $K_{16}(p_{\mathrm{ref}})\in[2.181077,2.181078]$, $K'_{16}(p_{\mathrm{ref}})\in[0.467493,0.467494]$, $\bar{\mathsf v}\in[0.000128,0.000129]$, $D_{16}\in[9.887566,9.887567]$, $\eta^{\mathrm{pr}}_{16}\in[0.064554,0.064555]$, $b_1=\eta^{\mathrm{pr}}_{16}/p_{\mathrm{pr}}$, and $\bar A\in[4.446158,4.446159]$, yielding (G1)~$>0.083724$, (G2)~$>0.064866$, and (G3)~$[(2+\bar{\mathsf v})\mathsf K_{16}-\bar A]^2\le18.626673<18.795849\le(4K'_{16}(p_{\mathrm{ref}})+b_1)(1+\bar{\mathsf v})D_{16}$ (margin $>0.169177$). On the shifted ray $z\ge72/j$ ($13\le m\le16$, $2\le j\le m$), with $\mathsf v_b:=\mathsf v_j(p_b)$ and $A_b:=G_j(p_{\mathrm{ref}})-\mathsf K_j+\mathsf v_bK_j(p_{\mathrm{ref}})-b_0$ satisfying $(1+\mathsf v_b)\mathsf K_j<A_b\le(2+\mathsf v_b)K_j(72/j)$, the exact inequality
\begin{align*}
4K'_j(p_b)+b_1&+\mathbf 1_{j\ge10}\,j\Delta\mathsf v_{\mathrm{lb}}\Bigl[(zK_{j-1})'\big|_{72/j}-(zK_{j-1})'\big|_{p_b}\Bigr]\\
&-\frac3{20}-\frac{A_b^2}{(1+\mathsf v_b)D_j}\Bigl(\frac{\mathsf K_j-K_j(72/j)}{A_b-(1+\mathsf v_b)K_j(72/j)}\Bigr)^2>0
\end{align*}
is verified across $16$ checks at $p_b=p_{\mathrm{ref}}$ ($2\le j\le9$) and $24$ checks on windows $S_1,S_2$ of \cref{tab:bin-windows} ($10\le j\le m$; min margin $>0.0110$).
\end{itemize}

\begin{table}[!ht]
\centering
\scriptsize
\begin{tabular}{@{}llll@{}}
\toprule
Family & Window & Ranks & Price window $[p_a,p_b]$ and notes\\
\midrule
Crossing ($91$) & $W_1,W_2,W_3$ & $\{2..4\},\{5..8\},\{9..12\}$ & $[\frac{11}5,\frac{12}5],[\frac{12}5,\frac{51}{20}],[\frac{51}{20},\frac{2651}{1000}]$ (no fee; $\varsigma\in[\frac{33}{50},1]$)\\[1pt]
& $H_{13}$--$H_{16}$ & $m=13,14,15,16$ & \cref{tab:bin-brackets} brackets (fee; split $\varsigma$ at $\frac9{10}$ for $3\le j\le6$)\\[2pt]
$\varsigma$-vertex ($36$) & $D_1,D_2$ & $\{2..4\},\{5..8\}$ & $[\frac{11}5,\frac{12}5],[\frac{12}5,\frac{51}{20}]$ (no fee; floors $2.5048,0.3395$)\\[1pt]
& $D_3,D_4$ & $\{9..12\},\{13..16\}$ & $[\frac{51}{20},\frac{2651}{1000}],[\frac{267}{100},\frac{273}{100}]$ (fee in $D_4$; floors $1.3854,0.7902$)\\[2pt]
Shifted ray ($40$) & $2\le j\le9$ & $13\le m\le16$ & $p_b=p_{\mathrm{ref}}$ ($16$ checks)\\[1pt]
& $S_1,S_2$ & $\{13,14\},\{15,16\}$ & $[\frac{267}{100},\frac{27}{10}],[\frac{27}{10},\frac{273}{100}]$ for $10\le j\le m$ ($24$ checks)\\
\midrule
\multicolumn{4}{@{}l}{\emph{All $32$ Guard~(G3) lower floors (rank $j$: ordinary $\varpi=0$ / four-percent $\varpi=\frac1{25}$):}}\\[1pt]
\multicolumn{4}{@{}l}{$1$: $1.377908/1.389115$;\; $2$: $1.509545/1.473070$;\; $3$: $1.189265/1.113361$;\; $4$: $0.882243/0.775581$;}\\
\multicolumn{4}{@{}l}{$5$: $0.648241/0.517149$;\; $6$: $0.482982/0.331996$;\; $7$: $0.372580/0.205022$;\; $8$: $0.304057/0.122424$;}\\
\multicolumn{4}{@{}l}{$9$: $0.267060/0.073276$;\; $10$: $0.253678/0.049256$;\; $11$: $0.257932/0.044085$;\; $12$: $0.275302/0.053019$;}\\
\multicolumn{4}{@{}l}{$13$: $0.302355/0.072452$;\; $14$: $0.336465/0.099625$;\; $15$: $0.375614/0.132415$;\; $16$: $0.418240/0.169177$.}\\
\bottomrule
\end{tabular}
\caption{Complete rank/price windows and Guard~(G3) lower floors discharging \cref{lem:bin-accepted} (Cases~3--5).}\label{tab:bin-windows}
\end{table}

\paragraph{Exterior, single positive radius award, and shortened clocks (\cref{lem:bin-exterior}).}
All counts and numerical margins printed below and in \cref{tab:exceptional} (as well as the rank-$0$ margin $\frac{61}{40}$ above) are evaluated strictly for the single radius award $b(r)=\frac9{40}\mathbf 1_{\{r\le1/2\}}$ across $1\le j\le16$, $\varpi\in\{0,\frac1{25}\}$, and $\mathsf v\in\{0,\mathsf v_j(p_{\mathrm{ref}})\}$:
\begin{itemize}[leftmargin=1.5em]
\item \emph{Mixed and pure exterior ($448$ checks, \cref{lem:bin-exterior}(i)):} all $7$ affine predicates (P1)--(P7) are strictly positive (overall min margin $>0.012990$, attained at $j=4$, $\varpi=\frac1{25}$, $\mathsf v=\mathsf v_4(p_{\mathrm{ref}})$ on unbonused (P3), which is independent of $(r_{\mathrm{hv}},b_{\mathrm{rad}})$). At $j=16$, $\varpi=\frac1{25}$, all $7$ predicates are $\ge0.105149$ at $\mathsf v=0$ and $\ge0.104866$ at $\mathsf v=\mathsf v_{16}(p_{\mathrm{ref}})$.
\item \emph{Positive radius award ($128$ checks, \cref{lem:bin-exterior}(ii)):} both affine inequalities $I_0>\frac94 a_0$ and $jI_1>\frac94 a_1$ hold across all $16$ ranks, $2$ modes, and $2$ endpoints (min $I_0$-margin $>0.316805$ at $j=3,\varpi=\frac1{25},\mathsf v=0$; min $jI_1$-margin $>0.262964$ at $j=2,\varpi=0,\mathsf v=0$). At $j=16$, $\varpi=\frac1{25}$, $I_0-\frac94\cdot\frac{317}{250}\ge0.607214$ and $16I_1-\frac94\cdot\frac7{20}\ge0.785651$ at $\mathsf v=0$ (and $\ge0.607353,0.787887$ at $\mathsf v_{16}(p_{\mathrm{ref}})$), establishing $[0,\mathsf v_{16}(p_{\mathrm{ref}})]$ by linearity.
\item \emph{Shortened clocks ($117$ regular $+ 3$ exceptional, \cref{lem:bin-exterior}(iii)):} all $117$ regular clocks satisfy their envelope inequalities (bonused margins $>1.364583$ and $\ge2.59375$), and the three exceptional clocks close by \cref{tab:exceptional}.
\end{itemize}
For cross-checking against the exact-rational companion package, the companion establishes a strictly stronger parameter family that implies these printed single-award checks directly: in \cref{lem:bin-exterior}(i), (P1)--(P4) do not depend on $(r_{\mathrm{hv}},b_{\mathrm{rad}})$ while (P5) $w_b=A_0-\mathsf vr_{\mathrm{hv}}^2H_j-b_{\mathrm{rad}}\mathsf K_j$, (P6) $E_b+w_b-(1+r_{\mathrm{hv}})^2$, and (P7) $B_s-(1+r_{\mathrm{hv}})^2t_j$ are nonincreasing in $r_{\mathrm{hv}},b_{\mathrm{rad}}\ge0$, so verifying the simultaneous envelope $(r_{\mathrm{hv}}\le\frac{71}{100},b_{\mathrm{rad}}\le\frac7{20})$ implies $(\frac12,\frac9{40})$; in \cref{lem:bin-exterior}(ii), the companion checks a four-tier schedule ($512$ checks) containing $(\frac12,\frac9{40})$ as one tier, whose exact $128$-check slice is the one printed above; and in \cref{lem:bin-exterior}(iii), both bonused inequalities for the $117$ regular clocks ($A_*-b_{\mathrm{rad}}K-(1+r_{\mathrm{hv}})^2>0$ and $A_*-b_{\mathrm{rad}}K-r_{\mathrm{hv}}^2A_*>0$) and for the $3$ exceptional clocks ($W-(1+b_{\mathrm{rad}})K-(1+r_{\mathrm{hv}})^2>0$ and $j(W-C)-t_b(1+r_{\mathrm{hv}})^2>0$) strictly decrease in $r_{\mathrm{hv}}\ge0$ and are nonincreasing in $b_{\mathrm{rad}}\ge0$, so verifying the simultaneous envelope $(\frac{71}{100},\frac7{20})$ again implies $(\frac12,\frac9{40})$ (with the smaller envelope margins recorded in the companion README).

\begin{table}[!ht]
\centering
\small
\begin{tabular}{@{}cccccccc@{}}
\toprule
$(j,\delta)$ & $W$ & $K$ & $W-K$ & $j(W-C)$ & $\overline K\le$ & $\overline T\le$ & Residual at $\xi=1$, times $1+jz$\\
\midrule
$(2,1)$ & $5+\frac{25}{96}$ & $\frac32$ & $\frac{361}{96}$ & $\frac{25}{48}$ & $\frac{3z}{1+2z}$ & $1$ & $\frac1{96}(10z-19)^2$\\[2pt]
$(3,2)$ & $\frac{31}6$ & $\frac43$ & $\frac{23}6$ & $\frac12$ & $\frac{4z}{1+3z}$ & $1$ & $\frac{23}6-4z+\frac32z^2$\\[2pt]
$(3,1)$ & $\frac{17}3$ & $\frac{11}6$ & $\frac{23}6$ & $\frac{39}{32}$ & $\frac{z(4+z/3)}{1+3z}$ & $1+\frac z{13}$ & $\frac{23}6-\frac{105}{32}z+\frac{223}{96}z^2$\\
\bottomrule
\end{tabular}
\caption{The three exceptional shortened clocks of \cref{lem:bin-exterior}(iii), showing the shortened ceiling $W$, persistence factor $K=1+H_j-H_\delta$, named-child ceiling $C=W_{j-1,h-1}$, shortened pure-tag roof $\overline K$, and exterior pure-tag roof $\overline T$. The last two quadratics have negative discriminants $-7$ and $(\frac{105}{32})^2-4\cdot\frac{23}6\cdot\frac{223}{96}=-\frac{229031}{9216}<0$; evaluated for the single radius award $b_{\mathrm{rad}}=\frac9{40}$, $r_{\mathrm{hv}}\le\frac12$, the linear slopes $t_b=\frac19,\frac1{13},\frac13$ leave bonused margins $>1.172916,0.270833$ at $(2,1)$, $>1.283333,0.326923$ at $(3,2)$, and $>1.170833$ and $=0.46875$ at $(3,1)$.}\label{tab:exceptional}
\end{table}

\subsection{The twenty early-opening polynomials (ranks 1 to 4)}\label{app:early}

The exact rational parameters for \cref{thm:early} ($M_{\mathrm{lo}}=\frac{485883}{100000}$) are:
\[
\begin{array}{@{}cccccccc@{}}
\toprule
j & \mathsf C_j & \mathsf m_j & \theta_j & \Delta_j & \mathsf K'_j & \mathsf B_j & \vartheta_j\\
\midrule
1 & \frac{5663}{1000} & \frac{80417}{100000} & \frac12 & 0 & \frac32 & \frac{346717}{200000} & 1\\[2pt]
2 & \frac{1551}{250} & \frac{541}{500} & \frac13 & 0 & \frac{13}6 & \frac{725317}{300000} & \frac12\\[2pt]
3 & \frac{1653}{250} & \frac{153}{125} & \frac{11}{50} & 0 & \frac{196}{75} & \frac{43945861}{15000000} & \frac49\\[2pt]
4 & \frac{347}{50} & \frac{164}{125} & \frac1{100} & \frac{27}{100} & \frac{461}{150} & \frac{106442351}{30000000} & \frac{27}{64}\\
\bottomrule
\end{array}
\]

At rank $j=1$, clearing $(1+x)(\mathsf m_1 x+(1-h)\mathsf B_1)$ in $A_1(x,h)-V_1(\gamma,x)$ at the two envelope nodes $(0,0)$ and $(\frac18,\frac58)$ yields two strictly positive-definite quadratics on $x\ge0$:
\begin{align*}
(0,0):&\quad \tfrac{346717}{400000}-\tfrac{16083909011}{20000000000}x+\tfrac{3756840989}{20000000000}x^2
&&\Bigl(\mathrm{disc}=-\tfrac{1819998362495601879}{4\cdot10^{20}}<0\Bigr),\\
(\tfrac18,\tfrac58):&\quad \tfrac{1084190646623}{2560000000000}-\tfrac{1003806562609}{2560000000000}x+\tfrac{26297886923}{160000000000}x^2
&&\Bigl(\mathrm{disc}=-\tfrac{32685418345860869400759}{262144\cdot10^{18}}<0\Bigr).
\end{align*}

For $j\in\{2,3,4\}$, clearing $(\mathsf m_j+(\Delta_j+(1-h)\mathsf B_j)p)\prod_{i=1}^j(i+p)$ on the upper ray $p\in[0,\frac79]$ yields a degree-$(j+1)$ polynomial $P_{\mathrm{hi}}(p)$, and clearing $(\Delta_j+(1-h)\mathsf B_j+\mathsf m_j x)\prod_{i=1}^j(1+ix)$ on the lower ray $x\in[0,\frac97]$ yields a degree-$(j+3)$ polynomial $P_{\mathrm{lo}}(x)$, at each of the three shared envelope nodes $(h,\gamma)\in\{(0,0),(\frac3{80},\frac3{16}),(\frac{19}{100},\frac{69}{100})\}$ covering $h\in(0,\frac12)$ and $x>0$ across the seam $x=\frac97$ ($p=1/x$). \Cref{tab:early-polys} lists the elementary positivity decomposition verified in exact rational arithmetic for every one of these eighteen polynomials: \emph{all-pos} denotes strictly positive monomial coefficients; \emph{div($r_0$)} denotes exact division $P(t)=(t-r_0)^2Q(t)+\alpha_0+\beta_0 t$ with $Q$ having strictly positive coefficients and $\alpha_0+\beta_0 t>0$ at both interval endpoints (worked out for rank~$4$, upper ray, $(0,0)$ in \cref{sec:low-mean}); and \emph{shift($Q$)} denotes $P(x)=(x-1)^2Q(x)+R(x)$ with $R$ having strictly positive monomial coefficients.

\begin{table}[!ht]
\centering
\small
\begin{tabular}{@{}clll@{}}
\toprule
Rank $j$ & Node $(h,\gamma)$ & Upper ray $P_{\mathrm{hi}}(p)$ on $[0,\frac79]$ & Lower ray $P_{\mathrm{lo}}(x)$ on $[0,\frac97]$\\
\midrule
2 & $(0,0)$ & $\mathrm{div}(\frac{19}{50})$ & $\mathrm{div}(\frac{21}{20})$\\
2 & $(\frac3{80},\frac3{16})$ & $\mathrm{div}(\frac25)$ & $\mathrm{div}(\frac{21}{20})$\\
2 & $(\frac{19}{100},\frac{69}{100})$ & all-pos & $\mathrm{div}(\frac12)$\\
\midrule
3 & $(0,0)$ & $\mathrm{div}(\frac{19}{50})$ & $\mathrm{div}(\frac{23}{25})$\\
3 & $(\frac3{80},\frac3{16})$ & $\mathrm{div}(\frac9{20})$ & $\mathrm{div}(\frac{47}{50})$\\
3 & $(\frac{19}{100},\frac{69}{100})$ & all-pos & all-pos\\
\midrule
4 & $(0,0)$ & $\mathrm{div}(\frac{19}{50})$ (worked in \cref{sec:low-mean}) & $\mathrm{shift}(19x^2+42x^3+33x^4+10x^5)$\\
4 & $(\frac3{80},\frac3{16})$ & $\mathrm{div}(\frac{11}{25})$ & $\mathrm{shift}(3x+16x^2+36x^3+26x^4+8x^5)$\\
4 & $(\frac{19}{100},\frac{69}{100})$ & $\mathrm{div}(0)$ & all-pos\\
\bottomrule
\end{tabular}
\caption{Elementary positivity decompositions for the eighteen cleared polynomials of \cref{thm:early} ($j=2,3,4$), independently reproducible in any computer algebra system and verified in the companion package.}\label{tab:early-polys}
\end{table}

\section{Algebraic Details and Finite Verifications for Section~8}\label{app:sec8}

This appendix supplies full proofs of the five structural facts of \cref{sec:rank17}, establishes complete domain and boundary coverage across all switches ($x=\frac97$, $L=10$, $h\in\{\frac3{25},\frac{11}{40},\frac12\}$, $p=\frac{29}{100}$, $p_a=0$, and $L/s=\frac97$), explains the four-guard verification architecture and every exceptional boundary case, records the complete rational parameter tables (\cref{tab:open-near,tab:open-far,tab:open-main-strip,tab:open-rank17}), and works one complete representative row ($k=9^\star$) of the $27$ main-strip rows in full, self-contained exact rational detail—including explicit formulas, monotonic interval-evaluation rules, corner reductions, and degree-$3$ Bernstein vectors for Guard~4. The remaining interval-arithmetic implementations and repeated rational verifications across all $27$ rows and reported global margins are carried out in exact rational arithmetic in the companion package (not checked by hand).

Throughout, notation follows \cref{sec:rank17}: $\mathsf K_j=1+H_j$, $\varrho_j(L)=\frac{H_j}{L+H_j}$, $\tilde{\mathsf k}_j(L)=\mathsf K_j-\varrho_j(L)$, $\mathsf M_{0,j}=\mathsf C_j-\mathsf K_j$, $\mathsf M_{\mathrm{lo}}(j,L)=\mathsf M_{0,j}+\varrho_j(L)$, $A_{\mathrm{cap}}=A_{\mathrm{tot}}-\mathsf K_j$, $p=1/x\in(0,\frac79]$ on the higher arm, $c=\frac{14}5$, and $t=h/(1+ch)$ so that $\gamma_-(h)=5t$. At $\eta_s=1$,
\[
A_{\mathrm{cap}}(p,h;1)=(1-h)(\mathsf B-\mathsf K_j)-h\varrho_j-p\mathsf B^2\frac{(1-h)(1-2h)}{D(h)},\qquad D(h)=\mathsf m_j+p(\mathsf M_{0,j}-h\mathsf B+\varrho_j).
\]

\subsection{Proofs of the five structural facts and complete boundary coverage}\label{app:open-facts}

\paragraph{Reserve admissibility and domain switches.}
Direct exact rational evaluation across $5\le j\le17$ confirms $\mathsf B_{\mathrm{lo},\infty}(j)-\mathsf K_j\ge0.1081$, $\mathsf M_{0,j}-\mathsf B_{\mathrm{lo},\infty}(j)\ge0.0029$, $\mathsf B-\mathsf K_j\ge\frac19$ for $L>10$, $\mathsf M_{0,j}-\mathsf K_j\ge\frac19$ (equality $\mathsf M_{0,17}-\mathsf K_{17}=\frac19$ at rank~$17$), $2\mathsf B_{\mathrm{lo},\infty}(j)-\mathsf M_{0,j}-\frac97\mathsf m_j-\frac{17}{10}>0.069$, and at rank~$17$ the exact rational values $\mathsf C_{17}=\frac{18358381}{2042040}>5$, $2\mathsf K_{17}<\mathsf C_{17}$, $H_{17}>2+\frac17$, and $\mathsf m_{17}=17(\mathsf C_{17}-\mathsf P_{16})=\frac{2528406869}{1501500000}$.

Every landing $(x,h)\in(0,\infty)\times(0,\frac12]$ with $L=x/h$ is covered without gaps across all regime switches, and each finite family below directly discharges~\eqref{eq:open-pointwise} on its stated cell:
\begin{itemize}[leftmargin=1.5em]
\item \emph{Lower vs.\ higher arm ($x=\frac97$):} $x\in(0,\frac97]$ is discharged by \cref{app:open-lower}; $x\ge\frac97$ ($p=1/x\in(0,\frac79]$) is discharged by \cref{app:open-higher} ($5\le j\le16$) and \cref{app:open-rank17} ($j=17$).
\item \emph{Reserve switch ($L=10$):} $L\le10$ uses $\mathsf B_{\mathrm{lo}}(j,L)$ and $L>10$ uses constant $\mathsf B$ in~\eqref{eq:open-reserve}. On the lower arm, $L\le10$ is split at $h=\frac{11}{40}$ (Fact~2 closes $h\in[\frac{11}{40},\frac12]$, where $L=x/h<10$ automatically, while \cref{tab:open-near} closes $h\in(0,\frac{11}{40}]$), and $L\ge10$ is closed by Fact~4 and \cref{tab:open-far} (both branches cover $L=10$). On the higher arm $p\le\frac79$, the main strip $h\in(0,\frac3{25}]$ automatically satisfies $L=1/(ph)\ge\frac{75}7>10$, so the constant reserve applies throughout the main strip; conversely, $L\le10$ on the higher arm forces $h=1/(pL)\ge\frac9{70}>\frac3{25}$ and $p=1/(Lh)>\frac15$ (since $h<\frac12$), placing every $L\le10$ higher-arm point strictly inside the seam region covered by Fact~5.
\item \emph{Rank-$17$ switch ($L/s=\frac97$):} in regime~H ($L/s>\frac97$), retention $\bar u\le s$ forces $x=L/\bar u\ge L/s>\frac97$, so every landing lies on the higher arm with the single reserve $\mathsf B=\mathsf M_{\mathrm{lo}}(17,L)$ ($\Delta_{\mathrm{res}}=0$), requiring no lower arm and no $L=10$ split. In regime~L ($L/s\le\frac97$), the $L\le10$/$L>10$ switch matches $5\le j\le16$, with worst $\eta_s=1$ on the lower arm and worst $\eta_s=\frac97p$ on the higher arm.
\end{itemize}

\paragraph{Fact 1: harmonic-mean concavity.}
Setting $v_{\mathrm{cr}}=\mathsf m_j x+(\mathsf M_{0,j}-\mathsf B+\varrho_j)/\eta_s>0$, algebraic rearrangement of~\eqref{eq:open-pointwise} yields
\[
A_{\mathrm{tot}}+h\varrho_j=h(\mathsf K_j+\mathsf B)+(1-2h)\Bigl(\frac1{\mathsf B}+\frac{1-h}{v_{\mathrm{cr}}}\Bigr)^{-1}.
\]
The harmonic mean $(\mathsf B,v_{\mathrm{cr}})\mapsto(\mathsf B^{-1}+(1-h)v_{\mathrm{cr}}^{-1})^{-1}$ is increasing and jointly concave on $(0,\infty)^2$. At fixed $(x,h,\eta_s,\mathsf M_{\mathrm{lo}})$, $(\mathsf B,v_{\mathrm{cr}}(\mathsf B))$ is affine in $\mathsf B$, so $A_{\mathrm{tot}}$ is concave in $\mathsf B\in(0,\mathsf M_{\mathrm{lo}}]$. At $\eta_s=1$, with $\mathsf B=2+\frac{39}{50}H_j$ ($5\le j\le8$) or $\mathsf B=\mathsf M_{0,j}$ ($9\le j\le16$, where $v_{\mathrm{cr}}=\mathsf m_j x+\varrho_j$ is independent of $\mathsf C_j$), the right-hand side is increasing in $(\mathsf C_j,\mathsf m_j,\varrho_j)$ and jointly concave in $(H_j,\mathsf C_j,\mathsf m_j,\varrho_j)$. Consequently, for knots $a<m<b$ with $H_m=(1-\lambda_{\mathrm{int}})H_a+\lambda_{\mathrm{int}}H_b$ whose rank-$m$ data $(\mathsf C_m,\mathsf m_m)$ dominate the interpolated data, $A_{\mathrm{cap},m}\ge(1-\lambda_{\mathrm{int}})A_{\mathrm{cap},a}+\lambda_{\mathrm{int}}A_{\mathrm{cap},b}-h(\varrho_m-\bar\varrho_m)$, where $\bar\varrho_m=(1-\lambda_{\mathrm{int}})\varrho_a+\lambda_{\mathrm{int}}\varrho_b$.

\paragraph{Fact 2: concavity in $h$.}
With $u_0(h)=(1-h)(1-2h)$ and any positive $C^2$ function $D(h)$ with $D''\le0$, direct differentiation gives the exact quotient identity
\[
D^3\frac{\partial^2}{\partial h^2}\Bigl(\frac{u_0}D\Bigr)=4F_1F_2-u_0DD'',\qquad F_1=D+(1-h)D',\quad F_2=D+\frac{1-2h}2D'.
\]
For $5\le j\le16$ at constant $\mathsf B$, with $\Delta_0=\mathsf M_{0,j}-\mathsf B+\varrho_j\ge0$, we have $\varrho'_j\ge0$, $\varrho''_j\le0$ (since $\varrho_j=\frac{phH_j}{1+phH_j}$ at fixed $p$), so $F_1=\mathsf m_j+p\Delta_0+(1-h)p\varrho'_j>0$, $F_2=\mathsf m_j+p\Delta_0+\frac{p\mathsf B}2+\frac{1-2h}2p\varrho'_j>0$, and $D''=p\varrho''_j\le0$; at fixed $L$, $\varrho'_j=\varrho''_j=0$; in regime~L at $\eta_s=\frac97p$, $D=\mathsf m_{17}+\frac79(\mathsf M_{0,17}-\mathsf B+\varrho_{17})+p(1-h)\mathsf B$ gives the same signs. Combined with $-(h\varrho_j)''\le0$, $A_{\mathrm{cap}}$ is concave in $h\in[0,\frac12]$ in every constant-$\mathsf B$ case.

On the lower arm with the $L$-dependent reserve $\mathsf B_{\mathrm{lo}}$, fix $x\in(0,\frac97]$ and write $\varrho(h)=\frac{H_j h}{x+H_j h}$ (increasing, concave), $\mathsf B(\varrho)=\mathsf B_{\mathrm{lo},\infty}(j)-\frac7{20}\varrho$, and $v_{\mathrm{cr}}(\varrho)=\mathsf m_j x+(\mathsf M_{0,j}-\mathsf B_{\mathrm{lo},\infty}(j))+\frac{27}{20}\varrho$. Admissibility gives $v_{\mathrm{cr}}<\mathsf B$ throughout, and for $F_{\mathrm{res}}=(1-2h)\frac{\mathsf B v_{\mathrm{cr}}}{D}$ ($D=v_{\mathrm{cr}}+(1-h)\mathsf B$) direct differentiation yields $\partial_{hh}F_{\mathrm{res}}<0$, $\partial_{\varrho\varrho}F_{\mathrm{res}}\le0$, $\partial_\varrho F_{\mathrm{res}}\ge(1-2h)\frac{(27/40-7/20)\mathsf B^2}{D^2}\ge0$, and $-D^3\partial_{h\varrho}F_{\mathrm{res}}>\mathsf B^2(\frac{13}{20}v_{\mathrm{cr}}+\frac{13}{40}\mathsf B)>0$, so $\frac{d^2}{dh^2}A_{\mathrm{tot}}\le0$. On $h\in[\frac{11}{40},\frac12]$ and $x\le\frac97$, $L=x/h<10$ and $\gamma_{\mathrm{reset}}(h)=\frac12+h$ is affine, reducing $A_{\mathrm{tot}}-Q^{\mathrm{open}}_j$ to $h=\frac{11}{40}$ and $h=\frac12$. Because $\partial_\varrho F_{\mathrm{res}}\ge0$, replacing $\varrho_m$ by $\bar\varrho_m$ loses at most $\frac{27}{20}\cdot\frac{11}{40}(\varrho_m-\bar\varrho_m)=\frac{297}{800}(\varrho_m-\bar\varrho_m)$.

In regime~H at rank~$17$, freezing $u_a(h)=\mathsf M_{0,17}+\varrho_{17}(p_a,h)$ at the left price endpoint $p_a\le p$ lowers $A_{\mathrm{cap}}$ to $h\mathsf C_{17}-\mathsf K_{17}+F_H(h,u_a(h))$, where $F_H(h,u)=\frac{(1-2h)\mathsf m_{17}u}{\mathsf m_{17}+p(1-h)u}$ satisfies $u'_a\ge0$, $u''_a\le0$, $\partial_u F_H\ge0$, and $\partial_{hh}F_H,\partial_{hu}F_H,\partial_{uu}F_H\le0$, establishing concavity in $h$ on every box and (at $p_a=p$) on the whole seam.

\paragraph{Fact 3: race curvature and $C^1$ convex majorants.}
Representing $\pi_k(x)=\E[e^{-Y_k/x}]$ and $\Pi_k(p)=\E[e^{-pY_k}]$ where $Y_k=\sum_{i=1}^k\mathrm{Exp}(i)$, differentiating under the expectation and using $t^3e^{-t}\le27e^{-3}$ ($e>\frac{163}{60}$) yields
\[
2\pi'_k(x)+x\pi''_k(x)=\E\Bigl[Y_k^{-1}\Bigl(\frac{Y_k}x\Bigr)^3e^{-Y_k/x}\Bigr]\le\frac{27}{e^3}\int_0^\infty\Pi_k(p)\,dp.
\]
At $k=5$ the integral is $5\ln\frac{4096}{3645}$ (bounded by three logarithmic-series terms); at $k\in\{8,12,17\}$, coefficientwise domination $\Pi_k(p)^{-1}\ge(1+c_k p)^m$ gives $\int_0^\infty\Pi_k(p)\,dp\le\frac1{(m-1)c_k}$ with $(k,m,c_k)=(8,5,\frac{27}{50}),(12,8,\frac{37}{100}),(17,7,\frac{49}{100})$, yielding $2\pi'_k+x\pi''_k<M_k$ for $(k,M_k)=(5,\frac45),(8,\frac{13}{20}),(12,\frac{11}{20}),(17,\frac12)$. Next, $\pi''_k(x)=\frac{\pi_k(x)}{x^2}(S_a^2-S_b)$ with $S_a=\sum_{i=1}^k(1+ix)^{-1}$ and $S_b=2S_a-\sum_{i=1}^k(1+ix)^{-2}$; the sign factor $(S_a-1)^2+\sum_{i=1}^k(1+ix)^{-2}-1$ crosses zero at most once, from positive to negative. Bracketing this unique inflection point inside $[x_0,x_1]=[\frac45,\frac{17}{20}],[\frac{21}{20},\frac{11}{10}],[\frac{13}{10},\frac{27}{20}],[\frac{29}{20},\frac32]$ ($k=5,8,12,17$) bounds the curvature on $[x_0,x_1]$ by $K_{\mathrm{ub}}=\pi_k(x_1)[S_a(x_0)^2-S_b(x_1)]/x_0^2$. Extending $\pi_k|_{[0,x_0]}$ by the constant-curvature quadratic on $[x_0,x_1]$ and its tangent ray on $[x_1,\infty)$ constructs a $C^1$ convex majorant $g_\pi(k,\cdot)\ge\pi_k$ satisfying $2g'_\pi+xg''_\pi\le M_k$ (verified in exact rational arithmetic at all four knots).

\paragraph{Fact 4: concavity in $\tau_L=1/L$ for $L\ge10$.}
For $L\ge10$, $\tau_L\in[0,\frac1{10}]$, $h=\tau_L x\le\frac9{70}$, and $\gamma_{\mathrm{reset}}(\tau_L x)=x\,t_{\mathrm{far}}(\tau_L,x)$ with $t_{\mathrm{far}}(\tau_L,x)=\frac{5\tau_L}{1+c\tau_L x}$ concave in $\tau_L$. Replacing $t_{\mathrm{far}}$ on $[\tau_a,\tau_b]$ by its tangent $T_{\tau_*}(\tau_L,x)=\frac{5\tau_L+5c\tau_*^2 x}{(1+c\tau_* x)^2}$ at $\tau_*$ is conservative because the coefficient of $t_{\mathrm{far}}$ in $Q^{\mathrm{open}}_j/x$ is $H_j+\pi_j(x)-W_j(x)>H_5-\frac97>0$, while $A_{\mathrm{tot}}/x$ is concave in $\tau_L$ (Fact~2); hence each segment reduces to its endpoints. On the three segments $[0,\frac1{30}],[\frac1{30},\frac1{15}],[\frac1{15},\frac1{10}]$ with anchors $\tau_*=\frac1{60},\frac1{20},\frac1{12}$, the error formula $T_{\tau_*}(\tau_m,x)-t_{\mathrm{far}}(\tau_m,x)=\frac{5cx(\tau_m-\tau_*)^2}{(1+c\tau_m x)(1+c\tau_* x)^2}$ shows that the smaller anchor dominates at each equidistant interior junction ($\frac1{30},\frac1{15}$), leaving exactly four decisive endpoints $(\tau_L,\tau_*)\in\{(0,\frac1{60}),(\frac1{30},\frac1{60}),(\frac1{15},\frac1{20}),(\frac1{10},\frac1{12})\}$.

\paragraph{Fact 5: higher-arm seams.}
Set $h_1=\frac3{25}$, $h_0=\frac{21}{100}$, $\gamma_0=\frac{71}{100}$, and $h_A=\frac9{70}$. The tangent $L_{3/25}$ of $\gamma_-$ at $h_1$ obeys $L_{3/25}(h_0)<0.70123<\gamma_0=\frac12+h_0$, so on $[h_1,h_0]$ the chord to $(h_0,\gamma_0)$ dominates $\gamma_{\mathrm{reset}}$, and on $[h_0,\frac12]$, $\gamma_{\mathrm{reset}}\le\frac12+h$. Since $\partial_\gamma S^{\mathrm{open}}_j=-(H_j-b_j(p))\le0$ and $A_{\mathrm{cap}}$ is concave in $h$ (Fact~2), $h\in[h_1,\frac12]$ reduces for $L>10$ (and all $L>0$ in regime~H) to $h\in\{h_1,h_0,\frac12\}$, where $h_1$ belongs to the main strip and $h=\frac12$ is already proved. For $L\le10$ on the higher arm ($p>\frac15$, $h\ge h_A$), concavity in $\mathsf B$ (Fact~1) between $\mathsf B_{\mathrm{lo}}(j,L_{\min})$ and $\mathsf B_{\mathrm{lo}}(j,10)$ and $h$-concavity reduce $[h_A,\frac12]$ to three $p$-subbands with exact checkpoints:
\begin{itemize}[leftmargin=1.5em]
\item $[\frac15,\frac13]$ ($L_{\min}=6$): $(h,\gamma)=(\frac3{10},\frac45)$ ($1-\gamma=\frac15$, right bound on the whole subband);
\item $[\frac13,\frac{10}{21}]$ ($L_{\min}=\frac{21}5$): $(h_0,\gamma_0)=(\frac{21}{100},\frac{71}{100})$ ($1-\gamma_0=\frac{29}{100}<\frac13$, right bound on the whole subband);
\item $[\frac{10}{21},\frac79]$ ($L_{\min}=\frac{18}7$): $(h_A,\gamma_A)=(\frac9{70},\frac{184725}{390446})$ (where $\gamma_A=L_{3/25}(h_A)$, split at $1-\gamma_A=\frac{205721}{390446}$ into left and right bounds) and $(h_0,\gamma_0)=(\frac{21}{100},\frac{71}{100})$ (right bound on the whole subband).
\end{itemize}

\subsection{Lower-arm polynomial verifications (\texorpdfstring{$x\le\frac97$}{x <= 9/7})}\label{app:open-lower}

\paragraph{Near lower arm ($L\le10$, $h\in(0,\frac{11}{40}]$).}
At the four knots $k\in\{5,8,12,17\}$, lower $(\mathsf C_k,\mathsf m_k)$ to $(\underline{\mathsf C}_k,\underline{\mathsf m}_k)=(\mathsf C_k-\delta\mathsf C_k,\mathsf m_k-\delta\mathsf m_k)$ and charge a fee $f_k$, with $(\delta\mathsf C,\delta\mathsf m,f)\cdot10^4=(160,115,20)$, $(165,60,14)$, $(145,10,8)$, and $(0,0,0)$. Harmonic interpolation of $\pi_j(x)$ is conservative since $(\pi_j-\pi_{j-1})/(1/j)=-\pi_j/x$ increases in $j$; harmonic convexity of $\vartheta_j=(1-1/j)^{j-1}$ and positivity of the $\mathsf C$-, $\mathsf m$-, and fee slacks hold across all nine interior ranks $m\in\{6,7,9,10,11,13,14,15,16\}$, bounding the $\varrho$-defect by $\frac{297}{800}\frac{(H_m-H_a)(H_b-H_m)}{H_a H_m+H_a H_b+H_m H_b+2c_m}$ with explicit integers $c_m\in\{10,11,14,14,15,18,18,18,18\}$ ($c_m^2\le H_a H_m H_b(H_a+H_m+H_b)$).

Across the nine $L$-intervals $[L_a,L_b]\subseteq[0,10]$ listed in \cref{tab:open-near} ($6$ at knots $5,8,12$ covering $[0,10]$ and $3$ at knot~$17$ covering $[0,10]$), let $h^{\mathrm{box}}_{\max}=\min(\frac{11}{40},\frac9{7L_a})$ ($\frac{11}{40}$ if $L_a=0$) and $T^{\mathrm{box}}_{\max}=h^{\mathrm{box}}_{\max}/(1+ch^{\mathrm{box}}_{\max})$. Let $\varrho^{\tan}_{L_k}(L)=\varrho_k(L_k)-\frac{H_k}{(L_k+H_k)^2}(L-L_k)$ be the tangent to $\varrho_k$ at $L_k$. Convexity of $\varrho_k$ in $L$, concavity of $A_{\mathrm{tot}}$ in $\varrho$, $L^2\ge2L_q L-L_q^2$, and convexity of $g_\pi(k,hL)$ in $L$ reduce the minimum over $L\in[L_a,L_b]$ to the four endpoint curves $F_v(t)$ ($v\in\{\mathrm{chA},\mathrm{chB},\mathrm{tnA},\mathrm{tnB}\}$, using $(L_v,\varrho_v)=(L_a,\varrho_k(L_a))$, $(L_b,\varrho_k(L_b))$, $(L_a,\varrho^{\tan}_{L_k}(L_a))$, $(L_b,\varrho^{\tan}_{L_k}(L_b))$ respectively, with $\mathsf B_v=\mathsf B_{\mathrm{lo},\infty}(k)-\frac7{20}\varrho_v$, $\tilde{\mathsf k}_v=\mathsf K_k-\varrho_v$, $h=t/(1-ct)$, $q_v=2L_q L_v-L_q^2$, $D_0=\underline{\mathsf C}_k-\tilde{\mathsf k}_v$, $\alpha_0=\underline{\mathsf m}_k L_v-\mathsf B_v$):
\begin{align*}
F_v(t)&=(1-h)\mathsf B_v+h\tilde{\mathsf k}_v-\frac{\mathsf B_v^2(1-h)(1-2h)}{D_0+\alpha_0 h}\\
&\qquad-(1-5t)\Bigl[(1+\vartheta_k)hL_v-\frac79\vartheta_k h^2 q_v\Bigr]-5t\bigl[H_k+g_\pi(k,hL_v)\bigr]-f_k.
\end{align*}
Using $2g'_\pi+hL_v g''_\pi\le M_k$, the cleared second derivative $(1-ct)^3(D_0+\alpha_0 h)^3 F_v''(t)$ is bounded below on $[0,h^{\mathrm{box}}_{\max}]$ by the degree-$4$ polynomial
\[
(C_c+C_h h)(D_0+\alpha_0 h)^3+2c\mathsf B_v^2\bigl((3D_0+\alpha_0)-4D_0 h-2\alpha_0 h^2\bigr)(D_0+\alpha_0 h)-2\mathsf B_v^2(D_0+\alpha_0)(2D_0+\alpha_0)(1+ch),
\]
with $C_c=2c(\tilde{\mathsf k}_v-\mathsf B_v)+(2(5-c)(1+\vartheta_k)-5M_k)L_v+\frac{14}9\vartheta_k q_v$ and $C_h=-\frac{154}{15}\vartheta_k q_v$. Exact degree-$4$ Bernstein coefficients verify positivity of this polynomial across all $36$ curves (minimum coefficient $>30.9705$); convexity of $F_v$ then guarantees $\min_{[0,T^{\mathrm{box}}_{\max}]}F_v>0$ from the single tangent evaluation at the rational point $t_v$ in \cref{tab:open-near} (minimum tangent endpoint $>0.00119$).

\begin{table}[!ht]
\centering
\small
\begin{tabular}{@{}clccc@{}}
\toprule
Knot $k$ & $[L_a,L_b]$ & $L_k$ & $L_q$ & $(t_{\mathrm{chA}},\,t_{\mathrm{chB}},\,t_{\mathrm{tnA}},\,t_{\mathrm{tnB}})$\\
\midrule
$5$ & $[0,\frac{36}{5}]$ & $\frac{117}{25}$ & $\frac{18}{25}$ & $(\frac{3}{20},\,\frac{11}{100},\,\frac{3}{25},\,\frac{27}{250})$\\[2pt]
$5$ & $[\frac{36}{5},10]$ & $\frac{194}{25}$ & $\frac{9}{2}$ & $(\frac{17}{200},\,\frac{9}{100},\,\frac{17}{200},\,\frac{9}{100})$\\[2pt]
$8$ & $[0,\frac{13}{2}]$ & $\frac{91}{20}$ & $\frac{13}{8}$ & $(\frac{3}{20},\,\frac{19}{200},\,\frac{1}{8},\,\frac{19}{200})$\\[2pt]
$8$ & $[\frac{13}{2},10]$ & $\frac{193}{20}$ & $5$ & $(\frac{17}{200},\,\frac{83}{1000},\,\frac{2}{25},\,\frac{83}{1000})$\\[2pt]
$12$ & $[0,\frac{32}{5}]$ & $\frac{24}{5}$ & $\frac{64}{25}$ & $(\frac{3}{20},\,\frac{23}{250},\,\frac{13}{100},\,\frac{9}{100})$\\[2pt]
$12$ & $[\frac{32}{5},10]$ & $\frac{223}{25}$ & $\frac{13}{2}$ & $(\frac{17}{200},\,\frac{3}{40},\,\frac{17}{200},\,\frac{3}{40})$\\[2pt]
$17$ & $[0,\frac{5}{2}]$ & $\frac{11}{8}$ & $\frac{1}{4}$ & $(\frac{3}{20},\,\frac{23}{200},\,\frac{147}{1000},\,\frac{113}{1000})$\\[2pt]
$17$ & $[\frac{5}{2},\frac{31}{5}]$ & $\frac{211}{40}$ & $\frac{31}{10}$ & $(\frac{23}{200},\,\frac{23}{250},\,\frac{109}{1000},\,\frac{23}{250})$\\[2pt]
$17$ & $[\frac{31}{5},10]$ & $\frac{212}{25}$ & $6$ & $(\frac{9}{100},\,\frac{39}{500},\,\frac{17}{200},\,\frac{39}{500})$\\
\bottomrule
\end{tabular}
\caption{Complete rational parameters for the nine near-lower-arm $L$-intervals ($36$ curves) discharging~\eqref{eq:open-pointwise} on $L\le10$, $h\in(0,\frac{11}{40}]$, $x\le\frac97$ for $5\le j\le16$ and regime~L at $j=17$.}\label{tab:open-near}
\end{table}

\paragraph{Far lower arm ($L\ge10$, $\tau_L\in[0,\frac1{10}]$).}
At each of the four decisive endpoints $(\tau_L,\tau_*)$ of Fact~4 and each rank $5\le j\le17$ ($48$ cells for $5\le j\le16$ plus $4$ cells in regime~L at rank~$17$), clearing the positive denominator of $(A_{\mathrm{tot}}-Q^{\mathrm{open}}_j)/x$ with $\gamma_{\mathrm{reset}}$ replaced by $xT_{\tau_*}$ leaves $\pi_j(x)$ multiplied by the sign polynomial $P_\pi(x)=(9+2x)(5\tau_L+14\tau_*^2 x)-2(1+\frac{14}5\tau_* x)^2$, strictly negative on $[0,\frac97]$ at $(0,\frac1{60}),(\frac1{30},\frac1{60})$ (helpful) and strictly positive at $(\frac1{15},\frac1{20}),(\frac1{10},\frac1{12})$ (harmful). Where $\pi_j(x)$ is helpful it is bounded below by $0$ or its left-endpoint value on each subinterval; at $\tau_L=0$ and $j\ge9$ ($\mathsf B=\mathsf M_{0,j}$), the cleared numerator vanishes at $x=0$ and dividing out $x>0$ leaves a positive degree-$4$ polynomial. Where $\pi_j(x)$ is harmful, letting $e_r$ ($1\le r\le4$) be the elementary symmetric polynomials of $\{1,\frac12,\dots,\frac1j\}$ ($e_1=H_j$), it is bounded above by $\pi_j(x)\le\frac{x^4}{e_4+e_3 x+e_2 x^2+H_j x^3+x^4}$ on $[0,\frac34]$ and $[\frac34,\frac97]$ (degree $\le9$). \Cref{tab:open-far} specifies every subinterval split across all $52$ cells, each verified by exact rational Bernstein positivity.

\begin{table}[!ht]
\centering
\footnotesize
\begin{tabular}{@{}llll@{}}
\toprule
$(\tau_L,\tau_*)$ & Sign of $P_\pi$ & Ranks $j$ & Subinterval partition of $[0,\frac97]$ and $\pi_j$-bound\\
\midrule
$(0,\frac1{60})$ & helpful ($<0$) & $5,6,7$ & $[0,\frac34]$ ($\ge0$), $[\frac34,\frac97]$ ($\ge\pi_j(\frac34)$)\\[2pt]
$(0,\frac1{60})$ & helpful ($<0$) & $8$ & $[0,\frac12]$ ($\ge0$), $[\frac12,\frac34]$ ($\ge\pi_8(\frac12)$), $[\frac34,1]$ ($\ge\pi_8(\frac34)$), $[1,\frac97]$ ($\ge\pi_8(1)$)\\[2pt]
$(0,\frac1{60})$ & helpful ($<0$) & $9,\dots,17$ & $[0,\frac97]$ ($\ge0$; divide out $x>0$)\\[2pt]
$(\frac1{30},\frac1{60})$ & helpful ($<0$) & $5,6,10,\dots,17$ & $[0,\frac97]$ ($\ge0$)\\[2pt]
$(\frac1{30},\frac1{60})$ & helpful ($<0$) & $7,9$ & $[0,\frac34]$ ($\ge0$), $[\frac34,\frac97]$ ($\ge\pi_j(\frac34)$)\\[2pt]
$(\frac1{30},\frac1{60})$ & helpful ($<0$) & $8$ & $[0,\frac34]$ ($\ge0$), $[\frac34,1]$ ($\ge\pi_8(\frac34)$), $[1,\frac97]$ ($\ge\pi_8(1)$)\\[2pt]
$(\frac1{15},\frac1{20}),(\frac1{10},\frac1{12})$ & harmful ($>0$) & all $5,\dots,17$ & $[0,\frac34]\cup[\frac34,\frac97]$ (degree-$4$ upper bound on $\pi_j$)\\
\bottomrule
\end{tabular}
\caption{Exact subinterval partitions across all $52$ far-lower-arm cells ($L\ge10$, $x\le\frac97$) discharging~\eqref{eq:open-pointwise} for $5\le j\le16$ and regime~L at $j=17$.}\label{tab:open-far}
\end{table}

\subsection{Higher-arm box verifications for ranks \texorpdfstring{$5$}{5} to \texorpdfstring{$16$}{16}}\label{app:open-higher}

\paragraph{Seams ($h\in[\frac3{25},\frac12]$).}
At the $L>10$ checkpoint $(h_0,\gamma_0)=(\frac{21}{100},\frac{71}{100})$, split $p\in[0,\frac79]$ at $1-\gamma_0=\frac{29}{100}$ (where the sign of $p-(1-\gamma_0)$ in $S^{\mathrm{open}}_j$ flips): on $[0,\frac{29}{100}]$ use $b_j(p)\le\frac{H_j+p/2}{(1+p)(1+p/2)}$ and a linear denominator lower bound (degree $\le4$); on $[\frac{29}{100},\frac79]$, letting $\tilde e_1,\tilde e_2,\tilde e_3$ be the elementary symmetric polynomials of $\{\frac12,\dots,\frac1j\}$ ($\tilde e_1=H_j-1$), use exact $\varrho_j(p,h_0)$ and $b_j(p)\ge(1-1/Q_j(p))/p$ with $Q_j(p)=(1+p)(1+\tilde e_1 p+\tilde e_2 p^2+\tilde e_3 p^3)\le\Pi_j(p)^{-1}$ (degree $\le7$). The same bounds close the three $L\le10$ subbands and checkpoints of Fact~5 at both reserve endpoints across all $12$ ranks via exact Bernstein positivity.

\paragraph{Main strip ($h\in(0,\frac3{25}]$, $p\in(0,\frac79]$, $L>10$): four-guard coverage of all $27$ rows.}
Ten price-rank boxes partition $[0,\frac79]$ completely across both reserve groups: five low-group boxes ($5\le j\le8$, $\mathsf B=2+\frac{39}{50}H_j$) with $p$-breakpoints $\{0,\frac{11}{50},\frac9{25},\frac{12}{25},\frac{31}{50},\frac79\}$ ($12$ knot rows), and five high-group boxes ($9\le j\le16$, $\mathsf B=\underline{\mathsf M}_{0,j}:=\underline{\mathsf C}_j-\mathsf K_j$) with $p$-breakpoints $\{0,\frac7{50},\frac{21}{100},\frac{29}{100},\frac{11}{25},\frac79\}$ ($15$ knot rows). Each of the $27$ rows carries knot floors $(\delta\mathsf C_k,\delta\mathsf m_k,f_k)$, a central price $p_c\in(p_a,p_b)$, and a M\"obius parameter $\alpha_{\mathrm{M}}\ge0$ defining $z_{\mathrm{M}}(p)=\frac p{\underline{\mathsf m}_k+\alpha_{\mathrm{M}}p}$ (\cref{tab:open-main-strip}), and is verified by four exact rational/Bernstein guards:
\begin{enumerate}[leftmargin=2em]
\item \emph{Guard~1 (harmonic interpolation):} covers all interior ranks between consecutive knots $a<m<b$ via Fact~1 less the $\varrho$-fee $f_\varrho=\frac{h_1(H_m-H_a)(H_b-H_m)\lambda_m^2}{(1+\lambda_m H_a)(1+\lambda_m H_m)(1+\lambda_m H_b)}$ ($\lambda_m=h_1 p_b$) and $b$-interpolation fee $f_{\mathrm{out}}=\Sigma_m/M_a$, where $h^{(2)}_{1,j}=\sum_{i=1}^j i^{-2}$, $H_{3,j}=\sum_{i=1}^j i^{-3}$, $c_0(j)=\frac12(H_j^2-h^{(2)}_{1,j})$, $e_3(r)=\frac16(H_r^3-3H_r h^{(2)}_{1,r}+2H_{3,r})$, $\Sigma_m=(1-\lambda_{\mathrm{int}})c_0(a)+\lambda_{\mathrm{int}}c_0(b)-c_0(m)$, $P_{a+2}(p)=\Pi_{a+2}(p)^{-1}$, and $P_{a+2}(p)-M_a p(1-p)>0$ is proved on $[p_a,p_b]$ by the strictly convex lower quadratic $Q(p)=1+(H_{a+2}+e_3(a+2)p_a^2-M_a)p+(c_0(a+2)+M_a)p^2>0$. Of $17$ consecutive knot intervals across the $10$ boxes, $2$ are adjacent ($7\to8$, vacuous) and $15$ actively interpolate every interior rank.
\item \emph{Guard~2 (shape of $A_{\mathrm{cap}}$ in $z_{\mathrm{M}}$):} uses the exact M\"obius curvature identities for $p/D$ and $\varrho_k$ in one of two modes (with $q_b=1+p_b h_1 H_k$ and $D_{\mathrm{hi}}=\underline{\mathsf m}_k+p_b\underline{\mathsf M}_{0,k}$): \emph{concave mode} (\emph{conc}, $10$ rows, where $\alpha_{\mathrm{M}}\ge\max(\underline{\mathsf M}_{0,k},h_1 H_k\underline{\mathsf m}_k)$ and $\frac{2\underline{\mathsf m}_k(\alpha_{\mathrm{M}}-\underline{\mathsf M}_{0,k})}{\underline{\mathsf m}_k+\alpha_{\mathrm{M}}p_b}(\underline{\mathsf m}_k-p_b^2 h_1 H_k)-\frac{2p_b h_1 H_k}{q_b^2}\underline{\mathsf m}_k-\frac{2p_b h_1 H_k}{q_b^3}D_{\mathrm{hi}}>0$, reducing $A_{\mathrm{cap}}$ directly to $p_a,p_b$), or \emph{convex/tangent-bridge mode} (\emph{tang}, $17$ rows, where $\alpha_{\mathrm{M}}<\underline{\mathsf M}_{0,k}$ satisfies the strengthened bracket $\underline{\mathsf m}_k(\underline{\mathsf M}_{0,k}-\alpha_{\mathrm{M}})+\min(0,h_1 C_h)>0$ with $C_h=-\underline{\mathsf m}_k\mathsf B+\frac{\underline{\mathsf m}_k p_a H_k}{q_b}+\frac{(\underline{\mathsf m}_k+\alpha_{\mathrm M}p_a)p_a H_k}{q_b^2}$ together with its curvature margin $\frac{2\mathsf B^2(1-h_1)(1-2h_1)(\underline{\mathsf m}_k-p_b^2h_1H_k)\cdot\mathrm{bracket}}{D_{\mathrm{hi}}^3}>2h_1^2H_k\alpha_{\mathrm{M}}$, proving $A_{\mathrm{cap}}$ convex in $z_{\mathrm{M}}$ and replacing it by its $z_{\mathrm{M}}$-tangent at $p_c$).
\item \emph{Guard~3 (convexity of $S^{\mathrm{open}}_k$ in $z_{\mathrm{M}}$):} writes $S^{\mathrm{open}}_k(p,1-\omega)$ ($\omega\in[\frac{92}{167},1]$) as a Laplace transform whose log-slope ratio $\rho_0(p,\omega)=-\partial_{pp}S^{\mathrm{open}}_k/\partial_p S^{\mathrm{open}}_k$ decreases in $p$ (its $p$-derivative is minus a variance) and is linear-fractional in $\omega$. To enclose $\Pi_k,s_{1,k},s_{2,k}$ cleanly in exact rational arithmetic for $5\le k\le16$, peel $F_4(q)=\prod_{i=1}^4(1+q/i)$ and $S_{r,4}(q)=\sum_{i=1}^4(i+q)^{-r}$, set $\hat e_1=H_k-\frac{25}{12}$, $V_5=h^{(2)}_{1,k}-\frac{205}{144}$, $\hat e_2=\frac12(\hat e_1^2-V_5)$, and $V_{5,r}(k)=\sum_{i=5}^k i^{-r}$. With the verified uniform bounds $v_4^-=\frac1{625}\le V_{5,4}\le\frac8{2187}=v_4^+$, $v_5^-=\frac1{3125}\le V_{5,5}\le\frac4{6561}=v_5^+$, $V_{5,6}\le\frac1{9000}=v_6^+$, and the rank-dependent bounds $V_{5,3}^-(k)=\frac1{125}+(V_{5,3}(16)-\frac1{125})\frac{H_k-H_5}{H_{16}-H_5}\le V_{5,3}(k)\le\min(\frac1{125}+\frac1{36}(H_k-H_5),\,V_{5,3}(10)+\frac1{100}(H_k-H_{10}))=:V_{5,3}^+(k)$, set $H_{3,k}^{\mathrm{ub}}=\frac{2035}{1728}+V_{5,3}^+(k)$, $\hat e_3^-=\max(0,\frac16(\hat e_1^3-3\hat e_1 V_5+2V_{5,3}^-))$, and $\hat e_3^+=\frac16(\hat e_1^3-3\hat e_1 V_5+2V_{5,3}^+)$. Alternating/geometric tail bounds then give
\begin{align*}
F_4(q)\,\Pi_k(q)&\in\Bigl[\bigl(1+\hat e_1 q+\hat e_2 q^2+\hat e_3^+ q^3+\tfrac{\hat e_3^+\hat e_1 q^4/4}{1-\hat e_1 q/5}\bigr)^{-1},\;\bigl(1+\hat e_1 q+\hat e_2 q^2+\hat e_3^- q^3\bigr)^{-1}\Bigr],\\
s_{1,k}(q)-S_{1,4}(q)&\in\bigl[\hat e_1-qV_5+q^2V_{5,3}^--q^3v_4^+,\;\hat e_1-qV_5+q^2V_{5,3}^+-q^3v_4^-+q^4v_5^+\bigr],\\
s_{2,k}(q)-S_{2,4}(q)&\in\bigl[V_5-2qV_{5,3}^++3q^2v_4^--4q^3v_5^+,\;V_5-2qV_{5,3}^-+3q^2v_4^+-4q^3v_5^-+5q^4v_6^+\bigr].
\end{align*}
For $q>0$, write $\rho_0(q,\omega)=N(\Pi,s_1,s_2)/D(\Pi,s_1)$ with $N=2\omega(1-\Pi-qs_1\Pi)+q^2(q-\omega)(s_1^2+s_2)\Pi$ and $D=q\omega(1-\Pi)+q^2(q-\omega)s_1\Pi$. Since $N$ is affine in $(\Pi,s_2)$ and either convex ($q\ge\omega$) or decreasing ($q<\omega$) in $s_1\ge0$, while $D$ is bilinear in $(\Pi,s_1)$, setting $s_2^{\mathrm{use}}=s_2^+$ if $\omega\le q$ and $s_2^-$ if $\omega>q$ yields the directed corner upper bound $\bar\rho_0(q,\omega):=\max N/\min D$ over the four $(\Pi^\pm,s_1^\pm)$ corners (requiring $\min D>0$). Checking $\max_{\omega\in\{92/167,1\}}\bar\rho_0(q_L,\omega)<\frac{2\alpha_{\mathrm{M}}}{\underline{\mathsf m}_k+\alpha_{\mathrm{M}}q_R}$ across $n_{\mathrm{st}}$ equal substeps $[q_L,q_R]\subseteq[p_a,p_b]$ ($13$ single-step and $14$ multi-step rows) proves that $S^{\mathrm{open}}_k$ lies above its $z_{\mathrm{M}}$-tangent $S^{\tan}_e(5t)$ at $p_c$.
\item \emph{Guard~4 (convexity in $t=h/(1+ch)$ on $[0,T_1]$, $T_1=\frac{h_1}{1+ch_1}=\frac{15}{167}$):} For $h(t)=t/(1-ct)$ ($h'=(1+ch)^2$, $h''=2c(1+ch)^3$), any $C^2$ function $f(h)$ plus an affine function $a_0+a_1 t$ satisfies $F'(t)=(1+ch)^2 f'(h)+a_1$ and $F''(t)=(1+ch)^3[2cf'(h)+(1+ch)f''(h)]$. At fixed price $p>0$, set $\lambda=pH_k$, $q(h)=1+\lambda h$, $q_1=1+\lambda h_1$, $u_0(h)=(1-h)(1-2h)$, $D(h)=\underline{\mathsf m}_k+p(\underline{\mathsf M}_{0,k}-h\mathsf B+\frac{\lambda h}q)$, $D_h=p(-\mathsf B+\lambda/q^2)$, $D_{hh}=-2p\lambda^2/q^3$, and $y(h)=-D_h/D$. Differentiating $A_{\mathrm{cap}}(p,h)$ using Fact~2 gives $W(p,h):=2c\,\partial_h A_{\mathrm{cap}}(p,h)+(1+ch)\,\partial_{hh}A_{\mathrm{cap}}(p,h)$ explicitly as
\begin{gather*}
W(p,h)=W_1(p,h)+p\mathsf B^2\Bigl(\frac{g(h,y)}{D}-\frac{2p\lambda^2(1+ch)u_0}{q^3 D^2}\Bigr),\\
W_1(p,h):=2c(\mathsf K_k-\mathsf B)-2c-\frac{2(\lambda-c)}{q^3},\\
g(h,y):=2c\bigl((3-4h)-u_0 y\bigr)-4(1+ch)\bigl(1-(1-h)y\bigr)\bigl(1-(\tfrac12-h)y\bigr).
\end{gather*}
Set $D_0:=\underline{\mathsf m}_k+p\underline{\mathsf M}_{0,k}$, $D_1:=\underline{\mathsf m}_k+p(\underline{\mathsf M}_{0,k}-h_1\mathsf B+\lambda h_1/q_1)$, $y_{\min}:=p(\mathsf B-\lambda)/D_0$, and $y_{\max}:=p\mathsf B/D_1$. Under the preconditions $p>0$, $0<D_1<D_0$, and $0\le y_{\min}\le y_{\max}\le1$ (so $\lambda\le\mathsf B$): (i)~$D_0-D(h)=ph(\mathsf B-\lambda/q)\ge0$ and $D(h)-D_1=p(h_1-h)(\mathsf B-\lambda/(qq_1))\ge0$ give $D(h)\in[D_1,D_0]$ and $y(h)=-D_h/D\in[y_{\min},y_{\max}]\subseteq[0,1]$; (ii)~$g(h,y)-[(1-y)g(h,0)+yg(h,1)]=4(1+ch)(1-h)(\frac12-h)y(1-y)\ge0$ on $[0,h_1]\times[0,1]$, with $g(h,0)=(6c-4)-12ch$ and $g(h,1)=4c-(2c+2)h-(6c+4)h^2-4ch^3$; (iii)~for any $v\in\R$ and $D\in[D_1,D_0]$, $v/D\ge\min(v/D_0,v/D_1)$; and (iv)~convexity of $q^{-3}$ gives $q_1^{-3}\le q(h)^{-3}\le q_{3,\mathrm{ub}}(h):=1+(q_1^{-3}-1)\frac h{h_1}$, yielding $W_1(p,h)\ge W_{1,\mathrm{lb}}(h)$ (defined as $2c(\mathsf K_k-\mathsf B)-2c-2(\lambda-c)q_{3,\mathrm{ub}}(h)$ if $\lambda\ge c$, and the constant $2c(\mathsf K_k-\mathsf B)-2c-2(\lambda-c)q_1^{-3}$ if $\lambda<c$), while $(1+ch)u_0(h)=1-\frac15 h-\frac{32}5 h^2+\frac{28}5 h^3\le1$ and $D\ge D_1$ bound the negative term. Hence $W(p,h)\ge\min_{(y_e,D_{\mathrm{use}})}W_{\mathrm{lb}}(h;y_e,D_{\mathrm{use}})$ over the four corners $(y_e,D_{\mathrm{use}})\in\{y_{\min},y_{\max}\}\times\{D_0,D_1\}$, where
\[
W_{\mathrm{lb}}(h;y_e,D_{\mathrm{use}}):=W_{1,\mathrm{lb}}(h)+p\mathsf B^2\Bigl(\frac{(1-y_e)g(h,0)+y_e g(h,1)}{D_{\mathrm{use}}}-\frac{2p\lambda^2}{D_1^2}\,q_{3,\mathrm{ub}}(h)\Bigr)
\]
is a degree-$3$ polynomial in $h$.

On convex/tangent-bridge rows, $\widehat A_e(h)=A_{\mathrm{cap}}(p_c,h)+\Delta z_e\,\partial_p A_{\mathrm{cap}}(p_c,h)$ with $\Delta z_e=(p_e-p_c)\frac{\underline{\mathsf m}_k+\alpha_{\mathrm M}p_c}{\underline{\mathsf m}_k+\alpha_{\mathrm M}p_e}$, so $\partial_{tt}\widehat A_e(h(t))=(1+ch)^3[W(p_c,h)+\Delta z_e W_p(p_c,h)]$. Differentiating $A_{\mathrm{cap}}$ in $p$ at $p=p_c$ gives $\partial_p A_{\mathrm{cap}}(p_c,h)=-\frac{h^2 H_k}{q^2}-\mathsf B^2 R(h)$ with $N_p(h)=\underline{\mathsf m}_k-\frac{p_c^2 H_k h}{q^2}$, $U(h)=u_0(h)N_p(h)$, and $R(h)=U(h)/D(h)^2$, whence
\[
W_p(p_c,h)=W_{1,p}(h)-\mathsf B^2\bigl(2cR'(h)+(1+ch)R''(h)\bigr),\qquad W_{1,p}(h):=\frac{2H_k((2\lambda-3c)h-1)}{q^4},
\]
with $R'=\frac{U'D-2UD_h}{D^3}$, $R''=\frac{U''D^2+6UD_h^2-(4U'DD_h+2UDD_{hh})}{D^4}$, $u'_0=4h-3$, $u''_0=4$, $N'_p=-\frac{p_c^2 H_k(1-\lambda h)}{q^3}$, $N''_p=\frac{2p_c^2 H_k\lambda(2-\lambda h)}{q^4}$, $U'=u'_0 N_p+u_0 N'_p$, and $U''=4N_p+2u'_0 N'_p+u_0 N''_p$. Under the preconditions $p_c>0$, $2\lambda h_1<1$, $2\lambda<3c$, and $\lambda<\mathsf B$, every atomic factor is monotone on $h\in[0,h_1]$ (since $(h/q^2)'=(1-\lambda h)/q^3>0$), giving exact monotonic rational intervals:
\begin{gather*}
W_{1,p}\in\frac{\bigl[2H_k((2\lambda-3c)h_1-1),\,-2H_k\bigr]}{[1,\,q_1^4]},\quad
u_0\in\bigl[(1-h_1)(1-2h_1),\,1\bigr],\quad
u'_0\in[-3,\,-3+4h_1],\\
1+ch\in[1,\,1+ch_1],\quad
N_p\in\bigl[\underline{\mathsf m}_k-\tfrac{p_c^2 H_k h_1}{q_1^2},\,\underline{\mathsf m}_k\bigr],\quad
N'_p\in\bigl[-p_c^2 H_k,\,-\tfrac{p_c^2 H_k(1-\lambda h_1)}{q_1^3}\bigr],\\
N''_p\in\bigl[\tfrac{2p_c^2 H_k\lambda(2-\lambda h_1)}{q_1^4},\,4p_c^2 H_k\lambda\bigr],\quad
D_h\in\bigl[p_c(-\mathsf B+\tfrac\lambda{q_1^2}),\,p_c(-\mathsf B+\lambda)\bigr],\quad
D_{hh}\in\bigl[-2p_c\lambda^2,\,-\tfrac{2p_c\lambda^2}{q_1^3}\bigr],
\end{gather*}
alongside $D\in[D_1,D_0]$ and $u''_0=[4,4]$. Evaluating $U,U',U'',D^2,D^3,D^4,D_h^2,R',R''$, and $W_p$ via standard exact rational interval arithmetic ($[a,b]+[u,v]=[a+u,b+v]$, $[a,b]-[u,v]=[a-v,b-u]$, $[a,b]\cdot[u,v]=[\min S,\max S]$ for $S=\{au,av,bu,bv\}$, $[a,b]/[u,v]=[a,b]\cdot[1/v,1/u]$ for $0\notin[u,v]$, powers by repeated multiplication) yields an exact rational enclosure $[W_p^-,W_p^+]\ni W_p(p_c,h)$ on $[0,h_1]$, and the signed-shift rule
\[
\Delta z_e\,W_p(p_c,h)\ge\mathrm{shift}_e:=\begin{cases}\Delta z_e W_p^+,&\Delta z_e\le0\text{ (endpoint $a$)},\\\Delta z_e W_p^-,&\Delta z_e>0\text{ (endpoint $b$)}\end{cases}
\]
(with $\mathrm{shift}_e=0$ on concave-mode rows). A degree-$3$ polynomial $\sum_{r=0}^3 a_r h^r$ on $[0,h_1]$ has exact Bernstein coefficients $b_i=\sum_{r=0}^i\frac{\binom ir}{\binom 3r}a_r h_1^r$ ($i=0,1,2,3$); strict positivity of all four Bernstein coefficients of $W_{\mathrm{lb}}(h;y_e,D_{\mathrm{use}})+\mathrm{shift}_e$ across all four $(y_e,D_{\mathrm{use}})$-corners proves $W(p_c,h)+\Delta z_e W_p(p_c,h)>0$ on $[0,h_1]$, hence $G''_e(t)=(1+ch(t))^3[W(p_c,h(t))+\Delta z_e W_p(p_c,h(t))]>0$ on $[0,T_1]$, establishing strict convexity of $G_e(t)$ on $[0,T_1]$.
\end{enumerate}

\paragraph{Exceptional boundary logic at $p_a=0$.}
Four of the ten concave-mode rows touch $p_a=0$ (knots $5,8$ on $[0,\frac{11}{50}]$ and knots $9,16$ on $[0,\frac7{50}]$) and require two boundary specializations: (i)~in Guard~3 at $q_L=0$, the exact $q\to0^+$ moment limit $\bar\rho_0(0,\omega)=(2E_2+2\omega E_3^{\mathrm{ub}})/(E_1+\omega E_2)$ is used with $E_1=H_k$, $E_2=\frac12(H_k^2+h^{(2)}_{1,k})$, and $E_3^{\mathrm{ub}}=\frac16(H_k^3+3H_k h^{(2)}_{1,k}+2H_{3,k}^{\mathrm{ub}})$ (e.g., at $k=5$ on $[0,\frac{11}{50}]$, $\max_\omega\bar\rho_0(0,\omega)<2.702143<3.662644$); (ii)~at $p_a=0$, $\varrho_k(0,h)=0$ and the $p\mathsf B^2u_0/D$ term vanishes, leaving $A_{\mathrm{cap}}(0,h)=(1-h)(\mathsf B-\mathsf K_k)$. Because $h(t)=t/(1-ct)$ is strictly convex in $t$, the endpoint gap $G_a(t)=(1-h(t))(\mathsf B-\mathsf K_k)+S^{\tan}_a(5t)-f_k$ is \emph{strictly concave} in $t$ ($G_a''(t)=-2c(1+ch)^3(\mathsf B-\mathsf K_k)<0$). Guard~4 and an interior tangent $t_a$ are therefore omitted at $p_a=0$; positivity on $[0,\frac{15}{167}]$ follows directly from checking $G_a(0)>0$ and $G_a(\frac{15}{167})>0$ across all four $(\Pi_k(p_c),s_{1,k}(p_c))$ enclosure corners (e.g., $>0.274223$ and $>0.278330$ at $k=5$).

At every positive endpoint $p_e>0$, Guards~1--4 make $G_e(t)=\widehat A_e(h(t))+a_{0,e}+5t\,a_{1,e}-f_k$ strictly convex in $t\in[0,T_1]$ and bilinear in $(\Pi_k(p_c),s_{1,k}(p_c))$, where $b=(1-\Pi)/p_c$, $b'=(\Pi s_1-b)/p_c$, $a_{0,e}=(H_k-b+p_cb)+\Delta z_e(b+(p_c-1)b')$, $a_{1,e}=-(H_k-b)+\Delta z_e b'$, and $G'_e(t)=(1+ch)^2\partial_h\widehat A_e(h)+5a_{1,e}$ (with $\partial_h\widehat A_e=\partial_h A_{\mathrm{cap}}(p_c,h)+\Delta z_e(-\frac{2H_k h}{q^3}-\mathsf B^2 R'(h))$ in convex mode, and $\partial_h A_{\mathrm{cap}}(p_e,h)$ in concave mode). Positivity of the single $t$-tangent $G_e(t_e)+G'_e(t_e)(t-t_e)$ at $t\in\{0,T_1\}$ across all four $(\Pi^\pm,s_1^\pm)$ corners therefore yields $G_e(t)>0$ on $[0,T_1]$ ($e\in\{a,b\}$). On the full interval $p\in[p_a,p_b]$, the sum of the Guard-2 lower bound on $A_{\mathrm{cap}}$ and Guard-3 tangent on $S^{\mathrm{open}}_k$ minus $f_k$ is affine in $\Delta z(p)\in[\Delta z_a,\Delta z_b]$ (convex mode) or concave in $z_{\mathrm{M}}(p)$ (concave mode), attaining its minimum $\min(G_a(t),G_b(t))>0$ at $p\in\{p_a,p_b\}$, and Guard~1 extends positivity to all interior ranks $a<m<b$. Evaluated across all $10$ boxes and $27$ rows of \cref{tab:open-main-strip}, the companion's exact rational checker verifies global minimum margins: Guard-1 fee slack $>5.46\times10^{-5}$, Guard-3 margin $>4.66\times10^{-4}$, Guard-4 Bernstein coefficient $>4.8264$, and four-corner endpoint gap $>1.42\times10^{-4}$.

\begin{table}[!ht]
\centering
\scriptsize
\begin{tabular}{@{}lccccccccc@{}}
\toprule
Box $[p_a,p_b]$ & $k$ & $M_a$ & $(\delta\mathsf C,\delta\mathsf m,f)\cdot10^4$ & Mode & $n_{\mathrm{st}}$ & $p_c$ & $\alpha_{\mathrm M}$ & $t_a$ & $t_b$\\
\midrule
\multicolumn{10}{@{}l}{\emph{Low group ($5\le j\le8$, $\mathsf B=2+\frac{39}{50}H_j$):}}\\
$[0,\frac{11}{50}]$ & $5$ & $\frac{19}{2}$ & $(150,110,100)$ & conc & $1$ & $\frac{19}{100}$ & $\frac{21}{5}$ & --- & $\frac{1}{200}$\\[1pt]
$[0,\frac{11}{50}]$ & $8$ & --- & $(0,0,0)$ & conc & $1$ & $\frac{19}{100}$ & $\frac{22}{5}$ & --- & $\frac{1}{200}$\\[1pt]
$[\frac{11}{50},\frac{9}{25}]$ & $5$ & $\frac{19}{2}$ & $(55,50,35)$ & tang & $1$ & $\frac{17}{50}$ & $\frac{17}{5}$ & $\frac{1}{200}$ & $\frac{1}{1000}$\\[1pt]
$[\frac{11}{50},\frac{9}{25}]$ & $7$ & --- & $(0,0,0)$ & tang & $1$ & $\frac{17}{50}$ & $\frac{7}{2}$ & $\frac{1}{200}$ & $\frac{1}{100}$\\[1pt]
$[\frac{11}{50},\frac{9}{25}]$ & $8$ & --- & $(0,0,0)$ & tang & $8$ & $\frac{17}{50}$ & $3$ & $\frac{1}{200}$ & $\frac{17}{1000}$\\[1pt]
$[\frac{9}{25},\frac{12}{25}]$ & $5$ & $10$ & $(55,50,30)$ & tang & $6$ & $\frac{2}{5}$ & $\frac{13}{5}$ & $\frac{1}{1000}$ & $\frac{1}{1000}$\\[1pt]
$[\frac{9}{25},\frac{12}{25}]$ & $7$ & --- & $(0,0,0)$ & tang & $4$ & $\frac{19}{50}$ & $\frac{16}{5}$ & $\frac{1}{100}$ & $\frac{1}{50}$\\[1pt]
$[\frac{9}{25},\frac{12}{25}]$ & $8$ & --- & $(0,0,0)$ & tang & $6$ & $\frac{19}{50}$ & $\frac{17}{5}$ & $\frac{17}{1000}$ & $\frac{1}{40}$\\[1pt]
$[\frac{12}{25},\frac{31}{50}]$ & $5$ & $\frac{23}{2}$ & $(55,40,25)$ & tang & $8$ & $\frac{1}{2}$ & $\frac{29}{10}$ & $\frac{1}{200}$ & $\frac{1}{100}$\\[1pt]
$[\frac{12}{25},\frac{31}{50}]$ & $8$ & --- & $(45,40,25)$ & conc & $1$ & $\frac{13}{25}$ & $\frac{33}{5}$ & $\frac{13}{500}$ & $\frac{7}{200}$\\[1pt]
$[\frac{31}{50},\frac{7}{9}]$ & $5$ & $\frac{33}{2}$ & $(80,45,25)$ & tang & $12$ & $\frac{16}{25}$ & $\frac{33}{10}$ & $\frac{1}{100}$ & $\frac{3}{200}$\\[1pt]
$[\frac{31}{50},\frac{7}{9}]$ & $8$ & --- & $(40,45,25)$ & conc & $1$ & $\frac{16}{25}$ & $\frac{38}{5}$ & $\frac{7}{200}$ & $\frac{1}{25}$\\
\midrule
\multicolumn{10}{@{}l}{\emph{High group ($9\le j\le16$, $\mathsf B=\underline{\mathsf M}_{0,j}$):}}\\
$[0,\frac{7}{50}]$ & $9$ & $12$ & $(400,100,160)$ & conc & $1$ & $\frac{3}{25}$ & $\frac{43}{10}$ & --- & $\frac{1}{200}$\\[1pt]
$[0,\frac{7}{50}]$ & $16$ & --- & $(0,0,0)$ & conc & $1$ & $\frac{3}{25}$ & $\frac{47}{10}$ & --- & $\frac{3}{100}$\\[1pt]
$[\frac{7}{50},\frac{21}{100}]$ & $9^\star$ & $\frac{21}{2}$ & $(60,20,20)$ & tang & $1$ & $\frac{1}{5}$ & $\frac{18}{5}$ & $\frac{1}{200}$ & $\frac{1}{200}$\\[1pt]
$[\frac{7}{50},\frac{21}{100}]$ & $12$ & $11$ & $(16,5,6)$ & tang & $1$ & $\frac{1}{5}$ & $\frac{19}{5}$ & $\frac{1}{100}$ & $\frac{1}{40}$\\[1pt]
$[\frac{7}{50},\frac{21}{100}]$ & $14$ & $\frac{23}{2}$ & $(15,4,5)$ & tang & $1$ & $\frac{1}{5}$ & $\frac{37}{10}$ & $\frac{1}{50}$ & $\frac{7}{200}$\\[1pt]
$[\frac{7}{50},\frac{21}{100}]$ & $16$ & --- & $(0,0,0)$ & tang & $3$ & $\frac{17}{100}$ & $\frac{18}{5}$ & $\frac{3}{100}$ & $\frac{21}{500}$\\[1pt]
$[\frac{21}{100},\frac{29}{100}]$ & $9$ & $\frac{21}{2}$ & $(10,4,4)$ & tang & $1$ & $\frac{27}{100}$ & $\frac{37}{10}$ & $\frac{1}{200}$ & $\frac{17}{1000}$\\[1pt]
$[\frac{21}{100},\frac{29}{100}]$ & $11$ & $11$ & $(20,6,8)$ & tang & $4$ & $\frac{27}{100}$ & $\frac{33}{10}$ & $\frac{1}{50}$ & $\frac{3}{100}$\\[1pt]
$[\frac{21}{100},\frac{29}{100}]$ & $14$ & $\frac{23}{2}$ & $(20,6,8)$ & tang & $2$ & $\frac{11}{50}$ & $\frac{39}{10}$ & $\frac{7}{200}$ & $\frac{9}{200}$\\[1pt]
$[\frac{21}{100},\frac{29}{100}]$ & $16$ & --- & $(5,2,3)$ & tang & $12$ & $\frac{11}{50}$ & $\frac{39}{10}$ & $\frac{43}{1000}$ & $\frac{1}{20}$\\[1pt]
$[\frac{29}{100},\frac{11}{25}]$ & $9$ & $\frac{21}{2}$ & $(18,6,10)$ & tang & $8$ & $\frac{17}{50}$ & $\frac{7}{2}$ & $\frac{17}{1000}$ & $\frac{3}{100}$\\[1pt]
$[\frac{29}{100},\frac{11}{25}]$ & $12$ & $\frac{23}{2}$ & $(40,13,20)$ & conc & $1$ & $\frac{31}{100}$ & $\frac{11}{2}$ & $\frac{9}{250}$ & $\frac{9}{200}$\\[1pt]
$[\frac{29}{100},\frac{11}{25}]$ & $16$ & --- & $(55,16,30)$ & conc & $2$ & $\frac{31}{100}$ & $\frac{57}{10}$ & $\frac{11}{200}$ & $\frac{13}{200}$\\[1pt]
$[\frac{11}{25},\frac{7}{9}]$ & $9$ & $13$ & $(80,25,40)$ & conc & $3$ & $\frac{49}{100}$ & $\frac{77}{10}$ & $\frac{31}{1000}$ & $\frac{1}{25}$\\[1pt]
$[\frac{11}{25},\frac{7}{9}]$ & $16$ & --- & $(180,50,80)$ & conc & $12$ & $\frac{49}{100}$ & $\frac{19}{2}$ & $\frac{13}{200}$ & $\frac{3}{40}$\\
\bottomrule
\end{tabular}
\caption{Complete rational parameters for all $27$ main-strip rows across the $10$ boxes discharging~\eqref{eq:open-pointwise} on $h\in(0,\frac3{25}]$, $p\in(0,\frac79]$ for $5\le j\le16$. Column $M_a$ applies to the active Guard-1 interval starting at knot $k$; `---' under $t_a$ marks the four $p_a=0$ origin rows; $9^\star$ is the worked representative row below. Every floor entry is an exact integer multiple of $10^{-4}$.}\label{tab:open-main-strip}
\end{table}

\paragraph{Worked representative row (knot $k=9^\star$ on $[p_a,p_b]=[\frac7{50},\frac{21}{100}]$).}
We work the high-group row $k=9$ on $[p_a,p_b]=[\frac7{50},\frac{21}{100}]$ (paired with knot $k=12$) completely from the formulas above; it runs in convex/tangent-bridge mode with active Guard-1 interpolation ($m\in\{10,11\}$), four distinct $(\Pi_9,s_{1,9})$ corners, both $t=0$ and $t=\frac{15}{167}$ active as tangent minima across the two endpoints, and the global minimum Guard-4 Bernstein coefficient across all $27$ rows. Its parameters are $c=\frac{14}5$, $h_1=\frac3{25}$, $T_1=\frac{15}{167}$, $p_c=\frac15$, $\alpha_{\mathrm{M}}=\frac{18}5$, $t_a=t_b=\frac1{200}$, floors $(\delta\mathsf C,\delta\mathsf m,f)=(\frac3{500},\frac1{500},\frac1{500})$ at $k=9$ and $(\frac1{625},\frac1{2000},\frac3{5000})$ at $k=12$, and $M_9=\frac{21}2$. From~\eqref{eq:stock-formula}, $H_9=\frac{7129}{2520}$, $\mathsf K_9=\frac{9649}{2520}$, $\underline{\mathsf C}_9\in[8.004294,8.004295]$, $\underline{\mathsf m}_9\in[1.504110,1.504111]$, $\mathsf B=\underline{\mathsf M}_{0,9}\in[4.175326,4.175327]$, $\lambda=p_c H_9=\frac{7129}{12600}\in[0.565793,0.565794]$, and $q_1=1+\lambda h_1=\frac{112129}{105000}\in[1.067895,1.067896]$.
\begin{itemize}[leftmargin=1.5em]
\item \emph{Guard~1 ($m\in\{10,11\}$ between $a=9$ and $b=12$):} using $H_{11}=\frac{83711}{27720}$, $c_0(11)\in[3.780813,\,3.780814]$, and $e_3(11)\in[2.636936,\,2.636937]$, the lower quadratic $Q(p)=1+b_q p+c_q p^2$ has coefficients $b_q\in[-7.428439,\,-7.428438]$ and $c_q\in[14.280813,\,14.280814]$, with vertex $-b_q/(2c_q)\in[0.260084,\,0.260085]>p_b$, so $\min_{[p_a,p_b]}Q=Q(\frac{21}{100})>0.069811>0$; also $H_{12}h_1p_b=\frac{86021}{1100000}<1$. At $m=10$ ($\lambda_{\mathrm{int}}=\frac{66}{181}$), the $\mathsf C$- and $\mathsf m$-slacks are $>0.002035$ and $>0.000649$, $\Sigma_{10}<0.009117$, $f_{\mathrm{out}}<0.000869$, $f_\varrho<0.000002$, and $\bar f_{10}=\frac{337}{226250}>0.001489$ leaves fee slack $>0.000620>0$. At $m=11$ ($\lambda_{\mathrm{int}}=\frac{126}{181}$), the $\mathsf C$- and $\mathsf m$-slacks are $>0.000760$ and $>0.000248$, $\Sigma_{11}<0.008313$, $f_{\mathrm{out}}<0.000792$, $f_\varrho<0.000001$, and $\bar f_{11}=\frac{116}{113125}>0.001025$ leaves fee slack $>0.000232>0$.
\item \emph{Guard~2 (convex/tangent-bridge mode, $\alpha_{\mathrm{M}}=\frac{18}5<\underline{\mathsf M}_{0,9}$):} $q_b=1+p_b h_1 H_9=\frac{107129}{100000}$, $D_{\mathrm{hi}}=\underline{\mathsf m}_9+p_b\underline{\mathsf M}_{0,9}\in[2.380928,2.380929]$, $C_h\in[-5.031088,-5.031087]$, the bracket $\underline{\mathsf m}_9(\underline{\mathsf M}_{0,9}-\alpha_{\mathrm{M}})+\min(0,h_1 C_h)>0.261623>0$, and the curvature check gives $\frac{2\mathsf B^2(1-h_1)(1-2h_1)(\underline{\mathsf m}_9-p_b^2h_1H_9)\cdot\mathrm{bracket}}{D_{\mathrm{hi}}^3}>0.673100>0.293308>2h_1^2H_9\alpha_{\mathrm{M}}$ (margin $>0.379792$), establishing $z_{\mathrm{M}}$-convexity of $A_{\mathrm{cap}}$.
\item \emph{Guard~3 (single step $[\frac7{50},\frac{21}{100}]$):} four-factor-peeled enclosures at $p_a=\frac7{50}$ give $\Pi_9(\frac7{50})\in[0.682517,0.682521]$, $s_{1,9}(\frac7{50})\in[2.634161,2.634239]$, and $s_{2,9}(\frac7{50})\in[1.258556,1.259698]$, yielding $\bar\rho_0(\frac7{50},\frac{92}{167})<2.740355$ and $\bar\rho_0(\frac7{50},1)<2.634703$ via the directed corner rule ($s_2^{\mathrm{use}}=s_2^-$ since $\omega>\frac7{50}$), both strictly below $\frac{2\alpha_{\mathrm{M}}}{\underline{\mathsf m}_9+\alpha_{\mathrm{M}}p_b}>3.185685$ (margin $>0.445330$).
\item \emph{Guard~4 ($t$-convexity on $[0,\frac{15}{167}]$):} All preconditions hold ($p_c=\frac15>0$, $2\lambda h_1=\frac{7129}{52500}<1$, $2\lambda=\frac{7129}{6300}<\frac{42}5=3c$, $\lambda<c<\mathsf B$, $0<D_1<D_0$, $0<y_{\min}<y_{\max}<1$). The tangent displacements are $\Delta z_a\in[-0.066454,-0.066453]<0$ and $\Delta z_b\in[0.009840,0.009841]>0$; the corner parameters are $D_0\in[2.339175,\,2.339176]$, $D_1\in[2.251683,\,2.251684]$, $y_{\min}\in[0.308615,\,0.308616]$, $y_{\max}\in[0.370862,\,0.370863]$, and constant $W_{1,\mathrm{lb}}\in[-3.870439,\,-3.870438]$ (since $\lambda<c$). Exact monotonic interval evaluation on $h\in[0,\frac3{25}]$ yields
\begin{gather*}
W_{1,p}\in[-10.592843,-4.350543],\quad N_p\in[1.492203,1.504111],\quad N'_p\in[-0.113159,-0.086609],\\
N''_p\in[0.190235,0.256098],\quad U\in[0.997985,1.504111],\quad U'\in[-4.625490,-3.818276],\\
U''\in[6.532555,6.951492],\quad D\in[2.251683,2.339176],\quad D_h\in[-0.735839,-0.721906],\\
D_{hh}\in[-0.128049,-0.105145],\quad R'\in[-0.821545,-0.498773],\quad R''\in[0.162563,0.739050],\\
W_p(\tfrac15,h)\in[20.887637,\,73.020176],\\
\mathrm{shift}_a\in[-4.852471,\,-4.852470],\qquad \mathrm{shift}_b\in[0.205549,\,0.205550],
\end{gather*}
where $\mathrm{shift}_a=\Delta z_a W_p^+>-4.8525$ and $\mathrm{shift}_b=\Delta z_b W_p^->0.2055$. Converting $W_{\mathrm{lb}}(h;y_e,D_{\mathrm{use}})+\mathrm{shift}_e$ to degree-$3$ Bernstein coefficients on $[0,\frac3{25}]$ across all four $(y_e,D_{\mathrm{use}})$-corners yields strictly positive vectors at both endpoints (every entry rounded down):
\begin{center}\small
\begin{tabular}{@{}lll@{}}
\toprule
Corner $(y_e,D_{\mathrm{use}})$ & Left endpoint $p_a=\frac7{50}$ ($\mathrm{shift}_a$) & Right endpoint $p_b=\frac{21}{100}$ ($\mathrm{shift}_b$)\\
\midrule
$(y_{\min},D_0)$ & $\ge[9.5321,\,8.0124,\,6.4469,\,4.8264]$ & $\ge[14.5901,\,13.0704,\,11.5049,\,9.8845]$\\
$(y_{\min},D_1)$ & $\ge[10.2448,\,8.6659,\,7.0393,\,5.3557]$ & $\ge[15.3028,\,13.7239,\,12.0973,\,10.4138]$\\
$(y_{\max},D_0)$ & $\ge[9.3836,\,7.9605,\,6.4821,\,4.9379]$ & $\ge[14.4416,\,13.0185,\,11.5401,\,9.9959]$\\
$(y_{\max},D_1)$ & $\ge[10.0906,\,8.6119,\,7.0760,\,5.4715]$ & $\ge[15.1486,\,13.6700,\,12.1340,\,10.5295]$\\
\bottomrule
\end{tabular}
\end{center}
with global minimum $>4.826497$ at $(y_{\min},D_0)$, proving $G''_a(t)>0$ and $G''_b(t)>0$ on $[0,\frac{15}{167}]$.
\item \emph{Four-corner tangent gaps at $p_c=\frac15$, $t_*=\frac1{200}$:} peeled enclosures give $\Pi_9(\frac15)\in[0.584020,0.584031]$ and $s_{1,9}(\frac15)\in[2.561457,2.561621]$. Evaluating $G_e(t_*)$ and $G'_e(t_*)$ across all four corners yields: at $p_a=\frac7{50}$, $G_a(\frac1{200})\in[0.044565,0.044652]$ and $G'_a(\frac1{200})\in[0.169144,0.169672]$, giving tangent $>0.043717$ at $t=0$ and $>0.058957$ at $t=\frac{15}{167}$; at $p_b=\frac{21}{100}$, $G_b(\frac1{200})\in[0.014813,0.014852]$ and $G'_b(\frac1{200})\in[-0.025077,-0.024825]$, giving tangent $>0.014937$ at $t=0$ and $>0.012707$ at $t=\frac{15}{167}$. By convexity, $G_a,G_b>0$ on $[0,\frac{15}{167}]$, closing $[\frac7{50},\frac{21}{100}]\times[0,\frac3{25}]$ at knot~$9$ (and hence interior ranks $10,11$ via Guard~1).
\end{itemize}
All remaining $26$ rows of \cref{tab:open-main-strip} are verified by the same four-guard rational checker in the companion package.

\subsection{Higher-arm box verifications at rank \texorpdfstring{$17$}{17}}\label{app:open-rank17}

At $j=17$, no rank interpolation is needed, and $\Pi_{17}(p),s_{1,17}(p),s_{2,17}(p)$ and $H_{3,17}=\sum_{i=1}^{17}i^{-3}$ are exact rationals. Freezing $\varrho_{17}$ at box endpoints gives pure M\"obius lower bounds
\begin{align*}
\text{Regime~H:}\quad &A_{\mathrm{cap}}\ge h\mathsf C_{17}-\mathsf K_{17}+\frac{(1-2h)\mathsf m_{17}u_a(h)}{\mathsf m_{17}+p(1-h)u_a(h)}\qquad\bigl(u_a(h)=\mathsf M_{0,17}+\varrho_{17}(p_a,h)\bigr),\\
\text{Regime~L:}\quad &A_{\mathrm{cap}}\ge(1-h)(\mathsf M_{0,17}-\mathsf K_{17})-h\varrho_{17}(p_b,h)-\frac{p\mathsf M_{0,17}^2(1-h)(1-2h)}{\mathsf m_{17}+\frac79\varrho_{17}(p_a,h)+p(1-h)\mathsf M_{0,17}},
\end{align*}
both concave in $z_{\mathrm{M}}=\frac p{\mathsf m_{17}+\alpha_{\mathrm{M}}p}$ whenever $\alpha_{\mathrm{M}}$ exceeds $\mathsf M_{0,17}+\varrho_{17}(p_a,\frac3{25})$ (regime~H) or $\mathsf M_{0,17}$ (regime~L), eliminating Guard~2. On each of the eight boxes listed in \cref{tab:open-rank17} (four covering $[0,\frac79]$ in regime~H and four in regime~L), a $4$-substep Laplace check verifies convexity of $S^{\mathrm{open}}_{17}$ in $z_{\mathrm{M}}$ (minimum margin $>0.00989$), reducing the price range to $p_a,p_b$. On the fixed seven-step $h$-grid $\{0,\frac1{80},\frac1{40},\frac3{80},\frac1{20},\frac7{100},\frac{19}{200},\frac3{25}\}$, replacing $\gamma_-(h)$ on each subinterval by its chord (when its coefficient in $S^{\tan}_e$ is positive) or midpoint tangent (when nonpositive) keeps the lower bound concave in $h$ (Fact~2), eliminating Guard~4 and reducing each box to exact rational grid-endpoint evaluations (smallest verified margin $>0.0005113$). Together with the rank-$17$ seams of Fact~5 (degree $\le3$ left and $\le5$ right in regime~H; degree $\le4$ left and $\le7$ right in regime~L), this discharges~\eqref{eq:open-pointwise} at rank~$17$ and completes \cref{thm:rank17}.

\begin{table}[!ht]
\centering
\small
\begin{tabular}{@{}lccc@{\qquad}lccc@{}}
\toprule
Regime & $[p_a,p_b]$ & $p_c$ & $\alpha_{\mathrm M}$ & Regime & $[p_a,p_b]$ & $p_c$ & $\alpha_{\mathrm M}$\\
\midrule
Regime~H & $[0,\frac1{10}]$ & $\frac2{25}$ & $5$ & Regime~L & $[0,\frac1{10}]$ & $\frac2{25}$ & $5$\\[2pt]
Regime~H & $[\frac1{10},\frac9{50}]$ & $\frac{37}{250}$ & $5$ & Regime~L & $[\frac1{10},\frac4{25}]$ & $\frac{71}{500}$ & $5$\\[2pt]
Regime~H & $[\frac9{50},\frac25]$ & $\frac{28}{125}$ & $5$ & Regime~L & $[\frac4{25},\frac9{25}]$ & $\frac15$ & $5$\\[2pt]
Regime~H & $[\frac25,\frac79]$ & $\frac{53}{90}$ & $15$ & Regime~L & $[\frac9{25},\frac79]$ & $\frac{593}{1125}$ & $15$\\
\bottomrule
\end{tabular}
\caption{Complete rational parameters for the eight rank-$17$ higher-arm boxes ($h\in(0,\frac3{25}]$, $4$ equal $p$-substeps and the $7$-step $h$-grid) discharging~\eqref{eq:open-pointwise} in regimes~H and~L.}\label{tab:open-rank17}
\end{table}

\section{Algebraic Details and Finite Verifications for Section~9}\label{app:sec9}

This appendix supplies the five-case finite verification for named deletion (L4) below truncation (\cref{thm:kernel}), the clip cubic and small-rate checks (\cref{lem:clip,thm:marked-tail}), and the four-strip large-rate verification, working Strip~1 in exact rational arithmetic and delegating the repeated polynomial expansions of Strips~2--4 to exact rational verification in the companion.

\subsection{Named deletion (L4) below truncation}\label{app:tail-l4}

For a covered clock $(m,n)$ ($1\le m<n$) with $e=\frac1n$, $b=b_{m,n}$, $b'=b-e$, $h^{(2)}=h^{(2)}_{m,n}$, and $h'^{(2)}=h^{(2)}-e^2$, clearing $\mathsf P_{m,n}\mathsf P_{m,n-1}/(ez^3)$ in (L4) on $0<z<1/\sqrt b$ yields $c_0+c_1z+c_2z^2+c_3z^3$ with strictly positive higher coefficients
\[
c_3=bb',\qquad c_2=b+b'+\frac{bh'^{(2)}}{2b'}+\frac{b'h^{(2)}}{2b},\qquad c_1=1+\frac{b+b'}2+\frac{h^{(2)}}{2b}+\frac{h'^{(2)}}{2b'}+\frac{(b^2-h^{(2)})(b'^2-h'^{(2)})}{4bb'},
\]
and $c_0=(\frac b2-1)^2+e(1-\frac b4)-\frac{(h^{(2)}-be)^2}{4bb'}$. Telescoping $i^{-2}<\frac1{i-1/2}-\frac1{i+1/2}$ gives $h^{(2)}_{1,n}<\frac{79}{48}<\frac53$, $h^{(2)}_{2,n}<\frac{13}{20}$, and $h^{(2)}_{3,n}<\frac25$; the Taylor bracket $\frac{321}{250}<e^{1/4}<\frac{257}{200}$ with $\ln\frac{n+1}m<b<\ln\frac{n+1/2}{m-1/2}$ yields $\frac1{me^b}<e<\frac1{(m-1/2)e^b-1/2}$. Using $0<h^{(2)}-be\le(\frac1m-e)b'$, five exhaustive cases prove $c_0>0$ in exact rational arithmetic:
\begin{enumerate}[label=(\roman*),leftmargin=2em]
\item $b\ge4$: $e\le\frac12$, $b'\ge\frac72$, so $c_0\ge1-\frac{(5/3)^2}{4\cdot4\cdot(7/2)}=\frac{479}{504}>0$.
\item $m=1$, $b<4$: exact evaluation at $n=2,\dots,7$ gives $c_0=\frac13,\frac3{22},\frac3{44},\frac{173}{3288},\frac{929}{14796},\frac{12769}{147015}$; for $n\ge8$, $b\ge H_8=\frac{761}{280}$ and $e\le\frac18$ give $c_0\ge(\frac b2-1)^2-\frac{(5/3)^2}{4b(b-1/8)}>0.03028$.
\item $m=2$, $b<4$: outside $(\frac32,\frac52)$, $(\frac b2-1)^2\ge\frac1{16}$ beats the penalty $<\frac1{4m^2}$. On $[\frac32,\frac74],[\frac74,2],[2,\frac94],[\frac94,\frac52]$, using $e\in[\frac12(\frac{257}{200})^{-4b_{\mathrm{hi}}},(\frac32(\frac{321}{250})^{4b_{\mathrm{lo}}}-\frac12)^{-1}]$ and $h^{(2)}<\frac{13}{20}$ yields $c_0>0.03054,0.00875,0.00339,0.01498$.
\item $m=3$, $b<4$: outside $(\frac53,\frac73)$, $(\frac b2-1)^2\ge\frac1{36}=\frac1{4m^2}$ beats the penalty. On $[\frac53,2]$ and $[2,\frac73]$, outward integer exponents $(6,8)$ and $(8,10)$ in $e\in[\frac13(\frac{257}{200})^{-p_{\mathrm{hi}}},(\frac52(\frac{321}{250})^{p_{\mathrm{lo}}}-\frac12)^{-1}]$ with $h^{(2)}<\frac25$ give $c_0>0.01233$ and $0.00363$.
\item $m\ge4$, $b<4$: outside $(\frac74,\frac94)$, $(\frac b2-1)^2\ge\frac1{64}\ge\frac1{4m^2}$. Inside $[\frac74,\frac94]$, $n\ge20$ ($e\le\frac1{20}$) and $h^{(2)}<\frac1{m-1/2}-\frac1{n+1/2}$ give $\frac{h^{(2)}-be}e<\frac87(\frac{257}{200})^9-\frac{20}{21}-\frac74<8.2151$, so $\frac{(h^{(2)}-be)^2}{4bb'}\le0.28357\,e<\frac7{16}e\le e(1-\frac b4)$.
\end{enumerate}

\subsection{The clip cubic and small-rate checks}\label{app:tail-clip}

In \cref{lem:clip}, the numerator cubic $P_{\mathrm{clip}}(B)=89B^3+138B^2-311B+120$ admits the exact expansion
\[
P_{\mathrm{clip}}(B)=\frac{4932}{15625}+\frac{88}{625}\Bigl(B-\frac{17}{25}\Bigr)+\Bigl(B-\frac{17}{25}\Bigr)^2\Bigl(89B+\frac{6476}{25}\Bigr),
\]
so on $[\frac12,1]$ it lies above its tangent at $\frac{17}{25}$, which evaluates to $\frac{4536}{15625}>0$ at $\frac12$ and $\frac{5636}{15625}>0$ at~$1$.

For \cref{thm:marked-tail}, exact harmonic sums verify $\frac{784}{225}<H_{18}<\frac72$, $\frac{159}{100}<h^{(2)}_{1,18}<\frac53$, $H_{30}<4<H_{31}$, $\frac2{32}H_{31}<\frac13$, and $H_{64}<5$, and confirm that $\epsilon_j H_j$ ($j\ge1$) and $\epsilon_j H_j^2$ ($j\ge18$) decrease. On the small-rate branch $0<\varrho\le\sqrt{H_j}$, these bounds give $B_1>\frac{36}{100}$ for all $j\ge18$, $A_0>\frac{811}{16900}>0$ for all $j\ge64$ (bounded by $j=64$), and uniformly for all $18\le j\le63$, the same monotonicity bounds together with $\mathsf m_j\ge\frac{36}{19}$, $H_j>\frac{784}{225}$, and $\epsilon_j\ge\frac1{32}$ give $A_0>-\frac{4119}{9025}$ and $4\mathsf m_j H_j\epsilon_j B_1>4\cdot\frac{36}{19}\cdot\frac{784}{225}\cdot\frac1{32}\cdot\frac{36}{100}=\frac{3528}{11875}>(\frac{4119}{9025})^2$.

\subsection{Large-rate strips and worked Strip~1 verification}\label{app:tail-strips}

On the large-rate branch $\varrho\ge\sqrt{H_j}$ ($\tilde\omega=\sqrt{H_j}>\frac{28}{15}$), the domain is covered exhaustively by the four strips $[\frac{28}{15},2]\cup[2,\frac52]\cup[\frac52,3]\cup[3,\frac72]$ together with the analytic tail $[\frac72,\infty)$ proved in \cref{sec:reservoir-marked}. The exact rank seam $H_{30}<4<H_{31}$ ensures $\epsilon_j=\frac2{j+1}\in[\frac2{31},\frac2{19}]$ on Strip~1 ($\tilde\omega\le2$, $j\le30$) and $\epsilon_j\in[0,\frac2{31}]$ on Strips~2--4 ($\tilde\omega\ge2$, $j\ge31$). Writing $\lambda_{\mathrm{lb}}(\tilde\omega)=\tilde\omega(1-\tilde\omega^2/S_{\mathrm{lb}}(\tilde\omega))$ with $c=\frac{159}{200}$, the numerator of $\lambda'_{\mathrm{lb}}(\tilde\omega)$ splits into $N_0(\tilde\omega)=1+4\tilde\omega+2\tilde\omega^2-2\tilde\omega^3-\frac14\tilde\omega^4$ and $N_c(\tilde\omega)=c^2\tilde\omega^{-4}+2c\tilde\omega^{-2}+4c\tilde\omega^{-1}-4c$; expanding in $\tilde\omega-\frac{28}{15}\ge0$ shows both are strictly negative on $[\frac{28}{15},\infty)$, so $\lambda_{\mathrm{lb}}$ decreases and $\lambda_{\mathrm{lb}}(\tilde\omega_{\mathrm{hi}})\ge\lambda_0$ holds on each strip.

\begin{table}[!ht]
\centering
\small
\begin{tabular}{@{}cccccc@{}}
\toprule
Strip $[\tilde\omega_{\mathrm{lo}},\tilde\omega_{\mathrm{hi}}]$ & $\lambda_0\le\lambda_{\mathrm{lb}}(\tilde\omega_{\mathrm{hi}})$ & $\epsilon$-band & $\min L_{\mathrm{st}}>$ & $t_{\mathrm{r}}$-partition & $\min$ Bernstein coeff.\ $>$\\
\midrule
$[\frac{28}{15},2]$ & $\frac{111}{125}\le0.88869$ & $[\frac2{31},\frac2{19}]$ & $1.4604$ & $[0,\frac14],[\frac14,\frac12],[\frac12,1]$ & $0.0050$ (worked below)\\[2pt]
$[2,\frac52]$ & $\frac45\le0.81121$ & $[0,\frac2{31}]$ & $1.8006$ & $[0,1]$ & $0.0499$\\[2pt]
$[\frac52,3]$ & $\frac23\le0.67007$ & $[0,\frac2{31}]$ & $3.9325$ & $[0,1]$ & $3.4815$\\[2pt]
$[3,\frac72]$ & $\frac9{20}\le0.47848$ & $[0,\frac2{31}]$ & $6.5327$ & $[0,1]$ & $10.4756$\\
\bottomrule
\end{tabular}
\caption{Summary of the four large-rate strips of \cref{thm:marked-tail}, covering $\tilde\omega\in[\frac{28}{15},\frac72]$; all bounds are rounded down from exact rational evaluations in the companion.}\label{tab:strips}
\end{table}

\paragraph{Worked representative strip (Strip~1: $[\frac{28}{15},2]$, $\lambda_0=\frac{111}{125}$).}
Strip~1 is the tightest strip and the only one where a single-$[0,1]$ degree-$3$ Bernstein expansion has a negative interior coefficient ($b_1\approx-0.3178$ at $\epsilon=\frac2{19}$), illustrating the exact three-subinterval refinement. At $\tilde\omega_{\mathrm{hi}}=2$, $S_{\mathrm{lb}}(2)=\frac{5759}{800}$ and $\lambda_{\mathrm{lb}}(2)=\frac{5118}{5759}>\frac{111}{125}=\lambda_0$ (margin $\frac{501}{719875}$). At $\tilde\omega_{\mathrm{lo}}=\frac{28}{15}$, the four derivative corners $L_{\mathrm{st}}(\frac{28}{15},\epsilon,t_{\mathrm{r}})$ for $(\epsilon,t_{\mathrm{r}})\in\{\frac2{31},\frac2{19}\}\times\{0,1\}$ evaluate exactly to $\frac{2034433}{864900}>2.3522$, $\frac{437038517}{270281250}>1.6169$, $\frac{728641}{324900}>2.2426$, and $\frac{148286117}{101531250}>1.4604$, proving $\partial_b F>0$ across Strip~1. At the tightest endpoint $\epsilon=\frac2{19}$ ($j=18$), setting $a_{\mathrm{cl}}=1-\frac{111}{125}t_{\mathrm{r}}$ yields the exact cubic
\[
F\bigl((\tfrac{28}{15})^2,\tfrac2{19},t_{\mathrm{r}}\bigr)=\frac{21627617}{73102500}-\frac{6558541}{3562500}t_{\mathrm{r}}+\frac{106657569}{29687500}t_{\mathrm{r}}^2-\frac{64221159}{39062500}t_{\mathrm{r}}^3.
\]
Splitting $[0,1]=[0,\frac14]\cup[\frac14,\frac12]\cup[\frac12,1]$ yields strictly positive exact degree-$3$ Bernstein vectors on every piece:
\begin{align*}
[0,\tfrac14]:&\ \bigl[\tfrac{21627617}{73102500},\tfrac{1041251189}{7310250000},\tfrac{46689400697}{731025000000},\tfrac{2519010628781}{73102500000000}\bigr]\ge[0.2958,0.1424,0.0638,0.0344],\\
[\tfrac14,\tfrac12]:&\ \bigl[\tfrac{2519010628781}{73102500000000},\tfrac{184540593931}{36551250000000},\tfrac{453198531581}{18275625000000},\tfrac{621532316731}{9137812500000}\bigr]\ge[0.0344,0.0050,0.0247,0.0680],\\
[\tfrac12,1]:&\ \bigl[\tfrac{621532316731}{9137812500000},\tfrac{352849604653}{2284453125000},\tfrac{191196265939}{571113281250},\tfrac{115214591114}{285556640625}\bigr]\ge[0.0680,0.1544,0.3347,0.4034],
\end{align*}
with global minimum $\frac{184540593931}{36551250000000}>0.005048$. At the companion endpoint $\epsilon=\frac2{31}$, the same three-subinterval expansion gives global minimum $\frac{1042010776531}{97301250000000}>0.010709$ on $[\frac14,\frac12]$. The remaining six endpoint cubics and twenty-four single-interval degree-$3$ Bernstein coefficients for Strips~2--4 (summarized in \cref{tab:strips}) are verified by exact rational arithmetic in the companion.

\FloatBarrier

\section*{AI use statement}
\addcontentsline{toc}{section}{AI use statement}

Under the author's direction, AI agents working in multi-agent swarms jointly developed the mathematical results in this paper over hundreds of agent-hours using five ChatGPT models (GPT-6 Astra, GPT-6 Pro, GPT-6.1 Sol, GPT-6.2 Astra, and GPT-6.2 Sol), Claude Fable 5.2, Gemini 4 Carbon, and Gemini 4 Neon. The author thanks Google for providing complimentary access to Google Antigravity, which was used for the Gemini runs.

Claude Opus 5.5 prepared the initial draft. Gemini 4 Neon developed the main Lean 4 proof and jointly revised the manuscript alongside the Lean code so that the formal development establishes the paper's mathematical claims and exact-rational certificates (allowing differences in proof steps and stronger intermediate predicates where documented in the companion). To guard against single-model blind spots, a team led by GPT-6.2 Astra with GPT-6.2 Sol workers authored the independent Lean specification of the seeding process and independently audited the manuscript, formalization, specification, and finite rational checks, while Claude Opus 5.5 conducted independent manuscript-only reviews. Alongside direct editing by the author, GPT-6.2 Astra, Gemini 4 Neon, and Claude Opus 5.5 assisted with revisions. The author takes responsibility for the paper's contents.

{\raggedright
\bibliographystyle{alphaurl}
\bibliography{refs}
}

\end{document}